# Characterization of debris disks observed with SPHERE

N. Engler[1], J. Milli[2], N. Pawellek[3,4], R. Gratton[5], P. Thébault[6], C. Lazzoni[5], J. Olofsson[7], H. M. Schmid[1], S. Ulmer-Moll[8], C. Perrot[6], J.-C. Augereau[2], S. Desidera[5], G. Chauvin[9], M. Janson[10], C. Xie[11], Th. Henning[9], A. Boccaletti[6], S. B. Brown-Sevilla[9], E. Choquet[12], C. Dominik[13], C. Ginski[14], A. Zurlo[15,16], M. Feldt[9], T. Fusco[17,12], J. H. Girard[18], D. Gisler[19], R. G. van Holstein[20], A.-M. Lagrange[6,2], M. Langlois[12,21], A.-L. Maire[2], D. Mesa[5], P. Rabou[2], L. Rodet[22], M. Samland[9], T. Schmidt[6], and A. Vigan[12]

*(Affiliations can be found after the references)*

**ABSTRACT**

*Aims.* This study aims to characterize debris disk targets observed with SPHERE across multiple programs, with the goal of identifying systematic trends in disk morphology, dust mass, and grain properties as a function of stellar parameters. By combining scattered-light imaging with photometric and parametric modeling, we seek to improve our understanding of the composition and evolution of circumstellar material in young debris systems and to place debris disks in the broader context of planetary system architectures.

*Methods.* We analyzed a sample of 161 young main-sequence stars using archival SPHERE observations at optical and near-infrared (IR) wavelengths. Disk geometries were derived from ellipse fitting and model grids, while dust mass and properties were constrained by modified blackbody (MBB) and size distribution (SD) modeling of spectral energy distributions (SEDs). We also carried out dynamical modeling to assess whether the observed disk structures can be explained by the presence of unseen planets.

*Results.* We resolve 51 debris disks, including four new detections where disks are resolved for the first time: HD 36968, BD-20 951, and the inner belts of HR 8799 and HD 36546. In addition, we find a second transiting giant planet in the HD 114082 system, with a radius of $1.29 \pm 0.05\ R_{\rm Jup}$ and an orbital distance of ~1 au, providing an important new benchmark for planet–disk interaction studies.

Beyond these new detections, we identify nine multi-belt systems, with outer-to-inner belt radius ratios of 1.5 − 2, and find close agreement between scattered-light and millimeter continuum belt radii with a mean ratio $R_{\rm belt}$(near-IR)/$R_{\rm belt}$(mm) of $1.05 \pm 0.04$. Belt radii scale weakly with stellar luminosity ($R_{\rm belt} \propto L_\star^{0.11\pm0.05}$), but show steeper dependencies when separated by CO and $CO_2$ freeze-out regimes, and also increase with age as $R_{\rm belt} \propto t_{\rm age}^{0.37\pm0.11}$.

Uniform image modeling yields vertical disk aspect ratios of 0.02 − 0.06, consistent with collisionally stirred belts, while gas-rich systems show unusually small values. Inner density slopes steepen with stellar luminosity, indicating more efficient dust removal around luminous stars.

Disk fractional luminosities follow collisional decay trends, declining as $t_{\rm age}^{-1.18\pm0.14}$ for A-type and $t_{\rm age}^{-0.81\pm0.12}$ for F-type stars. SD modeling yields minimum grain sizes consistently above the blowout limit, typically $> 0.8\,\mu$m, with a mean SD index of $q = 3.6$, assuming astrosilicate composition. The inferred dust masses span $10^{-5} - 1\ M_\oplus$ from MBB modeling (and $0.01 - 1\ M_\oplus$ from SD modeling for detected disks). These masses scale as $R_{\rm belt}^n$ with $n > 2$ in belt radius and super-linearly with stellar mass, consistent with trends seen in protoplanetary disks (PPDs).

Our detailed analysis of disk scattered-light non-detections indicates that they are mainly caused by low dust masses, unfavorable viewing geometries, or suboptimal observing conditions. SD modeling combined with Mie theory further shows that bulk albedos are consistently above 0.5 with little variation, making albedo differences an unlikely explanation. To explore this further, we introduced a new parametric approach based on scattered-light and polarized-light images, which provides independent estimates of dust albedo and maximum polarization fraction.

We find a correlation between measured disk polarized flux and IR excess, with a slope shallower than that of optical total-intensity fluxes measured with HST/STIS. The offset of ~1 dex between total-intensity and polarized fluxes arises because polarized flux represents only a fraction of the total scattered light which depends on both grain properties and disk inclination.

Finally, a comparison of planetary architectures shows that most benchmark systems resemble the Solar System, with multiple planets located inside wide Kuiper-belt analogues. Dynamical modeling further indicates that many observed gaps and inner edges can be explained by unseen planets below current detection thresholds, typically with Neptune- to sub-Jovian masses, underscoring the likely ubiquity of such planets in shaping debris disk morphologies.

## 1. Introduction

The field of exoplanet research, which has rapidly evolved in recent decades, has uncovered an immense diversity in planetary structure and composition, ranging from small rocky worlds to massive gas giants, orbiting their stars in periods spanning days to hundreds of years (Jontof-Hutter 2019; Winn & Fabrycky 2015; Dawson & Johnson 2018; Zhu & Dong 2021; Wordsworth & Kreidberg 2022, and references therein). The vast variety of exoplanets might arise from the distinct environments of circumstellar gas and dust in which planets form and evolve over millions, or even billions, of years. These environments undergo continuous transformations: beginning with the collapse of a molecular cloud that gives rise to a new star, progressing through a protoplanetary disk (PPD), where planets are born, and eventually becoming a debris disk as the star enters the main sequence after several million years. Studying these environments is essential to answering fundamental questions in exoplanet science.

The circumstellar material that provides the building blocks for future planets has different origins and properties in protoplanetary and debris disks. In PPDs, gas and dust are pristine, originating directly from the initial molecular cloud. In contrast, the primary mechanism of dust production in debris disks is the collisions between kilometer-sized rocky bodies. These collisions supply the disk and the forming planets with substantial amounts of dust grains of various sizes and small amounts of gas (e.g., Wyatt 2018; Hughes et al. 2018).

The evolution of dust particle properties from the protoplanetary to the debris disk phase can be studied using two distinct ranges of the electromagnetic spectrum, as dust particles interact with stellar light in two primary ways. Some stellar photons are scattered by dust grains in all directions, particularly at optical and near-infrared (IR) wavelengths. Meanwhile, other stellar photons are absorbed by the dust grains and re-emitted as thermal radiation predominantly in the IR to millimeter wavelength range.

In the past decades, space-based mid-IR and far-IR observations with the *Spitzer*, IRAS, and *Herschel* Space Observatory played a crucial role in advancing our understanding of debris disks. In particular, the Disc Emission via a Bias-free Reconnaissance in the IR/Submillimeter (DEBRIS; Sibthorpe et al. 2018) and DUst around NEarby Stars (DUNES; Eiroa et al. 2013) surveys provided comprehensive statistical studies of debris disks around nearby main-sequence stars. These surveys enabled precise measurements of IR excess and dust temperatures, revealing trends with stellar type and age. They also established a framework for estimating dust luminosity distributions and the incidence rate of debris disks, especially around solar-type and early-type stars.

In addition to mid-IR and far-IR observations, the *Hubble* Space Telescope (HST) made a groundbreaking contribution to the imaging of debris disks in scattered light. Using its high-contrast imaging (HCI) capabilities, HST provided the first resolved views of numerous debris disks, revealing their morphology and fine structures such as rings, warps, and asymmetries (e.g., Golimowski et al. 2006; Kalas et al. 2007). Systematic surveys of circumstellar environments led by Schneider et al. (2014, 2016) offered valuable complementary insights to thermal emission data, helping to constrain disk geometries and the scattering properties of dust grains.

The detailed studies of both scattered and thermal light from circumstellar disks using the ground-based telescopes became possible with the start of operation of high-contrast and high-resolution instruments such as the Spectro-Polarimetic High contrast imager for Exoplanets REsearch (SPHERE; Beuzit et al. 2019) at VLT, the Gemini Planet Imager (GPI; Macintosh et al. 2014) or the Subaru Coronagraphic Extreme Adaptive Optics (SCExAO; Jovanovic et al. 2015), along with interferometric facilities like the Atacama Large (sub)Millimeter Array (ALMA) which delivered unprecedented images of many protoplanetary and debris disks around young stars (e.g., Perrot et al. 2016; Andrews et al. 2018; Avenhaus et al. 2018; Boccaletti et al. 2020; Columba et al. 2024). These observations have targeted both individual disks (e.g., Garufi et al. 2016; Milli et al. 2017b; Olofsson et al. 2018; Ménard et al. 2020) and large disk samples (e.g., Ansdell et al. 2017; Ginski et al. 2024; Garufi et al. 2024; Matrà et al. 2025), facilitating the first demographic studies that address the morphology of these objects.

Most optical, near-IR and (sub)millimeter imaging campaigns have focused on studying PPD evolution and searching for forming planets within them (Benisty et al. 2023, and references therein). In contrast, only a few comparable studies have investigated direct imaging (DI) data for a large sample of debris disks (e.g., Schneider et al. 2014; Esposito et al. 2020; Crotts et al. 2024). One key reason is that debris disks are significantly older ($\gtrsim$ 7 Myr) and contain roughly three orders of magnitude less dust than PPDs (Wyatt 2008). As a result, they are much fainter and more challenging to image directly.

This study focuses on SPHERE observations of debris disks. SPHERE is an extreme adaptive optics (AO) instrument optimized for observing circumstellar environments. Since its commissioning in 2014, SPHERE has been extensively utilized and has proven to be one of the most productive HCI instruments. As part of the SPHERE guaranteed time observation (GTO) program, numerous debris disks have been observed and detected in the course of the dedicated disk program, and sometimes as a by-product of the SpHere INfrared survey for Exoplanets (SHINE; Chauvin et al. 2017; Desidera et al. 2021; Vigan et al. 2021; Langlois et al. 2021). To perform a comprehensive analysis of these observations, we compiled a sample of targets known to host debris disks from the archival datasets of GTO and various open-time programs, including all targets from the SPHERE High Angular Resolution Debris Disk Survey (SHARDDS, PI: J. Milli; Milli et al. 2017b; Dahlqvist et al. 2022).

This study aims to consistently characterize the structural and compositional properties of debris disks, focusing on both their radial and vertical extents, as well as the nature of their constituent dust. A primary goal is to explore how the architecture of debris disks, particularly the radial locations of planetesimal belts and their dust masses, which are key to understanding the evolution of debris disks and their interaction with planets (Krivov & Wyatt 2021), relates to the fundamental properties of their host stars. By analyzing a large sample of spatially resolved debris disks, we investigated how the belt radii and disk dust masses scale with stellar luminosity and mass, and how these relationships evolve over time. Additionally, we assessed the conditions that distinguish detected from non-detected disks in scattered light, accounting for both intrinsic disk properties and observational biases. We further investigated the architecture of planetary systems within the sample, analyzing how the presence, absence, or configuration of planetary companions correlates with the structure and detectability of debris disks.

The paper is organized as follows. In Sect. 2, we present the stellar parameters of the targets included in our sample. Section 3 describes the SPHERE observing modes and the data reduction techniques used throughout the study. In Sect. 4, we analyze the morphological structure of the detected disks, emphasizing the radial locations of the planetesimal belts. These constraints are then incorporated into spectral energy distribution (SED) modeling in Sect. 5, allowing us to derive the parameters of the dust grain size distributions (SDs) and to estimate plausible ranges for the scattering albedo.

A key part of our investigation, detailed in Sect. 6, addresses the question of why the majority of young debris disks remain undetected in scattered light imaging. In particular, we examine the role of the dust's optical properties and the observational biases associated with viewing geometry and disk structure. Special focus is placed on the evaluation methods for the dust albedo and polarization efficiency based on polarimetric imaging (Sect. 6.3).

In Sect. 7, we shift focus to planetary system architectures. We analyze systems where both exoplanets and debris disks are detected, as well as those with no detected planets, by estimating the locations and masses of planets that could dynamically shape the observed disk structures. This includes modeling scenarios in which unseen planets are responsible for clearing gaps or truncating the inner edges of planetesimal belts. The key findings of the study are summarized in Sect. 8.

## 2. Sample description

We compiled a sample of debris disks from archival SPHERE observations, selecting main-sequence stars with an IR excess above $10^{-6}$, based on data from the Jena debris disk database[1]. This sample comprises 161 stars spanning a broad range of spectral types and ages. The majority are young F-type (35%) and A-type (29%) stars (Fig. 1). These targets usually exhibit strong IR excesses, indicating significant amounts of dust, and are bright enough to serve as reference stars themselves for the AO system. For instance, ZIMPOL observations require a reference star with a G magnitude brighter than $9.5^m$ for a good AO correction in the optical. Additionally, the two surveys that contributed most to our sample, SHINE and SHARDDS, primarily focus on A-type and solar-type stars. The presence of an IR excess or debris disk was not a part of the selection criteria for the SHINE statistical sample (Desidera et al. 2021). However, in specific cases, the known presence of exoplanets or a disk, suggesting a higher likelihood of harboring young, directly imageable planets (e.g., Meshkat et al. 2017), led to a target being classified as a special object, thereby increasing its observational priority. In contrast, SHARDDS was a dedicated debris disk survey, with targets selected based on the predicted brightness of their disks ($f_{\rm disk} > 10^{-4}$)[2]. The SHARDDS survey included 55 main-sequence stars observable from the Southern hemisphere, covering spectral types A through M and stellar ages ranging from 10 Myr to 6 Gyr. Its aim was to provide a comprehensive overview of planetary system properties and their temporal evolution.

In addition to A- and solar-type stars, our sample includes eight B-type and eight M-type stars, with the latter group exhibiting the highest detection rate among all spectral types in our sample. However, this high detection rate is in part due to the unexpected discovery of a debris disk around the M1Ve star GSC 7396-0759 (Sissa et al. 2018), which was not previously known to exhibit an IR excess and was routinely observed within the SHINE program. The median stellar mass of the full sample is 1.43 $M_\odot$, while for the subsample of targets with detected disks it is slightly lower at 1.38 $M_\odot$ (Fig. 1).

The sample comprises 18 binary systems, including spectroscopic, visual, and astrometric binaries, as well as spectroscopic binary candidates (SBCs), four triple systems, three quadruple systems and two systems with higher-order multiplicity (N > 4) according to the Washington Double Star catalog (WDS; Mason et al. 2001) as of January 15, 2024. In two of the triple systems, debris disks are known around two different components, which were observed individually and listed separately in Table E.2. These include HD 216956 (Fomalhaut A) and GSC 06964-1226 (Fomalhaut C), as well as HD 181296 (A component), which shares a common proper motion with HD 181327 (B component). Among the quadruple systems, HD 20320 consists of a spectroscopic binary (SB) as its A component and an astrometric binary as its B component, while HD 98800 features a pair of SBs orbiting each other (Kennedy et al. 2019). Another quadruple star system in the sample is HD 102647 (Denebola). The components of multiple systems that host debris disks and were observed with SPHERE are specified in Col. 6 of Table E.2.

Our sample includes five chemically peculiar stars classified as Lambda Boo stars: HD 30422, HD 31295, HD 110411, HD 183324, and HD 218396. These stars exhibit surface deficiencies in iron-peak elements while maintaining nearly solar abundances of carbon, nitrogen, oxygen, and sulfur (e.g., Paunzen 2001; Gray et al. 2017). This anomaly may be explained by preferential gas accretion over dust from a dynamically evolving debris disk, possibly influenced by migrating planets or accretion from the atmospheres of hot Jupiters (Murphy & Paunzen 2017). The debris disk hypothesis is further supported by the high fraction (up to 77%) of Lambda Boo stars exhibiting IR excess, which is often linked to the presence of a debris disk. (Draper et al. 2016b). Some of these disks have been imaged with the Herschel Space Observatory at 70, 100 and 160 $\mu$m, including several debris disks analyzed in this study (Su et al. 2009; Draper et al. 2016b). In Sect. 4, we present a scattered light image of the inner belt surrounding a Lambda Boo star HD 218396 (HR 8799).

Stellar ages were compiled from the literature, with their lower and upper boundaries listed in Col. 11 of Table E.2. For some targets, particularly field stars, there are significant discrepancies, up to 3000 Myr, between ages reported in different studies. This large scatter arises from the use of diverse age-dating techniques, such as isochrone fitting, kinematic group membership, and indicators of stellar activity (e.g., Ca II H and K line strength or X-ray luminosity). For instance, published age estimates for HD 15115 include $12^{+8}_{-4}$ Myr (Moór et al. 2006), 100 Myr (Zuckerman & Song 2004), or $500^{+1500}_{-500}$ Myr (Holmberg et al. 2009).

In such cases, we adopted an age range that covers the full span of results derived from various methods. For bona fide members of moving groups (MGs) and targets lacking literature age estimates, upper and lower age limits were assigned based on the most probable MG membership. In Column 12 of Table E.2, we list the MG with the highest probability of association for each star, along with the corresponding probability percentage (in parentheses), as determined using the BANYAN $\sum$ tool (Gagné et al. 2018). According to this analysis, the majority of our sample consists of field stars (50%), followed by members of the $\beta$ Pictoris MG ($\beta$PMG, 9%).

The median age of our sample is 100 Myr, with approximately half of the targets estimated to be between 10 and 100 Myr old (Fig. 1). The debris disks around the youngest stars (< 10 Myr) such as Herbig Ae/Be stars HD 141569 (e.g., Perrot et al. 2016) and HD 156623, as well as T Tauri stars such as TWA 7 (e.g., Olofsson et al. 2018; Ren et al. 2021), exhibit structures with multiple rings and spiral arms (Figs. 2 and B.1), features typically associated with PPDs. The fractional IR luminosities of these young stellar objects are generally below 0.1, leading to their classification as debris disks. However, these systems may represent an intermediate stage between the protoplanetary and debris disk phases. We categorize such disks as transition disks due to their evolutionary status.

Furthermore, transition disks often contain high CO masses, comparable to those found in PPDs (Lieman-Sifry et al. 2016; Moór et al. 2017, 2019). Our sample includes several debris disk systems with a significant gas reservoir, commonly referred to as hybrid disks: HD 9672 (Moór et al. 2011; Choquet et al. 2017; Pawellek et al. 2019), HD 21997 (Kóspál et al. 2013), HD 121617 (Perrot et al. 2023), HD 131488 (Pawellek et al. 2024), HD 131835 (Hung et al. 2015; Feldt et al. 2017), and HD 141569 (Dent et al. 2005). In these hybrid systems, dust evolution may have progressed more rapidly than gas dissipation (Péricaud et al. 2017).

Similar to stellar ages, a wide range of metallicity values for the same star can be found in the literature. Depending on the method used to determine metallicity, discrepancies of up

[1] https://www.physik.uni-jena.de/21956/catalog-of-resolved-debris-disks

[2] In Appendix G, we list all the symbols used in this work and provide their definitions.

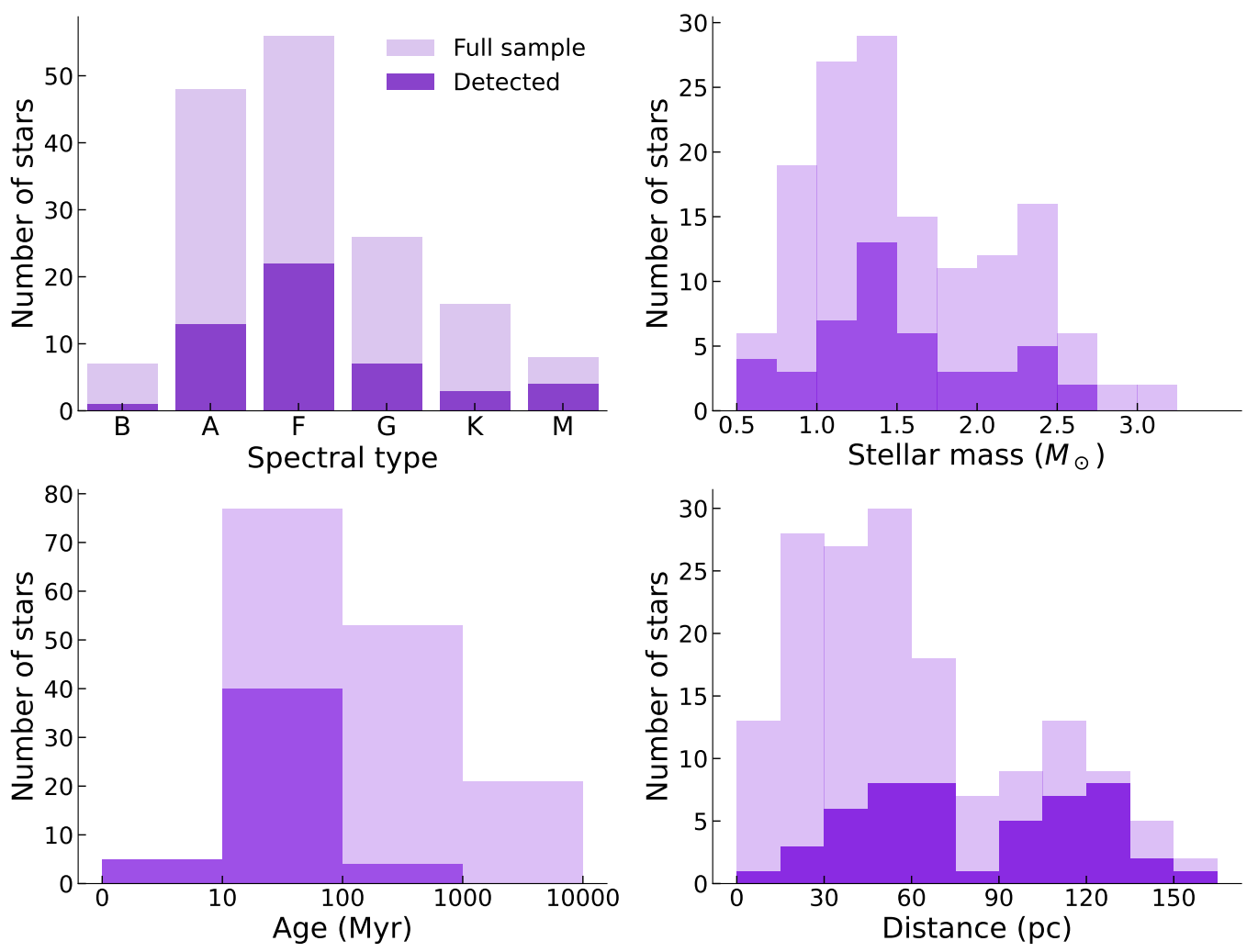

Fig. 1: Distributions of stellar parameters for the observed targets. Light-colored histograms show all targets (with detected and non-detected debris disks together), while the dark-colored histograms display the targets with detections only.

Table 1: SPHERE subsystem parameters[4].

| Instrument | Field of View | Science frame format in pixels | Pixel scale (mas/pixel) |
|---|---|---|---|
| IRDIS | $11'' \times 12.5''$ | $1024 \times 1024$ | $12.26 \pm 0.02^{(a)}$ |
| IFS | $1.73'' \times 1.73''$ | $291 \times 291$ | $7.46 \pm 0.02$ |
| ZIMPOL | $3.5'' \times 3.5''$ | $1024 \times 1024$ | $3.61 \pm 0.01$ |

**Notes.** $^{(a)}$ The IRDIS plate scale is evaluated in the H2 filter with the N_ALC_YJH_S coronagraph. For other IRDIS filters and coronagraphs the plate scale should be adjusted.

to 0.5 dex can arise between different studies. To ensure consistency, we opted to use the median metallicity value from all studies recorded in the SIMBAD database[3].

The target distances, listed in Col. 7 of Table E.2, were derived using stellar parallaxes from the Gaia DR3 catalog (Gaia Collaboration et al. 2023b). The closest object in our sample is HD 22049 ($\epsilon$ Eridani), located at 3.6 pc from the Sun, while the most distant target is HD 149914 at 154.3 pc. Two-thirds of all targets in the sample are located within 80 pc of the Sun, with a minor peak at ~100 pc, corresponding to the distance of the Scorpius-Centaurus OB association, a region rich in young stars (Fig. 1).

## 3. SPHERE observing modes and data reduction

The disk observations presented in this work were performed with different SPHERE subsystems (see Table 1 and the SPHERE User Manual[4]): the InfraRed Dual-beam Imager and Spectrograph (IRDIS; Dohlen et al. 2008), the Integral Field Spectrograph (IFS; Claudi et al. 2008) and the Zurich Imaging POLarimeter (ZIMPOL; Schmid et al. 2018). A variety of instrument modes were used, including pupil- and field-stabilized configurations, along with different filters ranging from optical to near-IR. Observations were conducted with classical or apodized pupil Lyot coronagraphs (Boccaletti et al. 2008; Carbillet et al. 2011; Guerri et al. 2011), or in some cases, without a coronagraph.

Many disks were observed only once using either classical or polarimetric imaging (de Boer et al. 2020; van Holstein et al. 2020) modes of IRDIS with the broadband H filter ($\lambda_c$ = 1.625 $\mu$m, $\Delta\lambda$ = 0.290 $\mu$m), or polarimetric imaging mode of ZIMPOL (Schmid et al. 2018), often employing the Very Broad Band filter (VBB or RI; $\lambda_c$ = 0.735 $\mu$m, $\Delta\lambda$ = 0.290 $\mu$m). The observations of all SHINE targets were performed in either the IRDIFS or IRDIFS_EXT modes (Langlois et al. 2021), which provide a simultaneous data acquisition with both IRDIS and IFS. With these instrument setups, the IRDIS is operated in the dual-band imaging mode (Vigan et al. 2010) with the filter pair H2H3 ($\lambda_{H2}$ = 1.593 $\mu$m, $\Delta\lambda_{H2}$ = 0.052 $\mu$m; $\lambda_{H3}$ = 1.667 $\mu$m, $\Delta\lambda_{H3}$ = 0.053 $\mu$m) for the IRDIFS mode, or with the filter pair K1K2 ($\lambda_{K1}$ = 2.110 $\mu$m, $\Delta\lambda_{K1}$ = 0.102 $\mu$m; $\lambda_{K2}$ = 2.251 $\mu$m, $\Delta\lambda_{K2}$ = 0.109 $\mu$m) for the IRDIFS_EXT mode, whereas the IFS is operated in the IRDIFS *Y-J* mode (0.95 − 1.35 $\mu$m, with a spectral resolution of $R_\lambda$ = 50), or IRDIFS_EXT *Y-H* mode (0.95 − 1.65 $\mu$m, with a spectral resolution of $R_\lambda$ = 35).

The IRDIS and IFS datasets were processed at the High-Contrast Data Center[5] (HC-DC, Delorme et al. 2017, formerly known as the SPHERE Data Center). For both instruments, the pre-processing steps are based on the SPHERE Data Reduction and Handling pipeline (Pavlov et al. 2008) to correct for bad pixels, flat-field non-uniformity, optical distortions, and telescope or sky background. In addition, for the IFS, the pre-processing includes a wavelength calibration and a correction for cross-talks between spectral channels. Coronagraphic images are centered via four satellite spots used to determine the accurate position of the star hidden behind the coronagraphic mask.

Pre-processed IRDIS and IFS datasets form spectral and temporal cubes of centered images, to which dedicated stellar subtraction algorithms can be applied. Such algorithms include classical Angular Differential Imaging (ADI; Marois et al. 2006), Principal Component Analysis (PCA; Soummer et al. 2012; Amara & Quanz 2012) or the Locally Optimized Combination of Images (LOCI; Lafrenière et al. 2007), implemented in the HC-DC in a template-oriented version (T-LOCI; Marois et al. 2014; Galicher et al. 2018). For several datasets, post-processing employing the reference-star differential imaging (RDI) technique was also applied, as described in Xie et al. (2022).

The IRDIS polarimetric datasets were processed using the IRDAP pipeline (van Holstein et al. 2020), while the ZIMPOL polarimetric datasets were reduced with a pipeline developed at ETH Zürich, as described in Engler et al. (2017) and Hunziker et al. (2020). Both pipelines are currently implemented in the HC-DC and include, as part of the pre-processing steps, subtraction of bias and dark frames, flat-fielding, and correction for instrumental polarization. Additionally, the ZIMPOL frames are corrected for modulation and demodulation efficiency (Schmid et al. 2018).

In both pipelines, the Stokes parameter $Q$ and $U$ images are computed from the calibrated and centered polarimetric frames using the double-difference method. These $Q$ and $U$ images are then transformed into the azimuthal Stokes parameter $Q_\varphi$ and $U_\varphi$:

$$Q_\varphi = -Q\cos 2\varphi - U\sin 2\varphi$$

$$U_\varphi = Q\sin 2\varphi - U\cos 2\varphi,$$

[3] https://simbad.cds.unistra.fr/simbad/
[4] https://www.eso.org/sci/facilities/paranal/instruments/sphere/doc.html
[5] https://hc-dc.cnrs.fr

where $\varphi$ is the polar angle measured east of north in a coordinate system centered on the star, and the sign convention $Q_\varphi = -Q_r$ and $U_\varphi = -U_r$ is adopted from Schmid et al. (2006) (see also Monnier et al. (2019)).

## 4. Morphology of resolved debris belts

Out of 161 targets, 51 debris disks were successfully detected with SPHERE in total intensity of scattered light (Fig. 2), linearly polarized intensity[6] (Fig. 5), or both. Four of these debris disks, BD-20 951 (Perrot et al. in prep.), HD 36968 and the inner belts of HD 218396 (HR 8799) and HD 36546 systems had not been imaged with any instrument before. Debris disks HD 38206, HD 36546, HD 38397 (Perrot et al. in prep.), HD 98800, HD 182681 are resolved in scattered light for the first time. Scattered light images of the HD 105, HD 377, TWA 25 (Langlois et al. in prep.), HD 30447, HD 92945, HD 145560, HD 192758 and HD 202917 debris disks have previously been obtained with instruments such as HST or GPI. However, the SPHERE images of these disks have not yet been published. The majority of detections were around F-type stars, with 23 discoveries, corresponding to a 45% detection rate in our sample (Fig. 1 upper left panel).

We determined the radii of the resolved debris belts by analyzing the $r^2$-scaled images of total or polarized intensities. The radial position of the peak surface brightness (SB) along the disk's major axis was measured, and the resulting disk radius, referred to as $R_{\rm belt}^{\rm mes}$, is listed in Col. 2 of Table 2. The inclination and position angle (PA) of each disk, provided in Table 2, were derived by fitting ellipses to the visible contours of the disk rims. Additionally, disk images with a higher signal-to-noise ratio (S/N) were modeled in more detail to obtain the fundamental geometrical and scattering parameters necessary for a comprehensive characterization of disk properties (Sect. 4.5.1).

In all images presented in Figs. 2 and 5, sky north is up and east to the left. The PA defines the orientation of the disk's major axis on the sky and is measured counterclockwise from sky north to east. The PA values of the eastern disk extensions are listed in Col. 4 of Table 2 ($0° \leqslant$ PA $\leqslant 180°$). The inclination of a debris disk is conventionally defined as the angle between the sky plane and the disk's minor axis, where a pole-on disk has an inclination of 0°, and an edge-on disk has an inclination of 90°. In this study, disk inclinations (Col. 3 of Table 2) follow the convention that an inclination is less than 90° when the brighter side of the disk is oriented southward, whereas an inclination is greater than 90° when the brighter side is oriented northward. In many debris disk studies, the inclination is reported as less than 90°, regardless of the orientation of the disk's brighter side. To enable comparison with such studies, the inclinations of disks with the brighter side oriented northward ($i > 90°$ in Table 2) can be converted using $i_{<90°} = 180° - i_{>90°}$.

### 4.1. Disk radii in SPHERE versus ALMA observations

We detect planetesimal belts in 33 debris disks which have also been resolved with ALMA and SMA at wavelengths of 0.856 – 1.34 mm as part of the REASONS survey (Matrà et al. 2025). The nature of dust emission observed in scattered light images (from optical to near-IR wavelengths) and thermal emission images (from mid-IR to millimeter wavelengths) is fundamentally different. In SPHERE images (both total and polarized intensities), we observe stellar photons scattered off dust grains into our line of sight. In contrast, the thermal emission detected by ALMA and SMA originates from the absorption of stellar photons by dust particles, which raises their temperature and leads to re-emission at longer wavelengths. Additionally, scattered light images trace predominantly dust particles with sizes smaller than a few microns, whereas (sub)millimeter imaging is more sensitive to (sub)millimeter particles. As a result, the disk morphology, particularly the radial position and extent of

[6] Hereafter, we refer to the total intensity of scattered light as scattered intensity or scattered light, and the linearly polarized intensity of scattered light as polarized intensity or polarized light.

Table 2: Parameters of spatially resolved debris disks.

| Debris belt | $R_{\rm belt}^{\rm mes}$ (au) | $i$ (deg) | PA (deg) | $T_{\rm bb}$ (K) |
|---|---|---|---|---|
| GSC 07396-0759 | 88 ± 10 | 83.0 ± 1.5 | 149.0 ± 2.0 | 18 |
| HD 105 | 87 ± 3 | 50.5 ± 3.5[a] | 13.9 ± 3.0 | 32 |
| HD 377 | 82 ± 5 | 95.5 ± 1.3 | 48.4 ± 1.7 | 32 |
| HD 9672 | 144 ± 10 | 79.0 ± 2.3 | 108.8 ± 2.0 | 47 |
| HD 15115 out | 98 ± 10 | 94.2 ± 2.7 | 278.9 ± 1.8 | 40 |
| HD 15115 inn | 64 ± 10 | 85.8 ± 5.8[a] | 278.9 ± 7.0 | 49 |
| HD 16743 | 149 ± 15 | 79.5 ± 2.9 | 169.5 ± 3.0 | 35 |
| HD 30447 | 89 ± 15 | 104.0 ± 5.0 | 33.0 ± 2.0 | 41 |
| HD 32297 | 117 ± 10 | 92.1 ± 1.3 | 47.7 ± 0.9 | 43 |
| HD 35841 | 66 ± 7 | 81.3 ± 1.9 | 165.9 ± 2.5 | 43 |
| HD 36546 out | 110 ± 30 | 79.3 ± 5.5 | 78.5 ± 5.2 | 55 |
| HD 36546 inn | 55 ± 20 | 79.3 ± 5.5 | 78.5 ± 5.2 | 77 |
| HD 36968 | 160 ± 24 | 102.0 ± 3.5 | 32.0 ± 1.7 | 31 |
| HD 38206 | 144 ± 15 | 86.7 ± 2.9 | 84.9 ± 2.5 | 53 |
| HD 38397 | 115 ± 15 | 55.3 ± 5.5[a] | 132.0 ± 9.0 | 28 |
| HD 39060 out | 110 ± 10 | 90.0 ± 1.0 | 23.0 ± 1.5 | 48 |
| HD 39060 inn | 65 ± 10 | 92.2 ± 1.0 | 27.0 ± 3.0 | 60 |
| HD 61005 | 67 ± 3 | 82.3 ± 1.3 | 71.0 ± 1.2 | 31 |
| HD 92945 out | 119 ± 10 | 64.0 ± 5.0[a] | 100.0 ± 2.0 | 20 |
| HD 92945 inn | 56 ± 10 | 64.0 ± 5.0[a] | 100.0 ± 2.0 | 30 |
| HD 98800 | 3.1 ± 0.4 | 35.0 ± 10.0[a] | 12.0 ± 5.0 | (...) |
| HD 106906 | 70 ± 6 | 94.7 ± 2.9 | 105.0 ± 1.4 | 54 |
| HD 109573 | 76 ± 2 | 102.7 ± 1.6 | 28.7 ± 1.0 | 71 |
| HD 110058 | 40 ± 12 | 85.0 ± 3.0 | 155.4 ± 2.7 | 76 |
| HD 111520 | 76 ± 10 | 92.0 ± 2.0 | 165.0 ± 2.5 | 41 |
| HD 112810 | 115 ± 6 | 76.0 ± 2.2 | 98.0 ± 2.5 | 35 |
| HD 114082 | 35 ± 2 | 83.2 ± 1.1 | 105.7 ± 1.4 | 66 |
| HD 115600 | 46 ± 3 | 104.4 ± 3.2 | 24.8 ± 1.7 | 62 |
| HD 117214 | 49 ± 3 | 72.8 ± 1.2 | 179.3 ± 0.2 | 61 |
| HD 120326 out | 119 ± 6 | 99.7 ± 3.5 | 86.0 ± 2.3 | 38 |
| HD 120326 inn | 50 ± 6 | 99.7 ± 3.5 | 86.0 ± 2.3 | 58 |
| HD 121617 | 82 ± 3 | 135.6 ± 1.5 | 61.0 ± 2.0 | 61 |
| HD 129590 out | 82 ± 6 | 99.2 ± 3.2 | 119.7 ± 2.9 | 41 |
| HD 129590 inn | 49 ± 6 | 99.2 ± 3.2 | 119.7 ± 2.9 | 53 |
| HD 131488 | 102 ± 10 | 94.5 ± 1.0 | 96.5 ± 1.9 | 53 |
| HD 131835 out | 105 ± 5 | 104.6 ± 1.7 | 58.5 ± 2.0 | 48 |
| HD 131835 inn | 70 ± 5 | 104.6 ± 1.7 | 58.5 ± 2.0 | 59 |
| HD 141011 | 129 ± 6 | 69.7 ± 1.5 | 155.5 ± 2.5 | 31 |
| HD 141943 out | 100 ± 10 | 99.0 ± 1.9 | 146.3 ± 3.5 | 34 |
| HD 141943 inn | 81 ± 10 | 99.0 ± 1.9 | 146.3 ± 3.5 | 38 |
| HD 145560 | 86 ± 10 | 47.5 ± 7.0 | 38.0 ± 4.0 | 41 |
| HD 146181 | 90 ± 20 | 107.5 ± 8.5 | 50.5 ± 5.0 | 37 |
| HD 146897 | 62 ± 15 | 84.4 ± 1.0 | 114.9 ± 0.8 | 48 |
| HD 156623 | 55 ± 10 | 33.0 ± 4.0 | 102.0 ± 7.0 | 71 |
| HD 157587 | 82 ± 9 | 110.5 ± 1.8 | 130.0 ± 2.0 | 42 |
| HD 160305 | 104 ± 10 | 81.8 ± 2.2 | 122.5 ± 1.5 | 31 |
| HD 172555 | 10 ± 3 | 105.0 ± 2.3 | 112.0 ± 2.4 | 147 |
| HD 181327 | 82 ± 3 | 151.0 ± 2.5 | 100.0 ± 1.4 | 40 |
| HD 182681 | 160 ± 10 | 75.9 ± 1.5[a] | 56.5 ± 2.9 | 51 |
| HD 191089 | 47 ± 4 | 120.7 ± 1.5 | 70.6 ± 1.6 | 53 |
| HD 192758 | 98 ± 12 | 130.0 ± 9.5 | 93.5 ± 5.0 | 43 |
| HD 197481 | 39 ± 1 | 88.6 ± 1.0 | 129.1 ± 0.6 | 30 |
| BD-20 951 | 122 ± 7 | 97.9 ± 2.1 | 31.5 ± 2.0 | 21 |
| TWA 7 out | 93 ± 8 | 10.0 ± 9.0 | 95.5 ± 10.0 | 17 |
| TWA 7 inn | 52 ± 4 | 10.0 ± 9.0 | 95.5 ± 10.0 | 23 |
| TWA 7 2 inn | 27 ± 4 | 10.0 ± 9.0 | 95.5 ± 10.0 | 31 |
| TWA 25 | 76 ± 2 | 101.7 ± 1.5 | 156.8 ± 1.8 | 23 |

**Notes.** [a] The image S/N is insufficient to determine the brighter side of the disk.
The columns list target IDs, measured disk radii ($R_{\rm belt}^{\rm mes}$), disk inclinations ($i$), PAs, and the BB temperature of the dust grains ($T_{\rm bb}$).

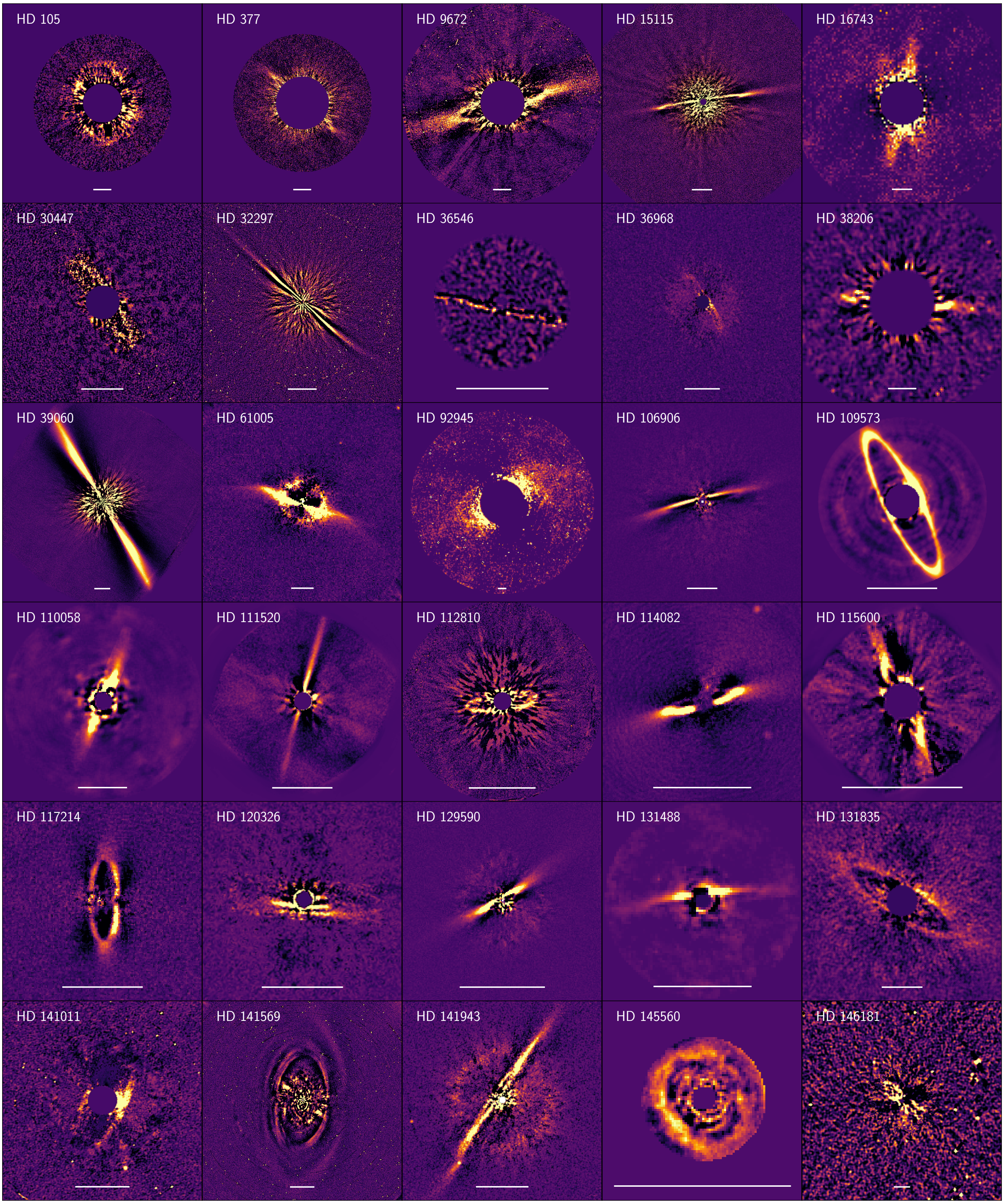

(a)

Fig. 2: Images of the total intensity of scattered light from debris disks detected with IRDIS, IFS, or ZIMPOL. The white bar at the bottom of each image corresponds to 1″. In all images, sky north is up and east is to the left.

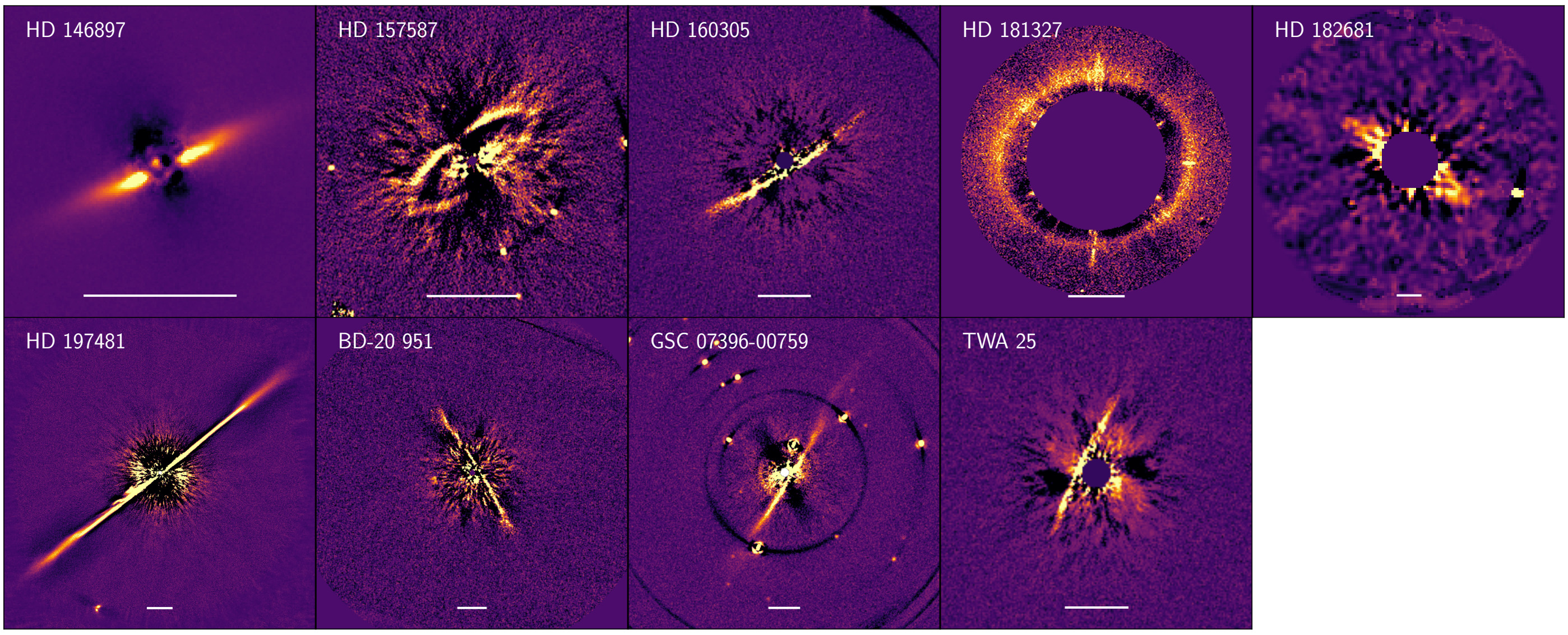

(b)

Fig. 2: Images of the total intensity of scattered light from debris disks detected with IRDIS, IFS, or ZIMPOL. The white bar at the bottom of each image corresponds to 1″. In all images, sky north is up and east is to the left. *(cont.)*

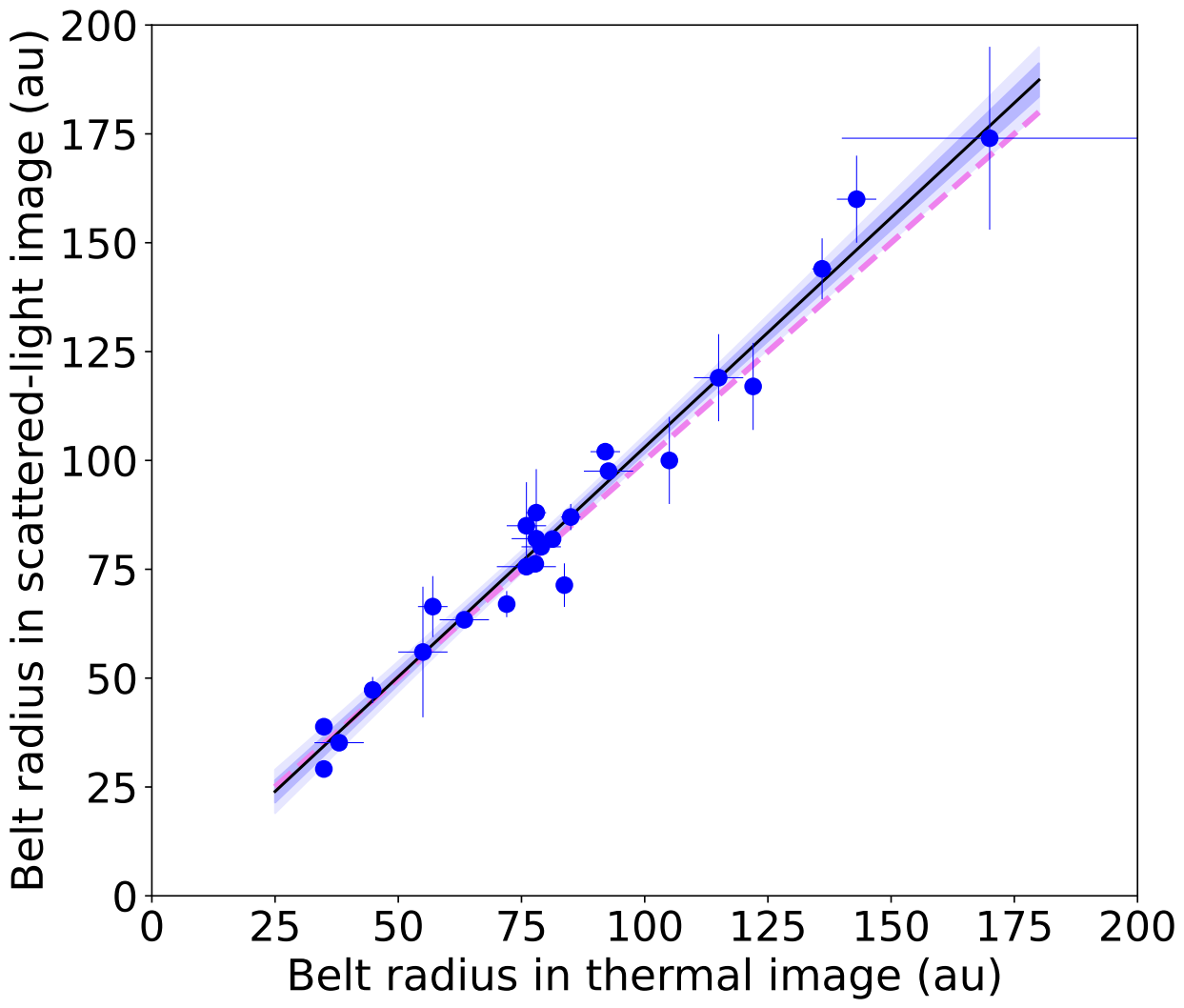

Fig. 3: Radii of planetesimal belts measured from the $r^2$-scaled scattered-light images (SPHERE) versus centroid radii of Gaussian distributions fitted to the thermal images (ALMA and SMA). The violet dashed line shows the 1:1 relation. The black solid line shows the empirical linear fit to the data, with a slope of $1.05 \pm 0.04$. The blue-shaded regions indicate the 68% and 95% confidence intervals for the fitted line.

the belt, can differ between scattered-light and thermal-emission images.

Since the spatial resolution of many millimeter observations is sufficiently high to examine the relationship between belt radii measured in both near-IR and millimeter wavelengths, we analyze this correlation and present our results in Fig. 3. For this comparison, we used disk radii measured from $r^2$-scaled SPHERE images (Col. 2 in Table 2), while the belt radii observed in thermal emission were obtained from the REASONS survey (Tables 1 and A.1 in Matrà et al. 2025). In that study, all targets are fitted with a single planetesimal belt model, where the radial surface density of particles is described by a Gaussian distribution. Consequently, the derived belt radii represent the centroid radii of this distribution.

To ensure consistency, we excluded from this comparison the REASONS targets that were only marginally resolved in millimeter observations or exhibited more than one planetesimal belt, with two exceptions:

- HD 15115: The radial locations of its two cold belts were taken from the two-belt model fit of the ALMA image presented by MacGregor et al. (2019).
- HD 92945: The SB profile of the disk, as shown in Fig. 2 by Marino et al. (2019), was used to determine the radial positions of its two belts in ALMA images.

Figure 3 demonstrates a good agreement between the belt radii measured from SPHERE and ALMA images, indicating a near 1:1 relationship between the locations of the radial SB peaks in near-IR scattered light and thermal emission images. A linear fit to the data (black solid line in Fig. 3) yields a slope of $1.05 \pm 0.04$, representing the ratio $R_{\rm belt}^{\rm mes}$(near-IR)/$R_{\rm belt}$(mm). This value is lower than the average ratio of 1.39 reported by Esposito et al. (2020) in a similar comparison. This finding highlights the need for higher-sensitivity and higher-resolution observations to better understand the connection between disk structures observed in scattered light and thermal emission.

### *4.2. Ratio of radii in multiple belt systems*

Observations across multiple wavelengths, from optical to millimeter, suggest that many young debris disks likely consist of multiple planetesimal belts (e.g., Golimowski et al. 2006; Bonnefoy et al. 2017; Marino et al. 2019). This is further supported by the fact that many disk SEDs are better modeled using two blackbody (BB) components with distinct equilibrium temperatures ($T_{\rm bb}$), requiring dust populations at different radial distances from the host star[7] (e.g., Chen et al. 2014).

[7] Note, however, that caution is required when interpreting SED-derived double belts. It has been demonstrated that when a significant population of submicron grains is present, as expected in bright and highly collisional debris disks, the SED of a single belt disk can mimic

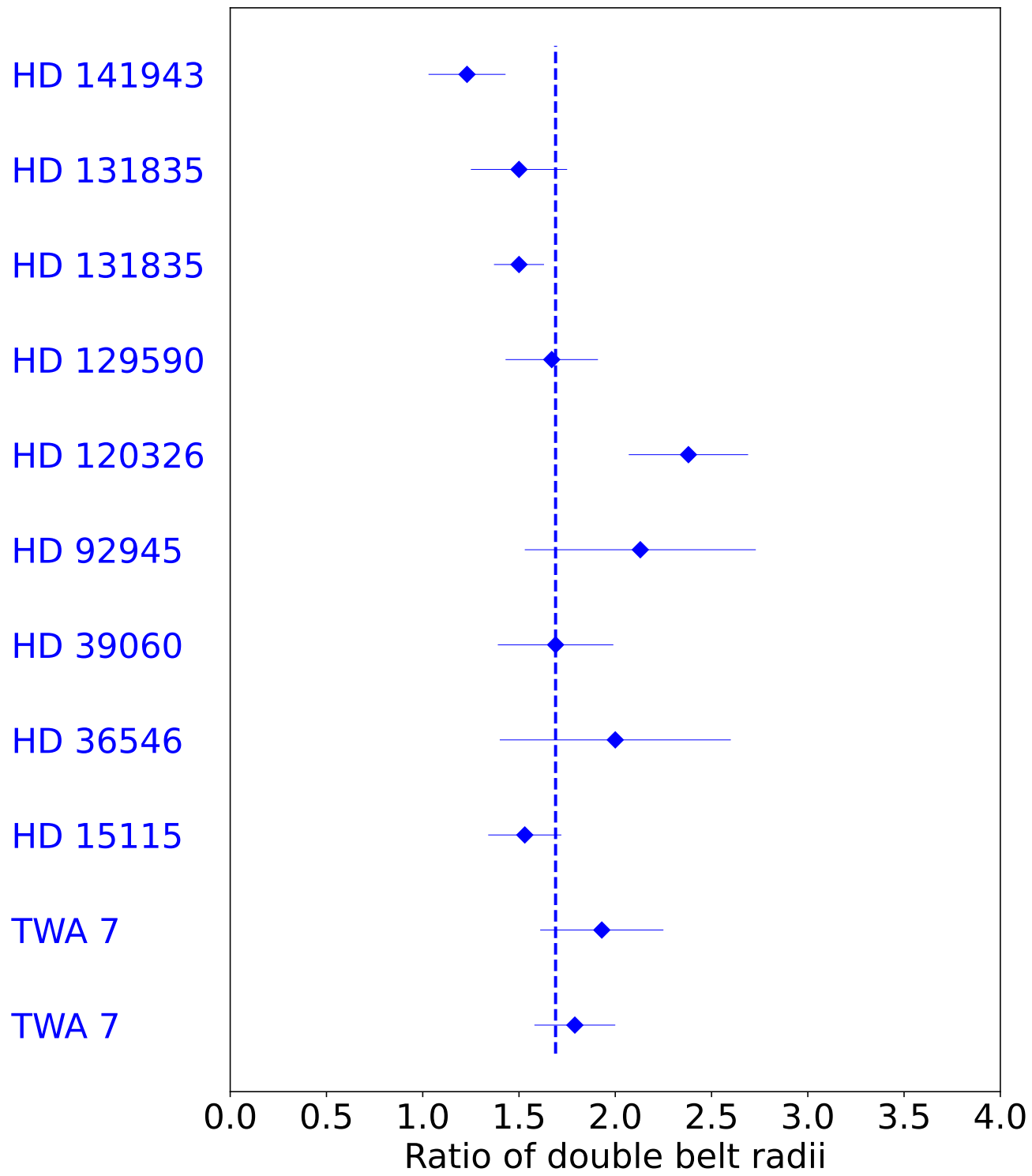

Fig. 4: Ratio of belt radii in double-belt systems. The radii of planetesimal belts were measured from the $r^2$-scaled scattered-light images. The two entries for TWA 7 and HD 131835 show the ratios between the intermediate and inner belts and between outer and intermediate belts. The blue dotted line indicates the median ratio value of 1.69.

Multiple belt configurations are rarely detected in DI, as disks with high inclinations are more easily resolved, as discussed in Sect. 6.2. Among the debris disks imaged with SPHERE, excluding transition disks such as HD 141569, seven systems exhibit spatially resolved double-belt structures, while HD 131835 (Feldt et al. 2017) and TWA 7 images reveal three distinct planetesimal belts. In these systems, both the inner and outer belts belong to the category of cold exo-Kuiper belts (Sect. 4.3) and are listed separately in Table 2. Interestingly, the ratio between the outer and inner belt radii is consistently around 1.5 or 2, with a median value of 1.69 across the nine resolved systems (Fig. 4). These similar belt spacing ratios may hint at similar evolutionary pathways of debris systems or could indicate the presence of mean-motion resonances, possibly due to unseen planets shaping these structures.

### 4.3. Empirical correlation between belt radius and star luminosity

As mentioned in Sect. 4.2, the SEDs of many debris disks require a two-BB model fit. In such two-temperature debris disks, the dust populations are typically classified into warm dust belts (belts with BB temperature between ~100 and 200 K) and cold dust belts (exo-Kuiper belts with BB temperature below 100 K). This bi-modal temperature distribution has been investigated in previous studies (e.g., Kennedy & Wyatt 2014) and is often explained by the preferential formation of planetesimals at the ice lines of volatile compounds such as water, ammonia, carbon dioxide, or carbon monoxide (Morales et al. 2011).

The ice line, also referred to as the frost or snow line, of a volatile compound marks the minimum radial distance from a star at which the temperature is sufficiently low for the compound to condense. Beyond this distance, gas condensation promotes the formation and growth of icy dust particles, thereby facilitating the development of planetesimals. Consequently, debris belts may be more likely to form just beyond the ice lines of common volatile compounds such as $H_2O$, CO or $CO_2$.

The positions of ice lines within a disk are not fixed throughout a star's lifetime, as they are influenced by the evolving stellar luminosity and the opacity of surrounding material. As a result, the disk's radial temperature profile and the condensation thresholds of different volatile substances change over time. This implies that the range of radial distances at which a specific volatile compound may condense into ice can be relatively broad. For example, in the solar nebula, the water snow line has been predicted to lie at 2.7 - 3.2 au, with grain temperatures between 170 and 143 K, depending on the model (Hayashi 1981; Podolak & Zucker 2004), In contrast, the current water snow line in the Solar System is estimated to be at ~5 au from the Sun (Jewitt et al. 2007). Moreover, the condensation temperature of volatile compounds is influenced by the properties of debris particles onto which the gases freeze. Kim et al. (2019), for instance, found that in the case of the young A-type star $\beta$ Pic (HD 39060), the water snow line could be located anywhere between 4.4 and 28.3 au, depending on the dust grain composition, grain size and ice phase (amorphous or crystalline).

To explore the correlation between the locations of ice lines and planetesimal belts within the subsample of debris disks spatially resolved with SPHERE, we estimated the temperature of their BB grains (Col. 5 in Table 2). These grains, being significantly larger than the peak wavelength of the emitted disk spectrum, allow their temperature to be determined using the following expression (Backman & Paresce 1993):

$$T_{\rm bb} = (278\,{\rm K}) \left(\frac{L_\star}{L_\odot}\right)^{0.25} \left(\frac{1\,{\rm au}}{R_{\rm belt}^{\rm mes}}\right)^{0.5},$$

where $L_\star$ is the stellar luminosity, and $R_{\rm belt}^{\rm mes}$ is the measured belt radius in au (Col. 2 in Table 2).

According to this estimation, all resolved debris belts fall into the category of exo-Kuiper belts containing cold dust ($T_{\rm bb}$ < 100 K), with the exception of the warm dust belt around HD 172555, which was detected with ZIMPOL (Engler et al. 2018). In Fig. 6, we present the derived BB temperatures of the belts as a function of their measured radii. The shaded regions in the plot indicate the upper temperature limits at which $H_2O$, $CO_2$ and CO may condense in young disks, depending on gas pressure and dust temperature (Harsono et al. 2015). For comparison, the plot also includes the locations of the Edgeworth–Kuiper belt at 40 au (Stern & Colwell 1997), with an estimated BB temperature of $T_{\rm bb}$ = 44 K, and the main asteroid belt in the Solar System at 3.5 au (Wyatt 2008), with $T_{\rm bb}$ = 150 K.

As shown in Fig. 6, the majority of planetesimal belts are located within the $CO_2$ and CO ice formation zones. This trend is also evident in seven double-belt systems, where both components reside within the same ice-species region. Notably, all three resolved planetesimal belts of HD 131835 lie within the $CO_2$ ice zone, suggesting that they originate from a common, broad debris disk in which gaps may have been sculpted by planetary bodies. The disk around HD 172555 is located in a region where water molecules can accumulate on grain surfaces. This analysis supports the hypothesis that planetesimals preferentially form beyond the ice lines of various gas species.

If this statement holds true, the radial distance of a debris belt should correlate with the luminosity of its host star. This relationship has recently been examined in samples of debris disks resolved at millimeter and far-IR wavelengths (Matrà et al. 2018; Marshall et al. 2021). The subsample of debris disks spatially that of a double belt system, with temperature ratios between the two belts reaching up to a factor of 2 (Thebault & Kral 2019).

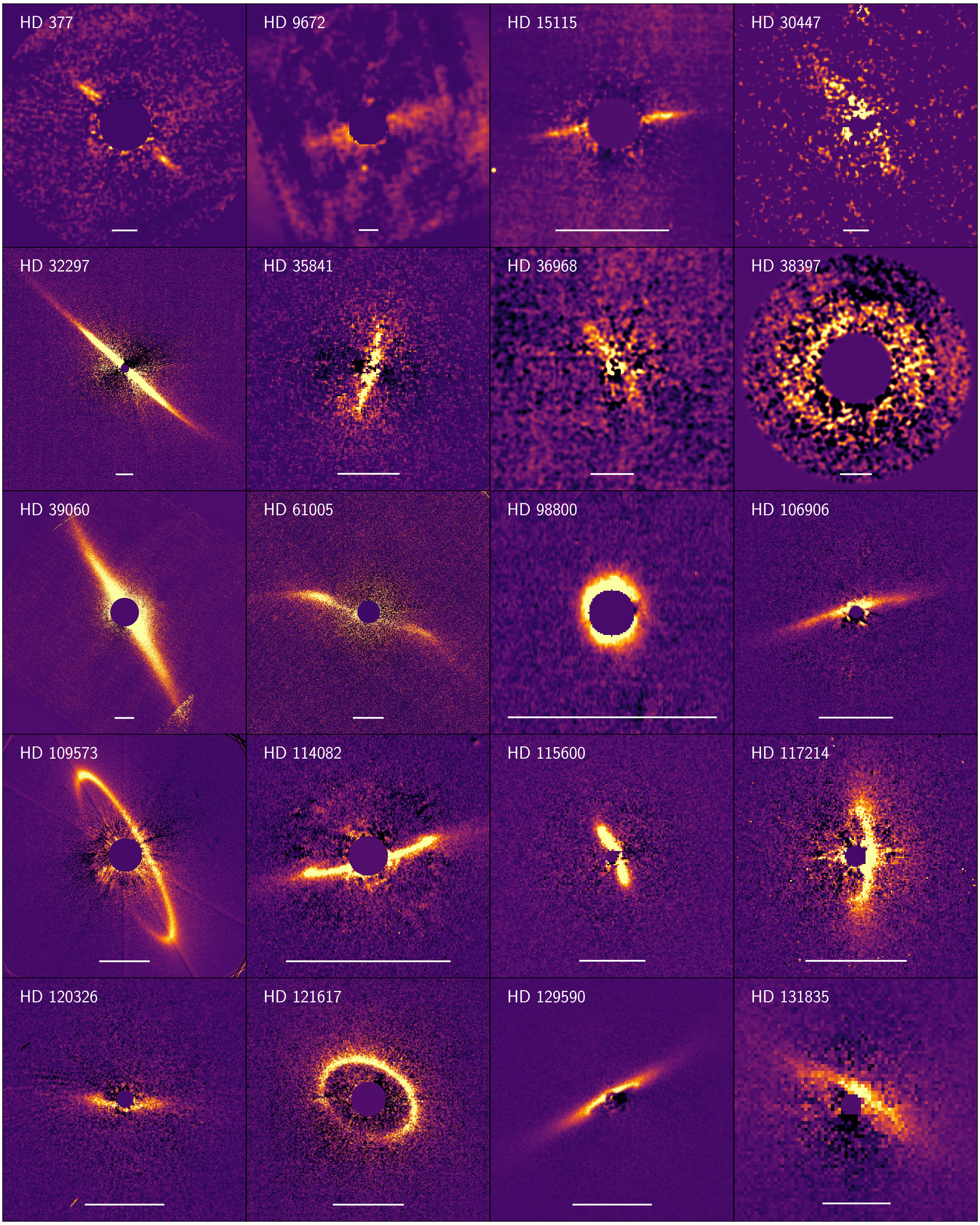

(a)

Fig. 5: Images of the polarized intensity of scattered light from debris disks detected with IRDIS, IFS, or ZIMPOL. The white bar at the bottom of each image corresponds to 1″, except for the HD 98800 image, where it represents 0.5″. In all images, sky north is up and east is to the left.

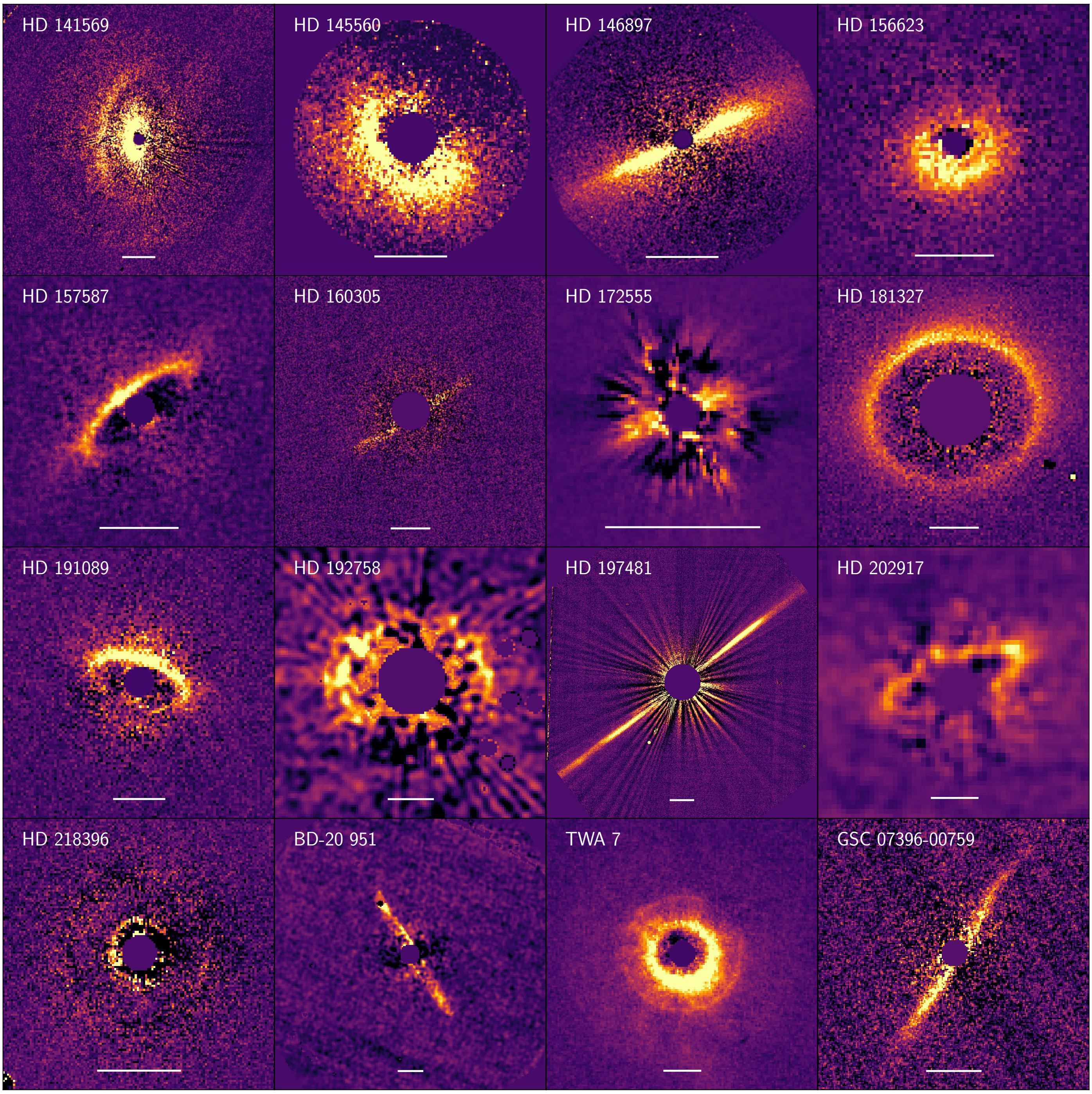

(b)

Fig. 5: Images of the polarized intensity of scattered light from debris disks detected with IRDIS, IFS, or ZIMPOL. The white bar at the bottom of each image corresponds to 1″, except for the HD 98800 image, where it represents 0.5″. In all images, sky north is up and east is to the left. *(cont.)*

resolved with SPHERE (Table 2) provides an opportunity to explore the correlation further. To quantify this relationship, we applied a power-law fit

$$R_{\rm belt} = R_{L_\odot}\left(\frac{L_\star}{L_\odot}\right)^{\alpha} \quad (1)$$

to the data points in Fig. 7, where we show the distribution of the exo-Kuiper belts in our subsample in the parameter space $[R_{\rm belt}^{\rm mes}, L_\star]$. The scaling factor $R_{L_\odot}$ is in au and represents the expected radial position of a planetesimal belt around a star with solar luminosity.

We obtained a relatively shallow linear dependence in logarithmic space $\log(R_{\rm belt}) = \alpha \log(L_\star/L_\odot) + \log(R_{L_\odot})$ (magenta dash-dotted line in Fig. 7) with $R_{L_\odot} = 74 \pm 7$ au and $\alpha = 0.11 \pm 0.05$. These values remain within the $1\sigma$ uncertainties of similar parameters reported in studies at millimeter and far-IR wavelengths (Matrà et al. 2018; Marshall et al. 2021).

The HD 172555 disk was excluded from this analysis, as it is the only system in our subsample that contains warm dust. How-ever, it is likely that the disks included in our subsample formed in connection with the ice lines of various volatile species, such as $CO_2$ and CO gases. If this is the case, analyzing the relationship between the radial distance of a belt and stellar luminosity requires categorizing the sample based on disk BB temperature, which may correspond to the freeze-out temperature of a specific gas specie.

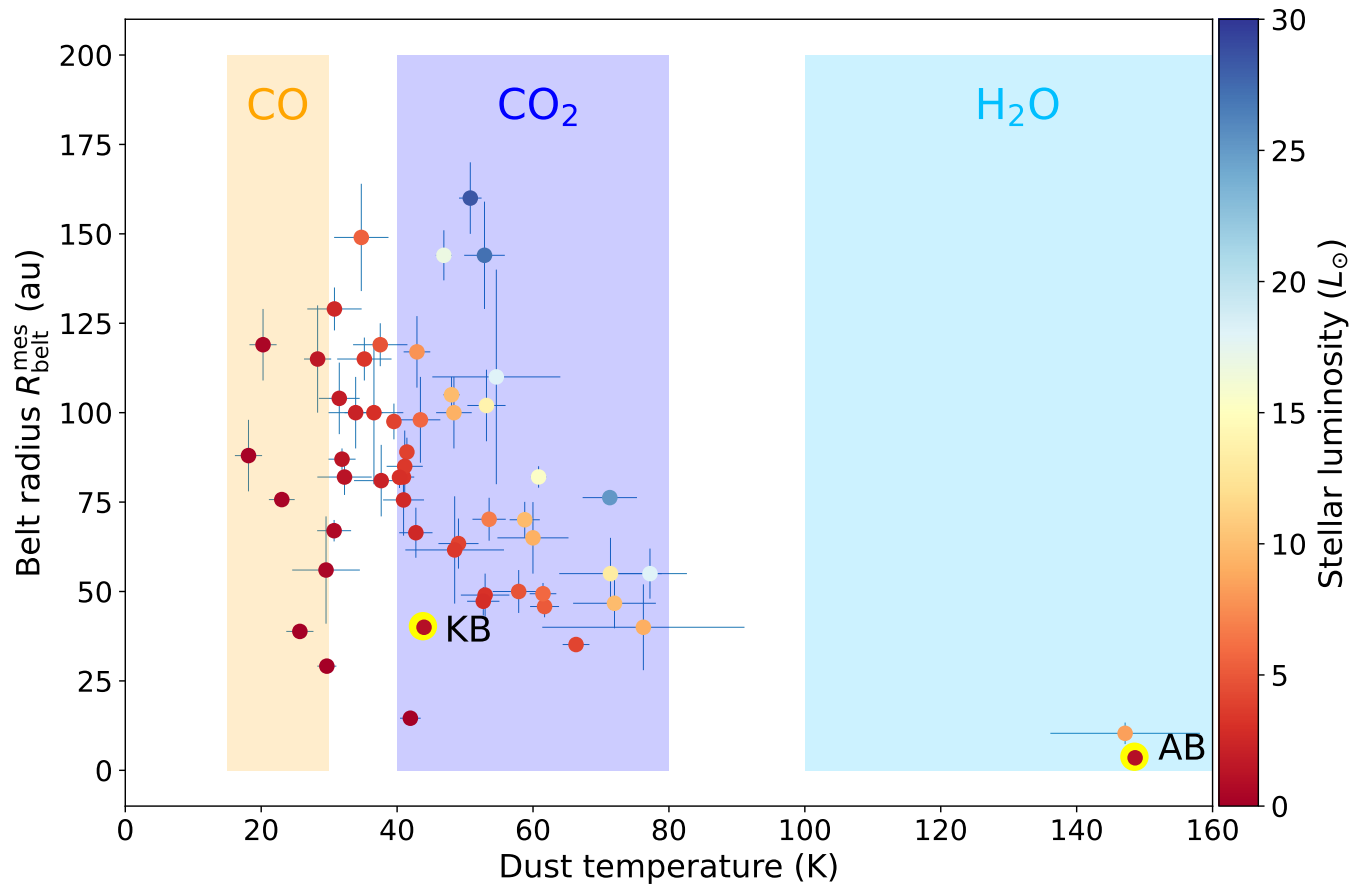

Fig. 6: Belt radii measured from $r^2$-scaled scattered-light images as a function of BB temperature $T_{\rm bb}$ for dust grains at the radial position of the planetesimal belt. The shaded areas indicate the upper temperature ranges where the volatile species $H_2O$ (light blue), $CO_2$ (violet) and CO (orange) begin to freeze out in the disk. KB and AB refer to the Kuiper belt and asteroid belt, respectively.

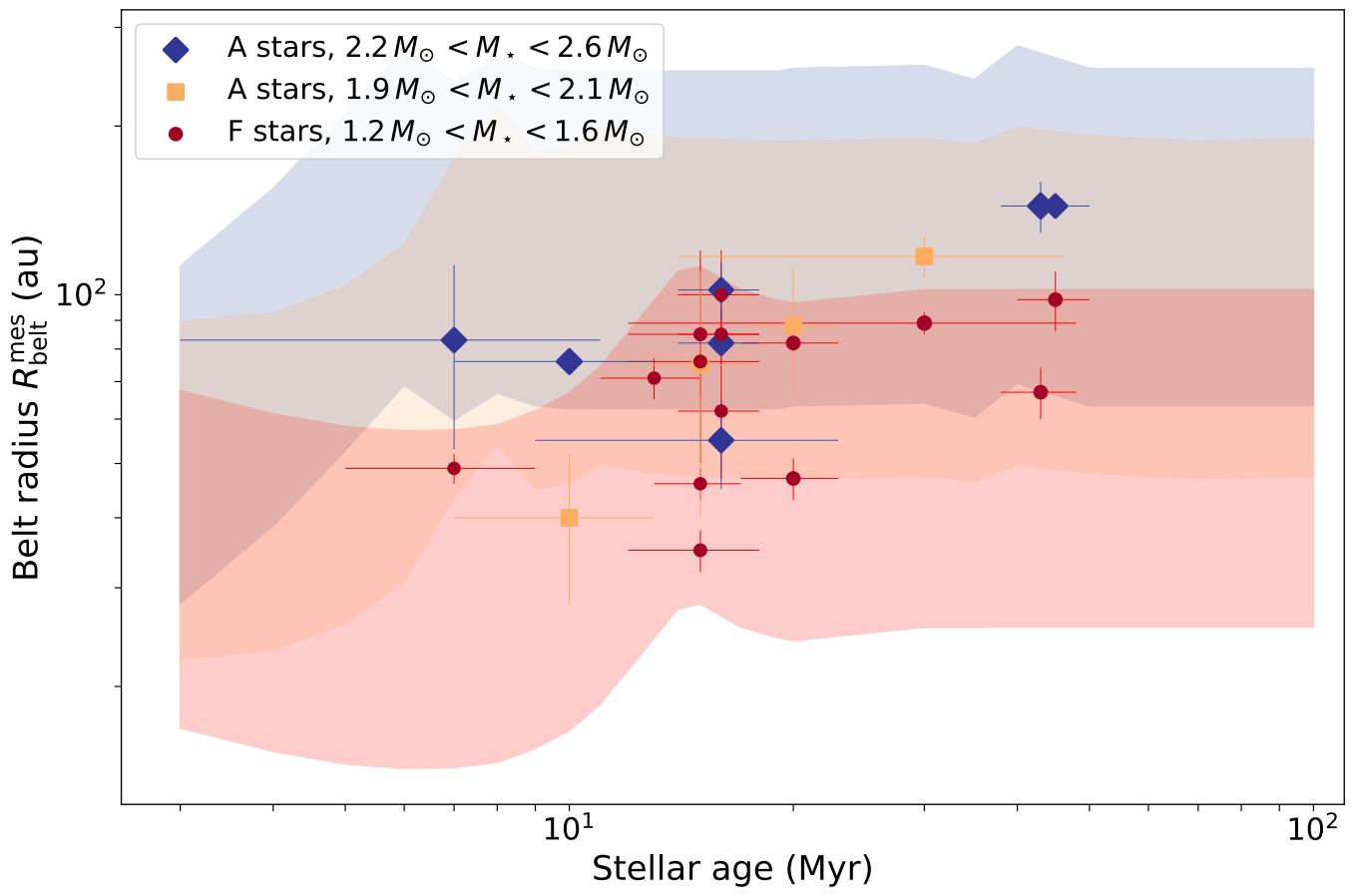

Fig. 8: Belt radii measured from $r^2$-scaled scattered-light images plotted as a function of stellar age for targets in the $CO_2$ subsample. Red, yellow and blue shaded regions indicate the temporal evolution of the $CO_2$ freeze-out zones for stars with masses of 1.5, 2 and 2.3 $M_\odot$, respectively. The upper and lower boundaries of the freeze-out zones correspond to the BB temperature of 40 and 80 K, respectively. The orange and gray shaded regions are the results of the overlap of the three regions.

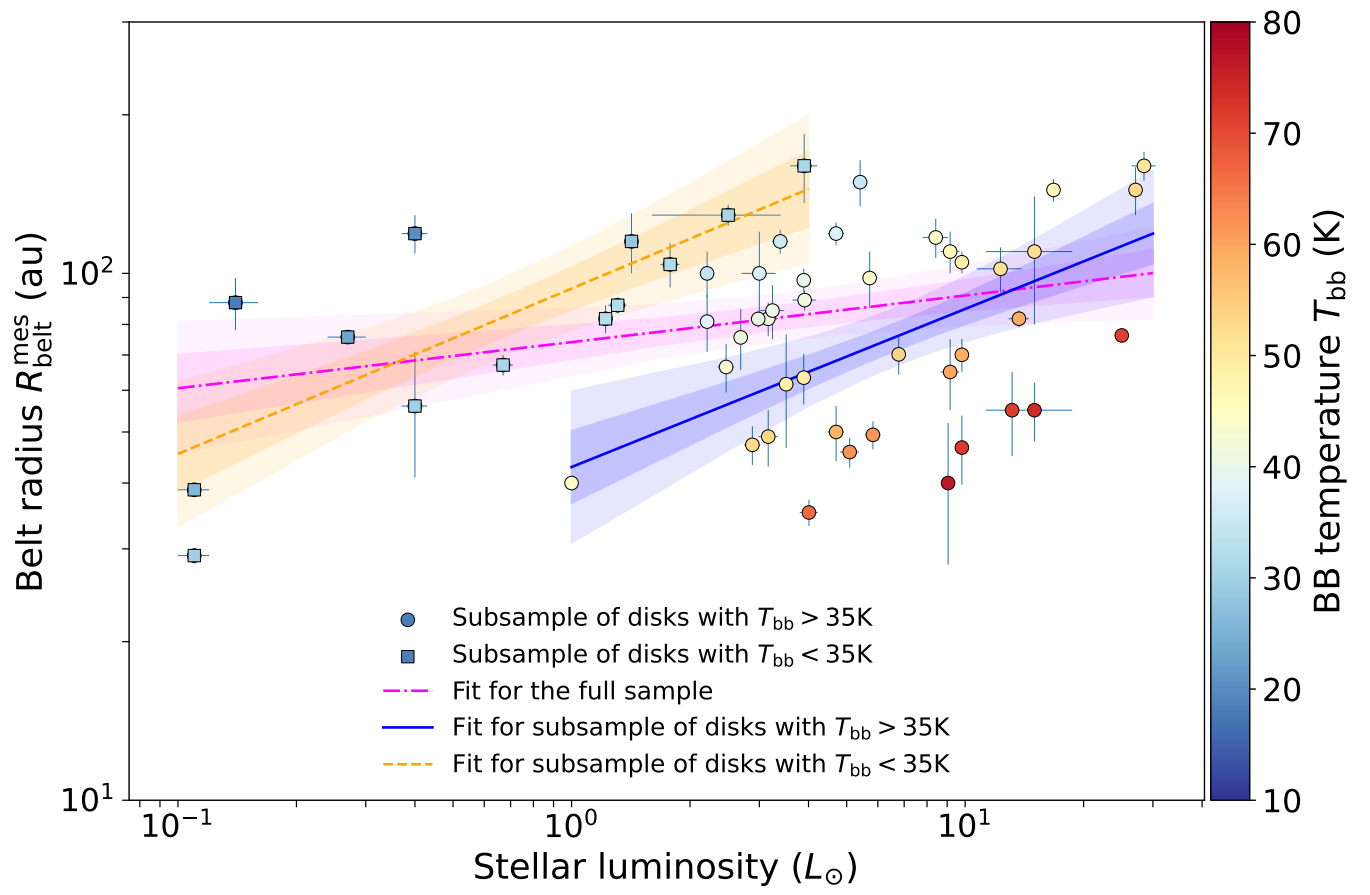

Fig. 7: Belt radii measured from $r^2$-scaled scattered-light images as a function of stellar luminosity. The magenta line represents the best-fit power-law relation for the full sample of resolved debris belts. The orange and blue lines show the fits for subsamples with BB dust temperatures below and above 35 K, respectively. The magenta-, orange- and blue-shaded regions indicate the 68% and 95% confidence bands for the corresponding fits.

Therefore we divided our subsample into two groups of disks based on their temperatures $T_{\rm bb}$. Taking into account the uncertainties in the estimated temperature, we set $T_{\rm bb}$ = 35 K as the upper limit for the coldest disks in the subsample, where the CO gas may freeze out (CO subsample). By fitting a power-law function (Eq. 1) to this group of disks, we obtained best-fit parameters of $R_{L_\odot} = 96 \pm 7$ au and $\alpha = 0.30 \pm 0.07$. This fit is represented by the orange dashed line in Fig. 7. For the group of disks with a local equilibrium temperature above 35 K ($CO_2$ subsample), the best-fit parameters are $R_{L_\odot} = 43 \pm 8$ au and $\alpha = 0.30 \pm 0.08$, as indicated by the blue solid line in Fig. 7.

As expected, the power-law functions for both disk groups are steeper than the function fitted to the entire sample. Notably, the parameter $R_{L_\odot}$ for the $CO_2$ subsample is found to be 43 au. This radial distance closely corresponds to the location of the Edgeworth–Kuiper belt in the Solar System.

As previously discussed in this section, the radial locations of the ice lines for volatile compounds vary over the course of stellar evolution. To investigate whether a corresponding evolution in the radial positions of planetesimal belts is observable, we plotted the measured belt radii as a function of stellar age for the systems in the $CO_2$ subsample. This subsample provides a relatively larger number of systems with stars of similar spectral type but different ages, allowing for a more meaningful comparison.

Given that stellar luminosity is a key parameter in this context, we examined three groups of stars categorized by spectral type and mass: (1) A-type stars with $2.2\,M_\odot < M_\star < 2.6\,M_\odot$, (2) A-type stars with $1.9\,M_\odot < M_\star < 2.1\,M_\odot$, and (3) F-type stars with $1.2\,M_\odot < M_\star < 1.6\,M_\odot$. For stars with multiple resolved belts, we adopted the mean belt radius, as all detected belts in these systems have $T_{\rm bb} > 35$ K.

We note, that the mean estimated age (Col. 11 in Table E.2) of most stars with detected disks (90% of detections) is below 50 Myr. According to stellar evolutionary models (e.g., Palla & Stahler 1999; Baraffe et al. 2015), such young objects are likely located either on the pre-main-sequence (PMS) or on the zero-age main sequence (ZAMS) in the Hertzsprung-Russell diagram.

Intermediate-mass stars ($1\,M_\odot < M_\star < 3\,M_\odot$) exhibit the most pronounced luminosity evolution during their PMS phase. These stars begin as fully convective objects with large radii and high luminosities, which decrease as the stars contract. This phase is followed by an increase in both temperature and luminosity as a radiative core begins to develop because it becomes hotter and denser during the contraction phase. This process leads to an increase in the rate of nuclear fusion, ultimately stabilizing the star and placing it on the main sequence.

The luminosity evolution during the PMS phase depends sensitively on the stellar mass and chemical composition (e.g., Tognelli et al. 2011). To illustrate this, Fig. 8 shows the evo-lution of the radial location of the $CO_2$ freeze-out zones (red, orange and blue shaded regions), corresponding to the $CO_2$ zone presented in Fig. 6, for stars with metallicity $Z = 0.028$ and helium abundance $Y = 0.304$, calculated for stellar masses of 1.5 $M_\odot$, 2 $M_\odot$ and 2.3 $M_\odot$ (Tognelli et al. 2011).

As shown in Fig. 8, stars in all three mass groups show a trend of increasing belt radius with stellar age within the region corresponding to the $CO_2$ freeze-out zone. This behavior may reflect the outward migration of the $CO_2$ ice line due to increasing luminosity as the star evolves. To better quantify this result, we performed a multiple regression analysis. We considered $\log t_{\rm age}$ (in Myr) and $\log M_\star$ (in solar masses) as independent variables, and $\log R_{\rm belt}$ (in au) as dependent variable. The best-fit relation is the following:

$$\log (R_{\rm belt}/{\rm au}) = (0.37 \pm 0.11) \log (t_{\rm age}/{\rm Myr}) + (0.59 \pm 0.23) \log (M/M_\odot) + (1.27 \pm 0.15).$$

Both coefficients are significant (with a 0.002 probability of being a chance result for the dependence on the age, and of 0.02 for the dependence on the mass). The dependence on age is then highly significant. This relation predicts $\log R_{\rm belt}$ for stars in this range of ages and masses with an accuracy of 0.12 dex.

### 4.4. Morphology of selected targets

In this section, we discuss debris systems that have been imaged in scattered light for the first time, as well as debris disks whose morphology exhibits notable features, such as multiple belts, that warrant further examination.

**HD 9672 (49 Ceti)** The A1V star HD 9672 is one of the youngest and brightest stars in the sample. It is likely a member of the ~40 Myr old Argus MG (99% membership probability; Zuckerman 2019). The debris disk surrounding HD 9672 is gas-rich, with an estimated CO mass exceeding $2.2 \times 10^{-4} M_\oplus$. The spatial distribution of CO gas closely resembles the structure of the outer debris disk, whereas no molecular gas has been detected within ~90 au of the star (Hughes et al. 2008).

The thermal emission of the HD 9672 disk is well characterized by two distinct dust populations: a warm component at 136−160 K and a cold component at 47−60 K (e.g., Wahhaj et al. 2007; Roberge et al. 2013; Chen et al. 2014, this work). Models reproducing the disk's emission at $\lambda = 12.5\,\mu$m and $\lambda = 17.9\,\mu$m suggest that the warm dust grains are located within 60 au (Wahhaj et al. 2007).

The debris disk has been observed multiple times with SPHERE using different filters (e.g., Pawellek et al. 2019), including the first detection of its scattered light (Choquet et al. 2017). The PCA-reduced data taken with the IRDIS B_Y filter, shown in Fig. 9a, reveal a broad debris ring extending to the image edges (∼ 6″). This structure may consist of multiple narrow rings. In the $r^2$−scaled image, the radial SB peak is located between 144 and 156 au. Additionally, image residuals hint at an inner planetesimal ring with a radius of ~105-110 au (see Fig. 9a). If this second cold debris ring is confirmed, it would contain dust grains with a blackbody temperature of $T_{\rm BB} = 54$ K and would not account for the warm dust excess observed in the HD 9672 SED.

Polarized scattered light observations of the disk, conducted with IRDIS in polarimetric mode using the B_Y filter, led to a detection, albeit with a relatively low S/N (Fig. 5). This may be attributed to an intrinsically low polarization fraction of the dust in this disk.

**HD 16743** The ALMA image of the debris disk around the F0 star HD 16743 was recently published by Marshall et al. (2023). While the authors report only a marginal detection of the disk in the IRDIS H-band data, our image, obtained with the IRDIS H23 filter (Fig. 2), provides the first clear detection of this debris disk in scattered light. This detection allows us to constrain the disk's geometrical parameters (Table 2). We determined that the PA of the disk is ~169°, which closely matches the value reported by Marshall et al. (2023). Additionally, our image reveals an extended feature at a PA of approx. 17° (see Fig. B.2). The origin of this feature remains unclear; it is most likely a residual PSF artifact, possibly caused by the telescope spider. However, the possibility of a scattered light signal cannot be entirely ruled out.

**HD 36546** HD 36546 is another A-type star in our sample (A0V-A2V; Lisse et al. 2017; Currie et al. 2017), located at a distance of 100.1 pc. It is a probable member of the Mamajek 17 group, with an estimated age between 3 and 10 Myr. For the first time, its debris disk has been spatially resolved using the Subaru/HiCIAO camera in the H band (Currie et al. 2017).

Our observations reveal a well-defined debris ring in the IFS data at a radial separation of 55 ± 10 au, as shown in Fig. 9c. This ring is also visible in the IRDIS image in the K band (Fig. 9b) albeit with a lower S/N. Additionally, the IRDIS image, along with some IFS data, reveals a more extended debris belt with a radius of 110 ± 30 au (outer ring in Fig. 9b). This belt appears both wider and brighter than the inner ring and may consist of multiple components. The presence of two cold debris rings at approx. 55 and 110 au aligns with the modeling results of Currie et al. (2017) and Lawson et al. (2021), who determined that the debris disk of HD 36546 extends between 60 and 115 au. However, due to the system's relatively high inclination (~80°), precisely determining the number of rings present remains challenging.

The residual pattern observed within the inner ring (Fig. 9b) resembles the structure of a smaller, distinct ring, particularly in its alignment along the major axis of the outer rings. If this feature represents a genuine ring rather than PSF residuals, an alternative that cannot be ruled out, its radial separation from the star would be ~30 au. Interestingly, Lisse et al. (2017) previously reported a debris belt at ~135 K, which is expected to be located between 20 and 40 au from HD 36546. Moreover, when fitting the disk spectrum and photometric data from multiple instruments, Lisse et al. (2017) also predicted the existence of an additional inner belt with a temperature of ~570 K, corresponding to hotter dust located between 1.2 and 2.2 au within the system.

**HD 36968** HD 36968 is a young (~20 Myr) F2V star located at 149 pc in Octans Association (Moór et al. 2011; Murphy & Lawson 2014). The debris disk surrounding this star exhibits a high IR excess of $1.34 \times 10^{-3}$ and has been detected for the first time in both total scattered intensity and polarized intensity (Figs. 2 and 5). The fundamental geometrical parameters of the disk are provided in Table 2.

**HD 92945** HD 92945 is a nearby K1V star located at a distance of 21.51 pc from the Sun. We resolved the inner debris belt with a radius of ~56 au and a part of the outer belt at ~119 au (Fig. 2), both of which were previously imaged with ALMA (Marino et al. 2019). The PCA reduction of the H-band data shows residual structures that suggest the presence of a third dust ring with a possible radius of ~38 au.

**HD 98800** HD 98800 is an intriguing quadruple system located 42.1 pc from the Sun (Gaia Collaboration et al. 2021; Boden et al. 2005). As a member of the TW Hydrae Association, it has an estimated age between 7 and 10 Myr (Ducourant et al. 2014). The system consists of two pairs of spectroscopic binaries which orbit each other with a semimajor axis of ~50 au and period of 246 years (Kennedy et al. 2019; Zúñiga-Fernández et al. 2021). A double-lined SB BaBb ($M_{\rm Ba}$ = 0.70 $M_\odot$, $M_{\rm Bb}$ = 0.58 $M_\odot$, $P$ = 315 days; Boden et al. 2005) is surrounded by a bright transitional debris disk, previously imaged at 1.3 mm with ALMA (Kennedy et al. 2019) and at 8.8 mm and 5 cm with the Very Large Array (VLA) (Ribas et al. 2018).

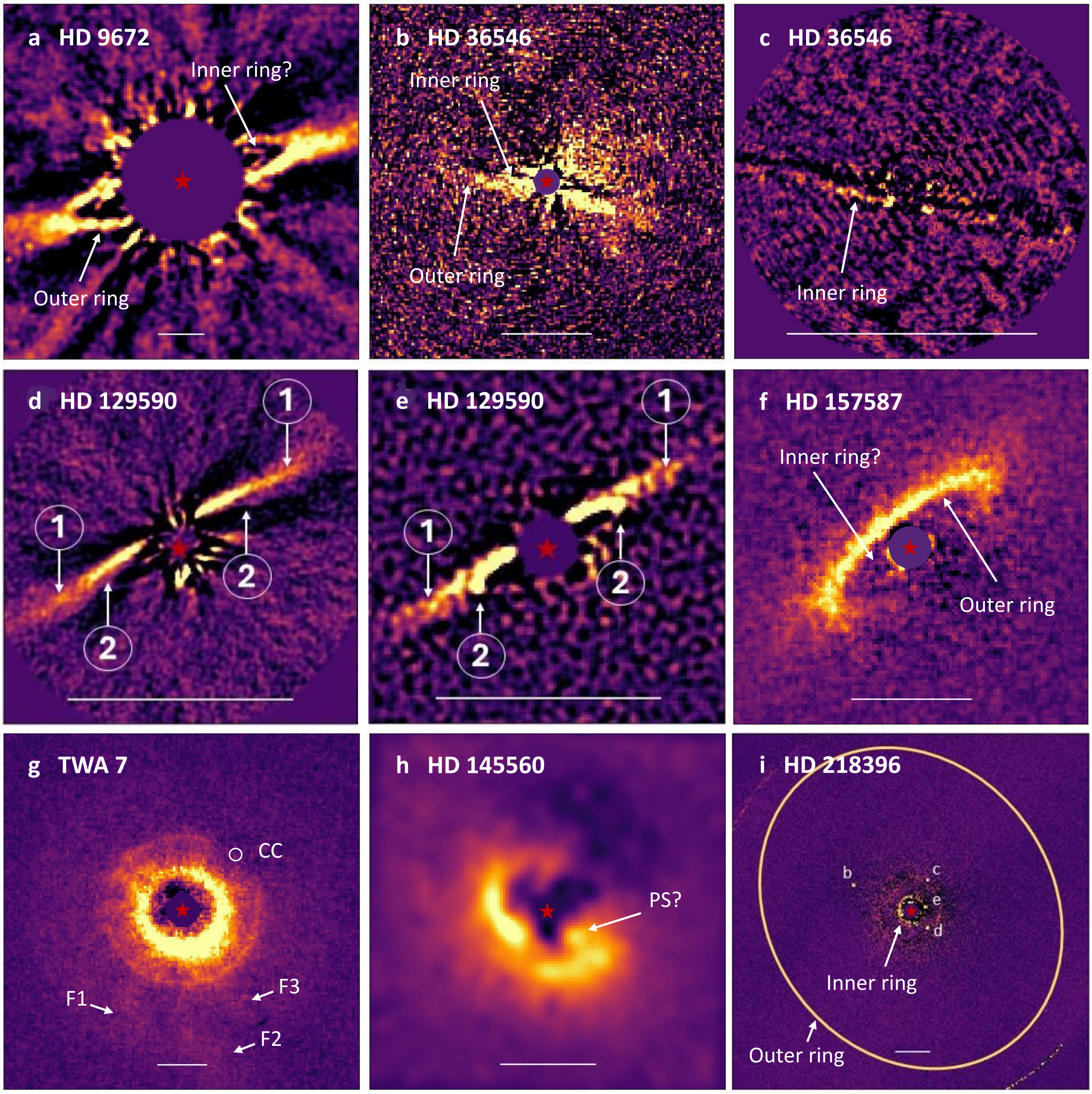

Fig. 9: Images of debris disks with signs of multiple rings. The white bars in the lower parts of the images correspond to 1 arcsec. The positions of stars are marked by red asterisks. In *panels d* and *e* the outer belt is indicated by a number "1" and the inner belt by a number "2". *Panel a*: The H2H3-filter total intensity image of debris disk HD 9672 (49 Ceti). *Panel b*: The H2H3-filter total intensity image of debris disk HD 36546. *Panel c*: The combined IFS total intensity image of inner belt around HD 36546. *Panel d*: The H-band total intensity image of debris disk HD 129590. *Panel e*: The H-band polarized intensity image of debris disk HD 129590. *Panel f*: The H-band polarized intensity image of debris disk HD 157587. *Panel g*: The H-band polarized intensity image of debris disk TWA 7. The position of the candidate planetary companion is labeled as "CC". The labels "F1", "F2" and "F3" indicate arc-like morphological features detected in the disk structure. *Panel h*: The VBB polarized intensity image of the HD 145560 debris disk. *Panel i*: The H-band $Q_\phi$ image of debris disk HD 218396 (HR 8799). The radial position of the outer belt at $r = 4.5''$ is schematically shown by the orange ellipse. The positions of planets HR 8799 b, c, d and e are taken from the total intensity image and overlaid over the $Q_\phi$ image.

Imaging this disk in scattered light presents a significant challenge due to its small angular size and the presence of a close central binary. The disk has a radius of less than 0.1″ and an inclination of less than 45°, making non-coronagraphic polarimetric differential imaging (PDI) with ZIMPOL the most suitable observing strategy for resolving it in scattered polarized light. Such observations were conducted on April 14, 2016 (ESO 097.C-0344, PI: Kennedy), and our data reduction successfully detected scattered light from the disk. Figure 5 presents the $Q_\phi$ image obtained using the ZIMPOL R_PRIME filter. The HD 98800 transitional disk is the smallest detected with SPHERE/ZIMPOL in both angular and physical size ($R_{\rm belt} \approx 0.07''$ or 3 au). In Fig. 5, the white bar in the HD 98800 panel showing the $Q_\phi$ image represents 0.5″ (20 au), whereas in all other panels it corresponds to 1″.

The Stokes $Q$ and $U$ signals, used to compute the $Q_\phi$ image, are partially reduced due to the significant PSF convolution effect caused by the small angular size of the disk (Engler et al. 2018). Additionally, the disk signal may be affected by residual flux from the central binary, as complete removal of stellar flux in the center of image is not possible even with PDI. These stellar residuals are always present at the image center and originate either from nonzero polarization of the star(s) or from slight mismatches in the PSF shapes of the two orthogonal polarization states, which do not perfectly align. These mismatches arise due to short coherence times and, particularly relevant for a binary system, differences between the PSF of a binary star and that of a single point source. For the HD 98800 disk, we estimated the extent of the stellar residuals from the BaBb binary in the $Q_\phi$ image by analyzing the residuals from the AaAb binary, which is located at ~0.7″ from BaBb, just outside of the frame in Fig. 5. These stellar residuals have been masked in the central region of the presented image.

The scattered polarized light from the disk is detected between 0.06″ (2.52 au) and 0.12″ (5.06 au). Two dips in SB are visible on the eastern ($PA$ = 109°) and western ($PA$ = 280°) sides of the disk, which may be attributed to stellar PSF effects. Based on the SB distribution in the $Q_\phi$ image, we derived the geometrical parameters of the disk, as listed in Table 2. Within uncertainties, these parameters are in good agreement with those obtained from VLA and ALMA images (Ribas et al. 2018; Kennedy et al. 2019).

**HD 111520** The strong brightness asymmetry in the scattered light of the nearly edge-on disk around HD 111520 (F5/6V star at $d$ = 108 pc) has been previously observed with HST (Padgett & Stapelfeldt 2016) and GPI (Draper et al. 2016a). In the SPHERE/IFS image, the northern extension of the debris disk appears significantly brighter than the southern extension, where a dip in SB is observed at approx. 0.5″. Within 0.8″, the disk morphology resembles that of AU Mic disk. The SB variations along the major axis in both systems may be explained by the presence of a spiral disk structure or a set of non-coplanar debris rings. Indeed, the SED of HD 111520 is best fitted with multiple dust populations at different temperatures, suggesting the existence of radially separated debris belts containing both warm and cold dust.

**HD 120326** The two distinct cold dust belts around the F0V star HD 120326 were first resolved in scattered light with SPHERE (Bonnefoy et al. 2017). We measure a radial distance of ~119 au (1.05″) for the larger planetesimal belt and ~50 au (0.44″) for the smaller one. The polarimetric data of HD 120326 reveal polarized light between 0.25″ and 0.7″ with a tentative SB peak at ~0.5″, potentially indicating the presence of an additional inner debris belt in this system.

**HD 129590** HD 129590 is a G3V star with one of the highest IR excesses in our sample ($f_{\rm disk}$ = (6.3 ± 1.8) × $10^{-3}$). The star is surrounded by two planetesimal belts, forming a structure reminiscent of a "moth" shape (Matthews et al. 2017; Olofsson et al. 2023) similar to that observable in the HD 61005 disk (Buenzli et al. 2010). The inner belt, located at ~49 au, is bright and exhibits an extended halo of small dust particles. In contrast, the outer planetesimal ring is significantly fainter but remains clearly visible in the PCA-reduced total intensity images (Fig. 9d). The region between the two belts does not appear to be completely cleared. The polarized intensity data show that although the outer ring is less pronounced in the halo, it remains detectable (Fig. 9e). This ring likely extends between 80 and 92 au, with a peak SB measured at ~82 au in the $r^2$-scaled polarized intensity image. Recently, CO gas was detected in the system (Kral et al. 2020), supporting the possibility of gas pileup as a contributing factor to the observed disk structure (Olofsson et al. 2023).

**HD 145560** We resolve the debris disk around HD 145560 (F5V star at 121.23 pc) in total intensity using the RDI technique (Xie et al. 2022) applied to the H2H3 dataset taken with IRDIS (Fig. 2), as well as in polarized intensity using the VBB filter of ZIMPOL (Fig. 9h). This disk has also been observed with ALMA (Lieman-Sifry et al. 2016; Matrà et al. 2025) and GPI (Esposito et al. 2020). Among all available data for this target, the ZIMPOL image provides the highest spatial resolution and appears to reveal a spiral-like structure on the southern side of the disk, as well as a point-source-like residual (denoted as "PS?" in Fig. 9h) on the western side. However, this image is affected by low-wind effects and a short coherence time during the observation, which lowered the S/N of the polarimetric data, making the detection of this structure uncertain. The apparent point source could, in reality, be a bright part of the disk.

The scattered light in both SPHERE images exhibits an elliptical structure, with a major axis PA = 39 ± 5.0° and a radius of $r$ = 87 ± 5 au, as measured from the $r^2$-scaled image, and an inclination of 48.7 ± 7.0°. These geometrical parameters are within 1$\sigma$ in good agreement with the GPI measurement (Esposito et al. 2020). However, there is a noticeable offset when comparing the disk's orientation on the sky as measured with ALMA: 20 ± 7.0° at 1.24 mm (Lieman-Sifry et al. 2016) and 28 ± 8.0° at 1.3 mm (Matrà et al. 2025). This discrepancy may be due to the different spatial and angular resolutions between the scattered light images (SPHERE, GPI) and the thermal emission images (ALMA). Since the HD 145560 disk is the only resolved disk in our sample that shows such a PA deviation compared to ALMA data, this offset may suggest a more complex disk structure than a simple ring. Possible explanations include a spiral structure or the presence of multiple planetesimal rings at different PAs and inclinations.

**HD 157587** The debris disk around F5V star HD 157587 is resolved with SPHERE instruments in both total and polarized scattered intensities. In the $Q_\phi$ image taken in the broadband H (Fig. 9f), the residual pattern inside the disk resembles the morphology of a smaller ring with a radius of ~50 au. These suspicious residuals are particular visible within the southeast extension of the disk and are also present in the Stokes $U$ image.

**HD 182681** HD 182681 is a B8.5V star located at 70.69 pc and a member of the $\beta$PMG. The debris disk surrounding this star was recently resolved with ALMA at 1.27 mm (Matrà et al. 2025). In the IRDIS H-band image, we detected extended scattered light emission from the debris belt, which has a radius of ~2.27″ (160 au). Additionally, there is evidence of a possible second belt at a radial distance of ~2.94″ (208 au).

**HD 218396 (HR 8799)** HD 218396, better known as HR 8799, is classified as an F0+VkA5mA5 Lambda Boo star (Gray et al. 2006) and is located at a distance of ~41 pc. It is surrounded by an extended exo-Kuiper belt, previously imaged at far-IR, submillimeter, and millimeter wavelengths using various facilities (e.g., Hughes et al. 2011; Faramaz et al. 2021, and references therein). Within the large disk cavity ($r \sim 100$ au), four giant planets with masses below 10 $M_{\rm Jup}$ have been discovered (Marois et al. 2008, 2010) and extensively studied (e.g., Esposito et al. 2013; Zurlo et al. 2016; Wang et al. 2018, see also Sect. 7). The relatively low inclination (∼30°) of the debris belt facilitates planet detection but makes the belt itself challenging to observe using DI with the ADI technique.

HD 218396 was observed in various modes with all SPHERE instruments. In the imaging modes, the cold debris belt, extending between ∼80 and 350 au (2″ − 7.5″) and peaking in SB at 180 − 200 au (∼4.5″, Faramaz et al. 2021), was not detected. Similarly, in the H-band polarized intensity image, the disk remains either undetectable or barely visible, likely due to its low SB in polarized light. However, the image reveals a bright ring-like structure near the coronagraph, at a radial distance of ∼0.4″ or ∼15 au (Fig. 9i).

The possibility that this structure results from stellar PSF residuals cannot be entirely excluded. HD 218396 was observed with IRDIS in polarimetric mode on two nights: October 11 and October 13, 2016. The ring-like structure is detected in the data from October 11, when the observing conditions were significantly better (seeing between 0.44″ and 0.78″, and coherence times between 3.5 and 6.1 ms) compared to those on October 13 (seeing between 0.92″ and 2.82″, and coherence times between 1.7 and 4.0 ms). Poor observing conditions, such as those during the second night, can completely prevent the detection of a debris disk (see discussion in Sect. 6). Therefore, the non-detection of the ring in the data from the second night does not rule out the presence of a warm planetesimal belt at the considered radial position. Additionally, PSF residuals of this kind, especially at locations farther from the coronagraph and AO ring, are uncommon in IRDIS polarimetric data. Thus, the imaged ring likely traces polarized scattered light from dust particles in a second, inner debris belt.

The idea that this feature originates from a warm dust belt is supported by flux measurements of HD 218396 obtained with IRAS, ISO, and Spitzer space telescopes. Based on these data, Su et al. (2009) modeled the disk SED with three distinct dust components: a warm belt, a cold belt, and an extended halo. Stronger evidence for the presence of a warm belt at ∼15 au comes from recent JWST/MIRI observations of HD 218396 at mid-IR wavelengths (Boccaletti et al. 2024) which provided spatially resolved signatures of the inner disk component.

If these interpretations are correct, the IRDIS H-band image resolves the inner warm dust belt in scattered polarized light for the first time. This belt has a radial distance of $r = 15.5 \pm 1.8$ au, an inclination of $i = (32 \pm 7)°$, and a PA of $(39 \pm 22)°$, and it may have an offset from the central star.

TWA 7  TWA 7 is a 4.4 ± 1.4 Myr old (Herczeg & Hillenbrand 2014) M2Ve star in the TW Hydrae association. The IRDIS polarimetric H-band image (Fig. 9g) reveals a nearly pole-on system consisting of three rings at approx. 27, 52 and 93 au (Olofsson et al. 2018; Ren et al. 2021). The disk in the region between the inner and middle rings exhibits a clumpy structure, with arc-like streamers particularly evident in the area extending from the middle ring to the outer boundary of the detected scattered-light emission at ∼96 au. These structural features are most clearly visible on the southern side of the disk, which is inclined toward the observer and exhibits enhanced SB due to the forward scattering of stellar light by dust grains. In the $Q_\phi$ image (Fig. 9g), we highlight three such features, labeled "F1", "F2", and "F3", all of which have been detected in at least three separate epochs of IRDIS polarimetric observations, albeit with varying S/N (see also Sect. B and Fig. B.1). The most pronounced of these, "F2", was also identified in HST/STIS and HST/NICMOS data (Ren et al. 2021).

These arc-like features may share a similar origin with the fast-moving clumps observed in the edge-on disk of AU Mic (Boccaletti et al. 2018). In both systems, sub-micron dust grains may be expelled by strong stellar winds from their active M-type host stars (Strubbe & Chiang 2006; Schüppler et al. 2015), potentially triggered by collisions in a secondary belt or in the vicinity of a planetary companion (e.g., Chiang & Fung 2017; Sezestre et al. 2017). Notably, a candidate Saturn-mass planet located at a projected distance of ∼ 52 au or 1.5″, coincident with the position of TWA 7's second planetesimal ring, has recently been detected with JWST/MIRI (Lagrange et al. 2025). This ring is both very narrow and flanked by two gaps, appearing underluminous at the planet's location relative to other azimuths (Fig. 9g). Such a morphology supports the scenario of a resonant planetesimal ring sculpted by the planet, which may be carving the adjacent gaps and generating a local void.

If confirmed, this planetary companion could be responsible for gravitational perturbations that locally enhance dust production. Once released, small grains are redistributed by interactions with stellar wind and radiation pressure, giving rise to asymmetric structures such as arcs, streamers, or clumps, depending on the disk inclination and viewing geometry. The nearly pole-on orientation of the TWA 7 disk may thus offer a complementary view of the dynamic processes that shape AU Mic's edge-on disk.

BD-20 951  The highly inclined circumbinary debris disk around the SB2 BD-20 951 (Torres et al. 2008) is resolved for the first time in both total and polarized scattered light with SPHERE in the H band (Perrot et al. in prep). The primary is a K1V(e) star, and the binary components have an estimated flux ratio of ∼0.25 (Elliott et al. 2014). The system may be a member of the Carina MG (28±11 Myr; Gratton et al. 2024) or Tucana-Horologium association (37±11 Myr; Gratton et al. 2024) as proposed by Torres et al. (2008), although the BANYAN Σ tool classifies it as a field star.

The IR excess was identified by Moór et al. (2016) who noted that the colder component may significantly contribute to the total near-IR flux of the system. The residuals in the PCA-reduced image suggest that the disk possesses sweap-back wings (Fig. 2). The geometrical parameters of belt are specified in Table 2.

### 4.5. Modeling of selected planetesimal belts

To investigate the morphology of the detected debris disks and potential correlations between disk parameters and host star properties, we fitted images of several debris belts (listed in Table E.1) using a grid of models that simulate single scattering of stellar photons by dust particles in an optically thin disk. A key advantage of our approach, compared to studies focused on individual disks, is the use of a uniform modeling framework for all systems in our sample. This consistency allows for a more direct and meaningful comparison of the derived disk parameters.

#### 4.5.1. Model for scattered light

To estimate the fundamental geometric parameters of the debris belts, we generated synthetic images of scattered (or polarized) light and compared them with the observed disk images. To create these synthetic images, we employed a 3D, rotationally symmetric model to describe the spatial distribution of grain number density, $n_{\rm gr}(r,h)$, within the disk. Following the approach of Augereau et al. (1999), we characterize this distribution as the product of a radial profile $R(r)$ and a Gaussian function $Z(r,h)$. The profile $R(r)$ defines the variation of grain number density in the disk midplane as a function of radial distance from the star $r$.

Meanwhile, the Gaussian function determines the vertical profile of $n_{gr}(r, h)$, shaping its distribution in the direction perpendicular to the disk midplane, as described by the height coordinate $h$:

$$n_{gr}(r, h) \sim R(r) \times Z(r, h) =$$
$$= \left( \left(\frac{r}{r_0}\right)^{-2\alpha_{in}} + \left(\frac{r}{r_0}\right)^{-2\alpha_{out}} \right)^{-1/2} \times \exp\left[ -\ln 2 \left(\frac{|h|}{H(r)}\right)^2 \right],$$

where $r_0$ is the reference radius of the debris belt, $\alpha_{\rm in} > 0$ and $\alpha_{\rm out} < 0$ are the exponents of the radial power laws for the dust distributions inside and outside of the belt, respectively. The scale height of the disk $H(r)$ is defined as a half-width at half-maximum (HWHM) of the Gaussian profile at radial distance $r$. In this work, the scale height is assumed to increase linearly with radial distance, following the relation $H(r) = H_0\,(r/r_0)^{\beta}$, where $H_0 = H(r_0)$ and the disk flaring index $\beta$ is fixed at $\beta = 1$.

In the model, each dust grain location within the disk is associated with a scattering angle $\theta$, defined as the angle between the incident stellar ray striking a dust particle and the observer's line of sight, where $\theta = 0$ corresponds to forward scattering. The scattering angle is a crucial model parameter, as it governs the fraction of incident stellar light scattered in a particular direction, described by the so-called scattering phase function (SPF). To generate synthetic disk images, we employed the Henyey-Greenstein (HG) function as the SPF (Henyey & Greenstein 1941), which provides a convenient parametrization of anisotropic scattering by dust grains:

$$SPF(\theta, g) = \frac{1 - g^2}{4\pi(1 + g^2 - 2\,g\,\cos\theta)^{3/2}},$$

where $g$ represents the HG scattering asymmetry parameter. This parameter quantifies the preferential direction of scattering, ranging from $g = -1$ (backward scattering) to $g = 1$ (forward scattering), with $g = 0$ corresponding to isotropic scattering.

Based on observations of scattered light from zodiacal and cometary dust in the Solar System (e.g., Leinert et al. 1976; Bertini et al. 2017), as well as laboratory experiments with dust analogs (e.g., Frattin et al. 2019; Muñoz et al. 2017), interplanetary dust grains are expected to preferentially scatter radiation in the forward direction. As a result, in disk images, the side of the disk that is closer to the observer appears brighter.

The forward-scattering behavior of dust particles can be described using a HG function with a positive asymmetry parameter ($g > 0$). However, real SPFs derived from observational data often exhibit a more complex structure: a pronounced diffraction peak at small scattering angles ($g >> 0$), a relatively flat mid-range ($g \sim 0$), and an enhanced backscattering component ($g < 0$). Consequently, a more accurate representation of an actual SPF would require a combination of three HG functions.

Nevertheless, we opted to model the data using a single HG function. This choice is justified by the fact that, in most cases, the inclination of resolved debris disks does not permit the measurement of scattering intensity across the full range of scattering angles (0° to 180°). Instead, we can only fit the portion of the SPF that is accessible in the data, which can be adequately approximated by a single HG function.

Another simplification adopted in our modeling approach is the assumption that the optical characteristics of dust grains, defined by their composition, shape, and size, are spatially uniform across the disk and can be represented by a single SPF. While this assumption simplifies the modeling process, it remains a coarse approximation, as the SPF is inherently dependent on dust properties that are expected to vary with radial distance. For instance, at the radial location of the peak grain number density, $R_{\rm belt}$, debris spanning a wide range of sizes is typically present, from submicron grains to kilometer-sized planetesimals (e.g., Wyatt 2008). In contrast, the outer disk is expected to form a halo of small grains that are collisionally produced within the main ring and subsequently placed on high-eccentricity orbits by stellar radiation pressure. This halo is anticipated to exhibit strong size segregation (Thebault et al. 2014), with the dominant grain size decreasing with increasing radial distance.

Depending on the quality of the available data, we fitted either an image of total intensity or polarized intensity. To generate a polarized intensity image ($Q_\phi$ image), we employed a polarized scattering phase function (pSPF) given by a functional form (e.g., Engler et al. 2017)

$$pSPF(\theta, g) = p_{\rm max} \frac{1 - \cos^2\theta}{1 + \cos^2\theta} \frac{1 - g^2}{4\pi(1 + g^2 - 2\,g\,\cos\theta)^{3/2}}, \quad (2)$$

where $p_{\rm max}$ is the maximum polarization fraction of dust particles.

In this equation, we employed a polarization fraction function $p(\theta)$ characteristic of Rayleigh scattering, in which the maximum polarization fraction is attained at a scattering angle of $\theta = 90°$ (Bohren & Huffman 1983):

$$p(\theta) = \frac{pSPF(\theta)}{SPF(\theta)} = p_{\rm max} \frac{1 - \cos^2\theta}{1 + \cos^2\theta}. \quad (3)$$

Rayleigh scattering describes the scattering of light by particles whose sizes are at least an order of magnitude smaller than the wavelength of the incident radiation. In near-IR observations of debris disks, this condition is generally satisfied in the outer disk regions, where the abundance of larger grains declines and the contribution of small particles to the scattered light becomes increasingly significant. The polarization fraction function of this small-particle population as well as that of micron-sized grains, typically traced in near-IR scattered-light images, can be reasonably approximated by the Rayleigh scattering function $p(\theta)$, as defined in Eq. 3 (see Appendix C).

For each disk specified in Table E.1, we generated a large set of models using a grid of fitting parameters. The parameter ranges were individually defined based on the findings of previous studies on these targets. We deliberately chose to use a model grid approach rather than a Markov Chain Monte Carlo (MCMC) algorithm, which, although widely employed in disk modeling, often yields unrealistically small uncertainties on the best-fitting parameters. Mazoyer et al. (2020) showed indeed that MCMC often leads to underestimated uncertainties, an effect probably due to the non-Gaussian statistics of the residual noise in coronagraphic images (Pairet et al. 2019).

To identify a family of models that adequately reproduce the observed disk images, we therefore adopted a more conservative threshold for the $\chi^2$ value than the one derived from the probability distribution of parameters, which assumes normally distributed errors:

$$\chi^2 < \chi^2_{\rm min} + \Delta\chi^2,$$

where $\chi^2_{\rm min}$ is the minimum chi-square value obtained from the fits, and $\Delta\chi^2 = \sqrt{2\nu}$ with $\nu$ denoting the number of degrees of freedom[8] (Thalmann et al. 2013). The mean values and standard deviations of the parameter distributions from the family of well-fitting models are adopted as the best-fitting parameters and their corresponding uncertainties, as reported in Table E.1. Note that these estimates can be affected by the parameter step size, as it determines the sample size and can influence the precision of the evaluated sample mean and standard deviation[9]. Also, the best-fitting parameters do not correspond to the single model with the minimum $\chi^2$, but rather reflect the statistical properties of the family of acceptable models.

[8] The degrees of freedom are defined as the difference between the number of data points and the number of free parameters.

[9] The sample size directly influences the standard error of a sampling distribution, such as that of the sample mean: as the sample size increases, the standard error decreases.

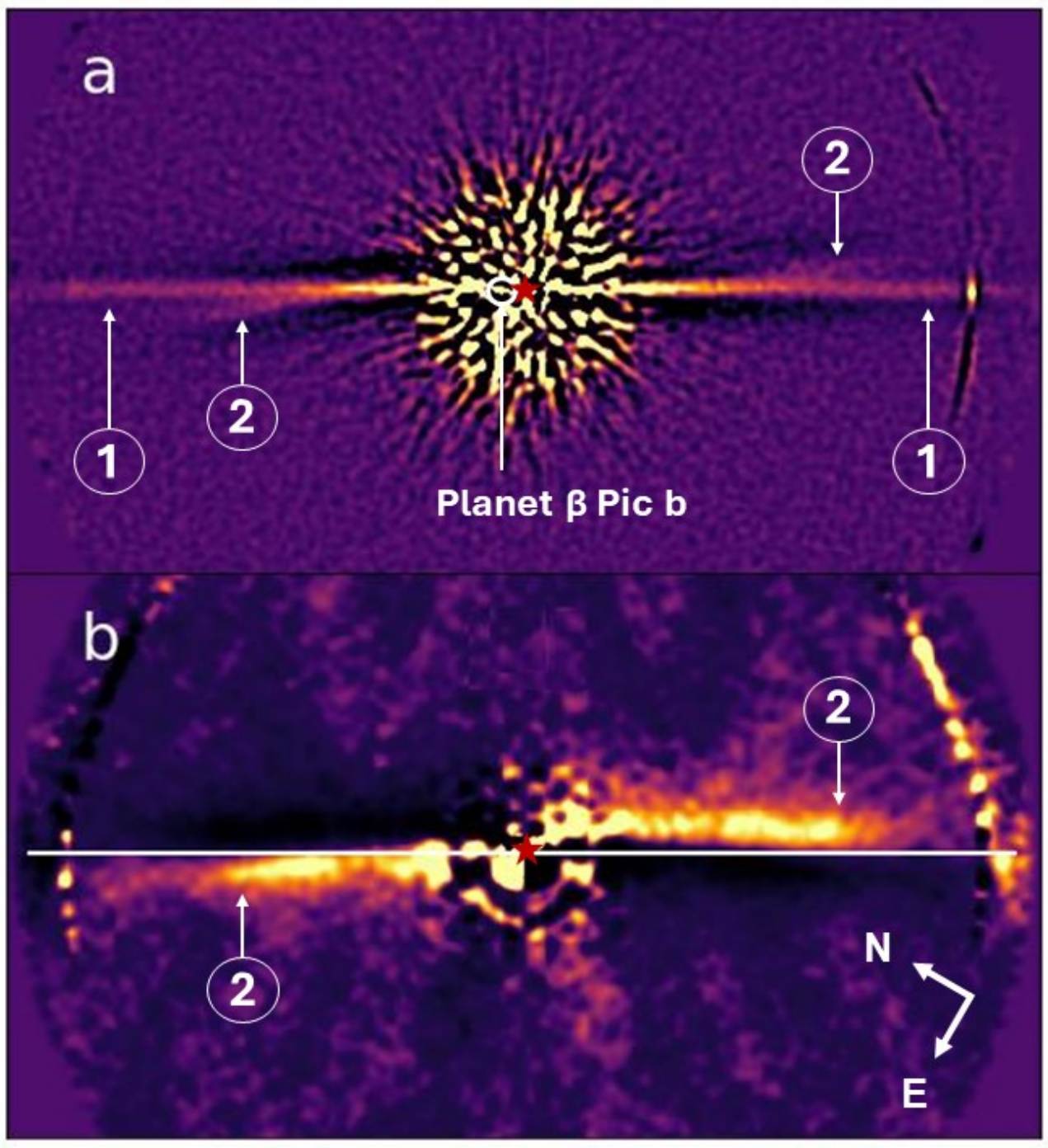

Fig. 10: H-band (IRDIS) images of HD 39060 debris belts: "1" indicates the outer belt and "2" the inner belt. The images are binned by 2×2 pixels. The position of star is marked by a red asterisk. The images are de-rotated by 60° to place the midplane of the outer belt in horizontal position. The field of view (FOV) of each displayed image is 6.27″ × 3.1″. *Panel a:* PCA data reduction of total intensity data. *Panel b:* Image showing the polarized flux from the inner belt. The image is obtained by subtraction of left (right) half of the $Q_\phi$ image from its right (left) half. The white solid line shows the position of outer belt which is invisible in this image. The length of this line is equal to 3″ or 118 au.

HD 39060 For the HD 39060 ($\beta$ Pic) debris disk, the model consists of outer and inner planetesimal belts and has, therefore, a double number of parameters. We chose a double-belt model for this particular target because IRDIS images in both total and polarized intensity reveal an inner disk located within the main outer belt (Fig. 10). While the inner belt shares a similar inclination with the outer belt, it has a slightly different PA. Consequently, its extensions become visible, producing the characteristic "butterfly pattern" in the scattered light distribution observed in HD 39060 disk images (e.g., Golimowski et al. 2006; Ahmic et al. 2009).

In the polarized intensity image of HD 39060 ($Q_\phi$ image), the inner disk becomes more distinct when the polarized flux from the outer belt is removed. To achieve this, we rotated the $Q_\phi$ image (Fig. 10b) counterclockwise by 60°, aligning the major axis of the outer belt horizontally. We then subtracted the left half of the $Q_\phi$ image from the right half and vice versa. The resulting image, shown in Fig. 10b, reveals the near side of the inner disk, which becomes visible in the lower left and upper right quadrants. In this image, the polarized flux from the outer belt is largely eliminated due to its symmetrical distribution with respect to the vertical axis. The polarized flux from the inner belt is partially reduced, particularly near the image center and in the upper left quadrant, due to the asymmetrical distribution of flux from the inner belt relative to the vertical axis of the image.

HD 39060 is the only target for which we applied a double-belt model. Other targets, such as HD 15115 (Engler et al. 2019; MacGregor et al. 2019), HD 120326 and HD 129590, also exhibit indications of a second planetesimal belt, though it is only marginally resolved (Sect. 4.4). The quality and spatial resolution of the available data do not allow for a reliable fit using a double-belt model to obtain robust constraints on the parameters of the secondary component. We modeled these systems using a single-belt approach despite their multiple-belt structure. In such cases, the fitted parameter values may be influenced by the presence of the second component, particularly affecting the derived belt radius or scale height, as discussed in the next section.

#### 4.5.2. Discussion of modeling results

*Comparison between measured and modeled radial distances of the planetesimal belts*

The reference radius $r_0$, in combination with the model parameters $\alpha_{\rm in}$ and $\alpha_{\rm out}$ obtained from disk image modeling, determines the modeled radial position of the debris belt $R_{\rm belt}^{\rm mod}$:

$$R_{\rm belt}^{\rm mod} = r_0 \left( -\frac{\alpha_{in}}{\alpha_{out}} \right)^{1/(2\,\alpha_{in} - 2\,\alpha_{out})} . \quad (4)$$

The modeled radius $R_{\rm belt}^{\rm mod}$ defines the location of the peak grain volume density in the radial profile of the disk midplane and determines the region where collisions between larger debris fragments or planetesimals generate dust particles. In the ADI-processed scattered-light images, the radial position of the SB peak measured along the disk's major axis may slightly deviate from the actual location of the planetesimal belt. This discrepancy arises from a combination of factors, including stellar illumination, spatial resolution of instruments, geometrical projection effects and the asymmetry of the SPF. For pole-on disks, for instance, the observed SB peak of the radial profile in the $r^2$-scaled images is expected to be at a radial position of the modeled peak surface density of grains, $R_{\max(\sigma)}^{\rm mod}$, which can be evaluated through the integration over the whole disk height in the vertical direction (Augereau et al. 1999):

$$R_{\max(\sigma)}^{\rm mod} = r_0 \left( -\frac{\alpha_{in} + \beta}{\alpha_{out} + \beta} \right)^{1/(2\,\alpha_{in} - 2\,\alpha_{out})} ,$$

where $\beta = 1$ is adopted in our model (see Sect. 4.5.1). With this value, the difference between the derived $R_{\rm belt}^{\rm mod}$ and $R_{\max(\sigma)}^{\rm mod}$ remains within 4% for all modeled debris disks listed in Table E.1 except for the HD 61005 disk, where the discrepancy reaches 10%.

In Fig. 11a, we compare the belt radii derived from disk image modeling (Col. 2 of Table E.1) with the radial locations of the SB peaks measured directly in the $r^2$-scaled images (Col. 2 of Table 2). The black line in Fig. 11a represents an empirical fit and coincides with the 1:1 relation, indicating that the modeled belt radii are in good agreement with the directly measured values. For most targets, the difference between $R_{\rm belt}^{\rm mes}$ and $R_{\rm belt}^{\rm mod}$ is within 10%, and within 12% between $R_{\rm belt}^{\rm mes}$ and $R_{\max(\sigma)}^{\rm mod}$. There are two noticeable exceptions: HD 120326 and HD 129590. Both of these targets are likely to host at least two distinct planetesimal belts (Sect. 4.4), which may explain the observed deviations.

*Asymmetry parameter*

The derived values for the HG asymmetry parameter $g$ range from 0.79 (for HD 106906 disk) to 0.5 (for HD 172555 disk). We observe a slight trend toward higher asymmetry parameters when modeling disks with higher inclinations (Fig. 11b). This trend can be attributed to the broader range of scattering angles accessible in highly inclined disks, which allows for a more pronounced forward-scattering peak to be observed. Consequently, this suggests that low-inclination disks may intrinsically exhibit higher asymmetry parameters, but their forward-scattering component remains less apparent due to the limited range of observable scattering angles. If these disks were viewed at higher inclinations, their asymmetry parameters might appear larger.

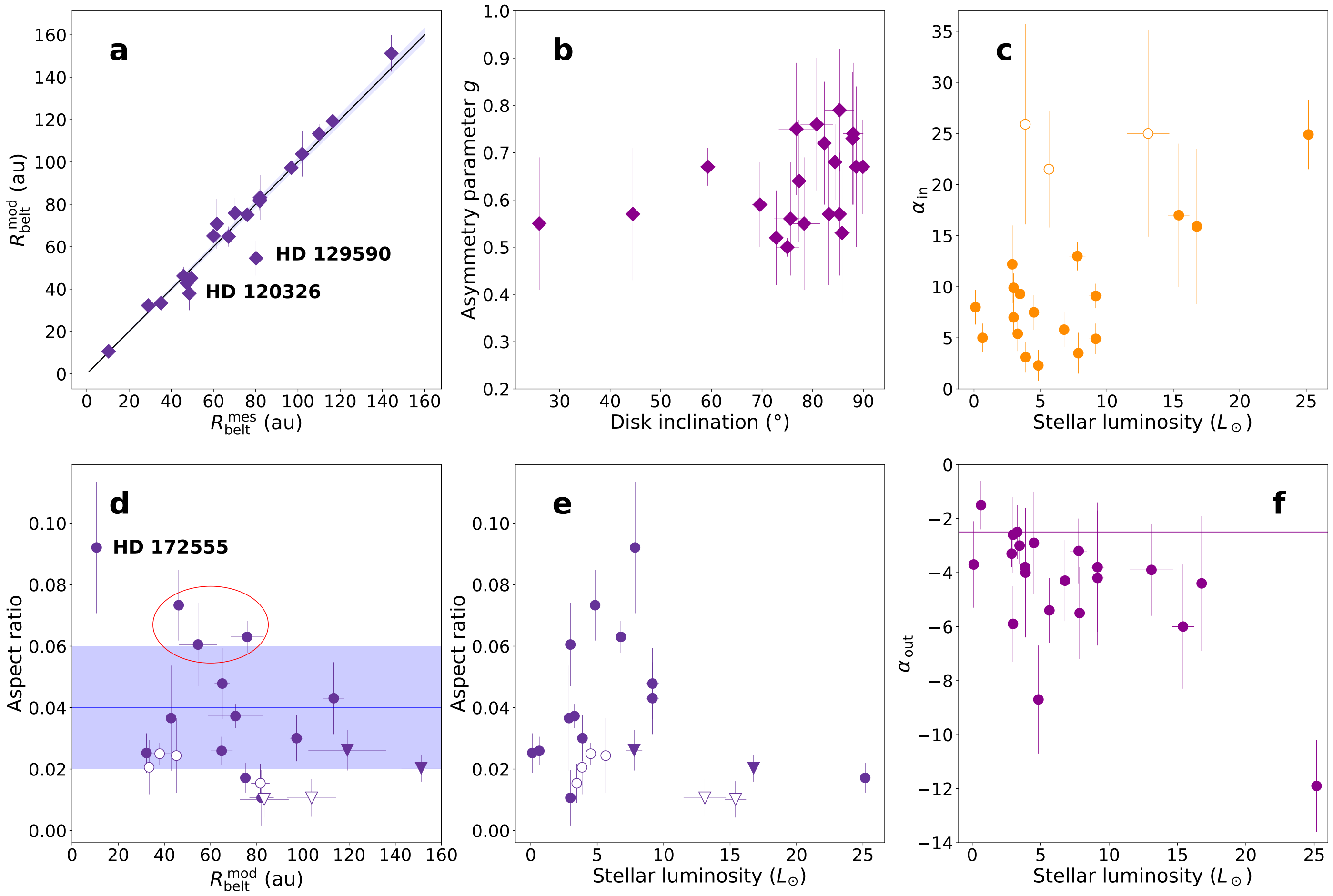

Fig. 11: Best-fit parameters for the debris disks listed in Table E.1. *Panel a:* Comparison of the measured ($R_{belt}^{mes}$) and modeled ($R_{belt}^{mod}$) radial distances of the planetesimal belts. The empirical fit is shown by the black solid line, which coincides with the 1:1 relation. The blue-shaded area represents the uncertainty on the slope of the fit. *Panel b:* HG asymmetry parameter $g$ versus disk inclination. *Panel c:* Exponent of the inner radial power law $\alpha_{in}$ as a function of stellar luminosity. Open circles show the exponents obtained by modeling the total intensity images of the HD 114082, HD 117214 and HD 131488 disks. *Panel d:* Disk aspect ratio versus radius of the planetesimal belt. Debris disks with unresolved FWHM are marked with open symbols. Gas-rich systems are indicated by triangles. The blue solid line represents the theoretical scale height value of 0.04 for a collisionally excited debris disk, while the blue-shaded area indicates its associated uncertainty (Thebault 2009). The red ellipse encloses targets HD 106906, HD 115600 and HD 129590. *Panel e:* Disk aspect ratio versus stellar luminosity. Marker symbols are the same as in *panel d*. *Panel f:* Exponent of the outer radial power law $\alpha_{out}$ as a function of stellar luminosity. The solid horizontal line indicates the value $\alpha_{out} = -2.5$, as theoretically predicted for the outer regions of debris disks (Strubbe & Chiang 2006; Thebault et al. 2023).

### *Exponents of radial power law for the radial distribution of grain number density*

The best-fit values for the exponent of the radial power law $\alpha_{in}$ span a relatively wide range, from 2.3 for the HD 115600 disk to 25.9 for the HD 114082 disk. However, it is important to note that three of the four highest exponents (> 20), shown as open circles in Fig. 11c, are not well constrained. Their distributions lack a clear peak indicating an optimal fit within the tested parameter space. Instead, these values consistently trend toward the upper limit of the parameter range, even when the maximum tested value is set as high as $\alpha_{in} = 80$. Notably, these unconstrained values were obtained from the fitting of total intensity images of the HD 114082, HD 117214 and HD 131488 disks, which may suggest a limitation in accurately determining this parameter using ADI forward-modeling, particularly for disks with high inclinations and small angular sizes.

Alternatively, these disks may indeed possess extremely sharp inner edges, a feature often interpreted as evidence of unseen planets clearing the space at the edges of the planetesimal belts. Milli et al. (2017b) tested methods to constrain large values of the $\alpha_{in}$ and $\alpha_{out}$ parameters and concluded that the modeling of these parameters is limited by the intrinsic steepness of the PSF. Specifically, belt edges that are steeper than the PSF wings are inherently blurred by convolution with the PSF, making it impossible to constrain $\alpha_{in}$ and $\alpha_{out}$ values steeper than approximately 30 in the IRDIS H band.

If the aforementioned targets are excluded, a trend emerges in which the $\alpha_{in}$ parameter increases with stellar luminosity (Fig. 11c). Regarding $\alpha_{out}$ (Fig. 11f), the derived values for most systems are consistent with the archival data presented in Ta-ble 1 of Thebault et al. (2023). As discussed in that study, the expected $\alpha_{out}$ value for a typical belt-like system with an outer halo composed of small grains placed on high-eccentricity orbits by radiation pressure is approx. −2.5. When considering error bars, we find that roughly half of the systems in Table E.1 have $\alpha_{out}$ estimates that are compatible with this reference value. For the remaining systems, we obtain steeper outer profiles, with $\alpha_{out}$ reaching values as low as -8 or even -12. In these cases, additional dynamical processes, such as perturbations by external planets or stellar companions, are likely responsible for clearing out the outer disk regions (Thebault et al. 2023).

*Vertical disk structure*

As discussed in Sect. 4.5.1, we adopted a Gaussian function to describe the vertical profile of each disk, accounting for its nonzero vertical width. In this study, we define the HWHM of the Gaussian profile as the scale height $H(r)$ of the disk at a given radial distance $r$. The ratio of the scale height to the radial distance, $A_{disk} = H(r)/r$, is referred to as the disk aspect ratio, also known as the disk half-opening angle. Since, in our model, the scale height $H(r)$ varies linearly with distance $r$, the aspect ratio remains constant throughout the disk and can be expressed as $A_{disk} = H_0/r_0$, where $H_0$ is the scale height at the reference radius $r_0$.

The scale height and aspect ratio of a disk serve as key indicators of the dynamical excitation within a debris system. Thebault (2009) numerically estimated a minimum aspect ratio of $0.04 \pm 0.02$ for a collisionally evolving disk observed at visible to mid-IR wavelengths. If the disk experiences additional dynamical perturbations from massive bodies such as giant exoplanets, its aspect ratio is expected to exceed 0.06.

In Figures 11d and 11e, we show the results of our analysis of the aspect ratios obtained for the modeled disks. We consider the disk to be resolved in direction perpendicular to the disk midplane, if the fitted FWHM of the vertical profile is larger than the FWHM of the stellar PSF. For four modeled disks (HD 114082, HD 117214, HD 120326, HD 131488) which are shown in Fig. 11d with open markers this condition is not fulfilled. We note that the parameters of these four disks are derived from the fitting of the total intensity images applying ADI forward-modeling approach. Therefore the value of the scale height might be underestimated. The other two disks (HD 109573, HD 181327) are only marginally resolved in vertical direction. Both disks have an inclination lower than 80°. In particular the HD 181327 disk has an inclination of ∼30° which is the lowest one we modeled. This might reflect the challenge of modeling the disk scale height when using images obtained with the ADI technique or images of low inclined disks.

Most of the modeled disks have a scale height between 0.02 and 0.06 (Fig. 11d). This relatively small vertical extent can be explained by the combined effects of radiation pressure acting on small dust particles and their mutual collisions, which naturally regulate the disk's thickness (Thebault 2009). In contrast, the HD 172555 disk exhibits a significant larger scale height of 0.1, which may indicate the influence of additional massive perturbers, although this result could be affected by the lower S/N of the image. This system is particularly intriguing, as it contains detected gas and a notable abundance of hot, small dust grains, features that may be the aftermath of a recent, violent collision between planetary bodies (Lisse et al. 2017; Riviere-Marichalar et al. 2012; Kiefer et al. 2014; Engler et al. 2018; Schneiderman et al. 2021; Samland et al. 2025).

Three targets (HD 106906, HD 115600, and HD 129590) with modeled scale heights ranging between 0.06 and 0.07 are enclosed by the red ellipse in Fig. 11d. All three disks are highly inclined and are suspected to host at least two cold belts. If so, one possible explanation for the relatively large scale heights inferred from the models is that the inner belts remain unresolved but lie in close projected proximity to the outer belts along the minor axis in the scattered-light images. This configuration could mimic the appearance of a geometrically thicker planetesimal belt. An alternative explanation, at least for HD 106906, is dynamical excitation of the disk by a massive substellar companion known to be present in the system (see Sect. 7). It is also worth noting that the HD 129590 disk contains small amounts of CO gas ($M_{CO} = 10^{-5} - 10^{-4} M_{\oplus}$; Kral et al. 2020). In contrast, no gas has been detected so far in HD 106906 and HD 115600, leaving the influence of gas on the observed disk scale heights in these systems uncertain.

However, there are four gas-rich systems with the estimated CO masses exceeding $10^{-2} M_{\oplus}$, namely HD 9672, HD 32297, HD 121617 (Moór et al. 2019), and HD 131488 (Moór et al. 2017; Pawellek et al. 2024). These systems are marked by triangles in Figs. 11d and 11e, and their aspect ratios span a range between 0.026 and 0.011. In these disks, the CO emission has been observed to be axisymmetric and co-located with the millimeter-sized dust particles (Hughes et al. 2017; Moór et al. 2017; MacGregor et al. 2018). The relatively low aspect ratios derived for these gas-rich systems may suggest that micron-sized dust grains are dynamically coupled to the gas, leading to their settling toward the disk midplane.

There appears to be a trend of decreasing disk aspect ratio with increasing stellar luminosity, as observed in Fig. 11e. This trend could also account for the low aspect ratios of gas-rich disks, as three of these systems are associated with high-luminosity A-type stars ($L_\star > 13\,L_\odot$), and, especially as we derive an aspect ratio of $A_{disk} = 0.017$ for the debris belt around the A0V star HD 109573 ($L_\star = 25.2\,L_\odot$), which is not known to be gas-rich, further supporting this possible correlation. However, confirming this trend would require a significantly larger sample of measured aspect ratios.

## 5. SED modeling

We applied SED modeling to characterize the thermal emission of debris disks in our sample, using two different approaches to fit the photometric data. The modified BB (MBB) approach (Backman & Paresce 1993) provides a uniform fitting method for all targets, making it particularly suitable for a statistical analysis of the sample. In contrast, the particle size distribution (SD) approach (e.g., Müller et al. 2010; Pawellek et al. 2021) allows for a more detailed characterization of dust grain properties, including the dominant grain size, the SD steepness, and the bulk optical properties of the dust. The SD model is only applicable to spatially resolved debris disks, as determining the disk radius is necessary to break the degeneracy between the location of dust particles and their sizes (Pawellek et al. 2014).

### 5.1. Modeling procedure

We utilized photometric data for our sample from published catalogs, such as 2MASS (Cutri et al. 2003), the WISE All-Sky Release Catalog (Wright et al. 2010), the AKARI All-Sky Catalog (Ishihara et al. 2010), the Spitzer Heritage Archive (Carpenter et al. 2008; Lebouteiller et al. 2011; Chen et al. 2014; Sierchio et al. 2014), and the Herschel Point Source Catalog (Marton et al. 2015; Marshall et al. 2021). In addition, we used data published in the literature (e.g., Chen et al. 2014; Matrà et al. 2017; Marshall et al. 2021). These data allow us to analyze the SEDs and assess the presence of IR excess emission beyond what is expected from the stellar photosphere.

To find excess emission, we fitted an SED model consisting of a star and a disk. Firstly, we fitted PHOENIX stellar photosphere models (Brott & Hauschildt 2005) for each target using the stellar luminosity and the stellar temperature as model parameters, and photometric data in the VIS/NIR where the stellar emission is supposed to dominate the SED and the disk emission is negligible. The resulting stellar luminosities and temperatures are listed in Table E.2. Secondly, knowing the stellar contribution to the mid- and far-IR data, we derived the excess emission in the appropriate wavelength bands between ~5 and ~1000 $\mu$m taking into account the uncertainties of the photometry and the photospheric model.

We followed the four criteria given in Ballering et al. (2013) and Pawellek et al. (2014) to check for the presence of a warm disk component in addition to the cold Kuiper belt analog. Firstly, the number of photometric data points must be large enough so that the data are not over-fitted. If that was the case in a second step we considered a warm component to be present, if there is a significant excess ($\geq 3\sigma$) in either the WISE/22 or MIPS/24 in excess of that which could originate in a single ring fitted to longer wavelength data. Thirdly, the fit of the two-component SED has to be much better than the one-component fit according to the Bayesian information criterion (BIC)

$$\mathrm{BIC} = \chi^2 + J \log_e (N_{\mathrm{data}}), \tag{5}$$

where $J$ represents the number of free parameters and $N_{\mathrm{data}}$ the number of data points. We used the classification given in Kass & Raftery (1995) to infer whether a one- or a two-component model is more likely (Pawellek et al. 2021). As a fourth criterion, we required the inferred ring containing the warm dust to be located outside the sublimation radius (assuming 1300 K as the sublimation temperature for astrosilicate). If all four criteria were met, we assumed the SED to consist of a two-component model.

The uncertainties of the fit parameters were inferred in the following way. We started at the position of the minimum $\chi^2$ in parameter space, e.g., from the best-fitting fractional luminosity, $f_{\mathrm{disk}}$, and BB radius, $R_{\mathrm{MBB}}$, in case of an MBB model (Sect. 5.2). A set of new parameter values was randomly generated from which we calculated the SED. This led to a new $\chi^2$ value that was compared to the previous minimum value. The $\chi^2$ parameter estimates how likely the set of parameter values fits the SED. If the probability was larger than a certain threshold value, the set was saved. In the end, it was counted how often the code reached a certain set of $f_{\mathrm{disk}}$ and $R_{\mathrm{MBB}}$. The closer the parameters get to the best-fitting values, the higher the probability. The resulting distribution in parameter space represents an estimate for the probability distribution of the parameters and thus, allows us to calculate the confidence levels for the parameters assuming that the values follow a normal distribution in parameter space (simulated annealing; e.g., Pawellek 2017).

### 5.2. Modified blackbody model

Every disk in the sample was fitted with a MBB model for which the thermal emission of the dust is described as

$$F_\lambda^{th} \sim B_\lambda(\lambda, T_{\mathrm{MBB}}) \left[ u(\lambda_0 - \lambda) + u(\lambda - \lambda_0) \left( \frac{\lambda}{\lambda_0} \right)^{-\beta_{\mathrm{op}}} \right], \tag{6}$$

where $F_\lambda^{th}$ is the spectral flux density of thermal emission, $B_\lambda$ is the Planck function and $u$ the Heaviside step function. The parameter $\lambda_0$ represents the characteristic wavelength, while $\beta_{\mathrm{op}}$ is the spectral opacity index.

From this model we derived the dust temperature, $T_{\mathrm{MBB}}$, and the resulting BB radius of the disk, $R_{\mathrm{MBB}}$, as well as the fractional luminosity, $f_{\mathrm{disk}}$. Here, $R_{\mathrm{MBB}}$ is the distance from the star that the temperature implies if the dust acted like BB in equilibrium with the stellar radiation. If a warm component was present (Sect. 5.1), we modeled it also with a MBB model, but assumed that the $\beta_{\mathrm{op}}$ parameter of the warm and cold component are similar to keep the number of free parameters as low as possible, and to avoid degeneracies between the component parameters.

For a single ring model the free modeling parameters are the fractional luminosity $f_{\mathrm{disk}}$, the BB radius $R_{\mathrm{MBB}}$, the characteristic wavelength $\lambda_0$, and the opacity index $\beta_{\mathrm{op}}$. Hence, at least five data points are needed to not over-fit the photometric data. In case of a two-component model we added the BB temperature, the fractional luminosity, and the characteristic wavelength of the inner ring as free parameters which required at least eight data points.

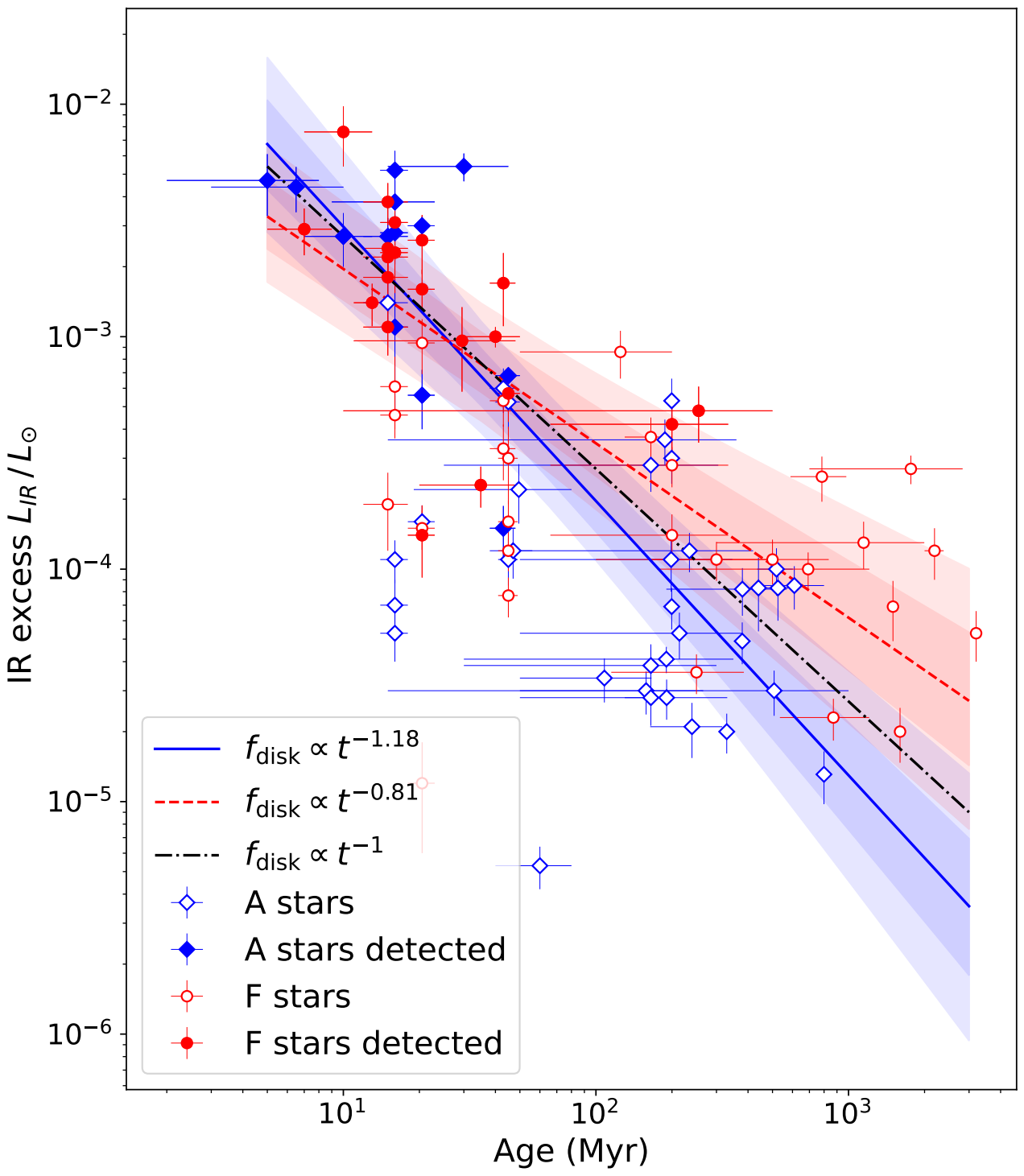

Fig. 12: Evolution of the IR excesses ($f_{\mathrm{disk}}$) for A- and F-type stars in our sample. Targets with debris disks detected using SPHERE instruments are shown as filled circles. The blue solid line represents a fit to the A-type star subsample, the red dashed line to the F-type star subsample, and the black dash-dotted line indicates the expected decline of IR excess for debris disks evolving in a steady-state collisional regime. The blue- and red-shaded regions indicate the 68% and 95% confidence bands for the fits to the A-type and F-type star subsamples, respectively.

We considered the SED model to be unreliable, i.e. it cannot be fitted to the data, if one of the following conditions is not fulfilled: (1) The number of available photometric data points is higher than a minimum number of data points needed to fit an SED, the number of needed data points varies with the modeling approach (MBB or SD); (2) There is a clear detection of the IR excess emission compared to the stellar photosphere, i.e. the total flux density exceeds the stellar photosphere by at least $3\sigma$; (3) The photometric data cover the peak of the thermal emission to constrain the model.

Following these criteria, the SEDs are counted as unreliable in case for the following targets: GSC 7396-0759, HIP 63942, HD 35114, HD 36968, HD 53842, HD 69830, HD 122705, HD 135379, HD 141011, HD 141943, HD 181869, and HD 274255.

Two targets (HD 17390 and TWA 25) fulfill points (2) and (3), but not point (1), as they have only three data points. For these targets we assumed a pure BB model where we only have two free parameters and thus, only needed three data points to achieve a proper fit.

### 5.3. Results of MBB modeling

The results of the one-component MBB fitting are specified in Table E.3. Based on the criteria outlined in Sect. 5.1, fourteen SEDs were fitted using a two-component model, incorporating both warm and cold dust components. The best-fit parameters from this modeling are provided in Table E.4. Figure D.1 (top row) shows examples of SEDs fitted with MBB models consisting of one or two components. Using the results of these fits, we investigated the evolution of the disk IR excess, the correlation between the estimated dust mass in the belt and its radial distance from the star for A-type and F-type stars in our sample, as well as the correlation between dust mass and host star mass for all targets.

The disk IR excess is one of the key factors influencing disk detectability (Sect. 6). Younger debris disks exhibit higher fractional luminosities, as their luminosity scales with dust mass, which is at its peak in early evolutionary stages. Various dust removal mechanisms, including the blowout of the smallest dust particles due to radiation pressure, the Poynting-Robertson drag, and photoevaporation, play a crucial role in shaping the evolution of disk mass and luminosity (e.g., Dominik & Decin 2003). Previous studies (e.g., Kalas et al. 2000; Rieke et al. 2005) have shown that the fractional luminosities of debris disks, $f_{\rm disk}$, and consequently their dust content, follow a power-law dependence on time

$$f_{\rm disk} \propto t_{\rm age}^{\alpha_t}.$$

Figure 12 shows the evolution of disk fractional luminosity for the debris disks around A- and F-type stars in our sample. By fitting the data points in the [$f_{\rm disk}$, $t_{\rm age}$] parameter space, we derived the power-law indices $\alpha_t = -1.18 \pm 0.14$ for A-type stars and $\alpha_t = -0.81 \pm 0.12$ for F-type stars. These exponents are in general agreement, though formally distinct within $1\sigma$ uncertainties with the steady-state collisional evolution theory of planetesimal belts, which predicts a disk luminosity decline proportional to $t_{\rm age}^{-1}$ (Dominik & Decin 2003). We note that the $\alpha_t = -1$ slope (black dash-dotted line in Fig. 12) is only valid for collisional systems in steady-state, where the disk age exceeds the collisional lifetimes of the largest planetesimals it contains. As pointed out by Löhne et al. (2008), this condition might be met only in very old systems (> 1 Gyr), or in very massive disks and/or disks at short radial distances from their host stars. Given that the median age of our sample is ~100 Myr, and the median belt distances are around 70 − 80 au, our results do not align with these expectations.

Our results show however that debris disks around A-type stars tend to decline more rapidly in brightness than those around F-type stars. This behavior can be interpreted in the context of differences in initial disk masses and their influence on collisional evolution timescales. Debris disks with higher masses produce dust more efficiently due to the increased frequency of planetesimal collisions (e.g., Wyatt 2008). However, these systems also deplete more rapidly, since both large planetesimals and the dust generated by their collisions are removed more quickly, either through further collisional grinding or via radiative forces acting on small grains (Löhne et al. 2008; Krivov 2010). A-type stars, being more massive and luminous than F-type stars, are expected to host more massive planetesimal belts, which evolve faster due to the enhanced dynamical excitation and stronger radiation pressure that efficiently clears small grains from the system. In contrast, the more gradual decay of IR excess observed in F-type systems suggests slower collisional processing, consistent with initially less massive disks and weaker dynamical stirring.

To further investigate this trend and quantify the effect of stellar mass on debris disk evolution, we estimated dust masses for our targets using the MBB best-fit parameters and equation

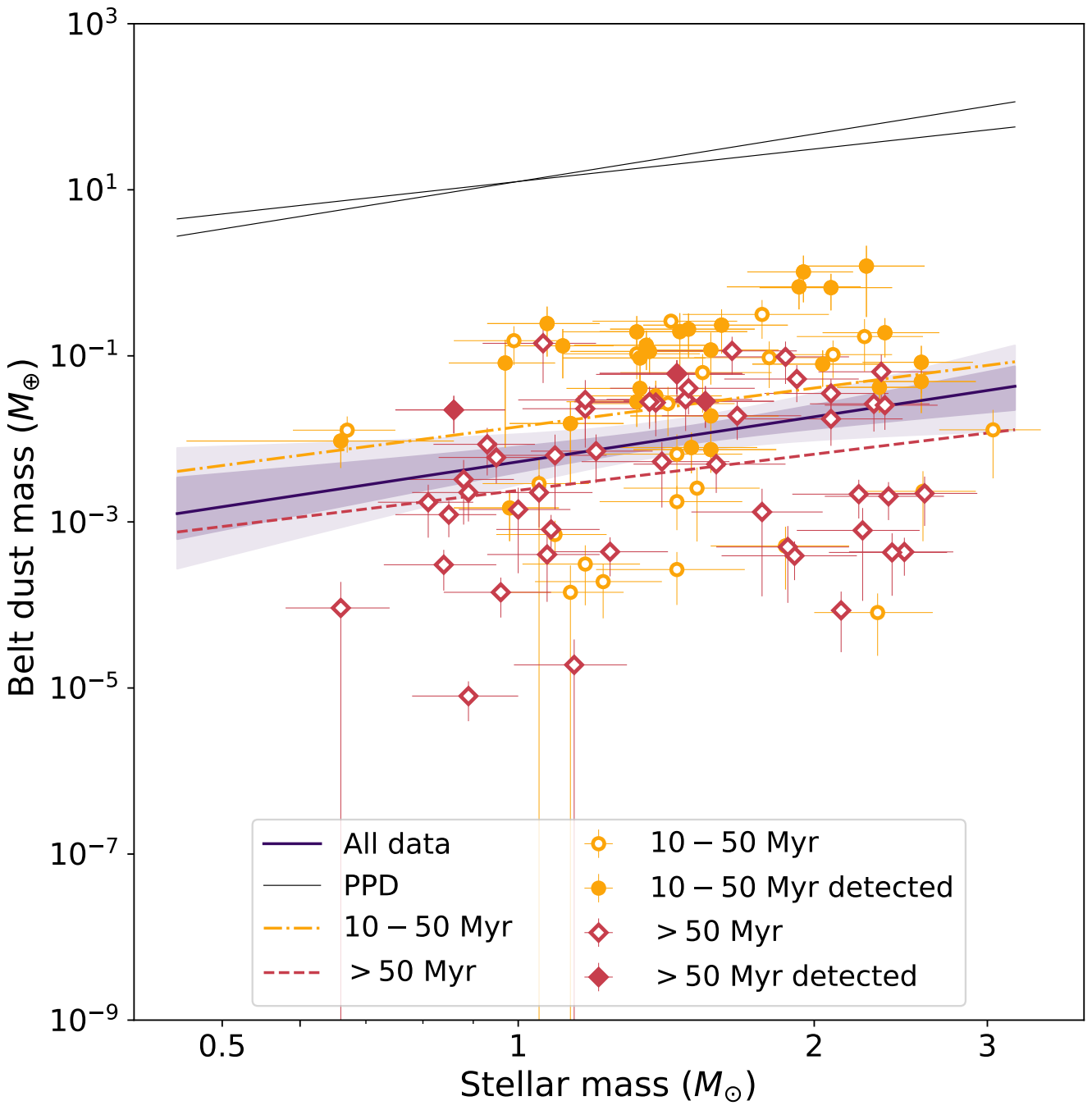

Fig. 13: Relation between dust mass ($M_{\rm dust}$) and stellar mass for all debris disks in our sample (violet solid line), a subsample of disks around stars aged between 10 and 50 Myr (orange dash-dotted line), and a subsample of disks older than 50 Myr (red dashed line). The violet-shaded regions indicate the 68% and 95% confidence bands for the fit to the entire sample. For comparison, two fits of the same relation derived for PPD in the 2 − 3 Myr old Chamaeleon I star-forming region by Pascucci et al. (2016) are shown as black solid lines. The differing slopes of the PPD fits reflect model-dependent uncertainties in the inferred scaling relations.

Table 3: Best-fit parameters for the relation between dust and stellar mass.

| Sample | $t_{\rm age}$ (Myr) | $\alpha_{\rm mass}$ | $\beta_{\rm mass}$ | Ref. |
|---|---|---|---|---|
| PPD, model 1[(a)] | 2 − 3 | 1.3 ± 0.2 | 1.1 ± 0.1 | (1) |
| PPD, model 2[(a)] | 2 − 3 | 1.9 ± 0.2 | 1.1 ± 0.1 | (1) |
| Debris disks | 10 − 50 | 1.6 ± 1.0 | −1.9 ± 0.2 | (2) |
| Debris disks | > 50 | 1.4 ± 0.9 | −2.6 ± 0.2 | (2) |
| Debris disks | > 10 | 1.7 ± 0.7 | −2.3 ± 0.2 | (2) |

**Notes.** [(a)] The PPD sample includes objects from Chamaeleon I star-forming region. Model 1 and model 2 correspond to two extremes of the possible relation between the average dust temperature and stellar luminosity.

**References.** (1) Pascucci et al. (2016); (2) this work.

for the disk dust mass following Wyatt (2008):

$$M_{\rm dust} = 12.6\, f_{\rm disk}\, R_{\rm MBB}^2 k_{850}^{-1} \left(\frac{850\,\mu m}{\lambda_0}\right)^{-\beta_{\rm op}}, \quad (7)$$

where $k_{850} = 45$ au$^2 M_\oplus^{-1}$, and $\lambda_0$ and $\beta_{\rm op}$ are the characteristic wavelength and spectral opacity index, respectively, obtained from the MBB fit (see Eq. 6).

The derived dust masses of the planetesimal belts, plotted as a function of stellar mass, are shown in Fig. 13. Since the major-ity of the debris disks detected in our sample are younger than 50 Myr (90% of all detections), we divided the sample into two groups based on stellar age: systems aged $10-50$ Myr and older systems. These are represented by orange dots and red diamonds in Fig. 13, respectively. The detected debris disks are indicated by filled markers.

To investigate the $M_{\rm dust} - M_\star$ relation and its potential evolution over time, we fitted a power-law function in log-log space $\log(M_{\rm dust}/M_\oplus) = \alpha_{\rm mass} \log(M_\star/M_\odot) + \beta_{\rm mass}$ to three datasets: disks in the $10-50$ Myr range, disks older than 50 Myr, and all disks older than 10 Myr. The best-fit parameters $\alpha_{\rm mass}$ and $\beta_{\rm mass}$ for the three subsamples are reported in Table 3 (rows 3-5) and the fits are visualized in Fig. 13, along with 68% and 95% confidence intervals for the full sample. In all cases, we find a steeper than a linear relation ($\alpha_{\rm mass} > 1$) between the estimated disk dust mass and the mass of the host star. Notably, the scaling relation for the younger group of debris disks (see row 3 of Table 3) is steeper than that of the older group (row 4 of Table 3). This decrease in steepness with age may indicate a more rapid dust mass depletion in initially more massive disks, a trend that is also apparent in Fig. 12.

It is well established that PPD masses correlate with stellar mass as well (e.g., Andrews et al. 2013; Barenfeld et al. 2016; Ansdell et al. 2016). For example, Pascucci et al. (2016) analyzed the scaling of disk masses with stellar mass in star-forming regions, using various assumptions for the dust temperature–luminosity relation, and similarly found super-linear trends. To place our findings in context, we included in Table 3 (rows $1-2$) the best-fit parameters reported by Pascucci et al. (2016) for the Chamaeleon I star-forming region and overplot the corresponding fits in Fig. 13 as black solid lines. As evident from this figure, the PPD relation is very similar to that we found for the debris disks and supports the idea that debris disks reflect the initial conditions set during the protoplanetary phase. This is particularly significant, as PPD masses represent the material reservoir available for planet formation. A super-linear scaling of disk mass with stellar mass therefore implies that the resulting planet populations, both in terms of typical masses and occurrence rates, are also expected to have a positive correlation with stellar mass.

Comparing the offset of the linear relations (Col. 4 in Table 3) between PPD and debris disk dust masses (Fig. 13) we find that the average dust mass decreases by $\sim 3$ dex within the first 50 Myr, and by 3.7 dex for the older stars in our sample. However, the dust masses exhibit a relatively large spread, which can be partially attributed to the variation in belt radii among the systems with the same mass of the host star, and scaling the disk mass with its radius.

A correlation between dust mass and radial location is consistent with previous studies suggesting that belts at larger distances from the star are likely to be more massive (e.g., Andrews et al. 2013; Matrà et al. 2025). This trend can be explained by the fact that a wider belt spans a larger volume, thereby allowing for a greater population of dust-producing planetesimals, assuming a roughly constant surface density or collisional activity per unit area. Moreover, the collisional timescales in outer regions of debris disks are longer due to lower orbital velocities and decreased dynamical stirring, allowing dust to persist for extended periods and accumulate to a higher levels (e.g., Wyatt 2008).

In Fig. 14, we show the derived belt dust masses as a function of the BB belt radius $R_{\rm MBB}$ for A-type stars (panel a) and F-type stars (panel b). The contour plots in this figure represent the bivariate probability density function (PDF) for subsamples of stars with estimated ages between 10 and 50 Myr (red contours) and stars older than 50 Myr (violet contours). A comparison of the PDF peak positions across different stellar age bins indicates that the dust mass declines by $1-1.5$ dex on average for older stars, particularly among A-type stars, in line with our results presented in Fig. 13.

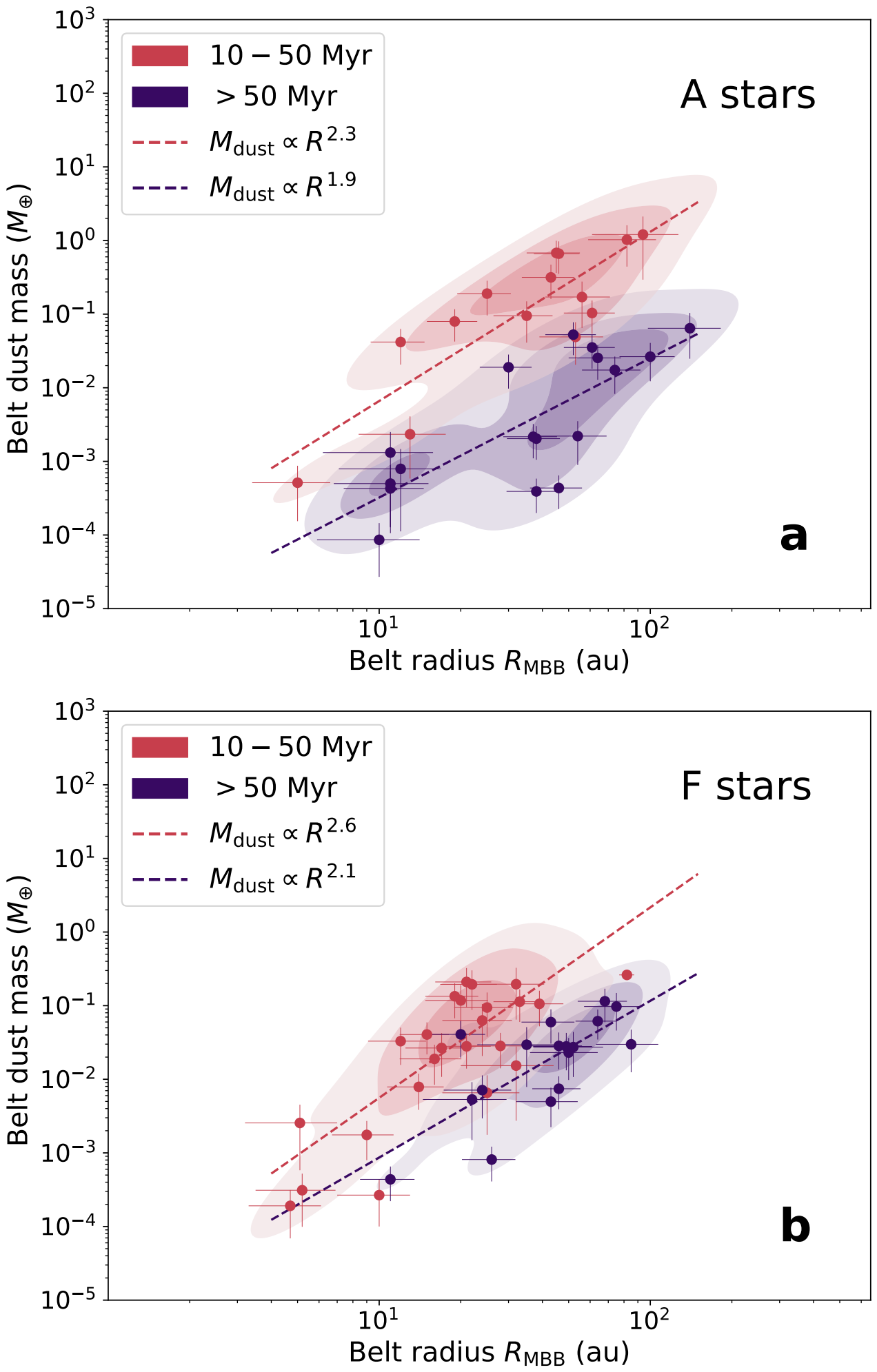

Fig. 14: Dust mass of debris belts derived using Eq. 7 versus BB belt radius $R_{\rm MBB}$ for A-type (*panel a*) and F-type stars (*panel b*). The red filled contours represent the probability density distribution of data consisting of stars with ages between 10 and 50 Myr (red circles), the violet contours of stars with ages above 50 Myr (violet circles). The contours contain 20%, 50% and 80% of the data points. The dotted lines show the power law fits $R_{\rm MBB}^{\alpha_{\rm R}}$ to these distributions.

We also observe a trend of increasing belt dust mass with growing radial distance from the star across all age bins. This trend follows the fitted radial power law $M_{\rm dust} \propto R_{\rm MBB}^{\alpha_{\rm R}}$, with $\alpha_{\rm R} = 2.3 \pm 0.4$ for A-type stars in the $10-50$ Myr age bin and $\alpha_{\rm R} = 1.9 \pm 0.4$ for A-type stars older than 50 Myr. Similarly, for F-type stars, we find $\alpha_{\rm R} = 2.6 \pm 0.4$ for the $10-50$ Myr bin and $\alpha_{\rm R} = 2.1 \pm 0.5$ for older stars. This result is expected, as it is consistent with the power-law form used in Eq. 7, within the given uncertainties. Therefore, we further investigate this relationship using belt dust masses derived from SD modeling (Sect. 5.5).

### 5.4. Size distribution model

In case of spatially resolved disks, we used the SONATA code (Pawellek et al. 2014; Pawellek & Krivov 2015) to model the SEDs with a dust SD. While for the MBB model we simply fitted a dust temperature and a fractional luminosity without consideration of dust properties, the SONATA code calculates the temperature and the thermal emission of dust particles at differ-ent distances to the star following

$$r(T_{\rm grain}) = \frac{R_\star}{2}\left[\frac{\int_0^\infty Q_\lambda^{\rm abs}(a)\,B_\lambda(T_\star)\,{\rm d}\lambda}{\int_0^\infty Q_\lambda^{\rm abs}(a)\,B_\lambda(T_{\rm grain}(r))\,{\rm d}\lambda}\right]^{1/2}, \qquad (8)$$

where $B_\lambda$ is the Planck function, $T_\star$ and $R_\star$ the stellar temperature and radius, and $T_{\rm grain}$ the grain temperature. The parameter $Q_\lambda^{\rm abs}(a)$ gives the absorption efficiency dependent on wavelength $\lambda$ and grain radius $a$. As seen in Eq. 8, this equation must be solved iteratively to determine the particle temperature as a function of distance from the star.

We assumed compact spherical grains and used Mie theory to compute the absorption efficiencies (Bohren & Huffman 1983). The dust composition was set to pure astronomical silicate (Draine 2003a), with a bulk density of $\varrho = 3.3$ g/cm$^3$. The SONATA code integrates the emission from particles over a range of sizes to generate the SED. As mentioned before, flux densities at wavelengths shorter than 5 $\mu$m were excluded from the dust disk fitting, as the stellar photosphere dominates the emission in this regime.

We applied a power law for the SD of the dust and assumed a Gaussian radial distribution for the spatially resolved ring using the surface number density $N_{\rm SED}(r, a)$ similar to Pawellek et al. (2021),

$$N_{\rm SED}(r, a) \sim a^{-q}\frac{1}{\sqrt{2\pi}\Delta R_{\rm belt}}\exp\left[-\frac{1}{2}\left(\frac{r - R_{\rm belt}^{\rm mes}}{\Delta R_{\rm belt}}\right)^2\right],$$

where $r$ represents the distance to the star, $R_{\rm belt}^{\rm mes}$ the belt radius measured from the $r^2$-scaled scattered-light images along the disk's major axis, and $\Delta R_{\rm belt}$ is taken to be $0.1R_{\rm belt}^{\rm mes}$, assuming a radial dust distribution width of approx. 20% of the belt radius. This assumption aligns reasonably well with the results from well-resolved debris belts exhibiting a wide range of eccentricities, such as HD 22049 (Booth et al. 2017), HD 109085 (Marino et al. 2017), HD 109573 (Milli et al. 2019), HD 181327 (Marino et al. 2016), HD 202628 (Faramaz et al. 2019), and HD 216956 (MacGregor et al. 2017; Kennedy 2020). The parameter $a$ represents the grain radius, while $q$ is the SD power-law index. The surface number density, $N_{\rm SED}(r, a)$, is directly related to the surface density, $\Sigma(r, a)$, by the equation

$$\Sigma(r, a)\,da = \pi a^2\,N_{\rm SED}(r, a)\,da.$$

We considered grain sizes ranging from a minimum value, $a_{\rm min}$, to a fixed maximum value of $a_{\rm max} = 5000\,\mu$m. Grains larger than this were assumed to contribute negligibly to the SED, as particles absorb and emit radiation efficiently only at wavelengths shorter than their size. For instance, the efficiency of interaction with radiation drops for weakly absorbing materials at wavelengths longer than $a/2\pi$ (Backman & Paresce 1993), while for moderately absorbing materials such as "dirty ice", this critical wavelength is approximately equal to the particle size (Greenberg 1978). Therefore, the adopted maximum size of 5 mm is sufficiently large to encompass all grain sizes that significantly contribute to the SED flux within the wavelength range covered by the available observations.

A one-ring model has three free parameters: the minimum grain radius, $a_{\rm min}$, the SD index, $q$, and the amount of dust, $M_{\rm dust}^{\rm SD}$, for particles between $a_{\rm min}$ and $a_{\rm max}$ assuming a bulk density $\varrho$. Hence, at least four photometric data points (and the disk radius) are needed to fit a SD model to the data.

If the scattered light images provided evidence for both an outer and an inner belt, and both rings were spatially resolved, we applied a two-component SD model to fit the data. To address the degeneracy in dust mass estimates between the two components, we determined the mass ratio as follows. First, we fitted the SED with a single-component model, considering a dust distribution spanning from the central radius of the inner ring to the central radius of the outer ring, allowing us to estimate the total dust mass of the entire disk. For simplicity, we assumed that the dust mass of a belt scales with the square of its radial distance from the star, such that it is given by

$$M_{\rm outer} = \frac{M_{\rm total}\times R_{\rm outer}^2}{R_{\rm inner}^2 + R_{\rm outer}^2} \qquad M_{\rm inner} = \frac{M_{\rm total}\times R_{\rm inner}^2}{R_{\rm inner}^2 + R_{\rm outer}^2},$$

where $M_{\rm inner}$ and $M_{\rm outer}$ are the dust masses of the inner and outer belt, $M_{\rm total}$ the estimated total dust mass, and $R_{\rm inner}$ and $R_{\rm outer}$ the central radii of the inner and outer belt.

We then fixed the masses of the individual belts and fitted the SD parameters $a_{\rm min}$ and $q$, assuming that both cold dust rings share the same SD. In this approach, only the dominant grain size and the SD index are the only free parameters, as the dust masses are fixed. Consequently, at least three photometric data points are required to fit the SED.

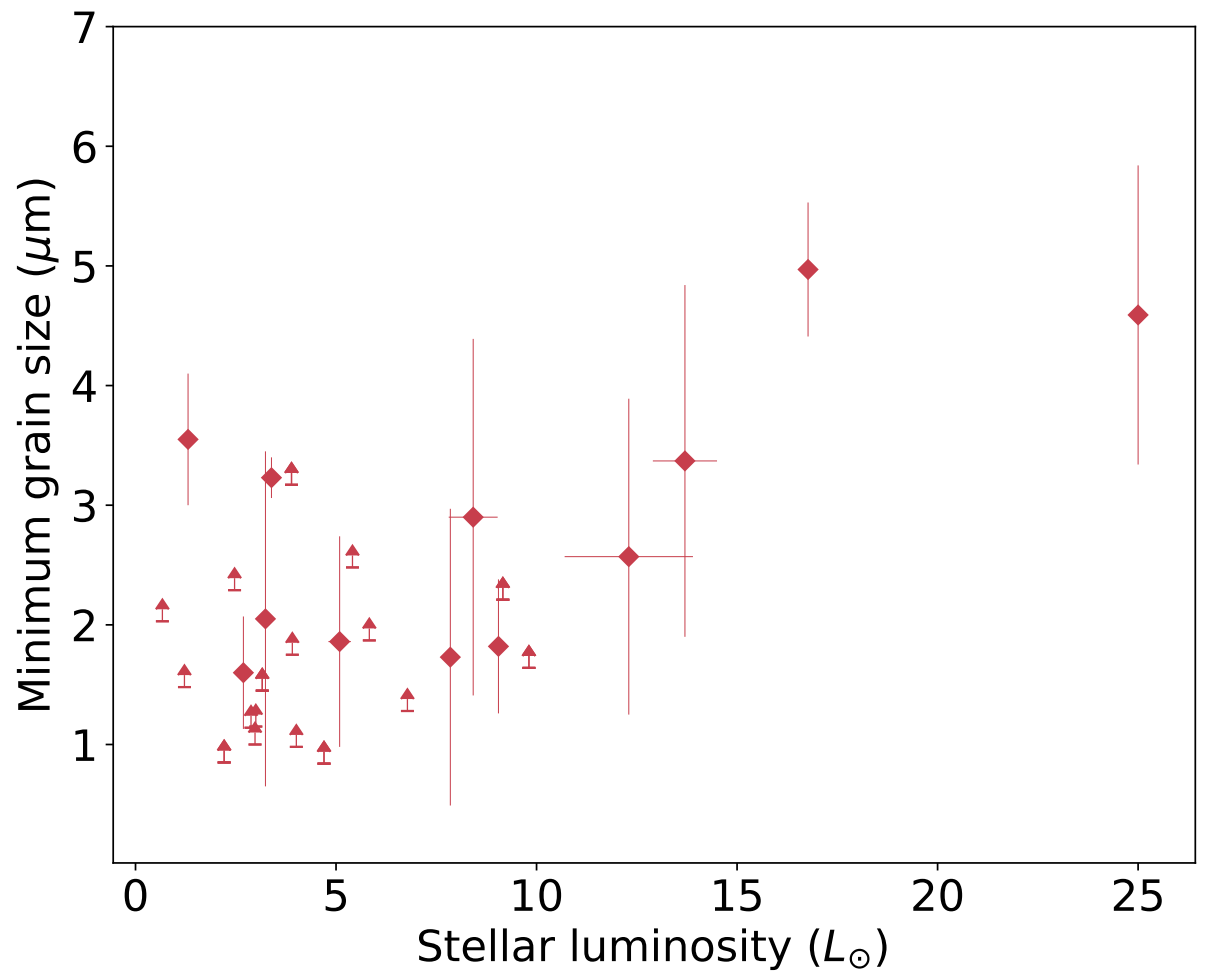

Fig. 15: Minimum grain size $a_{\rm min}$ derived from SD modeling for resolved debris disks, plotted as a function of stellar luminosity. For systems where fitting the warm component was not feasible, only lower limits on $a_{\rm min}$ are indicated.

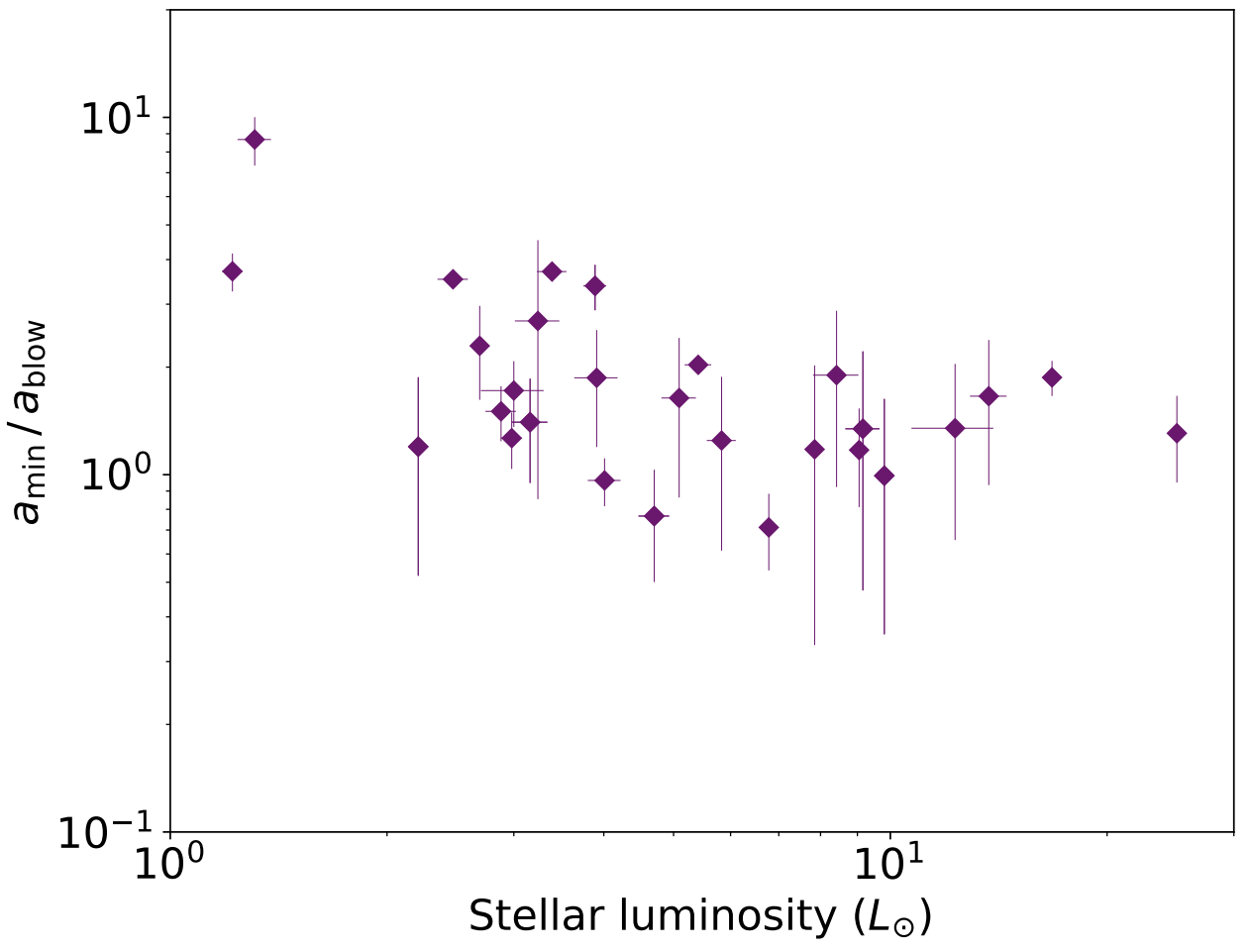

Fig. 16: Ratio of the minimum grain size $a_{\rm min}$, derived from SD modeling of resolved debris disks, to the blowout size $a_{\rm blow}$, plotted as a function of stellar luminosity. The ratio is shown for host stars with luminosities $L_\star > 1\,L_\odot$, where radiation pressure is expected to efficiently remove small dust grains from the system.

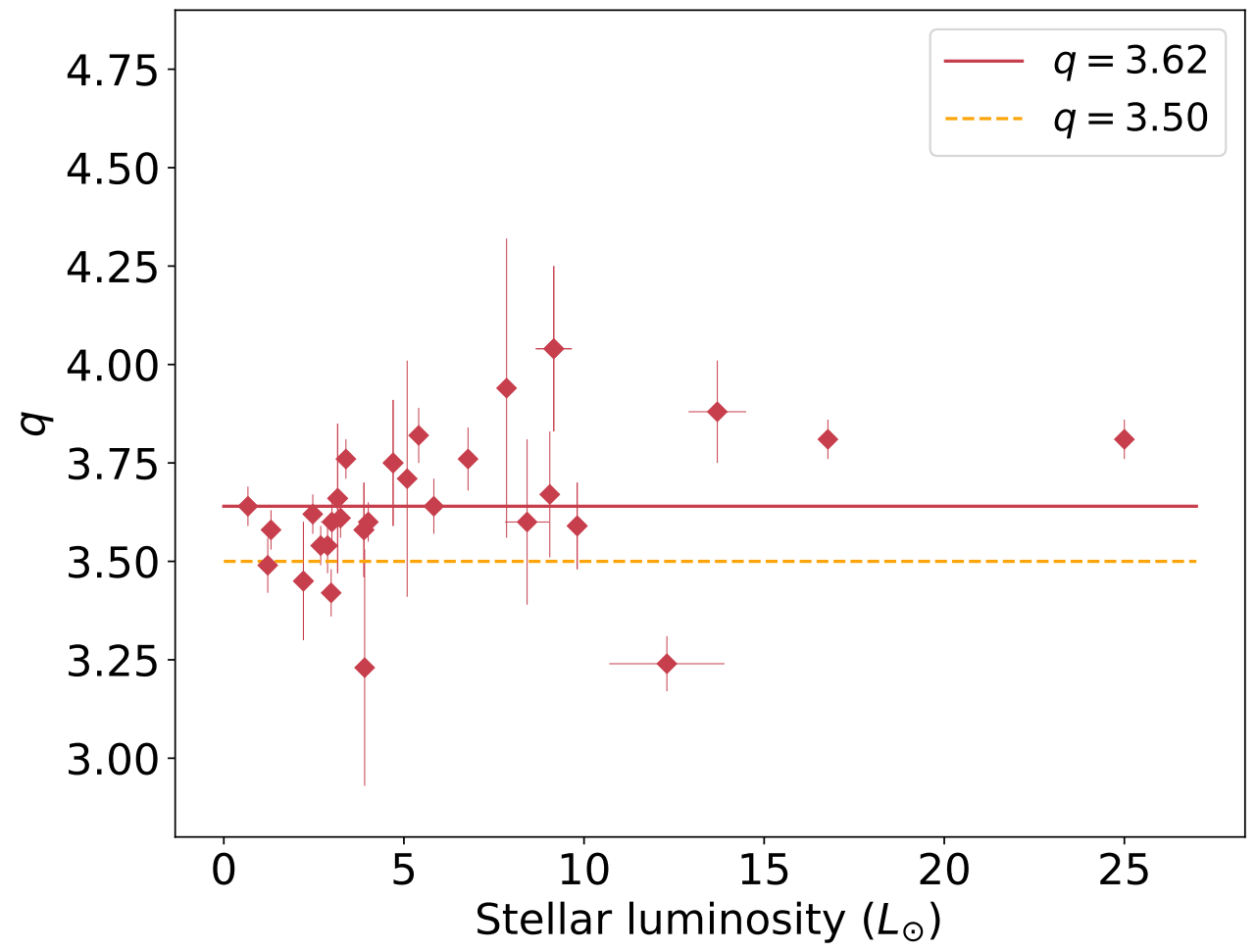

Fig. 17: SD power law index $q$, derived from SD modeling of resolved debris disks, plotted as a function of stellar luminosity. The red solid line indicates the mean value of $q = 3.62$, while the orange dotted line marks the canonical value $q = 3.5$ expected for a steady-state collisional cascade (Dohnanyi 1969).

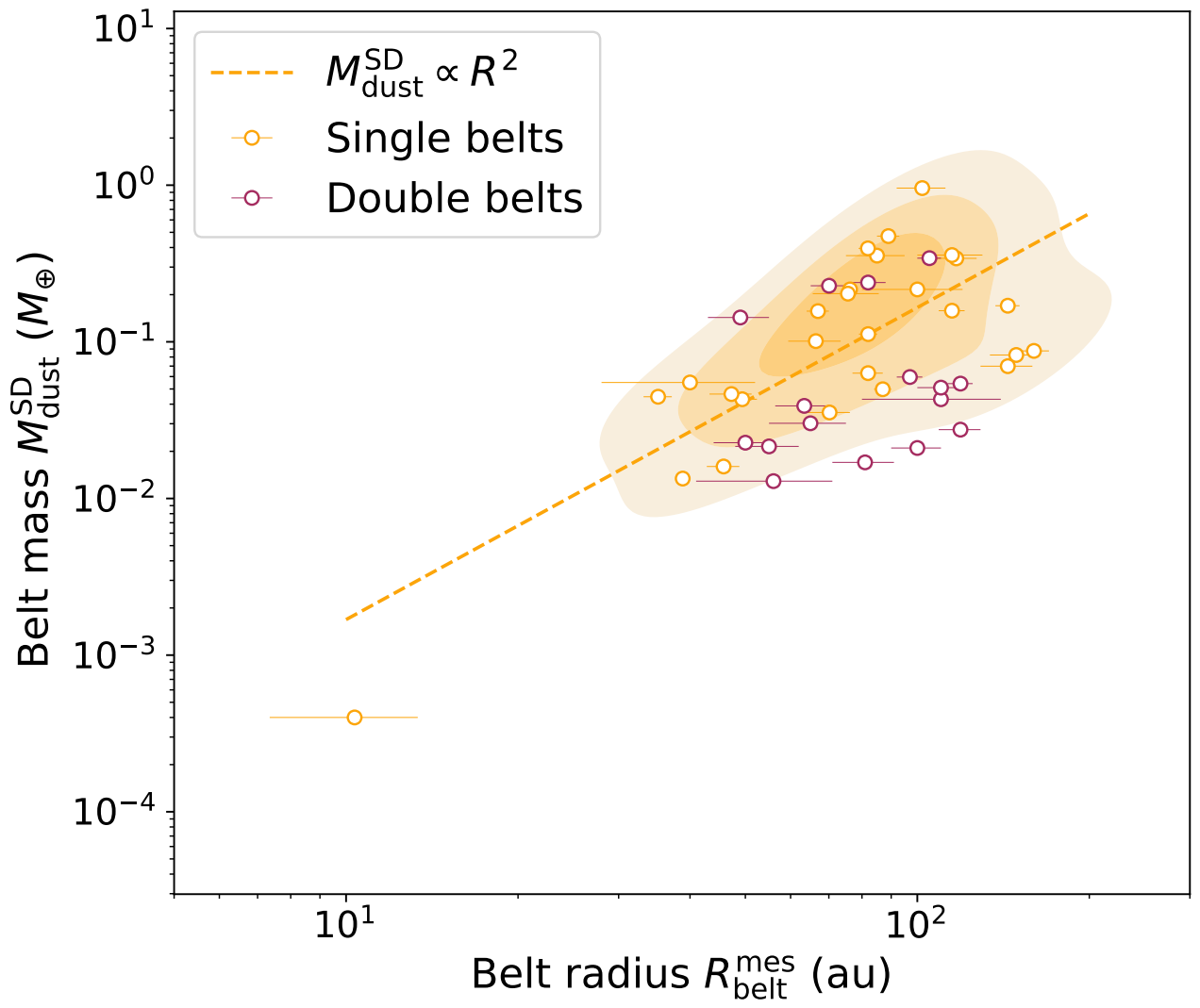

Fig. 18: Belt dust mass obtained from SD modeling versus measured belt radius (orange open circles). The filled contours represent the probability density distribution of single-belt data, containing 20%, 50% and 80% of the data points. The orange solid line represents the fit $R^{\alpha_R}$ to this distribution. Red open circles indicate the double-belt data, which are not included in the distribution fit.

With our chosen approach, we focused on spatially resolved data from scattered-light images. However, warm components that remain undetected in the images or are spatially unresolved are not included in the modeling process. As a result, the NIR/MIR data of such SEDs are not well-fitted, and part of the emission that could originate from a warm dust component is instead incorporated into the fit of the cold dust belt. In such cases, the fitted minimum grain size is underestimated.

To mitigate this issue, for targets with a sufficiently large number of available photometric points, we tested various fitting approaches. These included a fit with a warm component modeled as a BB, which provided an upper limit on $a_{\rm min}$, and a fit excluding the warm component, which yielded a lower limit on $a_{\rm min}$. These approaches allowed us to derive a mean value for the minimum grain size, along with its upper and lower boundaries. For targets, where a warm component fit was not feasible, only lower limits on the minimum grain size were obtained.

### 5.5. Results of SD modeling

Figure D.1 (bottom row) presents examples of SEDs fitted with SD models, featuring either one- or two-component configurations. The results of SD modeling are presented in Table 4, which provides the best-fitting parameters for the targets with reliable fits. These include the minimum grain radius, $a_{\rm min}$, the SD power-law index, $q$, the BB temperature corresponding to the peak of the fitted emission, $T^{SD}_{\rm dust}$, the IR excess, $f^{SD}_{\rm disk}$, of the resolved cold belts, and their dust masses, $M^{SD}_{\rm dust}$, integrated over grain sizes up to 5000 $\mu$m and assuming a pure astrosilicate composition. Additionally, the table lists the radiation pressure blowout size $a_{\rm blow}$, which represents the grain size threshold below which particles are expected to be expelled from the disk by stellar radiation pressure. Dust grains become unbound when the radiation pressure force, $F_{\rm rad}$, acting on them exceeds half of the gravitational force, $F_{\rm gr}$ (e.g., Krivov 2010). Based on the condition $F_{\rm rad}/F_{\rm gr} = 0.5$, the blowout grain size was calculated for a star with mass $M_\star$ and luminosity $L_\star$ following the formulation by Burns et al. (1979):

$$a_{\rm blow} = \frac{3L_\star\langle Q_{\rm pr}\rangle}{8\pi G M_\star c\varrho},$$

where $\langle Q_{\rm pr}\rangle$ is the mean radiation pressure coupling coefficient averaged over the stellar flux (e.g., Augereau et al. 1999), $G$ is the gravitational constant, and $c$ is the speed of light. We assume $\langle Q_{\rm pr}\rangle = 1$ which is close to the values of 1.1 and 1.4 for stars with luminosities of $L_\star = 10\,L_\odot$ and $L_\star = 1\,L_\odot$, respectively, for astrosilicate composition (Pawellek et al. 2014). For low-luminosity stars ($L_\star < 1\,L_\odot$), the blowout grain size is not calculated, as the radiation pressure is too weak to serve as an effective dust removal mechanism in such systems.

The derived minimum grain sizes, or their lower limits in cases where the warm component could not be fitted, are presented in Fig. 15, while the ratios of these values to the blowout grain sizes are shown in Fig. 16 as a function of stellar luminosity. Notably, nearly all derived $a_{\rm min}$ values are close to, but consistently larger than, the corresponding blowout sizes $a_{\rm blow}$. This is expected, as grains smaller than $a_{\rm blow}$ are efficiently removed from the systems on short timescales due to radiation pressure. Figure 16 reveals that the ratio $a_{\rm min}/a_{\rm blow}$ tends to increase with decreasing stellar luminosity, reaching values of up to ~9 in the case of HD 105. This trend corroborates earlier findings (see Pawellek et al. 2014; Pawellek & Krivov 2015) and could have several, not mutually exclusive, explanations: limitations in grain surface energy that inhibit the production of small collisional fragments (Krijt & Kama 2014; Thebault 2016) or lower dynamical excitation levels in cold disks, leading to a depletion of small grains (Thébault & Wu 2008).

For nearly all systems, the derived $q$ values (Col. 4 in Table 4) are close to the canonical value of 3.5, which is expected for an idealized infinite, self-similar collisional cascade (Dohnanyi 1969). The mean value of $q = 3.62$ is slightly higher (Fig. 17), which is consistent with expectations for more realistic collisional systems, where the critical specific energy required for fragmentation increases with decreasing particle size in the strength-dominated regime (O'Brien & Greenberg 2003).

Table 4: Results of SD modeling.

| Debris belt | $a_{\mathrm{blow}}$ ($\mu$m) | $a_{\mathrm{min}}$ ($\mu$m) | $q$ | $T^{SD}_{\mathrm{dust}}$ (K) | $f^{SD}_{\mathrm{disk}}$ ($10^{-4}$) | $M^{SD}_{\mathrm{dust}}$ ($10^{-2}M_\oplus$) |
|---|---|---|---|---|---|---|
| HD 105 | 0.41 | 3.55 ± 0.55 | 3.58 ± 0.05 | 46±2 | 2.95 | 4.98 |
| HD 377 | 0.40 | 1.48 ± 0.18 | 3.49 ± 0.07 | 52±6 | 4.64 | 6.31 |
| HD 9672 | 2.66 | 4.97 ± 0.56 | 3.81 ± 0.05 | 66±4 | 6.80 | 17.0 |
| HD 15115 out | 0.94 | 3.17 ± 0.46 | 3.58 ± 0.12 | 54±1 | 1.99 | 4.79 |
| HD 15115 inn | 0.94 | 3.17 ± 0.46 | 3.58 ± 0.12 | 68±1 | 3.05 | 2.04 |
| HD 16743 | 1.22 | 2.48 ± 0.14 | 3.82 ± 0.07 | 70±3 | 5.08 | 8.24 |
| HD 30447 | 0.94 | 1.75 ± 0.63 | 3.23 ± 0.30 | 49±11 | 10.5 | 47.3 |
| HD 32297 | 1.53 | 2.90 ± 1.49 | 3.60 ± 0.21 | 51±20 | 60.5 | 34.1 |
| HD 35841 | 0.65 | 2.29 ± 0.14 | 3.62 ± 0.05 | 66±3 | 16.1 | 10.1 |
| HD 39060 out | 1.64 | 2.21 ± 1.43 | 4.04 ± 0.21 | 107±7 | 11.4 | 4.16 |
| HD 39060 inn | 1.64 | 2.21 ± 1.43 | 4.04 ± 0.21 | 121±7 | 19.3 | 1.45 |
| HD 61005 | (...) | 2.03 ± 0.15 | 3.64 ± 0.05 | 59±4 | 27.8 | 15.7 |
| HD 106906 | 1.80 | 1.28 ± 0.31 | 3.76 ± 0.08 | 105±4 | 12.8 | 3.54 |
| HD 109573 | 3.51 | 4.59 ± 1.25 | 3.81 ± 0.05 | 96±10 | 45.5 | 21.6 |
| HD 110058 | 1.55 | 1.82 ± 0.56 | 3.67 ± 0.16 | 107±2 | 26.1 | 5.50 |
| HD 111520 | 0.70 | 1.60 ± 0.47 | 3.54 ± 0.05 | 73±6 | 21.7 | 20.3 |
| HD 112810 | 0.87 | 3.23 ± 0.17 | 3.76 ± 0.05 | 59±3 | 10.4 | 15.8 |
| HD 114082 | 1.02 | 0.98 ± 0.15 | 3.60 ± 0.05 | 111±4 | 36.3 | 4.46 |
| HD 115600 | 1.13 | 1.86 ± 0.88 | 3.71 ± 0.30 | 115±6 | 19.9 | 1.60 |
| HD 117214 | 1.50 | 1.87 ± 095 | 3.64 ± 0.07 | 121±8 | 23.6 | 4.30 |
| HD 120326 out | 1.10 | 0.84 ± 0.29 | 3.75 ± 0.16 | 96±7 | 5.30 | 4.61 |
| HD 120326 inn | 1.10 | 0.84 ± 0.29 | 3.75 ± 0.16 | 121±7 | 12.60 | 0.81 |
| HD 121617 | 2.03 | 3.37 ± 1.47 | 3.88 ± 0.13 | 87±15 | 43.6 | 11.2 |
| HD 129590 out | 1.03 | 1.45 ± 0.47 | 3.66 ± 0.19 | 76±6 | 24.00 | 19.45 |
| HD 129590 inn | 1.03 | 1.45 ± 0.47 | 3.66 ± 0.19 | 85±6 | 40.30 | 6.95 |
| HD 131488 | 1.90 | 2.57 ± 1.32 | 3.24 ± 0.07 | 60±8 | 18.3 | 95.8 |
| HD 131835 out | 1.65 | 1.64 ± 1.05 | 3.59 ± 0.11 | 76±5 | 14.7 | 27.35 |
| HD 131835 inn | 1.65 | 1.64 ± 1.05 | 3.59 ± 0.11 | 85±5 | 22.1 | 12.18 |
| HD 141943 out | 0.71 | 0.85 ± 0.53 | 3.45 ± 0.15 | 54±6 | 1.1 | 1.64 |
| HD 141943 inn | 0.71 | 0.85 ± 0.53 | 3.45 ± 0.15 | 60±6 | 1.3 | 1.07 |
| HD 145560 | 0.76 | 2.05 ± 1.40 | 3.61 ± 0.05 | 66±6 | 30.3 | 35.5 |
| HD 146181 | 0.67 | 1.15 ± 0.24 | 3.60 ± 0.07 | 76±5 | 24.6 | 21.6 |
| HD 172555 | 1.47 | 1.73 ± 1.24 | 3.94 ± 0.38 | 207±35 | 3.50 | 0.04 |
| HD 181327 | 0.79 | 1.00 ± 0.18 | 3.42 ± 0.06 | 62±5 | 25.3 | 39.5 |
| HD 191089 | 0.76 | 1.14 ± 0.20 | 3.54 ± 0.07 | 88±5 | 15.1 | 4.64 |
| HD 107146 out[a] | 0.34 | 2.79 ± 1.17 | 3.42 ± 0.04 | 34±2 | 3.33 | 17.30 |
| HD 107146 inn[a] | 0.34 | 2.79 ± 1.17 | 3.42 ± 0.04 | 54±2 | 8.01 | 7.21 |

**Notes.** [a] The planetesimal belts around HD 107146 were not resolved with SPHERE. The SD modeling for this target was conducted for comparison purposes.
The columns list target IDs, blowout grain sizes ($a_{\mathrm{blow}}$), minimum grain sizes ($a_{\mathrm{min}}$), the SD power-law index ($q$), the BB temperature corresponding to the peak of the fitted emission ($T^{SD}_{\mathrm{dust}}$), the disk IR excess ($f^{SD}_{\mathrm{disk}}$), and the belt dust masses ($M^{SD}_{\mathrm{dust}}$) derived from the SED fits using the SD model.

Figure 18 presents the belt dust masses derived from the SD model as a function of the measured disk radius, in order to examine the correlation between these two quantities. The contour plot in the figure represents the PDF for a subsample of resolved single belts around A-, F- and G-type stars with estimated ages between 10 and 200 Myr. The double-belt data points (red open circles in Fig. 18) are excluded from the PDF calculation due to the assumed relationship between the masses of two components in double-belt systems (Sect. 5.4).

We observe a tendency for increasing belt dust mass with growing radial distance from the star, following the fitted radial power law $M^{\mathrm{SD}}_{\mathrm{dust}} \propto R^{\alpha_{\mathrm{R}}}_{\mathrm{belt}}$ with $\alpha_{\mathrm{R}} = 2.1 \pm 0.4$. Although subject to small-number statistics bias, this trend is consistent with our findings for a larger sample of A- and F-type stars, based on the results of MBB modeling. The actual radial dependence may be less steep than our result suggests if single belts consist of multiple components that remain undetected due to insufficient spatial resolution. In such a case, the total cold dust mass would be distributed across multiple components, thereby reducing the mass assigned to a single belt and leading to a shallower mass distribution.

HD 107146 To investigate whether the non-detection of a disk could be attributed to the low reflectivity of its dust grains, we applied the SD model to fit the SED of the HD 107146 debris disk. This nearly pole-on ($i = 19°$) disk consists of two broad cold planetesimal belts located at ~50 and 120 au from its G-type host star, as previously observed with ALMA (Marino et al. 2018). The debris belts were not clearly detected in the IRDIS H-band polarimetric observations. To compare the dust optical properties inferred from SD modeling for this disk with those of a detected debris disk (see Sect. 6.3), we derived the best-fit parameters for HD 107146 as well. The results are listed in the last rows of Table 4.

## 6. Detections versus non-detections

To comprehensively understand the diversity of debris disk architectures, it is essential not only to analyze systems where belts are detected in scattered light, but also to interpret cases of non-detections, which provide valuable constraints. They may indicate disks that are intrinsically fainter, narrower, more pole-on, or more evolved, and thus consistent with lower dust masses or smaller planetesimal belts located closer to the star. Non-detections in scattered light may still host massive disks visible in the IR, and their lack of scattered-light visibility could be explained by viewing geometry or dust properties. Additionally, the observing conditions during the runs, such as atmospheric seeing, coherence time, or instrument stability, can significantly influence detection sensitivity, particularly for faint disks.

Two-thirds of the debris disks in our sample were not detected in SPHERE observations. Therefore, in the following sections, we use the results of disk and SED modeling to explore the potential causes of these non-detections in more detail. Particular attention is given to the optical properties of dust grains, focusing on two complementary approaches for deriving dust scattering characteristics, such as albedo and maximum polarization fraction: one based on theoretical predictions using Mie theory (Mie 1908), and the other on measurements of total and polarized scattered fluxes.

### 6.1. Disk luminosity and dust mass

Debris disks with an IR excess below $10^{-4}$ are very faint, making them challenging to image in scattered light with current instrumentation. The faintest debris disk successfully imaged with SPHERE is that of HD 141943 (Boccaletti et al. 2019), with a disk fractional luminosity of $f_{\rm disk} = 1.2 \times 10^{-4}$.

As illustrated in Fig. 12, where targets with detected disks are represented by filled circles, stellar age appears to be one of the most critical factors limiting the detectability of scattered light from debris disks around distant stars. Among all disks detected with SPHERE, 90% of the targets have a mean estimated age below 50 Myr. The stellar age histogram in Fig. 1 (lower left panel) also shows that debris disks are detected in more than 50% of systems younger than 100 Myr, whereas for older stars, the detection rate drops to just 5%. This declining detection rate with increasing stellar age can be attributed to the progressive reduction in dust mass over time, leading to a decrease in both scattered and thermal emission from dust particles (e.g., Wyatt 2008; Krivov 2010).

The relative small number of detected debris disks older than 50 Myr in our sample, which spans system ages from a few Myr to several Gyr, may provide constraints on the size of the largest planetesimals formed by the end of the PPD phase. To address the issue of overly high inferred debris disk masses, Krivov & Wyatt (2021) proposed that young debris disks may contain relatively small largest planetesimals (on the order of 1 km in size), suggesting that "planetesimals are born small". The detection statistics in our sample support this hypothesis, as disks formed with small planetesimals are expected to appear bright at young ages (tens of Myr) but fade rapidly within a few hundred Myr, whereas disks formed with larger planetesimals would maintain their brightness over several Gyr.

### 6.2. Disk geometry and observing techniques

In addition to system age, the viewing geometry of the disk and the observing technique significantly influence the detectability of debris disks through DI. In particular, disk inclination often plays a decisive role, as highly inclined (nearly edge-on) systems are generally easier to detect (e.g., Esposito et al. 2020). Such disks exhibit increased SB along the edges of the planetesimal belt, where the column density of dust particles is highest, and on the disk's front side, where forward-scattering enhances the intensity of scattered light. These effects make inclined disks more readily observable, whereas pole-on systems, which lack strong forward-scattering features and appear more diffuse, are inherently more challenging to detect.

To illustrate this effect, we generated model images of a typical debris disk observed at inclinations 0° (pole-on), 45° and 90° (edge-on). For this purpose we used the model described in Sect. 4.5.1 adopting parameters commonly found in disk studies. Figure 19 presents the corresponding scattered light images, with total intensity shown in the top row and polarized intensity in the bottom row. When viewed edge-on (top left panel), the disk exhibits the highest SB, significantly enhancing its detectability.

Additionally, the observing technique most commonly used for imaging debris disks in scattered light is the ADI (Marois et al. 2006). This method is particularly sensitive to edge-on systems, as they generate a signal that differs from the stellar PSF when the sky field rotates. For disks with lower inclinations ($i \lesssim 70°$), ADI is less effective, and for rotationally symmetric pole-on disks, it is inapplicable. As a result, despite their high IR excesses ($f_{\rm disk} > 10^{-3}$), the debris disks around HD 107146 and HD 95086 remained undetected in the SPHERE ADI datasets. Nonetheless, a large sky rotation angle during observations under good conditions can improve the detectability of low-inclination disks. This is demonstrated in the case of HD 105 debris disk (Fig. 2), which was successfully detected, even though it has a relatively low inclination of 50.5°.

For imaging debris disks at low inclinations, PDI is more suitable than ADI. PDI yields images of the polarized intensity of scattered light, referred to as polarimetric images, which complement total intensity images of the same disk. However, the SB distribution in polarimetric images differs from that in total intensity images, as the two are governed by distinct phase functions: the SPF and the pSPF, respectively. This distinction is illustrated in Fig. 19, which shows simulated images of total and polarized intensities for a debris disk generated using our model with a parametric representation of the phase functions: a HG function with an asymmetry parameter of $g = 0.6$ for the SPF, and the function given by Eq. 2 with a maximum polarization fraction $p_{\rm max} = 0.3$ for the pSPF.

The total flux, or integrated intensity, defined as the sum over all image pixels containing disk emission, also differs between total scattered and polarized intensity images and varies systematically with disk inclination. This effect is demonstrated in Fig. 20, which is based on the same model as in Fig. 19. The left panel of Fig. 20 shows the total scattered flux, $F_{\rm sca}$, while the middle panel displays the total polarized flux, $F_{\rm pol}$, both plotted as functions of the scattering asymmetry parameter $g$ and disk inclination. Each flux is normalized by a factor of $4\pi/L_{\rm sca}$, where $L_{\rm sca}$ denotes the scattered luminosity of the disk, or total intensity integrated over the full solid angle.

As expected, the total scattered flux reaches a maximum for an edge-on disk with $g = 0.9$, corresponding to strongly forward-scattering grains (Fig. 20 left panel). The maximum polarized flux is obtained for a pole-on disk with an isotropic scattering parameter of $g = 0$ (Fig. 20 middle panel). In this configuration, most scattering occurs at $\theta = 90°$, where, according to our model, the degree of linear polarization reaches its maximum (Eq. 3). However, the polarized scattered intensity represents only a fraction of the total scattered intensity (Fig. 20 right panel), and this fraction can decrease rapidly with lower disk inclination, particularly if the dust particles exhibit strong forward-scattering behavior. A comparison of the model images in Fig. 19 and Fig. 20 demonstrate that debris disks appear faintest in polarized intensity when viewed pole-on, and that they consistently exhibit lower brightness in polarized light than in total scattered light, irrespective of the dust's optical properties.

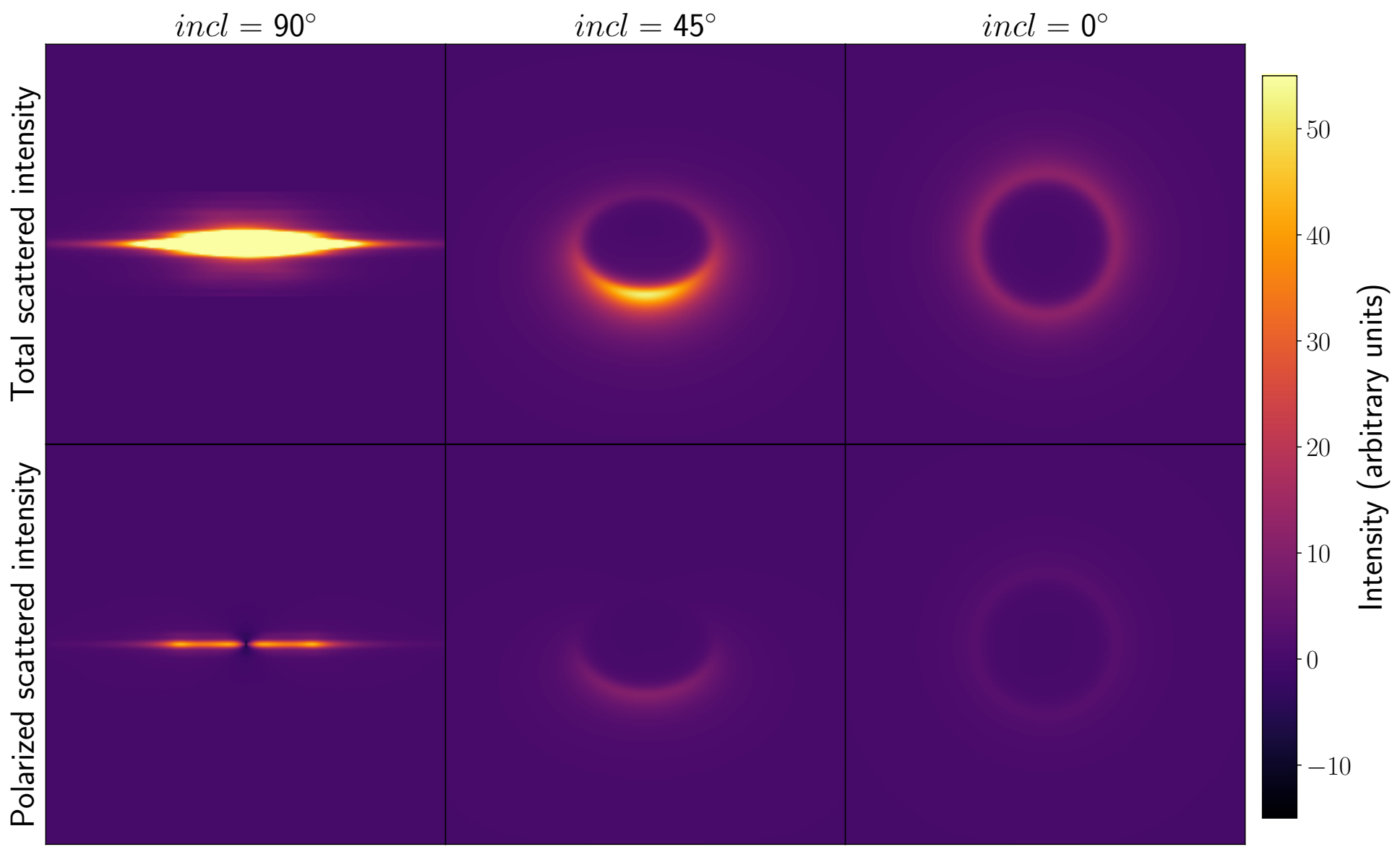

**Fig. 19.** Simulated images of a debris disk modeled using the parameters $r_0/H_0 = 0.01$, $\alpha_{\rm in} = 15$, $\alpha_{\rm out} = -3$, $g_{\rm sca} = 0.6$. The $p_{\rm max}$ was set to 0.3. The images were convolved with a typical IRDIS PSF. Forward-modeling to mimic the ADI data reduction was not applied.

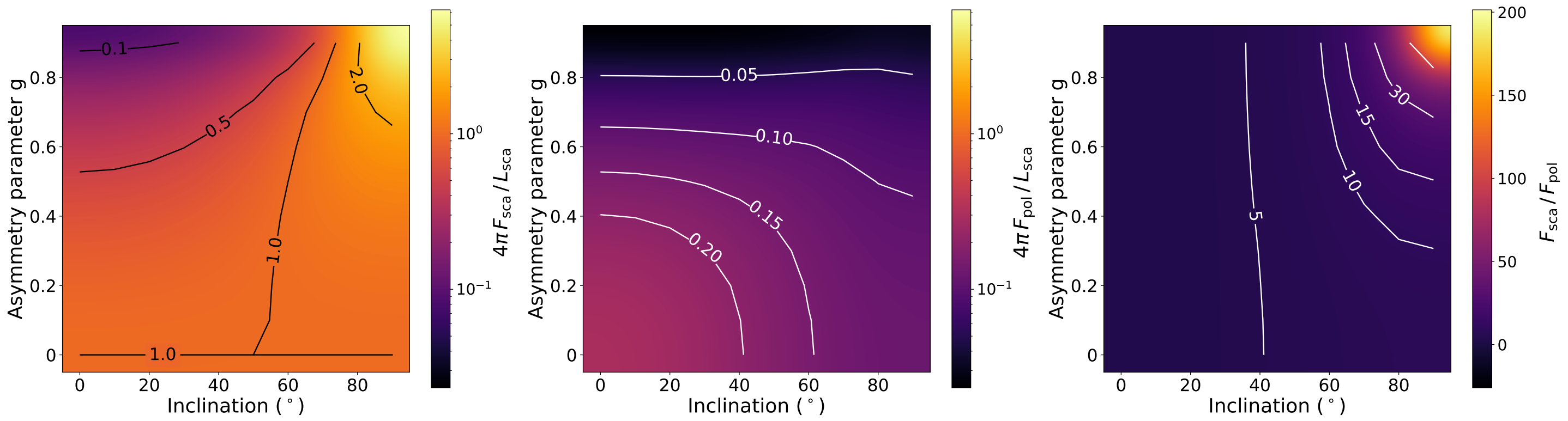

Fig. 20: Total fluxes measured in the model images convolved with IRDIS PSF. Left: Total scattered flux measured in the image of total intensity. Middle: Total polarized flux measured in the image of polarized intensity. Right: Ratio of scattered and polarized fluxes.

Despite this, thirty six debris disks in our sample were successfully detected using the PDI modes of IRDIS and ZIMPOL (Fig. 5). The PDI data processing methodology enables a more effective subtraction of the stellar PSF from the science frames compared to the ADI technique. As a result, polarimetric images can achieve higher contrast levels, reaching up to $10^{-8} - 10^{-7}$ with ZIMPOL (Hunziker et al. 2020; Tschudi et al. 2024). This allows for the easy detection of young, low-inclination disks with a high polarization fraction of scattered light, where dust particles are confined to a narrow, bright ring such as HD 181327 planetesimal belt (Fig. 5; Milli et al. 2024). Conversely, older low-inclination disks, which exhibit a broad radial dust distribution in some ALMA images and may consist of multiple faint planetesimal rings, are more challenging to resolve using PDI. Broad pole-on disks generally have a lower surface density resulting in lower SB in polarized light. This could explain the marginal detection of debris belts in HD 107146 using IRDIS H-band polarimetry, despite relatively favorable observing conditions.

### 6.3. Influence and derivation of dust albedo and polarization characteristics

The optical properties of dust particles, specifically their scattering and absorption efficiencies, can also be responsible for a non-detection of a debris disk. If the dust scattering efficiency is low at a particular wavelength, the SB of the debris disk is correspondingly low at that wavelength, decreasing the probability of disk detection.

The optical characteristics of dust grains are fundamental parameters in the study of circumstellar environments. They are intrinsically linked to the grains' composition, structure, and SD, and thus provide indirect constraints on the primordial solid-phase reservoir from which exoplanets may have formed. Understanding these properties is therefore a key objective in debris disk research.

One approach to constraining dust composition is through the evaluation of the dust albedo, which quantifies the relative efficiency of scattering versus total extinction (scattering plus absorption). A higher albedo than 0.5 indicates that scattering dominates over absorption, while a lower albedo suggests that absorption is more significant. If the dust grain SD within a disk is known, often inferred from modeling the SED (e.g., Pawellek et al. 2019) or scattered-light imaging (e.g., Olofsson et al. 2016), then theoretical predictions of albedo can be made for various dust compositions using Mie theory (Mie 1908) or more complex models accounting for grain porosity and non-sphericity (Draine & Flatau 1994; Min et al. 2005).

These theoretical predictions can then be compared to observational constraints, derived from combined measurements of scattered stellar light and thermal re-emission (SED). By jointly analyzing these datasets, the range of plausible dust compositions can be narrowed down. For example, grains composed primarily of astronomical silicates, carbonaceous materials, or ices each exhibit distinct scattering and absorption efficiencies, and thus different albedo values. This comparison allows to exclude certain grain compositions and structures, providing a more refined picture of the physical nature of dust in debris disks.

In the following section, we present the methodology used to calculate the albedo of dust grains in debris disks, employing Mie theory to derive the scattering and absorption efficiencies for particles of specified composition and size. We then describe how these theoretical predictions are compared with observational constraints, obtained from the analysis of scattered-light images of spatially resolved debris disks in our sample and their IR excess measurements.

#### 6.3.1. Albedo

The amount of stellar photons with a wavelength $\lambda$ which are scattered by a spherical dust particle of radius $a$ is determined by its spectral cross section for scattering, $C_\lambda^{\rm sca}$. This cross section quantifies the relationship between the intensity of the incident radiation, $I_{\star\,\lambda}$, and the scattered power or spectral radiant flux, $F_\lambda^{\rm sca}$ (Bohren & Huffman 1983):

$$C_\lambda^{\rm sca} = \frac{F_\lambda^{\rm sca}}{I_{\star\,\lambda}} = Q_\lambda^{\rm sca}\,A,$$

where $Q_\lambda^{\rm sca}$ is the scattering efficiency and $A = \pi a^2$ is the geometrical cross section of particle.

The fraction of stellar light attenuation caused by scattering, and thus the role of scattering in the overall extinction process, is characterized by the single scattering albedo $\omega_\lambda$, which describes the proportion of extinction resulting from scattering rather than absorption:

$$\omega_\lambda = \frac{C_\lambda^{\rm sca}}{C_\lambda^{\rm ext}} = \frac{C_\lambda^{\rm sca}}{C_\lambda^{\rm sca}+C_\lambda^{\rm abs}} = \frac{Q_\lambda^{\rm sca}}{Q_\lambda^{\rm sca}+Q_\lambda^{\rm abs}} = \frac{F_\lambda^{\rm sca}}{F_\lambda^{\rm sca}+F_\lambda^{\rm abs}},$$

where $C_\lambda^{\rm ext} = Q_\lambda^{\rm ext}A = (F_\lambda^{\rm sca}+F_\lambda^{\rm abs})/I_{\star\,\lambda}$ and $C_\lambda^{\rm abs} = Q_\lambda^{\rm abs}A = F_\lambda^{\rm abs}/I_{\star\,\lambda}$ are the spectral extinction and absorption cross sections, respectively, and $F_\lambda^{\rm abs}$ is the power of light absorbed by dust particle.

For a given dust composition characterized by a complex refractive index, Mie theory predicts the scattering efficiency $Q_\lambda^{\rm sca}$ and extinction efficiency $Q_\lambda^{\rm ext}$ as functions of the size parameter $x$ (Bohren & Huffman 1983):

$$x = \frac{2\pi\,a}{\lambda}. \tag{9}$$

For dust particles following a SD characterized by a differential grain number density $n(a)$[10], the effective size parameter can be defined as the ratio of the third to the second moment of the distribution[11] (Hansen & Travis 1974):

$$x_{\rm eff} = \frac{2\pi}{\lambda}\,a_{\rm eff} = \frac{2\pi}{\lambda}\frac{\int a^3\,n(a)\,da}{\int a^2\,n(a)\,da}.$$

[10] The differential grain number density $n(a)$, often expressed as $dn/da$, describes the number of grains per unit size interval. That is, $n(a)da$ represents the number of grains with sizes in the range $[a,\,a+da]$.

[11] This definition of $a_{\rm eff}$ is strictly valid only in the geometric optics limit, which applies to particle sizes larger than approximately 3 microns, assuming a wavelength of 1.6 microns.

The total cross sections for scattering $\sigma_\lambda^{\rm sca}$ and extinction $\sigma_\lambda^{\rm ext}$ are obtained by averaging over the distribution:

$$\sigma_\lambda^{\rm sca} = \frac{\int C_\lambda^{\rm sca}(a)\,n(a)\,da}{\int n(a)\,da} \qquad \sigma_\lambda^{\rm ext} = \frac{\int C_\lambda^{\rm ext}(a)\,n(a)\,da}{\int n(a)\,da}.$$

In this case the spectral albedo is given by

$$\omega_\lambda = \frac{\int C_\lambda^{\rm sca}(a)\,n(a)\,da}{\int C_\lambda^{\rm ext}(a)\,n(a)\,da}. \tag{10}$$

#### 6.3.2. Scattering albedo of various dust compositions

Using Eq. 10, we modeled the spectral albedo for a range of dust compositions, including astrosilicates, amorphous carbon, and silicate grains coated with either dirty or water ice, motivated by both observational evidence and theoretical considerations. Astrosilicates are widely used to represent the dominant silicate emission features observed in mid-IR spectra of circumstellar environments and reflect the mineralogical composition inferred from both debris disks and Solar System dust populations (e.g., Draine 2003a; Dorschner et al. 1995). Amorphous carbon is included to represent more absorbing, featureless materials, commonly invoked to model the continuum emission in disks and supported by the presence of carbonaceous material in interplanetary dust and meteorites (e.g., Zubko et al. 1996; Li & Greenberg 1997). To account for conditions in the outer, colder regions of disks, we also considered silicate grains coated with water ice, which are expected beyond the ice line and significantly modify scattering properties due to their high albedo and distinct optical constants (e.g., Donaldson et al. 2013; Xie et al. 2025). Additionally, dirty ice grains, incorporating refractory inclusions, provide a more realistic representation of ice mantles processed by collisions and irradiation (e.g., Preibisch et al. 1993; Li & Greenberg 1998). This set of compositions spans a physically plausible range and enables the exploration of how material properties influence key observables such as scattered light brightness.

For these four dust compositions, we computed the scattering albedo $\omega_\lambda$ and present the results as 2D maps in Fig. 21. The calculations are based on a grain SD model $n(a)da \propto a^{-q}da$, with grain sizes spanning the range $a_{\rm min} \leqslant a \leqslant a_{\rm max}$. We varied $a_{\rm min}$ between 0.9 and 5 $\mu$m while keeping $a_{\rm max}$ fixed at 5 mm, following our SED fitting procedure (Sect. 5.4). We considered three values for the SD power-law exponent, $q = 3.0$, 3.5 and 4.0, assuming that for most debris systems, the exponent falls within this range (see Col. 4 in Table 4).

The optical data for astrosilicates were taken from Draine (2003a), for water ice from Warren & Brandt (2008), and for amorphous carbon and dirty ice from Preibisch et al. (1993). The refractive indices of silicate grains coated with water or dirty ice were calculated assuming a volume fraction of 50% for each component. This corresponds to a mass fraction of 79% for silicates and 21% for water ice, based on their material densities of $\varrho_{\rm sil} = 3.5$ g cm$^{-3}$ and $\varrho_{\rm ice} = 0.92$ g cm$^{-3}$, respectively (Pollack et al. 1994). The optical constants of the dirty ice coating were derived for a mixture of $H_2O$- and $NH_3$-ices with a volume ratio of 3:1 polluted by amorphous carbon with a volume fraction of 10%. Adopting the material densities for carbon $\varrho_{\rm car} = 2.3$ g cm$^{-3}$ and for $NH_3$-ice $\varrho_{\rm NH_3\,ice} = 0.85$ g cm$^{-3}$, we obtained a mass fraction of 23% for the dirty ice mantle.

The left column in Fig. 21 presents the single scattering albedo for particles composed of pure astrosilicates. For $q = 3.0$ the albedo remains nearly constant (~0.56) for all SDs and wavelengths, corresponding to a gray disk, meaning that the reflectance spectrum of the disk does not vary with wavelength. For $q = 4.0$, the spectral variation of $\omega_\lambda$ is more significant, with the albedo increasing towards the lower right corner of the plot as the effective size parameter decreases $x_{\rm eff}$. This trend is due to the decreasing effective grain size $a_{eff}$ and the increasing wavelength $\lambda$. In the considered range of $x_{eff}$ the scattering cross section of astrosilicates is larger for smaller values of effective size parameter. Within the wavelength range covered by SPHERE filters (0.5 to 2.25 $\mu$m), the scattering cross section of dust grains increases with wavelength, resulting in a higher albedo and thereby an increase in the relative scattered flux (i.e., the disk flux normalized by the stellar flux). Consequently, a disk composed of such dust particles is expected to exhibit a red reflectance spectrum[12], which becomes more pronounced for dust compositions with a higher fraction of small particles (i.e., smaller $x_{eff}$ or larger $q$).

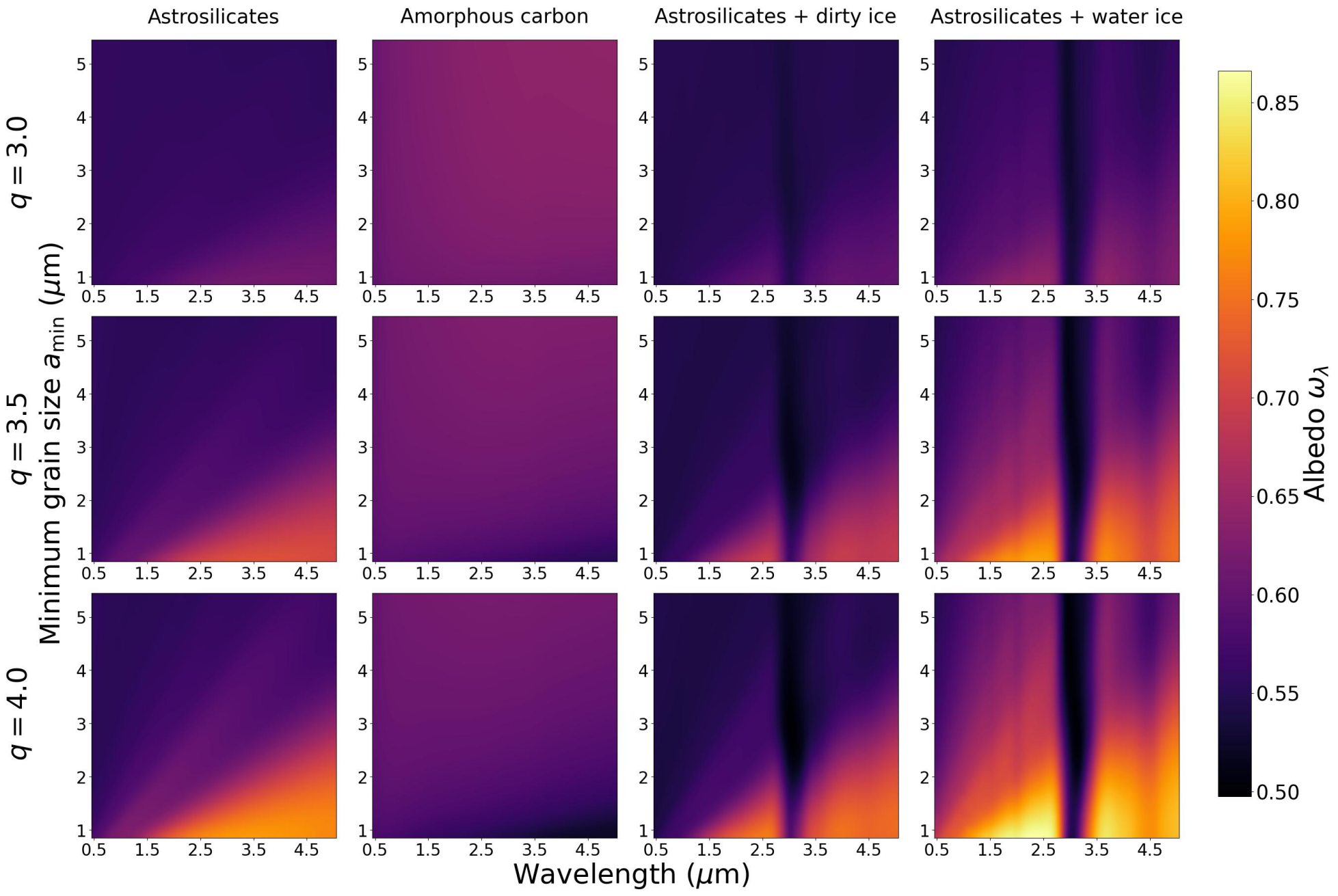

**Fig. 21.** Predictions of Mie theory for the single scattering albedo $\omega_\lambda$ of spherical dust particles exhibiting a SD $n(a) \propto a^{-q}$ with the grain radii in the range $a_{min} \leqslant a \leqslant 5$ mm. The albedo is calculated for $q = 3$ (top row), 3.5 (middle row) and 4 (bottom row). The dust is composed of astrosilicates (left column), amorphous carbon (second column) and grains with the astrosilicate core coated by dirty ice (third column) or pure water ice (right column), assuming a coating volume fraction of 50%.

For dust particles composed of amorphous carbon and following a SD $n(a) \propto a^{-3}$ (top panel of second column in Fig. 21) the reflectance spectrum color remains essentially unchanged regardless of variations in the minimum grain size or wavelength. In this case, the bulk albedo is ~0.63. In contrast to astrosilicates, the albedo decreases for steeper SDs, reaching a value of ~0.52 for a distribution with a minimum grain size of $a_{min} = 0.9$ $\mu$m (bottom panel of second column in Fig. 21). However, even for steeper SDs, the bulk albedo exhibits minimal spectral variation, meaning that a disk composed of such dust particles would appear gray to an observer using SPHERE, irrespective of the specific SD parameters.

The albedo maps of astrosilicate particles coated with either dirty or pure water ice reveal distinct water ice absorption features (third and fourth columns in Fig. 21), with the most prominent one at 3.1 $\mu$m attributed to O-H stretching vibrations of water ice. The spectral albedo of a mixture containing pure water ice is significantly higher, reaching up to 0.88, compared to that of pure astrosilicates. When observed using the broadband H filter with IRDIS, a disk composed of such icy grains would exhibit up to twice the scattered flux of a similar disk with identical viewing geometry and stellar irradiance but composed of amorphous carbon or even astrosilicate particles.

Based on the results of the SED fitting with the SD model (Cols. 3 and 4 in Table 4), we calculated the range of possible spectral albedo values for disks observed at $\lambda = 1.6$ $\mu$m (central wavelength of the IRDIS broadband H filter). For all resolved exo-Kuiper belts listed in Table 4, the albedo values lie within the range [0.54, 0.68], assuming an astrosilicate dust composition. Higher albedo values are obtained for SDs with smaller minimum grain sizes. Given the narrow range of derived albedo values, it is unlikely that dust albedo is the primary factor behind the disk non-detections.

#### 6.3.3. Detection of the HD 181327 debris belt versus non-detection of the inner belt around HD 107146

The difference in scattering efficiency of the dust material may explain the detection of the HD 181327 debris belt and the non-detection of the inner belt around HD 107146 in the H-band polarimetric observations with IRDIS. Both stars were observed using the same instrumental setup, under comparable observing conditions, and with nearly identical total exposure times.

The F6V star HD 181327 ($L_\star = 2.88\,L_\odot$, 18 − 23 Myr, $f_{disk} = (2.6 \pm 0.7) \times 10^{-3}$) hosts a debris belt at a radial distance of 82 au, inclined at 30° (Table 2). As mentioned in Sect. 5.5, the G2V star HD 107146 ($L_\star = 1.04\,L_\odot$, 50 − 2400 Myr, $f_{disk} = (1.1 \pm 0.3) \times 10^{-3}$) possesses two low-inclination ($i = 19°$) planetesimal belts located at ~50 and 120 au. The stellar illumination of the HD 181327 belt is comparable to that of the inner belt of HD 107146, as it is governed by the ratio $L_\star/R^2_{belt}$, which yields a value of 0.55 Wm$^{-2}$ in both cases. This is two orders of magnitude lower than the solar flux received by Jupiter in the Solar System. Additionally, the small difference in inclination between the two planetesimal belts is not expected to significantly impact the amount of observed polarized scattered light, as seen in the middle panel of Fig. 20. Thus, the stellar illumination and disk viewing geometry are not the primary factors responsible for the non-detection of the 50 au belt in the HD 107146 system, or at least, they do not play a decisive role.

Using the results of SD modeling, we estimated the scattering efficiencies of dust grains in both systems assuming they are composed of astrosilicates. Interestingly, for both systems we obtained the same SD power-law index of 3.42 but different minimum grain sizes: $a_{min} = 1.00 \pm 0.18\,\mu$m for the HD 181327 belt and $a_{min} = 2.79 \pm 1.17\,\mu$m for the HD 107146 belt. Such a difference in minimum grain sizes would lead to different bulk albedo values in the H band ($\lambda_c = 1.6\,\mu$m): $\omega_H = 0.61$ for HD 181327 and $\omega_{\rm H} = 0.56$ for HD 107146. Consequently, the polarized scattered flux from the HD 181327 belt could be 1.1 times higher than that from the HD 107146 inner belt, assuming a comparable number of scattering particles in both belts.

[12] The reflectance spectrum exhibits a red color when the disk's scattered flux, normalized by the stellar flux, increases with wavelength, a blue color when it decreases with wavelength, and a gray color when it remains approximately constant across the wavelength range.

However, it is entirely possible that the dust particles in the HD 107146 belt have a different composition, with lower scattering efficiency or a lower maximum polarization fraction of scattered light compared to the dust around HD 181327. Moreover, the dust spatial distribution and total mass may be the key factors contributing to the non-detection. Despite both systems exhibiting a high IR excess, the majority of dust in the HD 181327 system is confined to one relatively narrow debris belt, whereas in HD 107146, the total dust mass is distributed across at least two cold belts. According to our SD modeling results, the dust mass of the inner belt in HD 107146, and therefore the number of scattering particles, is estimated to be approx. 5.5 times lower than that of the HD 181327 debris belt (Col. 7 in Table 4). ALMA observations of both disks (Marino et al. 2016, 2018) show that the belt area of the HD 181327 disk ($R_{\rm belt} \times \Delta R_{\rm belt} = 86 \times 23.2 = 2 \times 10^3$ au$^2$) is larger than the area of the inner belt in HD 107146 ($R_{\rm belt} \times \Delta R_{\rm belt} \approx 50 \times 30 = 1.5 \times 10^3$ au$^2$). This implies that the dust surface density, and consequently the disk's SB in scattered light, may be up to four times higher for the HD 181327 belt, making its detection in polarized light with SPHERE significantly more likely.

#### 6.3.4. Parametric approach for deriving optical properties of dust grains

In the following sections, we introduce a new diagnostic approach for deriving the bulk albedo and maximum polarization fraction of dust grains based on scattered-light and polarized-light images combined with parametric modeling. This method provides independent albedo estimates that can be directly compared with the values obtained from Mie theory discussed in Sects. 6.3.1 and 6.3.2.

In most cases, the single scattering albedo of disk material cannot be directly retrieved from disk images. This is because the observer measures only a fraction of the total scattered flux, which is governed by the material's SPF or, equivalently, its differential scattering cross section $d\sigma_\lambda^{\rm sca}/d\Omega$. This quantity describes the fraction of incident light scattered into a specific direction per unit solid angle $\Omega$, so that

$$\sigma_\lambda^{\rm sca} = \int \frac{d\sigma_\lambda^{\rm sca}}{d\Omega}\, d\Omega = \int \sigma_\lambda^{\rm sca} \times SPF\, d\Omega.$$

Additionally, the total scattered flux, $F_{{\rm sca}\,\lambda}$, measured in disk images[13] is an integrated flux from disk regions with varying scattering angles $\theta$, and is further influenced by the disk's viewing geometry (Sect. 6.2). This geometry is characterized by parameters such as the disk inclination, $i$, the radius of the planetesimal belt, $R_{\rm belt}$, the disk opening angle, $H_0/r_0$, and the exponents of the radial power laws, $\alpha_{\rm in}$ and $\alpha_{\rm out}$ (see Sect. 4.5.1):

$$F_{{\rm sca}\,\lambda} = f\left(\frac{d\sigma_\lambda^{\rm sca}}{d\Omega},\, i,\, R_{\rm belt},\, H_0/r_0,\, \alpha_{\rm in},\, \alpha_{\rm out}\right).$$

Nevertheless, optical properties of the dust in debris disks can be constrained with a comparison between the amount of observed scattered radiation and the IR excess, which represents very roughly the relation between dust scattering and absorption. To examine this relation, several debris disk studies have estimated the so-called disk single scattering albedo $\omega_{{\rm disk}\,\lambda}$. This was done by computing the ratio of the total scattered flux derived from the disk image, $F_{{\rm sca}\,\lambda}$, to the disk's IR excess, $L_{\rm IR\,disk}/L_\star$, (e.g., Schneider et al. 2014; Choquet et al. 2018; Engler et al. 2023) according to equation

$$\omega_{{\rm disk}\,\lambda} = \frac{F_{{\rm sca}\,\lambda}/F_{\star\,\lambda}}{F_{{\rm sca}\,\lambda}/F_{\star\,\lambda} + L_{\rm IR\,disk}/L_\star}, \quad (11)$$

where $F_{\star\,\lambda}$ denotes the stellar flux at wavelength $\lambda$, and $L_{\rm IR\,disk}/L_\star$ is used as a proxy for the absorbed flux at that wavelength. The latter represents a rough approximation, as the IR excess reflects the total emission integrated over the entire IR spectrum, and the dust generating the thermal flux may be located not only at the position of planetesimal belt resolved in scattered light.

When calculated in this manner, the disk scattering albedo is proportional to the dust albedo $\omega_\lambda$ (Eq. 10). However, it also depends on the spatial distribution of dust particles in the disk, as discussed above:

$$\omega_{{\rm disk}\,\lambda} \propto \omega_\lambda \cdot f\left(\frac{d\sigma_\lambda^{\rm sca}}{d\Omega},\, i,\, R_{\rm belt},\, H_0/r_0,\, \alpha_{\rm in},\, \alpha_{\rm out}\right).$$

Therefore, this type of disk albedo can be useful for comparing the scattering properties of two debris disks with the same viewing geometry and stellar irradiance and, ideally, the same SPF. In any case, the measured scattered flux should always be corrected for flux losses introduced by the post-processing of ADI data to ensure accurate comparisons.

The single scattering albedo of dust material can be determined by considering the full angular distribution of scattered light. This requires disk modeling to estimate the scattered spectral luminosity of the disk $L_{{\rm sca}\,\lambda}$. Once this value is obtained, the dust albedo can be derived using the following relation:

$$\omega_\lambda \approx \frac{L_{{\rm sca}\,\lambda}/L_{\star\,\lambda}}{L_{{\rm sca}\,\lambda}/L_{\star\,\lambda} + L_{\rm IR\,disk}/L_\star} = \frac{\langle F_{{\rm sca}\,\lambda}\rangle/F_{\star\,\lambda}}{\langle F_{{\rm sca}\,\lambda}\rangle/F_{\star\,\lambda} + L_{\rm IR\,disk}/L_\star}, \quad (12)$$

where $\langle F_{{\rm sca}\,\lambda}\rangle = L_{{\rm sca}\,\lambda}/4\pi$ is the scattered disk flux averaged over the full solid angle.

The ratio between the observed disk flux $F_{{\rm sca}\,\lambda}$ and the averaged flux $\langle F_{{\rm sca}\,\lambda}\rangle$ can be determined if the shape of the SPF is known, for example, from the disk image model. To demonstrate it, we plot in Fig. 22a the ratio $F_{{\rm sca}\,\lambda}/\langle F_{{\rm sca}\,\lambda}\rangle$ for different HG functions and disk inclinations. By applying a correction factor for the disk inclination, hereafter referred to as the view factor for scattered flux and defined as $f_{{\rm sca}\,\lambda} = F_{{\rm sca}\,\lambda}/\langle F_{{\rm sca}\,\lambda}\rangle$, we obtain

$$\omega_\lambda \approx \frac{F_{{\rm sca}\,\lambda}/F_{\star\,\lambda}}{F_{{\rm sca}\,\lambda}/F_{\star\,\lambda} + f_{{\rm sca}\,\lambda} \cdot L_{\rm IR\,disk}/L_\star}. \quad (13)$$

The averaged scattered flux $\langle F_{{\rm sca}\,\lambda}\rangle$ is equal to the measured disk flux for all inclinations only in the case of isotropic scattering ($g = 0$). As shown in Fig. 22a (see also Schmid 2021), for each HG parameter $g$, there is a specific disk inclination where $f_{{\rm sca}\,\lambda} = 1$. In such a case, e.g., for a disk inclined at 60° with a HG parameter $g = 0.6$, the measured scattered flux does not require any correction, and the single scattering dust albedo can be approximated as $\omega_\lambda \approx \omega_{{\rm disk}\,\lambda}$.

The view factor $f_{{\rm sca}\,\lambda}$ depends only on the shape of the SPF and the disk inclination, making it independent of other disk geometrical parameters. If the scattered-light image is modeled using a single HG function (i.e., without a combination of multiple HG functions), the view factor $f_{{\rm sca}\,\lambda}$ can be directly obtained from Fig. 22a.

[13] Hereafter, we denote the scattered flux measured in the disk image, which represents the total scattered power received by an observer, as $F_{{\rm sca}\,\lambda}$ to distinguish it from the total flux scattered by a single particle, $F_\lambda^{\rm sca}$, defined in Sect. 6.3.1.

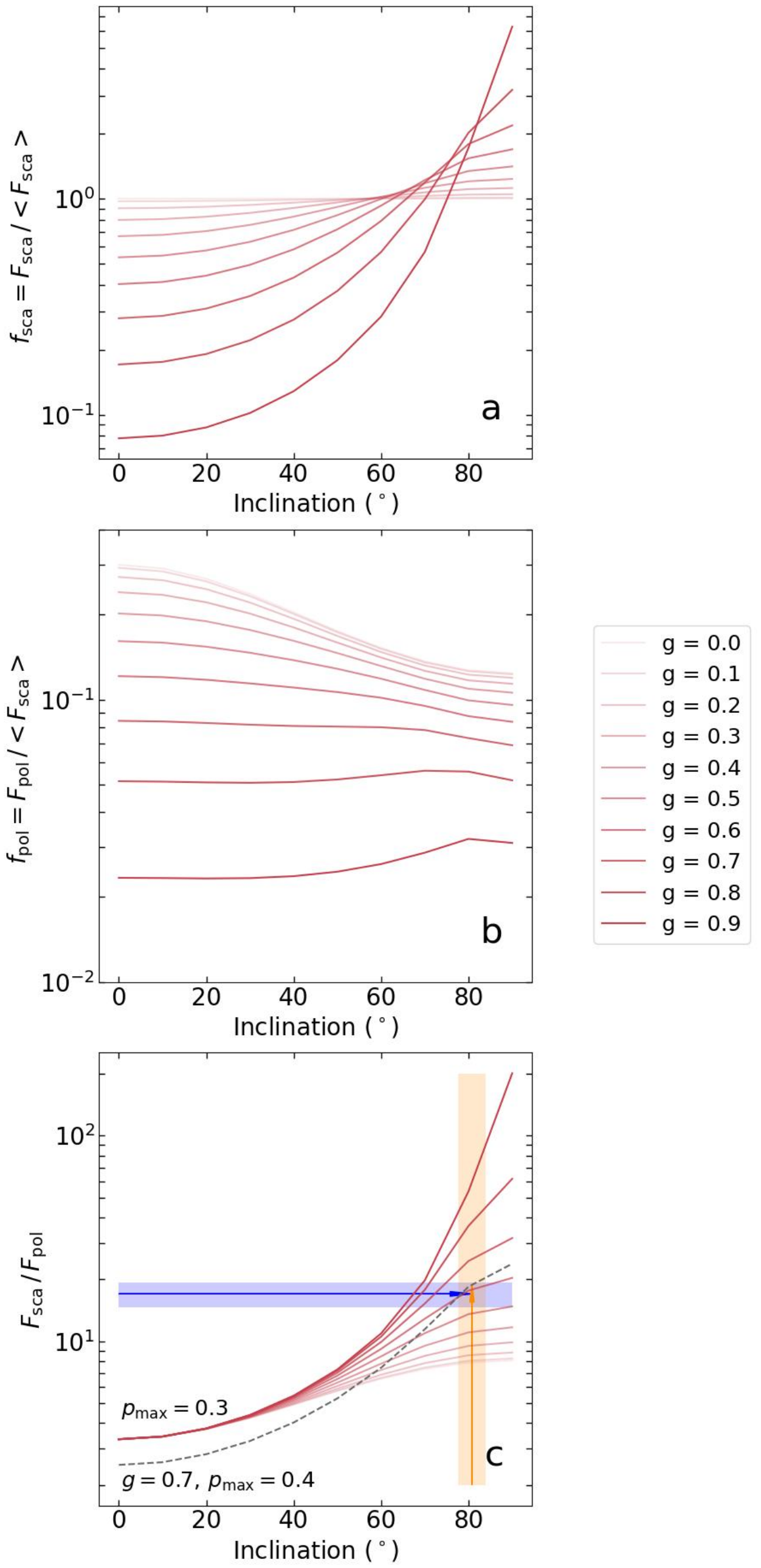

Fig. 22: Ratio of scattered (*panel a*) and polarized (*panel b*) flux to the disk scattered flux averaged over the $4\pi$ solid angle as a function of disk inclination and HG scattering asymmetry parameter. *Panel c:* Ratio of scattered to polarized flux for an optically thin debris disk. The ratios (*panel a* and *b*) define the view factors $f_{sca\,\lambda}$ and $f_{pol\,\lambda}$, respectively. The ratios presented in this figure are calculated using the HG function (as SPF) and Rayleigh-like function with $p_{max} = 0.3$ (as polarization fraction phase function). For a different value of $p_{max}$ the factor $f_{pol\,\lambda}$ should be linearly scaled.

### 6.3.5. Disk albedo with polarized flux. The Λ parameter

In Equation 13, the scattered flux $F_{sca\,\lambda}$ can be replaced by the measured polarized flux $F_{pol\,\lambda}$ using another multiplicative factor $f_{pol\,\lambda} = F_{pol\,\lambda} / \langle F_{sca\,\lambda} \rangle$ (see Sect. 6.3.6) to derive the scattering albedo of the disk:

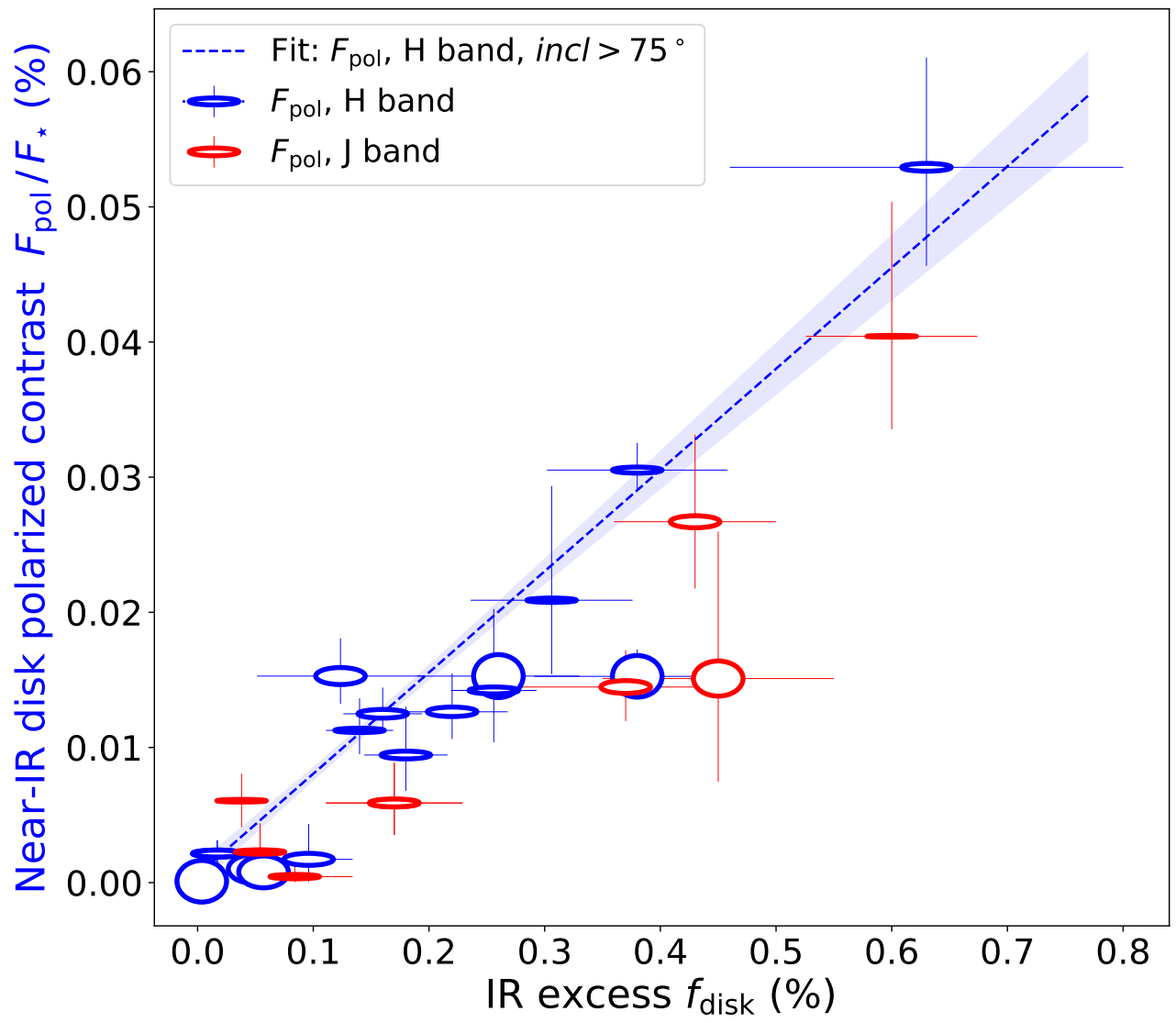

Fig. 23: Measured polarized contrast versus fractional IR luminosity for debris disks observed with SPHERE/IRDIS in broadband H (blue markers) and J (red markers). The axis ratio of each elliptical marker corresponds to the ratio of the minor to major axis of the respective disk, thereby visually representing the disk inclination. The blue dashed line denotes a linear fit to the H-band data for disks with inclinations higher than 75°, while the blue-shaded region indicates the 68% confidence interval for this fit.

$$\omega_\lambda \approx \frac{F_{pol\,\lambda}/F_{\star\,\lambda}}{F_{pol\,\lambda}/F_{\star\,\lambda} + f_{pol\,\lambda} \cdot L_{IR\,disk}/L_\star}. \quad (14)$$

Polarized flux is often easier to measure, particularly for debris disks with lower inclinations, and can be corrected for polarimetric signal losses more reliably (Engler et al. 2018).

Equation 14 can be reformulated in terms of the Λ parameter (see Eq. 17). This parameter quantitatively characterizes the relationship between the measured polarized flux and the fractional IR luminosity of a debris disk (Engler et al. 2017):

$$\Lambda_{pol\,\lambda} = \frac{F_{pol\,\lambda}/F_{\star\,\lambda}}{L_{IR}/L_\star}. \quad (15)$$

Similar to the disk single scattering albedo, $\omega_{disk\,\lambda}$, this observational parameter is proportional to the ratio of dust cross sections for scattering and absorption. Additionally, it depends on two key factors: the polarization fraction function, $p_\lambda(\theta)$, and the maximum polarization fraction of dust material, $p_{max\,\lambda}$.

Analogously to Eq. 15, we define the Λ parameters for the measured scattered flux $F_{sca\,\lambda}$ and disk-averaged flux $\langle F_{sca\,\lambda} \rangle$ as

$$\Lambda_{sca\,\lambda} = \frac{F_{sca\,\lambda}/F_{\star\,\lambda}}{L_{IR}/L_\star} \quad \text{and} \quad \langle \Lambda_{sca\,\lambda} \rangle = \frac{\langle F_{sca\,\lambda} \rangle / F_{\star\,\lambda}}{L_{IR}/L_\star}, \quad (16)$$

respectively. Using these definitions, the Eqs. 12 and 13 can be re-expressed in terms of the Λ parameters as well.

### 6.3.6. Disk polarized contrast versus IR excess for the studied debris disk sample

To derive the Λ parameter for debris disks detected with SPHERE in polarimetric modes, we measured their polarized contrast relative to the star, given by $F_{pol\,\lambda}/F_{\star\,\lambda}$. Most of these disks were observed with IRDIS using broadband filters H and J, indicated in Fig. 23 by blue and red markers, respectively. This figure displays the measured polarized contrast as a function of disk fractional IR luminosity. The elliptical shape of each marker reflects the disk inclination. However, in the case of polarized contrast, the dependence on inclination is relatively weak, significantly less pronounced than for the total scattered flux (Fig. 22a, b).

Figure 23 reveals a positive correlation between polarized contrast and fractional IR luminosity. This correlation is expected and can be attributed to the dust scattering and dust absorption opacities which both depend mainly on the amount and spatial distribution of dust particles.

Although the number of disks with similar inclinations is limited, we fitted a simple linear relation between polarized contrast and fractional IR luminosity to the H-band data, focusing on disks with inclinations greater than 75° (which represent the majority of detected disks). For these disks, the polarized contrast follows the fractional IR luminosity, $f_{\rm disk}$, according to the expression $0.074\, f_{\rm disk} - 8.1 \times 10^{-6}$, as shown by the blue dashed line in Fig. 23. This fit corresponds to a $\Lambda_{\rm disk\,H}$ parameter of $0.074 \pm 0.007$.

A few debris disks with inclinations lower than 75° observed in the H band, as well as most disks observed in the J band, exhibit a lower contrast and consequently a lower $\Lambda$ parameter. The scatter seen in Fig. 23 may reflect variations in the scattering properties of dust grains among disks with similar inclinations or may result from differences in viewing geometry.

The dependence of the data on viewing geometry can be eliminated by dividing the disk's polarized contrast by the view factor $f_{{\rm pol}\,\lambda} = F_{{\rm pol}\,\lambda} / \langle F_{{\rm sca}\,\lambda} \rangle$ (Fig. 22b). Similar to $f_{{\rm sca}\,\lambda}$, the view factor for polarized flux, $f_{{\rm pol}\,\lambda}$, is independent of the radial structure for axisymmetric dust distributions. However, its value is determined by the shape of the pSPF and scales linearly with the maximum polarization fraction of the dust material, $p_{\max\,\lambda}$.

Once the $\Lambda_{{\rm pol}\,\lambda}$ parameter and the view factor $f_{{\rm pol}\,\lambda}$ are determined, the single scattering albedo of the dust material can be estimated as follows:

$$\omega_\lambda \approx \frac{\Lambda_{{\rm pol}\,\lambda}}{\Lambda_{{\rm pol}\,\lambda} + f_{{\rm pol}\,\lambda}}. \tag{17}$$

For example, for disks with inclinations greater than 75° observed in the H band, we derived a value of $\Lambda_{\rm pol\,H} = 0.074 \pm 0.007$. Assuming a maximum polarization fraction $p_{\max\,\rm H} = 0.3$ and a scattering asymmetry parameter $g = 0.9$ for these disks, we obtain a view factor of $f_{{\rm pol}\,\lambda} = 0.033$ from Fig. 22b. Substituting these values into Eq. 17 yields a dust albedo of $0.69 \pm 0.02$. According to Fig. 21, such an albedo is consistent, for instance, with dust grains composed of astrosilicates, either pure or coated with a dirty ice mantle, and having a minimum size of $a_{\rm min} = 1\,\mu$m.

#### 6.3.7. Constraining the maximum polarization fraction from combined scattered and polarized light observations

The combined analysis of total and polarized intensity images enables a more comprehensive assessment of dust scattering properties. Specifically, the ratio of polarized to total scattered flux provides an observational constraint on the maximum polarization fraction, $p_{{\rm max},\lambda}$, of the dust grains at a given wavelength.

This parameter represents the peak of the polarization fraction phase function $p_\lambda(\theta)$ and is sensitive to grain composition, porosity, and SD. As such, it plays a critical role, alongside the single scattering albedo $\omega_\lambda$ and the shape of the SPF, in constraining the optical constants and morphology of the dust population (e.g, Graham et al. 2007; Kirchschlager & Wolf 2014). Because different grain materials and structures (compact vs. aggregate particles) produce distinct polarization signatures, the combination of albedo and $p_{\max\,\lambda}$ allows for a significantly narrower range of viable dust models than can be achieved using albedo alone.

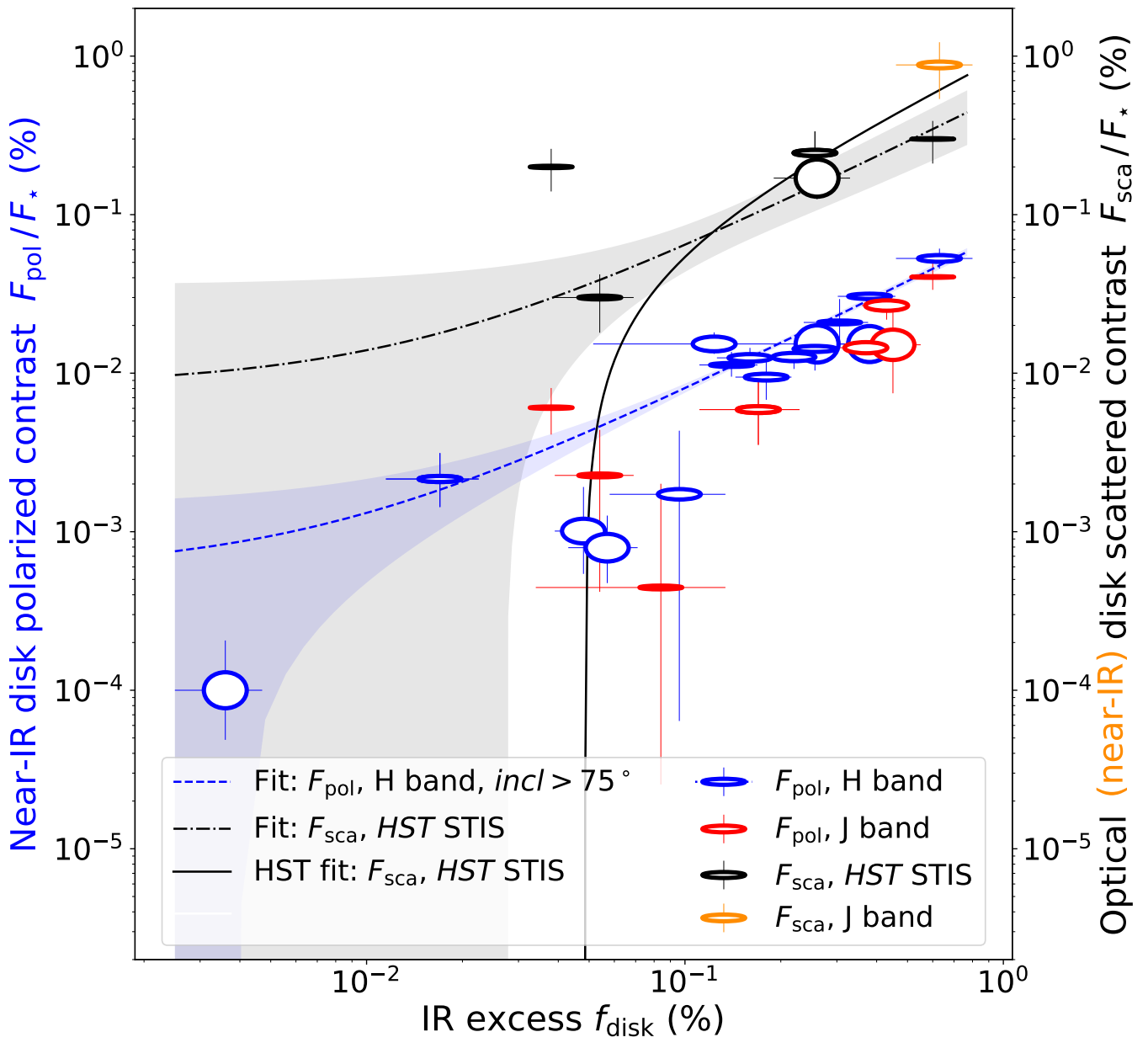

Fig. 24: Polarized disk contrast in the near-IR measured with SPHERE/IRDIS in the broadband H (blue markers) and J (red markers) filters, and total scattered-light contrast in the optical measured with HST/STIS (black markers) plotted against the disk fractional IR luminosity. The orange marker represents the scattered-light contrast for HD 129590 measured in the H band with IRDIS. The axis ratio of each elliptical marker corresponds to the ratio of the minor to major axis of the respective disk, representing its inclination. The black dash-dotted line shows a linear fit to the optical total scattered-light contrast measured with HST/STIS for five debris disks (Sect. 6.3.7). The gray-shaded region indicates the 68% confidence interval for the fit. The black solid line shows the fit to the optical total scattered-light contrast obtained by Schneider et al. (2014) for a set of ten debris disks. The blue dashed line denotes a linear fit to the H-band polarized contrast for disks with inclinations higher than 75°, while the blue-shaded region indicates the 68% confidence interval for this fit.

The analysis presented in this section requires accurately measured total scattered fluxes, which are best obtained from images processed with RDI, a technique that preserves the photometric fidelity of extended disk structures. Using this technique, we measured the total scattered flux of the HD 129590 debris disk (Olofsson et al. 2024) in the H band. This is the only IRDIS image in our sample from which the total scattered flux could be reliably extracted. To broaden the sample and enable a meaningful comparison, we complemented our analysis with scattered light measurements obtained from HST observations at optical wavelengths, where RDI processing is routinely applied and photometric calibration is robust. In the following, we compare total intensity contrasts derived from HST data (Schneider et al. 2014) to the polarized intensity contrasts measured with SPHERE.

Using broadband optical[14] images of ten debris disks obtained with the Space Telescope Imaging Spectrograph (STIS), Schneider et al. (2014) investigated the correlation between IR excess and the optical scattering fraction, analogous to the relation shown in our Fig. 23. Their result is reproduced in Fig. 24 as the black solid line (see also their Fig. 7). They derived a proportionality factor between the fractional scattered flux and the IR excess of $\Lambda_{\rm sca\,opt} \approx 1.05$ (according to Eq. 16), which is significantly higher than the $\Lambda_{\rm pol\,H}$ parameter derived from our polarized flux analysis in the near-IR (Sect. 6.3.6). This difference is expected, as the polarized flux represents only a fraction of the total scattered flux (Sect. 6.2).

[14] $\lambda_p = 575.2$ nm, FWHM of unfiltered passband = 433 nm.

We measured the polarized flux for five of the ten HST targets using SPHERE: in the J band for HD 15115, HD 32297, and HD 197481 (AU Mic), and in the H band for HD 61005 and HD 181327. The HST measurements for these targets are represented by black ellipses in Fig. 24. The SPHERE measurement of total scattered flux in the H band for the HD 129590 debris disk is marked with an orange ellipse in the same figure. In total, we thus have six targets for which both total and polarized scattered fluxes are available. However, for five of these, the measurements come from different instruments, HST for total intensity and SPHERE for polarized intensity, while HD 129590 is the only case where both measurements were obtained from the same dataset.

To provide a general comparison between HST/STIS and SPHERE/IRDIS measurements, we performed a linear fit to the STIS data points, yielding a best-fit slope of $\Lambda_{\rm sca\,opt} = 0.56 \pm 0.24$. This fit, shown as the black dash-dotted line in Fig. 24, is consistent within a $2\sigma$ confidence interval with the trend reported by Schneider et al. (2014). The result suggests that the fractional scattered fluxes measured in optical total intensity images are, on average, approx. 7.5 to 10 times higher than those derived from near-IR polarized intensity images.

A comparison of the five individual targets for which both optical scattered and near-IR polarized fluxes are available shows that the optical fluxes exceed the polarized fluxes by factors ranging from 7 (HD 32297) to 33 (HD 197481). In terms of magnitudes, this corresponds to differences of $2.18^m$ for HD 32297 (A0V, ~30 Myr) and $3.80^m$ for HD 197481 (M1V, 18-23 Myr), both of which are edge-on systems. For the other three debris disks, the magnitude difference falls within a comparable range: $2.61^m$ for HD 181327 (F6V, $i = 30°$, 18-23 Myr), $2.81^m$ for HD 15115 (F4V, $i = 85°$, 10-500 Myr), and $3.09^m$ for HD 61005 (G8V, $i = 82°$, 45-55 Myr). For the IRDIS measurement of the HD 129590 disk (G3V, $i = 81°$, 14-18 Myr), the ratio between the total scattered and polarized fluxes is $16.9 \pm 2.4$ ($3.07^m \pm 0.14^m$).

Using this ratio, we can estimate the value of maximum polarization fraction of the HD 129590 debris particles, as illustrated in Fig. 22c. In this figure, the blue arrow and shaded area represent the measured flux ratio and its uncertainty, while the orange arrow and shaded area indicate the measured disk inclination and its associated uncertainty. The intersection point of these two indicators should lie on the curve corresponding to the HG asymmetry parameter derived from the modeling of the disk image. In the case of HD 129590, the intersection falls on the curve with $g = 0.6$, which is the lower bound of the modeling result, suggesting a maximum polarization fraction of approx. 0.3. To test the compatibility of modeling result with higher values of $p_{\rm max}$, the plotted curves can be rescaled. As seen in Fig. 22c, the measured flux ratio and inclination are also consistent with a solution adopting $g = 0.7$ and $p_{\rm max} = 0.4$ (black dotted line), which is closer to the best-fit value for $g$ obtained for this target.

A similar conclusion can be drawn from Fig. 25, where we plot the ratio of the measured polarized to scattered flux as a function of the disk IR fractional luminosity for systems with available flux measurements from both HST/STIS and SPHERE/IRDIS images of total scattered intensity, as presented in Fig. 24. As discussed above, the data point for HD 129590 (orange marker in Fig. 25) is the only case where both the polarized and scattered fluxes were derived from the same IRDIS H-band dataset. For the remaining five data points (black markers in Fig. 25) the total scattered flux was measured from HST/STIS, while the polarized flux was obtained from IRDIS polarimetric observations in either the J or H band. Consequently, when estimating the polarization fraction in the near-IR, one should consider that the positions of the black markers in this figure may shift, if the contrast in scattered light differs between the optical and near-IR wavelengths.

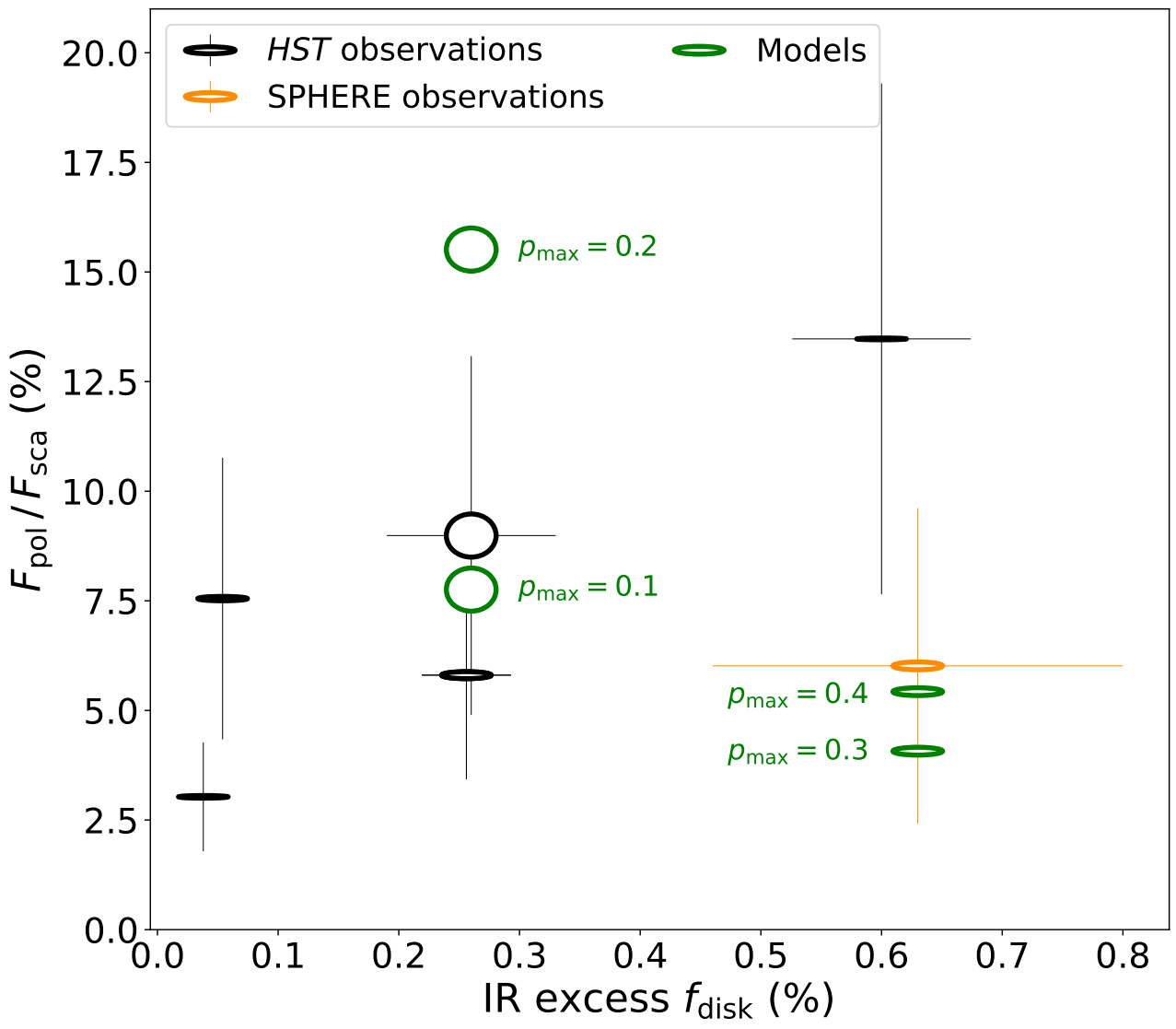

Fig. 25: Ratio of polarized and scattered flux as a function of disk IR excess for debris disks observed with HST/STIS (black markers) and for the HD 129590 disk measured with SPHERE/IRDIS in the H band (orange marker). Green markers indicate model-predicted positions for a disk inclined at 30° with $p_{\rm max} = 0.1$ and $p_{\rm max} = 0.2$ (for comparison with HD 181327 data point), and a disk inclined at 80° with $g = 0.7$ and $p_{\rm max} = 0.3$ or $p_{\rm max} = 0.4$ (for comparison with HD 129590 data point). The axis ratio of each elliptical marker corresponds to the ratio of the minor to major axis of the respective disk, thereby visually representing the disk inclination.

According to results presented in Fig. 25, the ratios of polarized to scattered flux are below 15% for all six considered debris disks, regardless of inclination or IR excess. These values are broadly consistent with the inverse of the modeled ratio $F_{\rm sca\,\lambda} / F_{\rm pol\,\lambda}$, assuming a maximum polarization fraction of $p_{\rm max} = 0.3$ (see right panel in Fig. 20). This means, that, if the fractional scattered flux, or contrast, in the near-IR is the same as at the optical wavelengths, the near-IR maximum polarization fraction of scattered light for these disks might be in the range $25 - 35\%$.

The largest difference to the modeling results adopting $p_{\rm max} = 0.3$ is observed for HD 181327. Two possible explanations may account for this deviation. First, if the scattered-light contrast is similar across optical and near-IR wavelengths, the HD 181327 disk may intrinsically exhibit a lower maximum polarization fraction than $p_{\rm max} = 0.3$. This scenario can be readily assessed, as the modeled ratio $F_{\rm sca\,\lambda} / F_{\rm pol\,\lambda}$ varies only marginally with $g$ for disks inclined at 30° (see Fig. 22c). By rescaling this ratio, we estimate the location of HD 181327 disk in Fig. 25 for assumed values of $p_{\rm max} = 0.1$ and $p_{\rm max} = 0.2$, indicated by annotated green markers. The marker for $p_{\rm max} = 0.1$ lies closest to the observed position of HD 181327, suggesting a maximum polarization fraction of ~12% for this disk. This value is about half of that estimated by Milli et al. (2024), who used HST/NICMOS F110W filter data to calibrate the total scattered flux inferred from the IRDIS total intensity image obtained using RDI technique, along with the same polarimetric dataset analyzed in this work. Therefore, an alternative explanation maybe valid, namely, that the optical scattered-light contrast measured by Schneider et al. (2014) is significantly higher than the near-IR scattered-light contrast. This system features a prominent halo of small grains, which is well resolved in the HST data due to the disk's low inclination. The contribution from this extended halo likely inflated the total scattered flux measured in the optical, resulting in a lower estimated flux ratio that does not accurately reflect the maximum polarization fraction in the near-IR.

An advantage of using low-inclination disks ($i < 40°$) in this diagnostic is that the ratio $F_{\mathrm{sca}\,\lambda}$ / $F_{\mathrm{pol}\,\lambda}$ remains nearly constant across all values of the HG asymmetry parameter (see Fig. 22c). Consequently, the positions of these disks in the diagnostic diagram are primarily sensitive to the assumed maximum polarization fraction $p_{\mathrm{max}}$ and are largely independent of $g$. This is not the case for higher-inclination systems, such as HD 129590, where the ratio $F_{\mathrm{sca}\,\lambda}/F_{\mathrm{pol}\,\lambda}$ shows a stronger dependence on both $g$ and $p_{\mathrm{max}}$. In Figure 25, for instance, the calculated position of the disk corresponding to $p_{\mathrm{max}} = 0.4$ and $g = 0.7$ lies closer to the observed value than the one with $p_{\mathrm{max}} = 0.3$ and the same $g$. However, a similar good agreement as for $p_{\mathrm{max}} = 0.4$ can also be obtained with $p_{\mathrm{max}} = 0.3$ and a lower asymmetry parameter of $g = 0.6$, since the modeled positions shift upward with decreasing $g$ or increasing $p_{\mathrm{max}}$. Although values of $p_{\mathrm{max}} > 0.4$ may appear to better match the data for HD 129590, given its best-fit asymmetry parameter of $g = 0.78$ (see Table E.1), the actual maximum polarization fraction in this system could be lower than 0.4. As discussed in Sect. 4.4, HD 129590 exhibits a double-belt structure, with the inner belt being significantly brighter than the outer one. For simplicity, we modeled this system using a single-belt model, resulting in a best-fit radius that lies between the measured radii of the two components. The dominance of the bright inner belt may have biased the modeling toward a higher apparent value of $g$. Therefore, the true asymmetry parameter could be lower than the derived value, which in turn could imply a maximum polarization fraction below 0.4.

It is important to note that results discussed in this section assume a simple HG function for the SPF and a Rayleigh-like function for the polarization fraction phase function. If alternative forms of the SPF and pSPF are used, the diagnostic relationships $F_{\mathrm{sca}\,\lambda}/F_{\mathrm{pol}\,\lambda}$ would need to be recalculated using a disk model.

#### 6.3.8. Averaged scattered flux for the studied debris disk sample

For each disk shown in Fig. 24, the averaged scattered flux $\langle F_{\mathrm{sca}\,\lambda}\rangle$ can be derived by dividing the measured scattered and polarized fluxes by the corrections factors $f_{\mathrm{sca}\,\lambda}$ and $f_{\mathrm{pol}\,\lambda}$, respectively, as described in Sect. 6.3.6. To illustrate this approach, we determined the view factors using Fig. 22 and assuming an asymmetry parameter of $g = 0.7$ (the average value from our disk modeling) and a maximum polarization fraction of $p_{\mathrm{max}} = 0.3$ for all disks, regardless of the observation wavelength. Figure 26 displays the positions of the disks in the parameter space $[\langle F_{\mathrm{sca}\,\lambda}\rangle/F_\star, f_{\mathrm{disk}}]$ following this flux correction.

If both the scattered and polarized fluxes for a given disk are measured at the same wavelength, as is the case for the HD 129590 disk (orange and blue markers on the far right in Figs. 24 and 26), and if the adopted values for $g$ and $p_{\mathrm{max}}$ are appropriate for this disk, then the corrected scattered and polarized flux markers should overlap. This overlap indicates a consistent estimate of the disk's averaged scattered flux $\langle F_{\mathrm{sca}\,\lambda}\rangle$. As shown in Fig. 26, a good match using the adopted view factors is achieved for two HST targets with the lowest IR contrast. For the remaining four targets, the corrections yield different values of the average scattered flux, although the discrepancy for HD 129590 remains within the error bars.

For the adopted values for $g$ and $p_{\mathrm{max}}$, the location of the data points in the diagram lies close to the line $\langle F_{\mathrm{sca}\,\lambda}\rangle/F_\star = L_{\mathrm{IR}}/L_\star$, indicating that comparable fractions of stellar radiation interacting with dust particles are either scattered or absorbed. This is equivalent to the statement that the typical scattering albedo is $\omega \approx 0.5$. This rough estimate relies on the assumption of a constant albedo across wavelengths, which may not be valid. For large particles, one would expect $\omega > 0.5$ because a value of $\omega \approx 0.5$ is already expected from diffraction. A scattering contribution from radiation interacting with the particle would then provide a total $\omega > 0.5$. For small particles, such as interstellar dust, one typically expects $\omega \lessapprox 0.5$ in the near-IR, along with a relatively low asymmetry parameter, $g \lessapprox 0.5$, but still a high scattering polarization $p_{\mathrm{max}}$ would be possible (Draine 2003b). This type of dust would not be consistent with the collected observational data of debris disks.

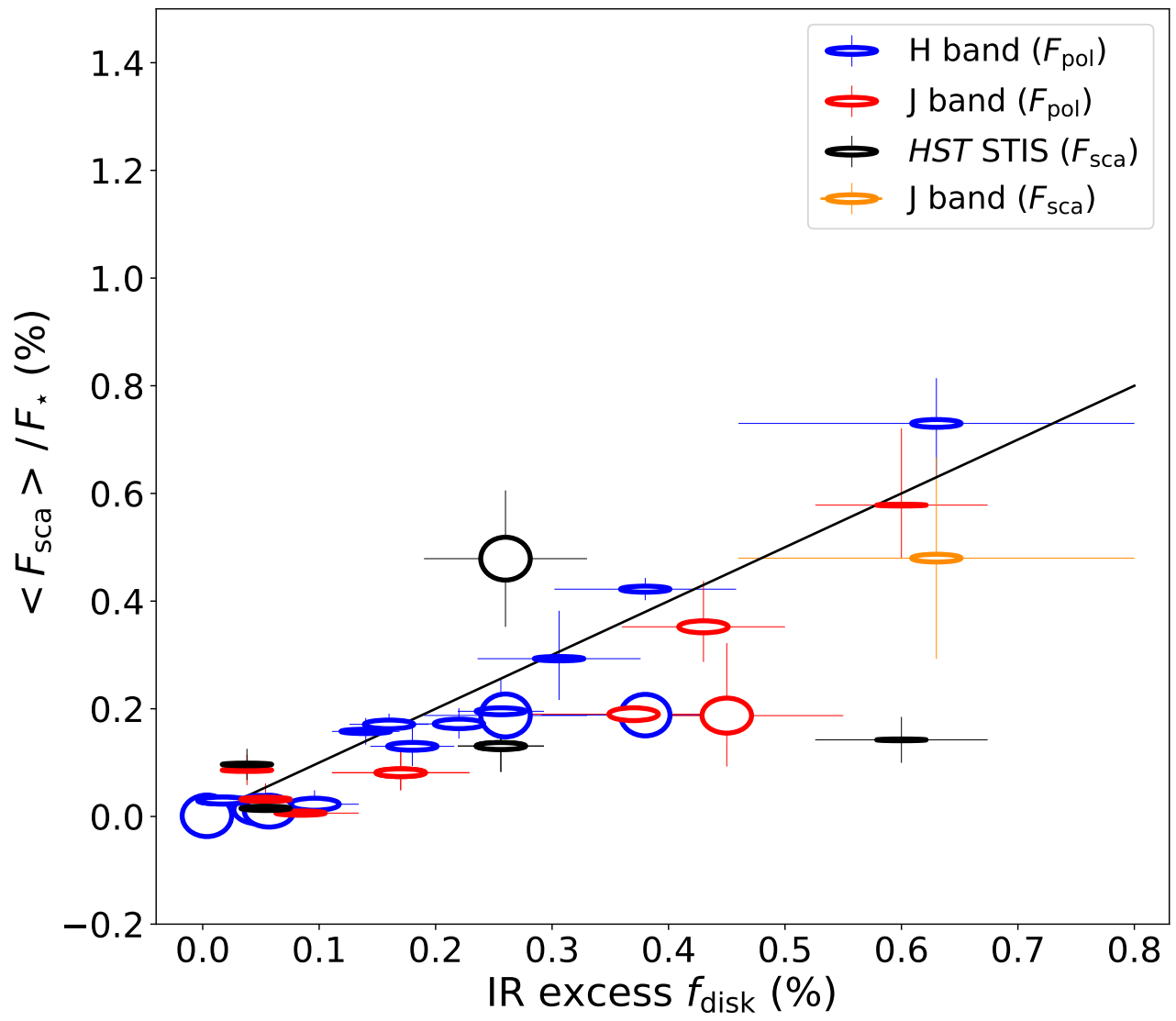

Fig. 26: Averaged scattered-light disk contrast plotted against disk fractional IR luminosity for measurements obtained with SPHERE/IRDIS in the broadband H (blue and orange markers) and J (red markers) filters, as well as with HST/STIS (black markers), as shown in Fig. 24. The averaged scattered-light contrasts were calculated from measured either polarized or total scattered fluxes, as indicated in the figure legend, and assuming an asymmetry parameter $g = 0.7$ and a maximum polarization fraction $p_{\mathrm{max}} = 0.3$ for all disks. Marker shapes reflect disk inclinations. The black solid line indicates the 1:1 relation.

As a final remark, we note that adopting higher values of the asymmetry parameter $g$ or lower values of the maximum polarization fraction $p_{\mathrm{max}}$ to derive the view factor for the polarized contrast shifts the data points upward in the diagram presented in Fig. 26, placing them above the line $\langle F_{\mathrm{sca}\,\lambda}\rangle/F_\star = L_{\mathrm{IR}}/L_\star$. This would imply a typical scattering albedo greater than 0.5, which is more consistent with the results obtained from our SD modeling (Sect. 6.3.1).

### 6.4. Observing conditions

Due to the faint and extended nature of debris disks in scattered light, their detection is strongly influenced by both ambient conditions (e.g., coherence time, seeing, wind speed) and observational parameters such as airmass, detector integration time (DIT or exposure time per frame), total exposure time, and, in the case of ADI, the total rotation angle of the sky field during the observation. For successful DI of debris disks with SPHERE instruments, the coherence time should ideally exceed 3 – 4 ms, while the seeing conditions should remain below 0.8″. If the coherence time is short (< 3 ms), the stellar position behind the coronagraph can deviate by more than one pixel from the intended cen-ter position. Misalignment in individual frames reduces the S/N in the final science image after frame combination. Similarly, high atmospheric turbulence (seeing > 0.8″) or strong variability in the PSF adversely impacts detection.

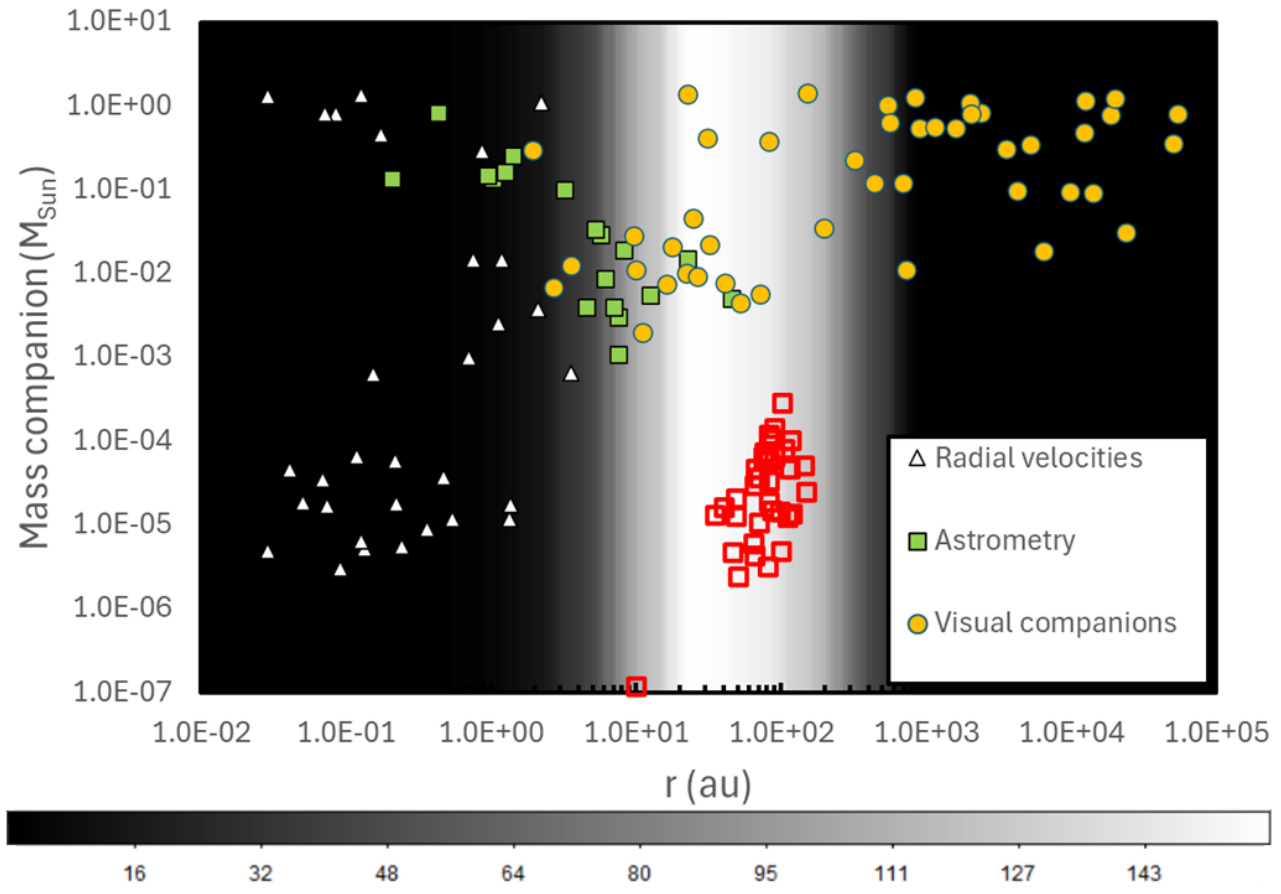

Fig. 27: Distribution of debris disks detected by SPHERE in the separation $a$ - mass plane (red open squares). For comparison, the distributions of stellar and planetary companions is also shown (see Tables F.1 and F.2). White triangles are companions detected using RV; orange filled circles are companions detected in DI; green filled squares are companions whose presence is deduced from astrometric perturbation. The masses of the debris disks have been multiplied by 100 to reduce the size of the plot. The gray background gives the number of program stars where the search of debris disks by SPHERE is sensitive (see scale on bottom of the figure).

This effect was notably observed during polarimetric observations of HD 117214 with ZIMPOL, where the degrading seeing conditions over the two-hour observing run significantly reduced the detectability of the debris disk (Engler et al. 2020). When the seeing exceeds 0.95 − 1″, the stellar PSF often becomes highly variable, making it unlikely to resolve a debris disk, even if the disk is intrinsically bright. However, in ADI observations, a sufficiently large sky rotation angle and a robust number of frames allow for selective frame combination, which can improve the S/N by applying thresholds on seeing (or equivalently PSF FWHM).

The quality of data obtained with SPHERE instruments is also affected by wind speed at Paranal Observatory. The optimal wind speed range for observations lies between 2 and 5 m s$^{-1}$. When wind speeds drop below 1 m s$^{-1}$, data quality is impacted by the "low wind effect" (LWE; Milli et al. 2018). To mitigate this issue, a special low-emissivity coating for the telescope spiders was introduced in August 2017. Before this modification, SPHERE observations conducted at wind speeds below 4 m s$^{-1}$ were often degraded by the LWE. Conversely, high-altitude turbulence with wind speeds exceeding 5 m s$^{-1}$ produces a wind-driven halo within the AO correction radius, thereby reducing achievable contrast (Cantalloube et al. 2018).

Furthermore, debris disks observed at high airmass (> 1.7) are essentially undetectable. The angular size of the disk also plays a crucial role; it must fit within the FOV of the instrument or exceed its inner working angle. This requirement is not met for some targets in our sample, such as HD 3003, where the suspected debris disk around the primary star of this binary system likely has an angular size smaller than 0.1″.

From the analysis of detections and non-detections of debris disks, we conclude that numerous observational requirements must be fulfilled for successful imaging of debris disks in scattered light with SPHERE. Under optimal conditions, debris disks with low IR excess (∼$10^{-4}$) can be imaged, whereas even bright disks ($f_{\rm disk} > 10^{-3}$) may remain undetected under suboptimal conditions. These considerations are critical for interpreting debris disk brightness in scattered light and estimating their scattering albedos, particularly when comparing different systems.

## 7. Debris disks as integral components of stellar systems

Along with stellar, substellar and planetary companions, debris disks represent key components of stellar systems, each tracing different aspects of their formation and dynamical evolution. Bright debris disks around young stars serve as visible signposts of ongoing or past planet formation, and their morphology offer insights into the dynamical environment of a system, often shaped by gravitational interactions with nearby massive bodies. Understanding how debris disks relate spatially and dynamically to both exoplanets and stellar companions is essential for building a more complete picture of young planetary systems.

In this section, we explore these relationships by examining debris disks detected with SPHERE in the context of the companions present around the stars in our sample. Among these systems are 23 young stars with confirmed exoplanets, discovered through various detection techniques, including DI, astrometry, transits, and radial velocity (RV) measurements. Table F.1 lists the parameters of these exoplanets, including only those with masses below 13 $M_{\rm Jup}$, the threshold below which companions are categorized as planets. All other types of companions, both confirmed and candidates, are described in Appendix F and listed in Table F.2.

### 7.1. Companions to the program stars

The debris disks detected with SPHERE occupy a specific region within stellar systems, clearly distinct from the regions where stellar and planetary companions are found. This distinction is likely due to both the intrinsic characteristics of debris disks and selection effects that lead to their detection. To explore these, we considered the separation - mass plane presented in Fig. 27. In this figure, we compared the distribution of debris disks with that of companions to the stars listed in Appendix F.

To interpret the locations of debris disks and compact companions around stars, we first briefly review the main selection effects at play. The FOV of SPHERE allows the detection of disks at projected separations of at least 0.1 − 0.15″ from the star. Disks located closer are obscured by the coronagraph and are difficult to detect due to the brightness of the stellar halo. Conversely, disks at separations beyond 5.5″ may fall outside the instrument's FOV, though they may still be detected if observed at a high inclination. Given that the distances to the program stars range from a few to over a hundred parsecs, the number of stars effectively surveyed for debris disks of a given physical size varies. This variation is illustrated by the background grayscale in Fig. 27. We note that the distribution of disk radii is narrower than the region where the survey is complete across the program stars, suggesting that an underlying physical cause, rather than observational bias alone, may be responsible.

Regarding the distribution of stellar companions, very few are found at separations similar to those of the debris disks in this sample. However, we know that the range from a few to hundreds of au is populated by many stellar companions and is the peak of the distribution of stellar companions for solar and A-type stars (Duquennoy & Mayor 1991; Raghavan et al. 2010; De Rosa et al. 2014; Gratton et al. 2023b). The lack of such companions in our sample may reflect the SPHERE observation strategy, as targets with known companions in this separation range were excluded from the SHINE survey, and only shallow observations were performed when such systems were inadvertently included (see Bonavita et al. 2022). However, stellar com-panions at these separations can destabilize debris disks, so it is reasonable to expect that they are absent in most of the stars with disks considered in this study. In an unbiased search for multiples among stars hosting debris disks, Rodriguez et al. (2015) found that the properties of disks in binary systems are not statistically different from those around single stars. However, their sample contained very few stellar companions with semimajor axes in the 10 – 100 au range, where the debris disks in our sample are found.

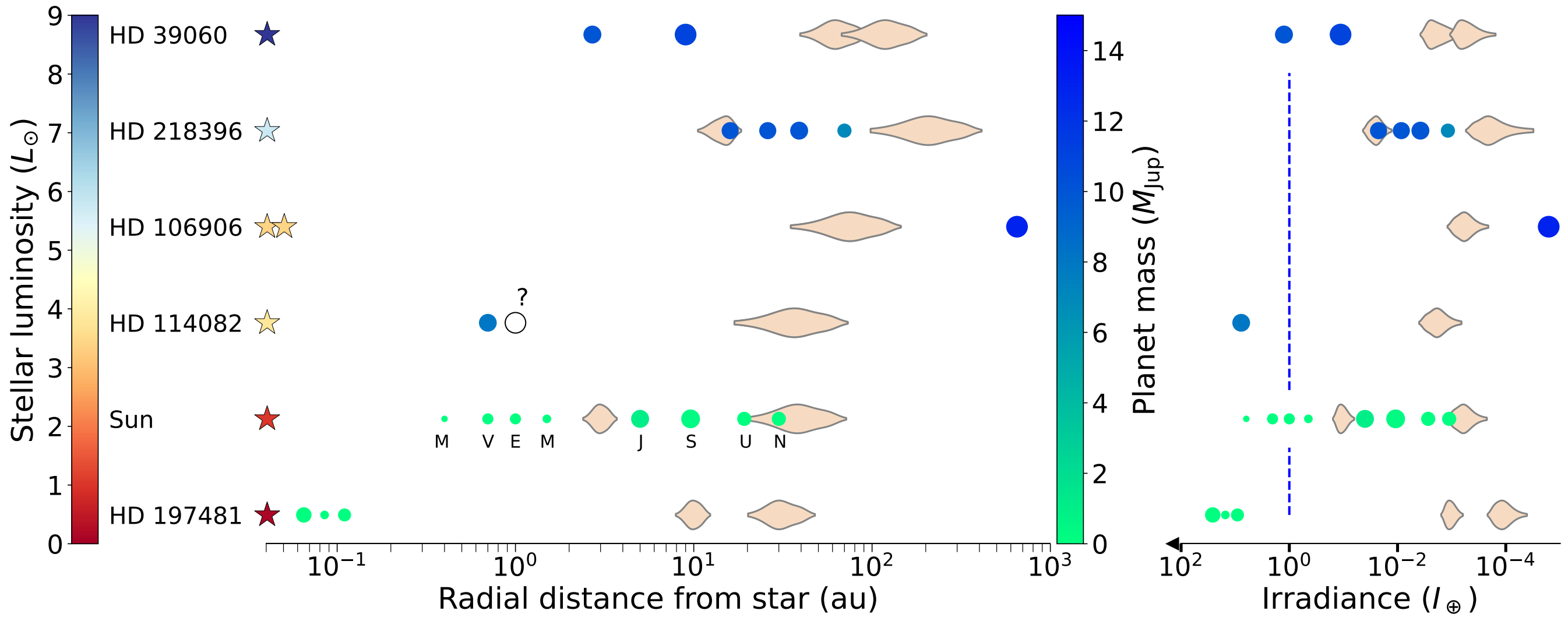

Fig. 28: Architecture of planetary systems in comparison with the Solar System. The sizes of stars and planets are schematically drawn and not to scale. *Left panel:* Planetary systems in which both exoplanets and debris disks have been detected. The letters correspond to the planets in the Solar System. Question mark indicates a candidate planet in the HD 114082 system. *Right panel:* The same planetary systems are shown, but with radial positions re-scaled in terms of irradiance. The blue vertical line marks the location where the stellar flux is $I_\star = 1\, I_\oplus$, corresponding to Earth's solar irradiance.

### *7.2. Architectures of individual planetary systems*

The spatial distribution of planetary companions relative to debris disks provides critical insights into the architecture and dynamical history of planetary systems. In particular, the location of planets within or near debris belts can indicate past migration processes, sculpting effects, and zones of dynamical stability or instability. In the sample of stars observed with SPHERE, the planetary companions ($M_\mathrm{p} < 13\, M_\mathrm{Jup}$) are located closer to their host stars than the resolved debris disks (Fig. 27). This separation is expected from both formation models and observational biases, but it also carries important implications for the interpretation of system dynamics.

Among the systems in our sample, only five (HD 39060, HD 106906, HD 114082, HD 197481, and HD 218396) have both debris disks resolved with SPHERE and confirmed planetary companions. In these cases, the relative radial positions of disks and planets are well constrained, allowing for a detailed investigation of system architecture and the study of how planets shape and interact with circumstellar debris.

Figure 28 presents the configurations of these five systems in comparison with the Solar System, highlighting both similarities and diversity in planet-disk arrangements. This comparative approach helps to identify trends and anomalies that can guide future searches and refine theoretical models. By analyzing these architectures, we can also place constraints on the potential locations and masses of additional, as-yet-undiscovered planets that may reside between known companions and debris structures.

HD 39060 ($\beta$ Pic) The HD 39060 disk is the most prominent debris disk among all targets, as it was the first to be directly imaged (Smith & Terrile 1984). It is highly asymmetric and extends beyond 1000 au (e.g., Janson et al. 2021). This vast disk consists of two cold exo-Kuiper belts, both observed edge-on but with slightly different PAs (Golimowski et al. 2006; Ahmic et al. 2009, this work). Additionally, several inner planetesimal belts are likely present (Okamoto et al. 2004; Wahhaj et al. 2003), though their exact radial positions and orientations remain uncertain; therefore they are not depicted in Fig. 28.

Two giant planets reside in the inner regions of the disk: HD 39060 b, a ~11 $M_\mathrm{Jup}$ planet located at ~9 au (Lagrange et al. 2010), and HD 39060 c, a ~10 $M_\mathrm{Jup}$ at ~3 au (Lagrange et al. 2019). The latter is positioned near the radial distance from the star where the stellar incident flux matches the solar incident flux on Earth, $I_\oplus$. This location is indicated by the blue vertical line in the right panel of Fig. 28, which compares the radial positions of exoplanets in terms of the incident flux or irradiance $I_\star$ from their host stars. For the two planets, a similar irradiance level suggests that their equilibrium temperatures, $T_\mathrm{eq}$ (or BB-equivalent temperatures), are comparable, given that $T_\mathrm{eq} \sim I_\star^{1/4}$.

HD 218396 (HR 8799) The planetary system of HD 218396 exhibits an even stronger resemblance to the Solar System in terms of stellar irradiance levels experienced by its giant exoplanets compared to the Jovian planets (Fig. 28 right panel). Its architectural structure, featuring a warm belt ($R_\mathrm{belt} \sim 15$ au), a cold belt ($R_\mathrm{belt} \sim 180$ au), and four giant planets residing in the space between them, closely parallels the Solar System's main asteroid belt, Edgeworth-Kuiper belt, and the orbits of the Jovian planets in between (Marois et al. 2010).

As discussed in Sect. 4, the polarized intensity image taken in the H band with IRDIS reveals the warm belt spatially resolved for the first time (Fig. 9i). Its radial position, indicated in left panel of Fig. 28, is close to that of planet HR 8799 e, which has a projected separation of 0.39″ ± 0.01″ (~16 au) in the same dataset. This spatial coincidence hints at a possible role for the planet in shaping the inner belt's edge or maintaining its structure through dynamical shepherding.

Table 5: Parameters of planets HD 114082 b and c.

| Planet | $R_p$ $R_{Jup}$ | $a_p$ (au) | $P$ (days) | $i$ (deg) | $b$ | $T_{tr}$ (h) | Reference |
|---|---|---|---|---|---|---|---|
| HD 114082 b | $0.98^{+0.03}_{-0.03}$ | $0.7^{+0.4}_{-0.3}$ | $197^{+171}_{-109}$ | $89.78^{+0.10}_{-0.25}$ | $0.42^{+0.23}_{-0.18}$ | $14.58^{+0.06}_{-0.06}$ | Engler et al. (2023) |
| HD 114082 c | $1.29^{+0.05}_{-0.05}$ | $1.0^{+0.4}_{-0.4}$ | $317^{+199}_{-158}$ | $89.72^{+0.09}_{-0.16}$ | $0.77^{+0.03}_{-0.20}$ | $13.20^{+0.08}_{-0.08}$ | This work |

**Notes.** The columns list planet IDs, modeled planet radii ($R_p$), orbital semimajor axes ($a_p$), orbital periods ($P$) and inclinations ($i$), transit impact parameters ($b$), and transit durations ($T_{tr}$).

HD 106906 The HD 106906 is another prominent planetary system hosting a debris disk around a spectroscopic binary, composed of two F5V stars with nearly equal masses (Rodet et al. 2017). The edge-on disk is oriented ~21° away from a planetary-mass companion, which is located at a large projected separation of 650 au from HD 106906 AB (e.g., Bailey et al. 2014; Lagrange et al. 2016; Kalas et al. 2015). The disk appears symmetric in SPHERE near-IR images (Lagrange et al. 2016), in contrast to optical-wavelengths images, which reveal a needle-like structure. A similar morphological difference is observed in the near-IR and optical images of the HD 15115 debris disk (Kalas et al. 2007; Engler et al. 2019) suggesting that the asymmetric, needle-like appearance of edge-on disks is predominantly shaped by submicron-sized dust particles.

HD 114082 HD 114082 is a young F3V star belonging to the Lower Centaurus Crux subgroup of the Sco-Cen association. Its debris disk, with a fractional luminosity of $f_{disk} = 3.8 \times 10^{-3}$, ranks among the brightest disks observed in scattered light to date (Wahhaj et al. 2016; Engler et al. 2023). A transiting super-Jovian planet was detected orbiting the star at a radial distance of 0.5 – 0.7 au using the RV and transit techniques (Zakhozhay et al. 2022b; Engler et al. 2023). If the predicted orbital parameters of this planet are accurate, its orbit is not co-planar with the midplane of the debris disk, which has an inclination of ~83°.

A second transiting candidate has been detected around HD 114082 in photometric data from the Transiting Exoplanet Survey Satellite (TESS; Ricker et al. 2015). The system was observed by TESS in Sectors 38, 64, and 65, with a second distinct transiting event clearly identified in Sector 64. However, this event does not correspond to HD 114082 b, as the observed transit duration and depth do not match those previously reported by Zakhozhay et al. (2022b). This second single-transit event is not associated with an asteroid crossing or centroid shifts, suggesting that it may be caused by a second planetary body.

To analyze this candidate, we download the TESS light curves produced by the Science Processing Operation Center (Jenkins et al. 2016) from the Mikulski Archive for Space Telescopes. For photometric modeling, we selected the pre-search data conditioning simple aperture flux (PDCSAP) and associated errors. Using the software package `juliet` (Espinoza et al. 2019) with the `dynesty` sampling method, we modeled the TESS light curve of Sector 64, incorporating a transiting planet model and a Gaussian process to account for stellar variability. A log-uniform prior ranging from 33 to 1000 days was set for the orbital period of the candidate planet. Additionally, we used the stellar parameters derived in Zakhozhay et al. (2022b) to impose a normal prior on the stellar density, which, when combined with the transit model, constrains the planetary transit speed across the stellar disk.

For this second planetary candidate, we obtain a large radius of $1.29^{+0.05}_{-0.05}$ $R_{Jup}$ and orbital parameters listed in Table 5 along with the orbital parameters of planet HD 114082 b for a comparison. The TESS light curve with the median transit model is shown in Fig. 29.

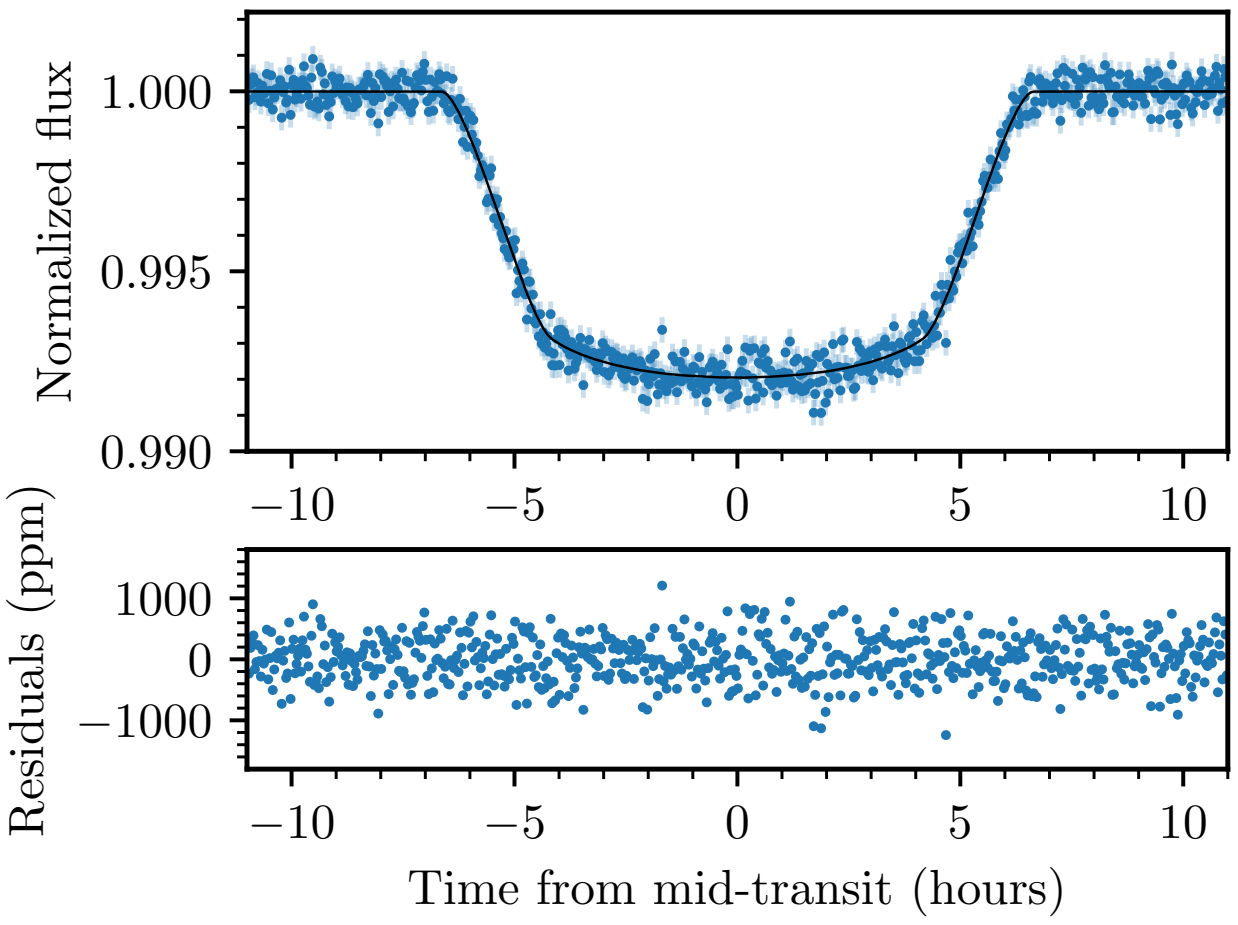

Fig. 29: De-trended TESS light curve from Sector 64 (blue dots) showing the second transiting candidate around HD 114082 with the median transit model (black line).

The system HD 114082 offers a compelling example of a planetary system where giant planets orbit close to the host star, while a spatially extended debris disk lies much farther out and appears misaligned with the inner planetary orbits, so that their planes are inclined by at least 6-7 degrees relative to the disk plane.

Such orbital misalignment between planets and debris disks have important implications for the dynamical history of the system. Several mechanisms have been proposed to explain such configurations. One possibility is planet–planet scattering, where gravitational interactions between giant planets lead to the ejection of one or more bodies and a reconfiguration of the survivors onto eccentric and inclined orbits. This process can disrupt the coplanar architecture established during the PPD phase and has been invoked to explain the high eccentricities and inclinations observed in many giant exoplanets (e.g., Chatterjee et al. 2008; Jurić & Tremaine 2008; Raymond et al. 2011).

Another mechanism is secular perturbations from additional, possibly undetected, massive companions. If a distant planet is present on an inclined orbit, it can induce long-term precession in the orbits of inner planets or in the disk itself, leading to gradual misalignment over time. This mechanism, particularly the Kozai-Lidov effect, has been shown to cause significant orbital inclination variations under specific conditions (e.g., Nagasawa et al. 2008; Naoz et al. 2013).

In the case of HD 114082, the wide separation between the planetary system and the debris belt, more than a factor of 30 in radius, creates a dynamical environment in which such long-term perturbations are plausible. Furthermore, the large radial cavity separating the planets from the debris disk may itself be sculpted by additional low-luminosity companions or represent the aftermath of dynamical clearing by now-ejected bodies.

The HD 114082 system offers a compelling case for investigating the early dynamical evolution of tightly packed planetary systems accompanied by outer debris structures. Although TESS has detected single transit events for both planets, their orbital geometries remain poorly constrained due to the limited transit coverage, and the PAs of their orbital planes are still unknown. Continued photometric and spectroscopic monitoring of HD 114082 is essential to refine the orbital parameters of the planets, including their mutual orientation and orbital eccentricities.

In particular, precisely determining the transit timing would enable targeted RV observations during the planet's passage across the stellar disk. This could allow for the detection of the Rossiter–McLaughlin effect (Triaud 2018), a spectroscopic signature that occurs when a transiting planet temporarily blocks part of the rotating stellar surface. Measuring this effect provides the sky-projected angle between the planetary orbital axis and the stellar spin axis, offering insight into the mutual alignment (or misalignment) between the planetary orbital planes and the host star's equatorial plane. This, in turn, helps to reconstruct the three-dimensional architecture of the system and test scenarios of planet migration or dynamical perturbation.

HD 197481 (AU Mic) HD 197481 is an active flaring M1Ve star belonging to the $\beta$PMG and hosts a highly dynamical debris system. A monitoring campaign of AU Mic conducted with SPHERE between 2015 and 2017 revealed multiple dust clumps appearing above and below the disk midplane (Boccaletti et al. 2018), seemingly moving in non-Keplerian orbits around the star. The origin of these fast-moving dust structures is still debated. One possibility is that they are generated in a collisional avalanche, where the main planetesimal belt intersects a debris stream resulting from the catastrophic disruption of a large asteroid-like body (Chiang & Fung 2017). Another explanation suggests that these dust clumps are expelled from the system by AU Mic's strong stellar wind, having been emitted by a parent body, such as a planet or a disk substructure within the main debris belt (Sezestre et al. 2017). Indeed, AU Mic is known to host three confirmed planets that follow close orbits, possibly in a 4:6:9 mean-motion resonance (see Table F.1; Plavchan et al. 2020; Martioli et al. 2021; Wittrock et al. 2023). However, they are too close to the star to be responsible for the observed dust clumps (Boccaletti 2023). In Fig. 28, these planets are shown inside two dust belts, positioned at the locations of radial SB peaks as measured along the disk's major axis in the $r^2$-scaled H-band image from May 20, 2017. The AU Mic's morphology, however, varies over time and resembles an edge-on spiral structure more than a stable belt system.

The comparative architectures shown in Fig. 28 indicate that our Solar System is not unusual but instead falls within the general distribution of planetary systems hosting both debris disks and planets. In most benchmark systems, the giant planets are confined to the region inside the exo-Kuiper belt, typically spanning orbital radii from a few to a few tens of au, while the planetesimal belts extend from several tens to more than one hundred au. An exception is HD 106906, where a massive (13 $M_{\rm Jup}$) companion resides well outside the belt. When scaled by stellar irradiance (right panel in Fig. 28), the exo-Kuiper belt locations cluster around the equivalent solar-system value, underscoring the analogy. This alignment demonstrates that multi-planet arrangements inside wide Kuiper-belt analogs are common, and that the Solar System's configuration, with its giant planets located interior to an outer cold belt, represents a recurring outcome of planetary system formation and evolution.

### 7.3. Theoretical predictions for the masses of undetected planets

To investigate the potential planetary architectures in other systems with detected debris disks in our sample, we applied a set of analytical prescriptions under the assumption that observed gaps and edges of planetesimal belts have a dynamical origin. The adopted method depends on whether the system hosts a single resolved belt or multiple belts. For all systems, we assumed a radial belt width equal to 20% of its resolved central radius in consistency with our SED modeling approach (Sect. 5.4).

For systems with only one resolved belt, we estimated the range of planet masses and semimajor axes for a single planet on a circular orbit capable of sculpting the inner edge of the belt. We assumed that such a planet dynamically clears a region interior to the belt by gravitationally ejecting dust particles. To relate the width of the cleared zone, $\Delta a$, to the planet's mass, $M_{\rm p}$, we adopted the expression derived by Morrison & Malhotra (2015):

$$\Delta a = 1.7 \left(\frac{M_{\rm p}}{M_\star}\right)^{0.31} a_{\rm p},$$

where $a_{\rm p}$ is the semimajor axis of the planet and $M_\star$ is the mass of the host star. Planet masses were explored in the range of $0.1 - 13\,M_{\rm Jup}$, with corresponding semimajor axes calculated to match the observed belt edge. The resulting minimum and maximum allowed semimajor axes, corresponding to the assumed mass bounds, are reported in Table 6.

For systems with two or more resolved belts, we examined the gap between each pair of belts, testing dynamical configurations involving one, two, or three planets on circular orbits capable of maintaining the observed separations. This analysis follows the framework of Lazzoni et al. (2018). In multi-planet scenarios, we assumed equal-mass planets arranged in a maximally packed configuration, thus with the minimum orbital spacing required for long-term dynamical stability, in order to reduce degeneracy in the possible solutions. For each configuration, the resulting planet masses and semimajor axes are reported in Table 7. For HD 141943, only the single-planet scenario is reported, as the resulting mass required to account for the observed belt structure is already extremely low ($5 \times 10^{-6}\,M_{\rm Jup}$), making the consideration of multi-planet solutions unnecessary.

The results of our dynamical analysis suggest that many debris disk systems may host planetary companions whose gravitational perturbations shape the observed disk morphology. This is particularly evident in the case of HD 218396 (HR 8799). In a three-planet configuration with equal-mass planets of 6.78 $M_{\rm Jup}$, our model yields a configuration that closely approximates the known planetary architecture of this system (Sect. 7.2), albeit without reproducing it in full detail.

For three other systems, HD 36546, HD 92945 and HD 120326, we find that a single planet located in the gap between two resolved belts must possess a mass greater than 3.3 $M_{\rm Jup}$, to account for the observed structure. This mass range lies within the detection capabilities of current HCI instruments. In the case of HD 92945, contrast limits achieved in IRDIS and IFS observations on 27 January 2018 were analyzed by Mesa et al. (2021). Using AMES-COND evolutionary models (Allard et al. 2012) to convert IRDIS contrast limits into mass constraints, they derived mass limits between 1 and 2 $M_{\rm Jup}$ at the gap radial position. This comparison indicates that the observed gap is more likely carved by multiple lower-mass planets, although Marino et al. (2019) and Mesa et al. (2021) found that a single planet with a mass of $0.3 - 0.6\,M_{\rm Jup}$ could reproduce the structure, assuming a narrower gap width of 20 au. In contrast, our analysis adopts a gap width of 45.6 au (see Table 7), which would require a more massive perturber. According to our model, the masses of planets in a multi-planet configuration for the HD 92945 system would fall below the SPHERE detection limit of 1 $M_{\rm Jup}$.

Table 6: Planets shaping the inner edges in systems with single belt.

| Single belt | Belt edge (au) | $a_{\rm p\,max}$ 0.1 $M_{\rm Jup}$ (au) | $a_{\rm p\,min}$ 13 $M_{\rm Jup}$ (au) |
|---|---|---|---|
| GSC 7396-0759 | 79.2 | 71.0 | 51.8 |
| HD 105 | 78.3 | 71.6 | 55.1 |
| HD 377 | 73.8 | 67.4 | 51.7 |
| HD 9672 | 129.6 | 120.5 | 96.6 |
| HD 16743 | 134.1 | 123.7 | 97.1 |
| HD 30447 | 80.1 | 73.8 | 57.7 |
| HD 32297 | 105.3 | 97.6 | 77.7 |
| HD 35841 | 59.8 | 54.9 | 42.7 |
| HD 36968 | 144.0 | 132.5 | 103.4 |
| HD 38206 | 129.6 | 120.9 | 97.8 |
| HD 38397 | 103.5 | 94.7 | 72.8 |
| HD 61005 | 60.3 | 54.9 | 41.9 |
| HD 106906 | 63.2 | 58.1 | 45.1 |
| HD 109573 | 68.6 | 64.0 | 51.7 |
| HD 110058 | 36.0 | 33.4 | 26.7 |
| HD 111520 | 68.0 | 62.5 | 48.7 |
| HD 112810 | 103.5 | 95.2 | 74.1 |
| HD 114082 | 31.6 | 29.1 | 22.7 |
| HD 115600 | 41.2 | 38.0 | 29.9 |
| HD 117214 | 44.4 | 40.9 | 31.8 |
| HD 121617 | 73.8 | 68.7 | 55.3 |
| HD 131488 | 91.8 | 85.4 | 68.6 |
| HD 141011 | 116.1 | 106.9 | 83.5 |
| HD 145560 | 76.5 | 70.5 | 55.2 |
| HD 146181 | 90.0 | 83.0 | 65.3 |
| HD 146897 | 55.5 | 50.9 | 39.3 |
| HD 156623 | 49.5 | 46.1 | 37.1 |
| HD 160305 | 93.6 | 85.6 | 65.9 |
| HD 172555 | 9.3 | 8.6 | 6.8 |
| HD 181327 | 73.7 | 67.7 | 52.7 |
| HD 182681 | 144.0 | 134.3 | 108.6 |
| HD 191089 | 42.5 | 39.1 | 30.4 |
| HD 192758 | 88.2 | 81.4 | 63.9 |
| HD 197481 | 26.2 | 24.6 | 20.2 |
| BD-20 951 | 109.8 | 97.6 | 77.7 |
| TWA 25 | 68.1 | 61.2 | 45.1 |

**Notes.** The columns list the target ID, the radial location of the inner edge of a single belt, the maximum constrained semimajor axis $a_{\rm p\,max}$ for a planet with a mass of 0.1 $M_{\rm Jup}$, and the minimum constrained semimajor axis $a_{\rm p\,min}$ for a planet with a mass of 13 $M_{\rm Jup}$.

The SPHERE data for HD 120326 were investigated by Bonnefoy et al. (2017), who discovered the double-belt structure around this star and identified ten candidate companions. Based on their positions in the color–magnitude diagram and comparison with earlier HST/STIS observations (Padgett & Stapelfeldt 2016), all candidates were classified as background objects. By converting the 5$\sigma$-detection limits into mass constraints, the authors ruled out the presence of giant planets with masses greater than 2 $M_{\rm Jup}$ at radial separations between 55 and 100 au, corresponding to the edges of the gap in our model (see Table 7). This excludes our single-planet model but still allows for a two-planet configuration with $M_{\rm p}$ = 1.49 $M_{\rm Jup}$, which remains below the detection threshold of available data.

In all other cases investigated, the predicted range of planet masses capable of sculpting the inner edges of single belts or clearing the gaps between two belts extends to values below the current detection limits of HCI surveys. This implies that a significant fraction of planetary systems could contain sub-Jovian or Neptune-like planets in wide orbits, which remain elusive to existing instruments but play a central role in shaping circumstellar dust distributions.

The inferred masses and orbital distances of these hypothetical planets in our analysis align well with the population of exoplanets identified through RV and transit surveys, especially in the super-Earth to sub-Saturn mass range (Winn & Petigura 2020). These studies have shown that such planets are common around a wide range of stellar types and may often occur in multi-planet configurations. The orbital distances we derive, typically tens to over a hundred au, complement these findings by probing an otherwise underexplored region of radial separations. Our results further support the notion that debris disk morphology can serve as an indirect tracer of planetary companions (e.g., Raymond et al. 2011; Lee & Chiang 2016) and highlight the need for combined approaches incorporating disk modeling, DI, and indirect detection methods, to fully characterize the architectures of planetary systems.

## 8. Summary

In this work, we performed a demographics study of debris disks around young stars observed with SPHERE at optical and near-IR wavelengths. By analyzing a large sample of 161 targets with IR excess, we addressed morphological, photometric, and polarimetric properties of debris disks across different stellar types and ages. From this sample, compiled from archival GTO and open-time program observations, we resolve 40 disks in scattered light and 36 in polarized light, identifying seven systems with two planetesimal belts and two with three distinct belts (HD 131835 and TWA 7). Newly resolved structures include the disks around HD 36968 and BD-20 951, as well as the inner belts of HR 8799 and HD 36546.

Using SPHERE images, we measured geometrical parameters of detected disks through ellipse fitting, and applied a grid of models for higher-quality data to derive disk radii, aspect ratios, radial density slopes, and scattering asymmetry parameters. This uniform modeling enables a consistent comparison of structural disk properties and reveals systematic dependencies of geometrical parameters on stellar luminosity. The inner slopes of grain density distributions steepen with increasing stellar luminosity, while disk vertical aspect ratios tend to decrease. For most systems, aspect ratios are between 0.02 and 0.06, consistent with expectations for collisionally active belts, though gas-rich disks show unusually small values.

A direct comparison between the radii measured in SPHERE scattered-light images and those derived from ALMA and SMA thermal-emission observations demonstrate a close spatial agreement, with a mean radius ratio of 1.05±0.04. In double-belt systems, the outer belts are typically 1.5 – 2 times larger than the inner ones.

Almost all resolved planetesimal belts contain cold dust with BB temperatures below 100 K, with HD 172555 being the only exception. A weak correlation between belt radius and stellar luminosity is found, following $R_{\rm belt} \propto L_\star^{0.11\pm0.05}$. When dividing belts according to dust temperatures associated with CO and $CO_2$ freeze-out, this correlation becomes steeper: with $\alpha$ = 0.30 ± 0.08 for CO subsample ($T_{\rm bb}$ < 35 K) and $\alpha$ = 0.30 ± 0.07 for $CO_2$ subsample ($T_{\rm bb}$ > 35 K). We also investigated how the locations of debris belts evolve with stellar age in relation to the possible migration of ice lines. For disks in the $CO_2$ subsample, we find that belt radii increase systematically with stellar age, following $R_{\rm belt} \propto t_{\rm age}^{0.37\pm0.11}$, indicating that debris architectures evolve alongside stellar luminosity growth during the PMS phase.

Complementary SED modeling with MBB and SD approaches allowed us to connect disk morphology with photo-

Table 7: Planets clearing the gaps between edges of systems with multiple belts.

| Multiple belts | Edge inner (au) | Edge outer (au) | One planet | | Two planets | | | Three planets | | | |
|---|---|---|---|---|---|---|---|---|---|---|---|
| | | | $a_{\rm p}$ (au) | $M_{\rm p}$ $M_{\rm Jup}$ | $a_{\rm p,1}$ (au) | $a_{\rm p,2}$ (au) | $M_{\rm p}$ $M_{\rm Jup}$ | $a_{\rm p,1}$ (au) | $a_{\rm p,2}$ (au) | $a_{\rm p,3}$ (au) | $M_{\rm p}$ $M_{\rm Jup}$ |
| HD 15115 | 69.7 | 87.3 | 78.2 | 0.275 | 74.5 | 81.9 | 0.0446 | 71.9 | 78.2 | 85.0 | 0.0026 |
| HD 36546 | 60.5 | 99.0 | 78.2 | 6.32 | 69.6 | 86.4 | 0.898 | 64.4 | 77.6 | 93.5 | 0.047 |
| HD 39060 | 71.5 | 99.0 | 84.8 | 1.288 | 78.6 | 90.3 | 0.196 | 74.6 | 84.3 | 95.2 | 0.011 |
| HD 92945 | 61.6 | 107.1 | 82.4 | 3.382 | 72.0 | 92.0 | 0.472 | 66.0 | 81.5. | 100.6 | 0.024 |
| HD 120326 | 55.0 | 107.1 | 78.6 | 10.94 | 66.3 | 89.5 | 1.49 | 59.6 | 77.0 | 99.3 | 0.075 |
| HD 129590 | 53.9 | 73.8 | 63.3 | 0.627 | 59.0 | 67.5 | 0.096 | 56.2 | 63.2 | 71.1 | 0.0053 |
| HD 131835 out | 77.0 | 94.5 | 85.4 | 0.287 | 81.8 | 89.1 | 0.047 | 79.2 | 85.5 | 92.2 | 0.0028 |
| HD 131835 inn | 51.4 | 63.0 | 57.0 | 0.282 | 54.6 | 59.4 | 0.0464 | 52.9 | 57.0 | 61.4 | 0.0027 |
| HD 141943 | 89.1 | 90.0 | 89.4 | $5\times10^{-6}$ | (...) | (...) | (...) | (...) | (...) | (...) | (...) |
| HD 218396 | 17.1 | 180.0 | >80 | (...) | 33.8 | 109.8 | 71.95 | 23.3 | 56.7 | 137.7 | 6.78 |
| TWA 7 out | 57.2 | 83.7 | 69.6 | 0.516 | 63.8 | 75.2 | 0.0764 | 60.1 | 69.3 | 80.0 | 0.0041 |
| TWA 7 inn | 29.7 | 46.8 | 37.7 | 0.934 | 33.8 | 41.2 | 0.135 | 31.5 | 37.4 | 44.4 | 0.0071 |

**Notes.** The columns list the target ID, the radial locations of the inner and outer edges of the gap between two belts, and the constrained semimajor axis ($a_{\rm p}$) of a planet with mass $M_{\rm p}$ in planetary configurations with one (Cols. 4 – 5), two (Cols. 6 – 9) and three (Cols. 10 – 13) equal-mass planets within the gap.

metric properties across both single- and multi-belt systems estimating their luminosities, dust masses and grain SDs. From this analysis, we find that disk fractional luminosities evolve approximately as $t_{\rm age}^{-1.18\pm0.14}$ for A-type stars and $t_{\rm age}^{-0.81\pm0.12}$ for F-type stars, supporting collisional evolution as the main driver of long-term decay.

In addition, we examined the scaling of disk dust masses with stellar properties and compared them with relations established for PPD. The dust masses derived with MBB approach, scale super-linearly with stellar mass following $M_{\rm dust} \propto M_\star^{\alpha_{\rm mass}}$ with $\alpha_{\rm mass} = 1.6 \pm 1.0$ for systems aged 10–50 Myr and $\alpha_{\rm mass} = 1.4 \pm 0.9$ for older disks, similar to the relations observed in 2 – 3 Myr old star-forming regions (e.g., Pascucci et al. 2016). This continuity suggests that the initial conditions set during the protoplanetary phase strongly influence debris disk evolution. Furthermore, typical debris disk masses decrease by about three orders of magnitude within the first 50 Myr and by nearly four orders at later ages, consistent with collisional depletion. The $M_{\rm dust} - R_{\rm belt}$ relation also follows a power law with index larger than 2, implying that more extended belts tend to host proportionally larger dust reservoirs. Together, these trends demonstrate that debris disks preserve imprints of their primordial PPD phase while revealing the efficiency of collisional evolution in regulating dust content over time.

To model the SEDs of the detected planetesimal belts with a SD approach, we adopted belt radii measured from SPHERE scattered-light images. The fits yield minimum grain sizes consistently larger than 0.8 $\mu$m and an average power-law slope of $q = 3.62$, slightly steeper than the canonical collisional cascade value of 3.5. The resulting dust masses of exo-Kuiper belts, integrated over particle sizes from $a_{\rm min}$ to 5 mm and assuming astrosilicate composition, lie in the range $0.01 - 1\,M_\oplus$ and agree with those inferred from MBB modeling. Moreover, these dust masses scale approx. as $R_{\rm belt}^{2.1}$ with radial distance in the subsample of A-, F-, and G-type stars with ages between 10 and 200 Myr.

Building on the SD modeling results, we estimated bulk dust albedo values using Mie theory for four different grain compositions. The derived values are consistently higher than 0.5, but the variation between compositions is relatively small, indicating that dust albedo is unlikely to be the primary factor behind disk non-detections. In addition, we introduced in this work a parametric approach based on image modeling and flux measurements from scattered-light and polarized-light images, demonstrating how both the dust albedo and the maximum polarization fraction can be derived with this method.

Analysis of polarized-light images reveals a correlation between polarized fluxes in H and J bands and modeled IR excesses. The slope of this relation is shallower than that found for total-intensity optical images from HST, consistent with the fact that polarized flux traces only a fraction of total scattered light, dependent on dust properties and disk inclination.

The analysis of non-detections further shows that 90% of the detected disks have estimated ages below 50 Myr, indicating that many undetected systems can be explained by intrinsically low dust masses resulting in faint scattered-light emission. Nevertheless, even bright disks ($f_{\rm disk} > 10^{-3}$) may remain undetected in cases of unfavorable viewing geometry or suboptimal observing conditions, whereas under optimal conditions disks with excesses as low as $10^{-4}$ are detectable. These findings establish practical detection thresholds and observational biases that are critical for interpreting the demographics of debris disks in current and future HCI surveys.

The connection between debris disks and planetary architectures was probed through dynamical modeling. In systems with resolved disks, we estimated the masses and semimajor axes of planets that could sculpt gaps or belt inner edges. For single-belt systems, planets capable of shaping the inner edges span masses from 0.1 to 13 $M_{\rm Jup}$, with orbital radii consistent with observed disk edges. In double-belt systems, multiple planets with sub-Jovian masses can reproduce the gaps while remaining below current detection thresholds. These findings imply that Neptune- to sub-Saturn-mass planets at tens to hundreds of au may be common but remain undetected. The inferred planet populations align with those found by RV and transit surveys in the super-Earth to sub-Saturn range, though at larger orbital separations.

Among the systems in our sample, only five, HD 39060 ($\beta$ Pic), HD 106906, HD 114082, HD 197481 (AU Mic), and HD 218396 (HR 8799), host both resolved debris disks and confirmed planets. In HD 114082, we identify a second transiting giant planet candidate in TESS data, with a radius of $1.29 \pm 0.05\,R_{\rm Jup}$ and orbit near 1 au. Its misaligned orbit relative to the debris disk and the known planet, along with their close proximity to the host star, well inside the typical formation region for giant planets, suggests a history of dynamical evolution involving planet–planet scattering or planet migration. Continued monitoring of HD 114082 is required to refine the orbital parameters of both planets, particularly their eccentricities and the orientation of their orbital planes. These constraints are critical for assessing the system's long-term dynamical stability and for distinguishing between potential migration and scattering scenarios.

Finally, we considered stellar companions. Few are found at separations overlapping the debris disks in our sample, consistent with selection biases excluding known multiples from SPHERE surveys. Nonetheless, stellar companions at 10−100 au are known to destabilize debris disks, likely explaining their scarcity in our resolved sample.

This study presents the largest homogeneous analysis of debris disks imaged in scattered light with SPHERE, complementing previous surveys conducted with HST, *Herschel*, GPI, and ALMA. It is important to note, however, that our results are based on a sample that is inherently biased, as it primarily includes targets selected for their high IR excess or previously known disk structures. This selection effect should be carefully considered when interpreting trends in disk properties and dust evolution. A more comprehensive and unbiased statistical assessment would require a broader sample, including fainter and lower-mass disks that remain undetected with current HCI techniques.

Future studies will benefit from advancements in observational capabilities, such as those provided by JWST, which can probe debris disks at mid-IR wavelengths (e.g., Boccaletti et al. 2024; Mâlin et al. 2024; Su et al. 2024), as well as next-generation ground-based instruments including ELT/METIS (Brandl et al. 2024). Additionally, combining high-resolution scattered-light imaging with ALMA millimeter observations will allow for a more complete picture of disk morphology, grain composition, and spatial segregation of dust populations. Expanding such multiwavelength approaches, together with advancements in disk modeling techniques, is essential for refining our understanding of debris disk evolution and the architectures of planetary systems.

*Acknowledgements.* We would like to thank the anonymous referee for many thoughtful comments which helped to improve this paper.
SPHERE is an instrument designed and built by a consortium consisting of IPAG (Grenoble, France), MPIA (Heidelberg, Germany), LAM (Marseille, France), LESIA (Paris, France), Laboratoire Lagrange (Nice, France), INAF Osservatorio di Padova (Italy), Observatoire de Genève (Switzerland), ETH Zurich (Switzerland), NOVA (Netherlands), ONERA (France) and ASTRON (Netherlands) in collaboration with ESO. SPHERE was funded by ESO, with additional contributions from CNRS (France), MPIA (Germany), INAF (Italy), FINES (Switzerland) and NOVA (Netherlands). SPHERE also received funding from the European Commission Sixth and Seventh Framework Programmes as part of the Optical Infrared Coordination Network for Astronomy (OPTICON) under grant number RII3-Ct-2004-001566 for FP6 (2004-2008), grant number 226604 for FP7 (2009-2012) and grant number 312430 for FP7 (2013-2016). This work has made use of the High Contrast Data Centre, jointly operated by OSUG/IPAG (Grenoble), PYTHEAS/LAM/CeSAM (Marseille), OCA/Lagrange (Nice), Observatoire de Paris/LESIA (Paris), and Observatoire de Lyon/CRAL, and supported by a grant from Labex OSUG@2020 (Investissements d'avenir – ANR10 LABX56).
This work has been supported by the DDISK ANR contract number ANR-21-CE31-0015.
This research has made use of the NASA Exoplanet Archive, which is operated by the California Institute of Technology, under contract with the National Aeronautics and Space Administration under the Exoplanet Exploration Program.
This work has made use of data from the European Space Agency (ESA) mission *Gaia* (https://www.cosmos.esa.int/gaia), processed by the *Gaia* Data Processing and Analysis Consortium (DPAC, https://www.cosmos.esa.int/web/gaia/dpac/consortium). Funding for the DPAC has been provided by national institutions, in particular the institutions participating in the *Gaia* Multilateral Agreement.
This paper includes data collected with the TESS mission, obtained from the MAST data archive at the Space Telescope Science Institute (STScI). Funding for the TESS mission is provided by the NASA Explorer Program. STScI is operated by the Association of Universities for Research in Astronomy, Inc., under NASA contract NAS 5–26555. JM and JCA acknowledge funding from the "Programme National de Planétologie" (PNP) of CNRS-INSU in France through the EPOPEE project (Etude des POussières Planétaires Et Exoplanétaires). A.Z. acknowledges support from ANID – Millennium Science Initiative Program – Center Code NCN2024_001 and Fondecyt Regular grant number 1250249.

[1] ETH Zurich, Institute for Particle Physics and Astrophysics, Wolfgang-Pauli-Strasse 27, CH-8093 Zurich, Switzerland
e-mail: englern@phys.ethz.ch
[2] CNRS, IPAG, Université Grenoble Alpes, IPAG, 38000 Grenoble, France
[3] Institut für Astrophysik Universität Wien, Türkenschanzstraße 17, 1180 Vienna, Austria
[4] IKonkoly Observatory, Research Centre for Astronomy and Earth Sciences, Konkoly-Thege Miklós út 15-17, H-1121 Budapest, Hungary
[5] INAF – Osservatorio Astronomico di Padova, Vicolo dell'Osservatorio 5, 35122 Padova, Italy
[6] LIRA, Observatoire de Paris, Université PSL, CNRS, Sorbonne Université, Université de Paris, 5 place Jules Janssen, 92195 Meudon, France
[7] European Southern Observatory, Karl Schwarzschild St, 2, 85748 Garching, Germany
[8] Leiden Observatory, University of Leiden, Einsteinweg 55, 2333CA Leiden, The Netherlands
[9] Max-Planck-Institut für Astronomie, Königstuhl 17, 69117 Heidelberg, Germany
[10] Department of Astronomy, Stockholm University, AlbaNova University Center, SE-10691 Stockholm, Sweden
[11] Johns Hopkins University: Baltimore, Maryland, USA
[12] Aix Marseille Université, CNRS, LAM - Laboratoire d'Astrophysique de Marseille, UMR 7326, 13388, Marseille, France
[13] Anton Pannekoek Astronomical Institute, University of Amsterdam, PO Box 94249, 1090 GE Amsterdam, The Netherlands
[14] University of Galway, University Road H91 TK33 Galway, Ireland
[15] Instituto de Estudios Astrofísicos, Facultad de Ingeniería y Ciencias, Universidad Diego Portales, Av. Ejército Libertador 441, Santiago, Chile
[16] Millennium Nucleus on Young Exoplanets and their Moons (YEMS), Chile
[17] DOTA, ONERA, Université Paris Saclay, F-91123, Palaiseau, France
[18] Space Telescope Science Institute, 3700 San Martin Drive, Baltimore, MD 21218, USA
[19] Kiepenheuer-Institut für Sonnenphysik, Schneckstr. 6, D-79104 Freiburg, Germany
[20] European Southern Observatory, Alonso de Córdova 3107, Casilla 19001, Vitacura, Santiago, Chile
[21] Centre de Recherche Astrophysique de Lyon, CNRS/ENSL Université Lyon 1, 9 av. Ch. André, 69561 Saint-Genis-Laval, France
[22] Center for Astrophysics and Planetary Science, Department of Astronomy, Cornell University, Ithaca, NY 14853, USA

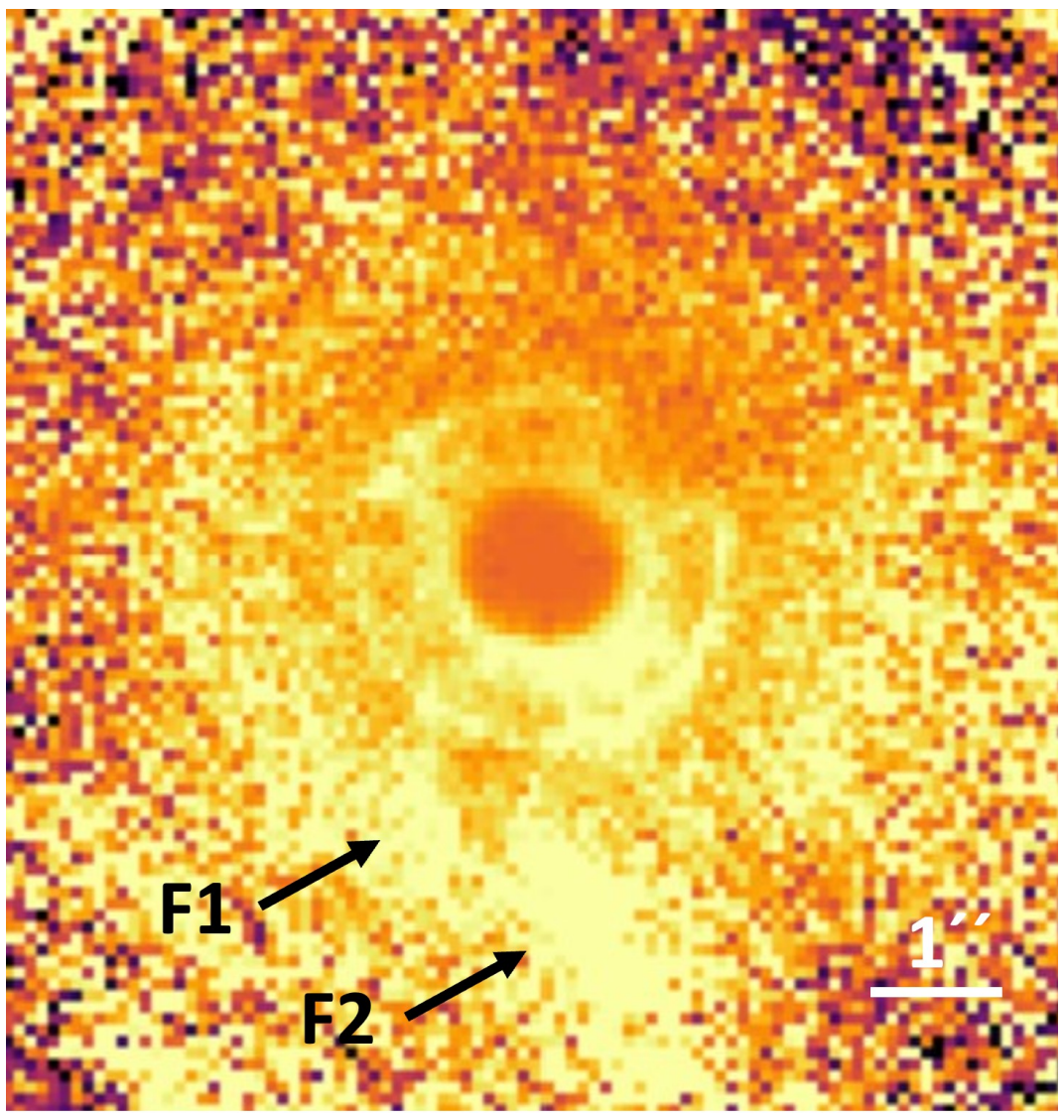

Fig. B.1: Polarized intensity ($Q_\phi$) image of the TWA 7 debris disk, obtained by combining H-band data from three observing epochs. The image has been binned by 8×8 pixels and smoothed using a Gaussian kernel with $\sigma$ = 2 px. Features labeled "F1" and "F2" correspond to structures discussed in Sect. 4.4. In the image, sky north is up and east is to the left.

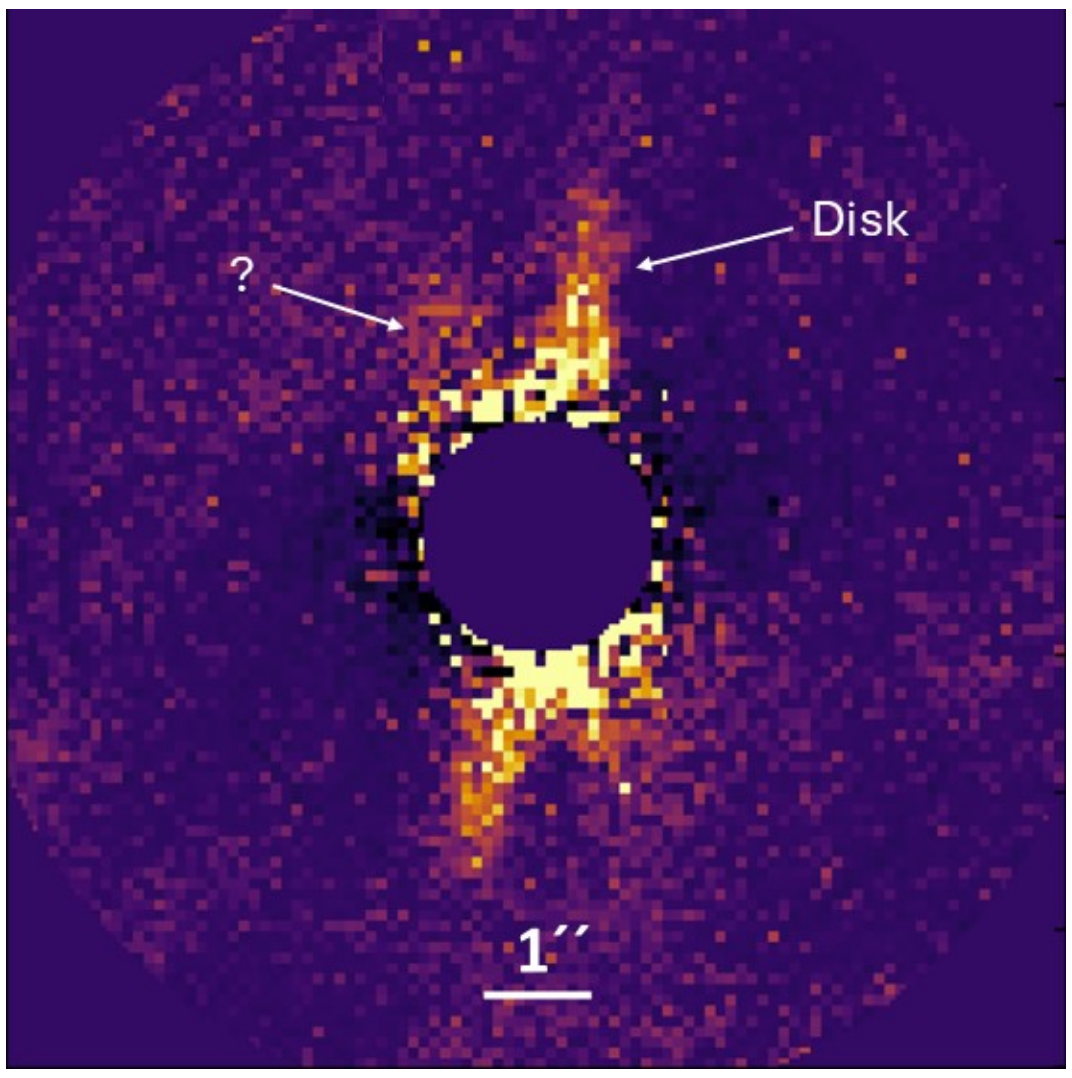

Fig. B.2: Total intensity image of the HD 16743 debris disk with the H2H3 filter. The image is binned by 8×8 pixels and smoothed using a Gaussian kernel with $\sigma$ = 2 pixels. An extended feature of uncertain origin is labeled with a question mark. In the image, sky north is up and east is to the left.

## Appendix A: ESO program IDs

This study is based on data collected at the European Southern Observatory in Chile under programs 095.C-0192, 095.C-0273, 096.C-0388, 097.C-0394, 096.C-0640, 097.C-0344, 098.C-0505, 098.C-0686, 098.C-0790, 099.C-0147, 0100.C-0548, 0101.C-0016, 0101.C-0015, 0101.C-0128, 0101.C-0420, 0101.C-0422, 0101.C-0502, 0102.C-0916, 0102.C-0861, 0104.C-0436, 0104.C-0456, 105.20GP.001, 1104.C-0416, 0102.C-0453, 1100.C-0481, and 198.C-0209.

## Appendix B: SPHERE images of disks TWA 7 and HD 16743

To complement the findings discussed in Sect. 4.4, Figs B.1 and B.2 present supplementary SPHERE/IRDIS images of the TWA 7 and HD 16743 systems, highlighting additional structural features within their respective debris disks. Figure B.1 shows the polarized intensity image of the TWA 7 disk, combining H-band data from three separate observing epochs. In this composite image, all three planetesimal belts are discernible, with the outer belt particularly prominent. Also visible are arc-like features connecting the middle and outer belts, labeled "F1" and "F2" in Figs. 9g and B.1. The most extended of these structures, "F2", reaches the outer edge of the SPHERE FOV and was previously detected in both HST and ALMA observations (Ren et al. 2021; Bayo et al. 2019).

Figure B.2 displays the total intensity image of the HD 16743 disk, showing a planetesimal belt oriented at a PA of ~170°, as well as an extended feature at PA = 17° marked with a question mark. The origin of this feature is uncertain; it may represent either a residual PSF artifact or genuine scattered light from disk material.

## Appendix C: Polarization fraction function of micron-sized dust particles

To model the disk images of polarized scattered light, we adopted the pSPF incorporating a polarization fraction function derived under the Rayleigh scattering assumption (Eq 3). Rayleigh scattering describes the interaction of light with particles significantly smaller than the wavelength of the incident radiation, and it exhibits a pronounced angular dependence, with maximum polarization occurring at scattering angle of 90°. This approximation provides a useful analytic form for the polarization behavior of small and compact grains in debris disks, although it becomes less accurate for larger or more complex dust particles.

To illustrate this, we present in Fig. C.1 the polarization fraction functions measured for Mg-rich olivine samples with three different particle SDs alongside the theoretical Rayleigh scattering polarization fraction function for comparison. We selected olivine as a representative material because its presence is commonly inferred from cometary and debris disk spectra (e.g., Kolokolova & Jockers 1997; Chen et al. 2006). The measurements for samples with effective grain radii of 2.6 $\mu$m and 3.8 $\mu$m were conducted using a laser source at 633 nm (Muñoz et al. 2000), while the data for the sample with an effective radius exceeding 20 $\mu$m were obtained using a white light source and the spectral response of imaging polarimeter cameras in the wavelength range of 1.5 – 1.6 $\mu$m (Renard et al. 2014)[15].

The scaled Rayleigh scattering function $p(\theta)$ is plotted in Fig. C.1 with a maximum polarization fraction $p_{\max}$ of 9%, chosen to match the measurement for the sample with an effective grain radius of $a_{\rm eff}$ = 3.8 $\mu$m. As shown, the overall shape of the Rayleigh polarization function reproduces the observed trend reasonably well. The largest discrepancy arises at large scattering angles ($\theta > 150°$), corresponding to small phase angles (phase angle = 180° – scattering angle), where the measured polarization fraction (< 4%) becomes negative, forming a so-called negative polarization branch. This sign inversion indicates a change in the orientation of the polarization vector from azimuthal to radial in the image of polarized intensity, meaning that the scattered light becomes preferentially polarized parallel to the scattering plane[16] rather than perpendicular to it.

[15] The experimental datasets are publicly available from the following databases: `https://old-scattering.iaa.csic.es/` (for $a_{\rm eff}$ = 2.6 $\mu$m and $a_{\rm eff}$ = 3.8 $\mu$m) and `https://www.icare.univ-lille.fr/progra2-en/banque-de-donnees/`.

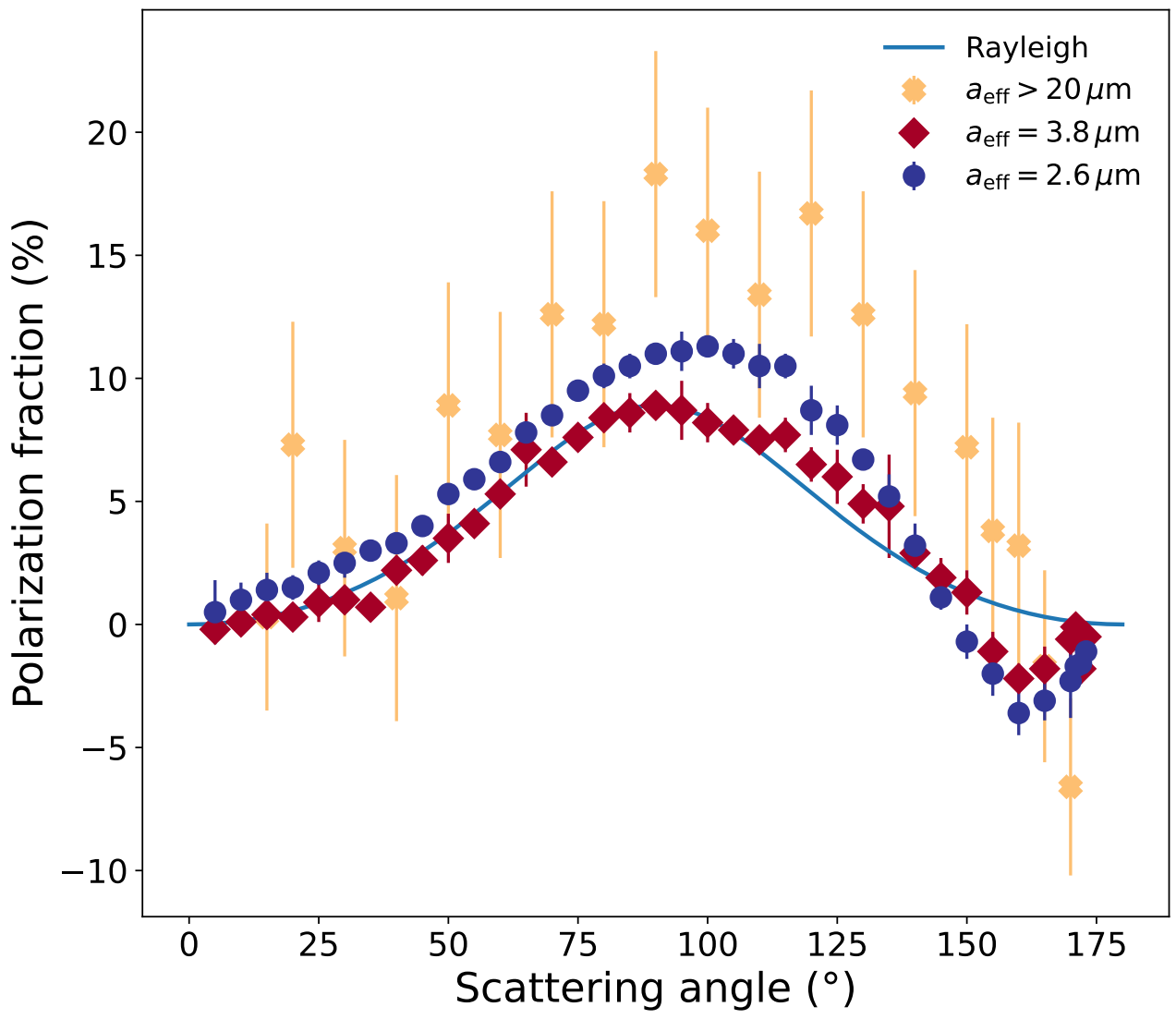

Fig. C.1: Polarization fraction measured for olivine dust samples with particle SDs characterized by different effective radii ($a_{eff}$), shown as a function of scattering angle. For comparison, a Rayleigh scattering polarization fraction function with a maximum polarization $p_{max}$ of 9% is overplotted (blue solid line). Measurements for samples with $a_{eff}$ = 2.6 $\mu$m (blue circles) and $a_{eff}$ = 3.8 $\mu$m (red diamonds) were obtained using a laser source at 633 nm. Data for the sample with $a_{eff} > 20\mu$m (yellow crosses) were acquired using a white light source and camera systems with a spectral response in the 1.5–1.6 $\mu$m range.

The maximum polarization fraction of dust particles may also occur at scattering angles different from 90°, though still typically close to it. For example, in Fig. C.1, the sample with an effective grain size of $a_{eff}$ = 2.6 $\mu$m exhibits a peak polarization fraction at a scattering angle of approx. 100°. In practice, the polarization fraction function of a debris disk, along with the SPF and pSPF, is expected to be smoothed due to averaging over a distribution of particle sizes and a mixture of dust populations across different disk regions. We therefore conclude that the Rayleigh scattering function $p(\theta)$ offers a reasonable, though simplified, approximation that can be applied to model polarimetric images of debris disks. If a more accurate representation is required, the location of the maximum polarization fraction can be parametrized using a scaled version of the beta distribution, as proposed by Ren et al. (2023). However, this approach introduces two additional degrees of freedom into the model.

## Appendix D: Examples of the SED fits using MBB and SD models

Figure D.1 shows several examples of the SED best-fitting models applying various approaches. Top row presents the MBB models consisting of one or two planetesimal rings. Bottom row shows the SD models based on the belt radii measured from the $r^2$-scaled disk images.

[16] The scattering plane is defined as the plane containing the incident light source (e.g., the star), the scattering particle, and the observer.

## Appendix E: Sample stars and results from disk and MBB modeling

Table E.2 lists the sample stars along with their stellar parameters. Tables E.1, E.3, and E.4 present the results of disk modeling and the one- and two-component MBB modeling, respectively.

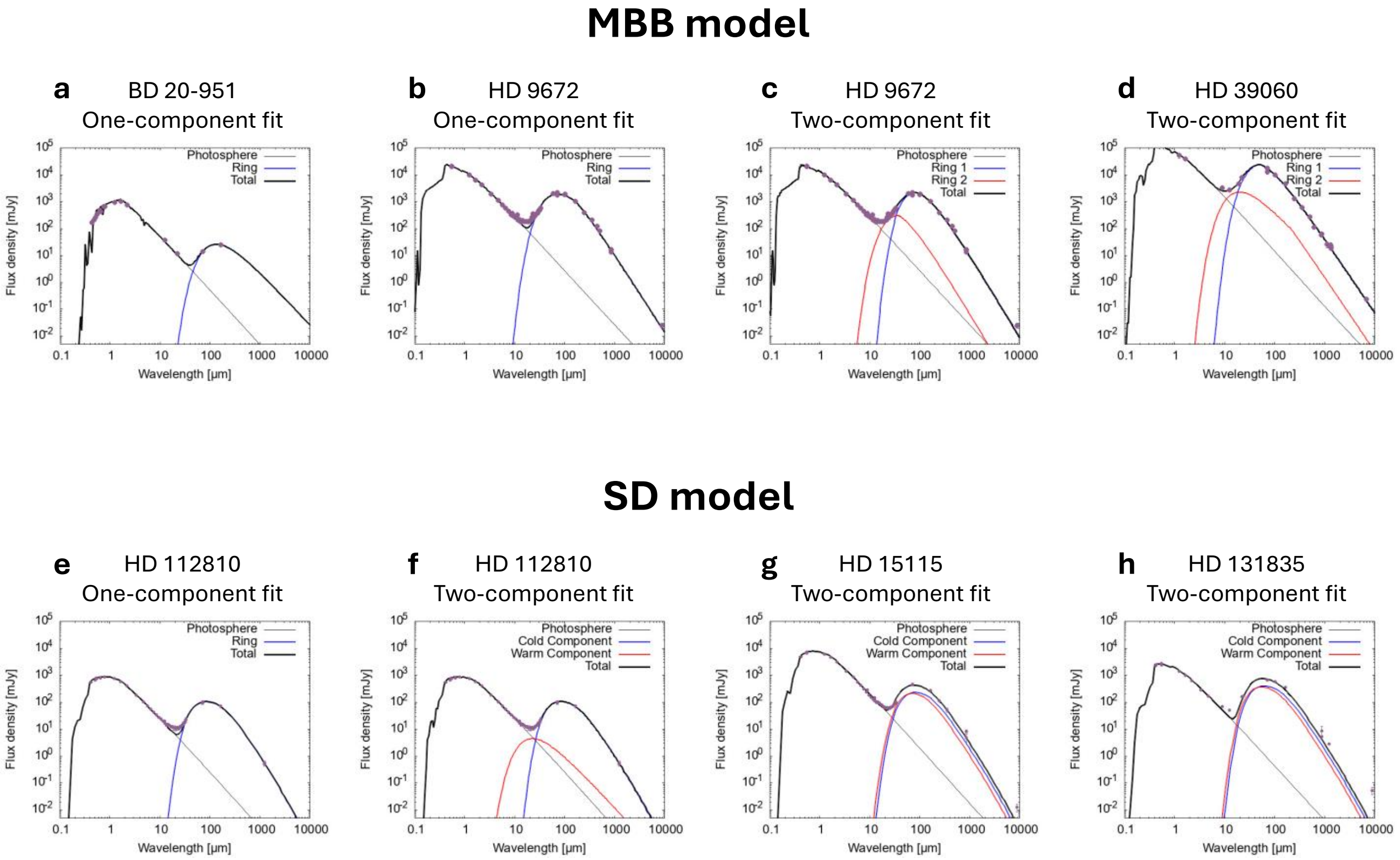

Fig. D.1: Examples of best-fit SED models using various approaches (see Sect. 5). The top row illustrates MBB models in which the ring radius is a free parameter. In all cases, the disk SEDs were fitted with a single planetesimal ring, as shown in *panel a* for BD 20-951 and *panel b* for HD 9672. For disks exhibiting a warm component, an additional BB component (Ring 2) representing warm dust was included in the fit, as demonstrated in *panel c* for HD 9672 and *panel d* for HD 39060. The bottom row shows fits using the SD model, where the ring radius is fixed to the value measured from the $r^2$-scaled scattered-light images. If only one ring is resolved, the SED is fitted with a single SD component, as in *panel e* for HD 112810. For systems with a warm dust component, an additional BB component (Warm Component) was added, as illustrated in *panel f* for HD 112810. In cases where two cold belts are resolved, the SED is fitted with two SD rings, as shown in *panel g* for HD 15115 and *panel h* for HD 131835.

Table E.1: Modeled parameters of debris disks resolved with SPHERE.

| Target ID | $R_{\rm belt}^{\rm mod}$ (au) | $r_0$ (″) | $H_0$ (″) | $i$ (°) | PA (°) | $\alpha_{\rm in}$ | $\alpha_{\rm out}$ | $g$ |
|---|---|---|---|---|---|---|---|---|
| Single belt model | | | | | | | | |
| HD 9672 | 151 ± 19 | 2.56 ± 0.15 | 0.052 ± 0.008 | 78.3 ± 3.2 | 108.5 ± 2.8 | 15.9 ± 7.6 | −4.4 ± 2.5 | 0.55 ± 0.14 |
| HD 15115 | 97 ± 3 | 2.03 ± 0.04 | 0.061 ± 0.015 | 94.2 ± 1.6 | 98.9 ± 1.2 | 3.1 ± 1.5 | −4.0 ± 2.4 | 0.53 ± 0.15 |
| HD 32297 | 119 ± 17 | 0.88 ± 0.13 | 0.023 ± 0.005 | 92.1 ± 1.3 | 47.7 ± 0.9 | 13.0 ± 1.4 | −3.2 ± 1.2 | 0.73 ± 0.14 |
| HD 61005 | 65 ± 5 | 1.62 ± 0.13 | 0.042 ± 0.005 | 82.3 ± 1.3 | 71.0 ± 1.2 | 5.0 ± 1.4 | −1.5 ± 0.9 | 0.72 ± 0.13 |
| HD 106906 | 76 ± 7 | 0.73 ± 0.07 | 0.046 ± 0.002 | 94.7 ± 2.9 | 105.0 ± 1.4 | 5.8 ± 1.7 | −4.3 ± 1.5 | 0.79 ± 0.13 |
| HD 109573 | 75 ± 2 | 1.05 ± 0.03 | 0.018 ± 0.005 | 102.7 ± 1.6 | 28.7 ± 1.0 | 24.9 ± 3.4 | −11.9 ± 1.7 | 0.64 ± 0.13 |
| HD 114082 | 33 ± 1 | 0.34 ± 0.01 | 0.007 ± 0.003 | 83.2 ± 1.1 | 105.7 ± 0.7 | 25.9 ± 9.8 | −4.6 ± 1.3 | 0.56 ± 0.15 |
| HD 115600 | 46 ± 4 | 0.45 ± 0.04 | 0.033 ± 0.005 | 104.4 ± 5.2 | 24.8 ± 1.7 | 2.3 ± 1.5 | −8.7 ± 2.0 | 0.56 ± 0.12 |
| HD 117214 | 45 ± 1 | 0.41 ± 0.01 | 0.010 ± 0.005 | 72.8 ± 1.2 | 179.3 ± 0.2 | 21.5 ± 5.7 | −5.4 ± 1.2 | 0.52 ± 0.10 |
| HD 120326 | 38 ± 8 | 0.32 ± 0.07 | 0.008 ± 0.001 | 76.8 ± 3.5 | 86.5 ± 1.8 | 7.5 ± 1.7 | −2.9 ± 1.9 | 0.75 ± 0.14 |
| HD 121617 | 82 ± 3 | 0.69 ± 0.09 | 0.007 ± 0.004 | 136.1 ± 1.5 | 60.5 ± 2.8 | 17.5 ± 7.0 | −6.0 ± 2.3 | 0.57 ± 0.14 |
| HD 129590 | 55 ± 8 | 0.38 ± 0.06 | 0.023 ± 0.005 | 99.2 ± 3.2 | 119.7 ± 2.9 | 7.0 ± 1.7 | −2.6 ± 1.4 | 0.76 ± 0.18 |
| HD 131488 | 104 ± 11 | 0.66 ± 0.07 | 0.007 ± 0.003 | 94.7 ± 1.2 | 96.5 ± 0.3 | 24.2 ± 10.1 | −4.0 ± 1.7 | 0.57 ± 0.13 |
| HD 146897 | 67 ± 12 | 0.48 ± 0.12 | 0.019 ± 0.005 | 84.4 ± 1.0 | 114.9 ± 0.8 | 5.3 ± 1.7 | −2.5 ± 1.0 | 0.68 ± 0.08 |
| HD 157587 | 82 ± 4 | 0.78 ± 0.04 | 0.012 ± 0.005 | 110.4 ± 1.5 | 129.5 ± 1.4 | 9.3 ± 2.6 | −3.0 ± 0.7 | 0.59 ± 0.09 |
| HD 172555 | 11 ± 3 | 0.38 ± 0.06 | 0.019 ± 0.011 | 105.0 ± 2.3 | 112.0 ± 7.4 | 3.9 ± 2.0 | −5.5 ± 1.7 | 0.50 ± 0.04 |
| HD 181327 | 82 ± 5 | 1.69 ± 0.11 | 0.018 ± 0.005 | 151.0 ± 1.4 | 100.0 ± 1.4 | 9.9 ± 1.4 | −5.9 ± 1.4 | 0.55 ± 0.14 |
| HD 191089 | 50 ± 2 | 0.82 ± 0.03 | 0.030 ± 0.014 | 120.7 ± 1.5 | 70.6 ± 1.6 | 12.2 ± 3.8 | −3.3 ± 0.5 | 0.67 ± 0.08 |
| HD 197481 | 32 ± 2 | 3.21 ± 0.14 | 0.081 ± 0.017 | 88.6 ± 1.0 | 129.1 ± 0.6 | 8.0 ± 1.7 | −3.7 ± 1.6 | 0.67 ± 0.17 |
| Double belt model | | | | | | | | |
| HD 39060 inn | 65 ± 6 | 3.22 ± 0.17 | 0.154 ± 0.026 | 92.0 ± 1.8 | 26.5 ± 3.5 | 9.1 ± 1.2 | −4.2 ± 2.5 | 0.74 ± 0.15 |
| HD 39060 out | 113 ± 10 | 5.69 ± 0.23 | 0.245 ± 0.035 | 90.0 ± 0.9 | 23.0 ± 1.2 | 4.9 ± 1.5 | −3.8 ± 2.4 | 0.67 ± 0.10 |

**Notes.** The columns list target IDs, modeled disk radii according to Eq. 4 ($R_{\rm belt}^{\rm mod}$), reference radii ($r_0$), scale heights ($H_0$), inclinations ($i$), PAs of the disks, power law exponents for the radial distribution of grain number density in the disk midplane ($\alpha_{\rm in}$ and $\alpha_{\rm out}$), and the scattering asymmetry parameter ($g$).

Table E.2: Sample stars and their parameters.

| HD ID | HIP ID | Alias | SpType | $M_\star$ ($M_\odot$) | Comp.[a] | $d$ (pc) | $L_\star$ ($L_\odot$) | $T_{\rm eff}$ (K) | $L_{\rm IR}/L_\star$ ($10^{-4}$) | Age (Myr) | MG[b] (%) | Detected?[c] |
|---|---|---|---|---|---|---|---|---|---|---|---|---|
| (...) | 11437 | AG Tri | K7V | 0.67 ± 0.08 | A | 40.94 ± 0.03 | 0.21 ± 0.03 | 4410 | 12.00 ± 2.80 | 18-23[1] | $\beta$PMG (99) | N |
| (...) | (...) | CPD-72 2713 | K7Ve | 0.62 ± 0.08 | S | 36.72 ± 0.02 | 0.19 ± 0.01 | 4065 | 10.00 ± 3.70 | 18-23[1] | $\beta$PMG (99) | N |
| (...) | 74995 | GJ 581 | M3V | 0.31 ± 0.02 | S | 6.30 ± 0.00 | 0.01 ± 0.005 | 3609 | 0.72 ± 0.14 | 2000-8000[2] | Field (100) | N |
| (...) | 83043 | GJ 649 | M2V | 0.51 ± 0.02 | S | 10.39 ± 0.00 | 0.05 ± 0.015 | 3644 | (...) | (...) | Field (100) | N |
| (...) | (...) | GSC 06964-1226 | M4.0Ve | 0.20 ± 0.02 | C | 7.68 ± 0.00 | 0.005[39] | 3200[39] | 1.70 ± 0.40 | 400-480[3] | Field (100) | N |
| (...) | (...) | GSC 07396-0759 | M1Ve | 0.56 ± 0.09 | S | 71.84 ± 0.10 | 0.14 ± 0.02[45] | 3630[45] | (...) | 18-23[1] | $\beta$PMG (99) | I, P |
| 105 | 490 | (...) | G0V | 1.12 ± 0.15 | S | 38.83 ± 0.03 | 1.31 ± 0.07 | 5937 | 2.80 ± 0.55 | 34-66[4] | THA (100) | I |
| 166 | 544 | HR 8 | G8/K0Ve | 0.96 ± 0.12 | A | 13.77 ± 0.01 | 0.69 ± 0.05 | 5465 | 0.75 ± 0.14 | 100-300[5] | Field (100) | N |
| 203 | 560 | HR 9 | F3V | 1.45 ± 0.24 | S | 39.74 ± 0.05 | 4.26 ± 0.15 | 6830 | 1.50 ± 0.29 | 18-23[1] | $\beta$PMG (96) | N |
| 377 | 682 | (...) | G2V | 1.07 ± 0.13 | S | 38.40 ± 0.04 | 1.22 ± 0.04 | 5835 | 8.40 ± 7.70 | 40-250[6, 20] | Field (100) | I, P |
| 1466 | 1481 | (...) | F8V | 1.17 ± 0.16 | S | 42.82 ± 0.03 | 1.65 ± 0.08 | 6073 | 0.77 ± 0.15 | 41-49[7] | THA (100) | N |
| 3003 | 2578 | $\beta^3$ Tuc | A0V | 2.32 ± 0.32 | A | 46.13 ± 0.15 | 22.5 ± 4.3 | 9400 | 1.10 ± 0.25 | 41-49[7] | THA (100) | N |
| 3670 | (...) | (...) | F5V | 1.32 ± 0.21 | S | 77.07 ± 0.10 | 2.52 ± 0.03 | 6273 | 5.30 ± 1.20 | 38-48[7] | COL (56) | N |
| 9672 | 7345 | 49 Cet | A1V | 2.21 ± 0.30 | S | 57.23 ± 0.18 | 16.8 ± 0.1 | 8867 | 6.80 ± 0.41 | 45-55[8] | ARG (99) | I, mP |
| 10472 | 7805 | (...) | F2IV/V | 1.45 ± 0.24 | S | 71.32 ± 0.08 | 3.64 ± 0.01 | 6698 | 3.00 ± 1.10 | 41-49[7] | COL (35) | N |
| 10638 | 8122 | (...) | A3 | 1.80 ± 0.29 | S | 67.77 ± 0.16 | 7.61 ± 0.56 | 7753 | 2.20 ± 0.63 | 20-150[10] | Field (93) | N |
| 10647 | 7978 | $q^1$ Eri | F9V | 1.17 ± 0.16 | S | 17.35 ± 0.01 | 1.65 ± 0.06 | 5972 | 2.70 ± 0.38 | 700-2830[26] | Field (98) | N |
| 10700 | 8102 | $\tau$ Cet | G8V | 0.92 ± 0.11 | S | 3.65 ± 0.00 | 0.48 ± 0.02 | 7731 | 0.24 ± 0.05 | 2900-6700[5] | Field (100) | N |
| 10939 | 8241 | $q^2$ Eri | A1V | 2.30 ± 0.32 | S | 62.28 ± 0.27 | 34.73 ± 1.25 | 9027 | 0.83 ± 0.23 | 448-603[11] | Field (100) | N |
| 13246 | 9902 | (...) | F7V | 1.22 ± 0.18 | A | 45.40 ± 0.03 | 1.72 ± 0.02 | 6140 | 1.60 ± 0.37 | 41-49[7] | THA (100) | N |
| 14082 | 10679 | (...) | G2V | 1.09 ± 0.14 | B | 39.63 ± 0.03 | 1.06 ± 0.04 | 5847 | 2.40 ± 0.51 | 18-23[1] | $\beta$PMG (99) | N |
| 15115 | 11360 | (...) | F4IV | 1.45 ± 0.24 | S | 48.77 ± 0.07 | 3.89 ± 0.15 | 6612 | 4.80 ± 1.30 | 8-2000[10, 12] | Field (100) | I, P |
| 15257 | 11486 | 12 Tri | F0III | 1.65 ± 0.27 | S | 48.99 ± 0.23 | 15.30 ± 0.92 | 7195 | 1.20 ± 0.30 | 2000-2380[13] | Field (100) | N |
| 16743 | 12361 | (...) | F0/2III/IV | 1.55 ± 0.27 | A | 57.81 ± 0.06 | 5.41 ± 0.23 | 6859 | 4.20 ± 1.20 | 10-334[11, 47] | ARG (98) | I |
| 17390 | 12964 | (...) | F3IV/V | 1.50 ± 0.26 | S | 48.26 ± 0.06 | 4.75 ± 0.13 | 6799 | 2.30 ± 0.60 | 100-500[11] | Field (100) | N |
| 17848 | 13141 | $\nu$ Hor | A2V | 2.08 ± 0.38 | S | 51.92 ± 0.17 | 18.00 ± 0.42 | 8331 | 0.49 ± 0.10 | 30-300[9] | Field (100) | N |
| 17925 | 13402 | EP Eri | K1V | 0.89 ± 0.11 | SBC | 10.36 ± 0.00 | 0.44 ± 0.04 | 5130 | 0.89 ± 0.19 | 30-00[5, 17] | Field (100) | N |
| 20320 | 15197 | $\zeta$ Eri | kA4hA9mA9V | 1.76 ± 0.29 | AaAb | 36.50 ± 0.18 | 13.28 ± 1.60 | 7399 | 0.21 ± 0.04 | 800[14] | Field (100) | N |
| 20794 | 15510 | 82 Eri | G6V | 0.94 ± 0.12 | S | 6.04 ± 0.00 | 0.66 ± 0.01 | 5500 | 0.02 ± 0.02 | 3050-9150[5] | Field (100) | N |
| 21997 | 16449 | (...) | A3IV/V | 2.09 ± 0.31 | S | 69.69 ± 0.14 | 11.50 ± 0.62 | 8425 | 6.00 ± 1.30 | 38-48[7] | COL (99) | N |
| 22049 | 16537 | $\epsilon$ Eri | K2V | 0.85 ± 0.10 | S | 3.22 ± 0.00 | 0.33 ± 0.01 | 5029 | 1.30 ± 0.28 | 165-835[11] | Field (100) | N |
| 22179 | (...) | (...) | G5IV | 1.05 ± 0.13 | S | 70.30 ± 0.10 | 1.14 ± 0.02 | 5865 | 2.90 ± 1.80 | 16 − 63[15, 20] | Field (79) | N |
| 23484 | 17439 | (...) | K2V | 0.88 ± 0.11 | S | 16.17 ± 0.00 | 0.44 ± 0.04 | 5215 | 0.92 ± 0.19 | 500 − 900[16] | Field (100) | N |
| 24636 | 17764 | (...) | F3IV/V | 1.45 ± 0.25 | S | 57.11 ± 0.05 | 3.69 ± 0.12 | 6723 | 1.20 ± 0.22 | 41-49[7] | THA (100) | N |
| 25457 | 18859 | (...) | F7V | 1.20 ± 0.17 | S | 18.71 ± 0.02 | 2.09 ± 0.06 | 6212 | 1.00 ± 0.18 | 30-1209[9], [17] | ABDMG (100) | N |
| 27290 | 19893 | $\gamma$ Dor | F1V | 1.59 ± 0.28 | S | 20.43 ± 0.07 | 6.95 ± 0.30 | 6865 | 0.23 ± 0.05 | 535-1207[11] | Field (100) | N |
| 29391 | 21547 | 51 Eri | F0IV | 1.65 ± 0.27 | A | 29.91 ± 0.07 | 5.82 ± 0.16 | 7190 | 0.12 ± 0.06 | 18-23[1] | Field (91) | N |
| 30422 | 22192 | EX Eri | A7VkA3mA3 | 1.91 ± 0.30 | S | 57.13 ± 0.07 | 9.06 ± 0.60 | 7868 | 0.49 ± 0.10 | 380[18] | Field (100) | N |
| 30447 | 22226 | (...) | F3V | 1.46 ± 0.25 | S | 80.31 ± 0.14 | 3.91 ± 0.21 | 6667 | 9.60 ± 3.80 | 38-48[7] | COL (99) | I, P |
| 31295 | 22845 | 7 Ori | A0Va_lB | 2.34 ± 0.32 | S | 37.03 ± 0.25 | 16.19 ± 0.80 | 8474 | 0.41 ± 0.05 | 30-350[18], [19] | Field (100) | N |
| 31392 | 22787 | (...) | G9V | 0.93 ± 0.11 | S | 25.76 ± 0.01 | 0.57 ± 0.01 | 5398 | 1.20 ± 0.24 | 66-334[11] | Field (99) | N |
| 32297 | 23451 | (...) | A0V | 1.93 ± 0.30 | S | 129.73 ± 0.55 | 8.41 ± 0.65 | 7846 | 54.00 ± 7.40 | 15-45[11] | Field (69) | I, P |
| 35114 | 24947 | (...) | F6V | 1.23 ± 0.19 | S | 47.34 ± 0.03 | 2.18 ± 0.12 | 6159 | 0.53 ± 0.13 | 3200[12] | Field (74) | N |
| 35650 | 25283 | (...) | K6V | 0.66 ± 0.08 | S | 17.46 ± 0.002 | 0.13 ± 0.01 | 4334 | 1.10 ± 0.27 | 50-200[7], [21] | ABDMG (100) | mI |
| 35841 | (...) | (...) | F3V | 1.33 ± 0.22 | S | 103.08 ± 0.14 | 2.47 ± 0.12 | 6343 | 17.00 ± 5.90 | 38-48[22] | Field (73) | P |
| 36546 | 26062 | (...) | A0V-A2V | 2.37 ± 0.34 | S | 100.18 ± 0.42 | 15.25 ± 3.70 | 9093 | 44.00 ± 9.70 | 3-10[22] | 118TAU (99) | I |
| 36968 | (...) | (...) | F2V | (...) | S | 148.56 ± 0.26 | 4.46 ± 0.07 | 6750 | 10.00 ± 4.00 | 30-50[46] | OCT (100) | I, P |
| 37484 | 26453 | (...) | F4V | 1.42 ± 0.24 | S | 58.84 ± 0.06 | 3.51 ± 0.02 | 6695 | 3.30 ± 0.89 | 38-48[7] | COL (100) | N |
| 38206 | 26966 | HR 1975 | A0V | 2.57 ± 0.35 | S | 70.69 ± 0.23 | 27.1 ± 0.90 | 9779 | 1.50 ± 0.37 | 38-48[7] | COL (100) | I |
| 38207 | (...) | (...) | F2V | 1.45 ± 0.25 | S | 109.96 ± 0.22 | 4.07 ± 0.07 | 6670 | 10.00 ± 3.70 | (...) | Field (84) | N |
| 38397 | 26990 | (...) | G0V | 1.11 ± 0.14 | S | 53.59 ± 0.05 | 1.42 ± 0.05 | 5962 | 4.80 ± 0.88 | 35-50[23] | COL (100) | P |

Table E.2: continued.

| HD ID | HIP ID | Alias | SpType | $M_\star$ ($M_\odot$) | Comp.$^{(a)}$ | $d$ (pc) | $L_\star$ ($L_\odot$) | $T_{\rm eff}$ (K) | $L_{\rm IR}/L_\star$ ($10^{-4}$) | Age (Myr) | MG$^{(b)}$ (%) | Detected?$^{(c)}$ |
|---|---|---|---|---|---|---|---|---|---|---|---|---|
| 38678 | 27288 | $\zeta$ Lep | A2IV-V(n) | 2.13 ± 0.30 | S | 22.32 ± 0.12 | 16.03 ± 0.29 | 8300 | 1.10 ± 0.30 | 50-347[24] | Field (73) | N |
| 38858 | 27435 | (...) | G2V | 1.02 ± 0.12 | S | 15.21 ± 0.01 | 0.87 ± 0.03 | 5752 | 0.80 ± 0.17 | 1600-4800[5] | Field (100) | N |
| 39060 | 27321 | $\beta$ Pic | A6V | 1.95 ± 0.24 | S | 19.63 ± 0.06 | 9.16 ± 0.50 | 8014 | 10.00 ± 3.00 | 18-23[1] | $\beta$PMG (100) | I, P |
| 40540 | 28230 | (...) | A8IV(m) | 1.66 ± 0.27 | S | 88.34 ± 0.24 | 5.86 ± 0.09 | 7391 | 5.30 ± 1.30 | 200[9] | Field (94) | N |
| 43989 | 30030 | (...) | G0V | 1.13 ± 0.15 | S | 51.67 ± 0.06 | 1.72 ± 0.17 | 6006 | 0.49 ± 0.46 | 38-48[7] | COL (100) | N |
| 48370 | (...) | (...) | K0V(+G) | 0.99 ± 0.13 | S | 35.95 ± 0.03 | 0.83 ± 0.06 | 5608 | 4.70 ± 0.88 | 20 − 50[25] | Field (84) | N |
| 50571 | 32775 | HR 2562 | F5VFe+0.4 | 1.38 ± 0.22 | S | 33.93 ± 0.02 | 3.54 ± 0.20 | 6448 | 1.10 ± 0.21 | 100-500[11] | Field (100) | N |
| 52265 | 33719 | (...) | G0V | 1.11 ± 0.15 | S | 29.92 ± 0.02 | 2.39 ± 0.15 | 6015 | 0.24 ± 0.05 | 2500[12] | Field (100) | N |
| 53143 | 33690 | (...) | G9V | 0.95 ± 0.12 | S | 18.34 ± 0.001 | 0.62 ± 0.04 | 5421 | 2.80 ± 0.56 | 500-800[16] | Field (100) | N |
| 53842 | 32435 | (...) | F5V | 1.30 ± 0.21 | S | 57.66 ± 0.06 | 2.79 ± 0.07 | 6433 | 1.26 ± 0.20 | 1500[12] | THA (97) | N |
| 54341 | 34276 | (...) | A0V | 2.60 ± 0.34 | S | 101.46 ± 0.86 | 26.12 ± 2.92 | 9604 | 2.80 ± 0.65 | 25-305[27] | Field (100) | N |
| 60491 | 36827 | (...) | K1V | 0.84 ± 0.11 | S | 23.46 ± 0.01 | 0.31 ± 0.01 | 5091 | 2.00 ± 0.51 | 70-500[28] | Field (100) | N |
| 61005 | 36948 | (...) | G8Vk | 0.97 ± 0.12 | S | 36.45 ± 0.02 | 0.67 ± 0.04 | 5509 | 19.00 ± 0.99 | 45-55[8] | ARG (100) | I, P |
| 69830 | 40693 | HR 3259 | G8+V | 0.93 ± 0.12 | S | 12.58 ± 0.01 | 0.60 ± 0.05 | 5490 | (...) | 5670-6100[5, 14] | Field (100) | N |
| 71155 | 41307 | 30 Mon | A0Va | 2.47 ± 0.30 | S | 39.46 ± 0.32 | 42.27 ± 1.02 | 9381 | 0.30 ± 0.06 | 50-266[24] | Field (100) | N |
| 71722 | 41373 | (...) | A0V | 2.31 ± 0.32 | S | 69.27 ± 0.12 | 16.54 ± 0.28 | 8848 | 1.10 ± 0.23 | 14-18[29] | ABDMG (96) | N |
| 73350 | 42333 | (...) | G8/K0(IV) | 1.05 ± 0.14 | S | 24.35 ± 0.01 | 1.03 ± 0.03 | 5808 | 1.30 ± 0.41 | 110-300[17] | ABDMG (96) | N |
| 75416 | 42637 | $\eta$ Cha | B8V | (...) | S | 98.35 ± 0.65 | 119.1 ± 1.0 | 12946 | 1.10 ± 0.60 | 8-14[7] | ETAC (83) | N |
| 76582 | 44001 | 63 CnC | F0IV | 1.87 ± 0.30 | S | 48.89 ± 0.13 | 10.42 ± 0.58 | 7675 | 2.50 ± 0.55 | 589-980[11] | Field (100) | N |
| 80950 | 45585 | (...) | A0V | 2.58 ± 0.34 | S | 76.39 ± 0.18 | 30.8 ± 0.92 | 10000 | 1.20 ± 0.29 | 38-56[7] | CAR (99) | N |
| 82943 | 47007 | (...) | F9VFe+0.5 | 1.08 ± 0.13 | S | 27.69 ± 0.02 | 1.63 ± 0.11 | 5959 | 1.10 ± 0.24 | 165-835[11] | Field (100) | N |
| 84075 | 47135 | (...) | G2V | 1.09 ± 0.14 | S | 63.65 ± 0.05 | 1.33 ± 0.03 | 5934 | 2.20 ± 0.64 | 45-55[8] | ARG (100) | N |
| 90905 | 51386 | (...) | G1V | 1.14 ± 0.15 | S | 30.76 ± 0.02 | 1.51 ± 0.09 | 5976 | 0.36 ± 0.09 | 1900[12] | Field (100) | N |
| 92945 | 52462 | V419 Hya | K1V | 0.86 ± 0.11 | S | 21.51 ± 0.01 | 0.40 ± 0.03 | 5171 | 7.90 ± 1.70 | 100[9] | Field (100) | I |
| 95086 | 53524 | (...) | A8III | 1.77 ± 0.33 | S | 86.46 ± 0.14 | 6.94 ± 0.35 | 7509 | 14.00 ± 3.10 | 12-18[29] | LCC (80) | N |
| 98800 | 55505 | TV Crt | K5V(e) | (...) | BaBb | 42.10 ± 0.70 | 0.67 ± 0.36 | 4156 | 990 ± 300 | 9-10[30] | TWA (99) | P |
| 102647 | 57632 | $\beta$ Leo | A3Va | (...) | A | 11.00 ± 0.00 | 13.6 ± 0.12 | 6300 | 0.28 ± 0.05 | 50-331[24] | ARG (87) | N |
| 104600 | 58720 | HR 4597 | B9V | 3.04 ± 0.36 | S | 103.78 ± 0.44 | 64.08 ± 5.40 | 11055 | 0.85 ± 0.15 | 12-18[29] | LCC (99) | N |
| 105850 | 59394 | (...) | A1V | 2.26 ± 0.32 | S | 61.21 ± 0.30 | 22.37 ± 1.02 | 8831 | 0.30 ± 0.07 | 8-1000[31] | Field (100) | N |
| 106906 | 59960 | (...) | F5V | 1.32 ± 0.20 | AB | 102.38 ± 0.19 | 6.78 ± 0.20 | 6382 | 14.00 ± 2.90 | 11-15[29] | LCC (100) | I, P |
| 107146 | 60074 | (...) | G2V | 1.06 ± 0.14 | S | 27.47 ± 0.02 | 1.04 ± 0.10 | 5827 | 11.00 ± 2.90 | 50-299[11] | Field (100) | mP |
| 107301 | 60183 | (...) | B9V | 2.73 ± 0.35 | S | 97.29 ± 0.29 | 38.16 ± 0.06 | 10599 | 1.20 ± 0.23 | 12-90[32] | Field (96) | N |
| 107649 | 60348 | (...) | F5V | 1.32 ± 0.21 | S | 108.37 ± 0.20 | 2.79 ± 0.05 | 6725 | 1.24 ± 0.70 | 12-18[29] | LCC (99) | N |
| 109085 | 61174 | $\eta$ Crv | F2V | 1.48 ± 0.25 | S | 18.24 ± 0.05 | 5.39 ± 0.29 | 6553 | 0.20 ± 0.05 | 1600[12] | CAR (65) | N |
| 109573 | 61498 | HR 4796 | A0V | 2.49 ± 0.33 | A | 70.77 ± 0.24 | 25.16 ± 0.18 | 9650 | 27.00 ± 7.00 | 7-13[7] | TWA (100) | I, P |
| 110058 | 61782 | (...) | A0V | 2.04 ± 0.31 | S | 130.08 ± 0.53 | 9.05 ± 0.16 | 8162 | 27.00 ± 5.40 | 12-18[29] | LCC (93) | I |
| 110411 | 61960 | $\rho$ Vir | A0Va_lB | 2.22 ± 0.32 | S | 38.92 ± 0.19 | 15.00 ± 0.80 | 8670 | 0.69 ± 0.14 | 200[18] | Field (100) | N |
| 111520 | 62657 | (...) | F5/6V | 1.35 ± 0.21 | S | 108.05 ± 0.21 | 2.69 ± 0.03 | 6214 | 24.00 ± 5.50 | 12-18[29] | LCC (95) | I |
| 112810 | 63439 | (...) | F3/5IV/V | 1.36 ± 0.23 | S | 133.66 ± 0.29 | 3.39 ± 0.16 | 6466 | 11.00 ± 2.70 | 12-18[29] | LCC (95) | I |
| (...) | 63942 | BD+21 2486 | K4V | (...) | A | 19.68 ± 0.01 | 0.11 ± 0.01 | 4123 | (...) | 3500-5000[33] | Field (100) | N |
| 113766 | 63975 | (...) | F3/5V | (...) | A | 108.90 ± 0.31 | 5.93 ± 1.31 | 5987 | (...) | 399-1400[12] | LCC (72) | N |
| 114082 | 64184 | (...) | F3V | 1.38 ± 0.23 | S | 95.06 ± 0.20 | 4.01 ± 0.21 | 6504 | 38.00 ± 7.80 | 12-18[29] | LCC (98) | I, P |
| 115600 | 64995 | (...) | F2IV/V | 1.57 ± 0.27 | S | 109.04 ± 0.25 | 5.09 ± 0.28 | 6771 | 22.00 ± 4.80 | 12-18[29] | LCC (98) | I, P |
| 115617 | 64924 | 61 Vir | G6.5V | 0.98 ± 0.12 | S | 8.53 ± 0.01 | 0.8 ± 0.06 | 6640 | 0.27 ± 0.06 | 3050-9150[5] | Field (100) | N |
| 117214 | 65875 | (...) | F6V | 1.36 ± 0.22 | S | 107.35 ± 0.25 | 5.83 ± 0.27 | 6202 | 29.00 ± 6.60 | 5-9[34] | LCC (98) | I, P |
| 120326 | 67497 | (...) | F0V | 1.50 ± 0.25 | S | 113.27 ± 0.38 | 4.70 ± 0.23 | 6821 | 18.00 ± 3.60 | 12-18[29] | LCC (53) | I, P |
| 120534 | (...) | (...) | A5V+(F) | 1.67 ± 0.27 | AB | 84.88 ± 0.25 | 9.26 ± 0.04 | 7342 | 3.60 ± 0.80 | 360[10] | Field (72) | N |
| 121617 | (...) | (...) | A1V | 2.36 ± 0.32 | S | 117.89 ± 0.45 | 15.41 ± 0.90 | 9021 | 45.0 ± 10.3 | 14-18[29] | UCL (87) | P |
| 122652 | 68593 | (...) | F8 | 1.17 ± 0.17 | S | 39.58 ± 0.03 | 1.77 ± 0.01 | 6149 | 1.30 ± 0.30 | 300-1995[9, 20] | Field (100) | N |
| 122705 | 68781 | (...) | A2V | 2.21 ± 0.32 | S | 122.82 ± 0.49 | 11.25 ± 0.45 | 8459 | 0.70 ± 0.20 | 14-18[29] | UCL (87) | N |
| 128311 | 71395 | HN Boo | K3V | 0.81 ± 0.09 | S | 16.32 ± 0.01 | 0.32 ± 0.03 | 4980 | 0.26 ± 0.06 | 172-276[17] | Field (100) | N |

Table E.2: continued.

| HD ID | HIP ID | Alias | SpType | $M_\star$ ($M_\odot$) | Comp.[a] | $d$ (pc) | $L_\star$ ($L_\odot$) | $T_{\rm eff}$ (K) | $L_{\rm IR}/L_\star$ ($10^{-4}$) | Age (Myr) | MG[b] (%) | Detected?[c] |
|---|---|---|---|---|---|---|---|---|---|---|---|---|
| 129590 | 72070 | (...) | G3V | 1.07 ± 0.14 | S | 136.32 ± 0.44 | 3.16 ± 0.18 | 5841 | 63.00 ± 18.00 | 14-18[29] | UCL (100) | I, P |
| 131488 | (...) | (...) | A1V | 2.26 ± 0.33 | S | 152.24 ± 0.85 | 13.9 ± 1.62 | 8950 | 11.00 ± 6.70 | 14-18[29] | UCL (98) | I |
| 131835 | 73145 | (...) | A2IV | 2.08 ± 0.32 | S | 129.74 ± 0.47 | 9.81 ± 0.16 | 8281 | 28.00 ± 7.50 | 14-18[29] | UCL (100) | I, P |
| 133803 | 73990 | (...) | F2IVm-2 | 1.52 ± 0.24 | ABC | 110.11 ± 0.35 | 5.58 ± 0.23 | 6965 | 4.60 ± 0.95 | 14-18[29] | UCL (97) | N |
| 135379 | 74824 | $\beta$ Cir | A3Va | 2.14 ± 0.25 | S | 29.57 ± 0.22 | 17.57 ± 0.05 | 8452 | 0.53 ± 0.12 | 50-378[24] | Field (97) | N |
| 135599 | 74702 | (...) | K0V | 0.89 ± 0.11 | S | 15.82 ± 0.01 | 0.41 ± 0.01 | 5308 | 1.10 ± 0.32 | 650-1950[5] | Field (100) | N |
| 136246 | 75077 | (...) | A1V | 2.91 ± 0.44 | S | 114.67 ± 0.45 | 15.71 ± 0.03 | 8715 | 0.53 ± 0.13 | 14-18[29] | UCL (92) | N |
| 138965 | 76736 | HR 5792 | A1V | 2.25 ± 0.33 | S | 78.60 ± 0.17 | 14.72 ± 0.34 | 8775 | 4.50 ± 1.40 | 40-50[8] | ARG (99) | N |
| 139664 | 76829 | g Lup | F3/5V | 1.40 ± 0.24 | S | 17.40 ± 0.04 | 3.63 ± 0.26 | 6415 | 1.40 ± 0.32 | 66-334[11] | Field (100) | N |
| 140840 | 77317 | (...) | B9/A0V | (...) | S | 144.08 ± 0.66 | 25.45 ± 4.95 | 9784 | 1.80 ± 0.60 | 14-18[29] | UCL (99) | N |
| 141011 | 77432 | (...) | F5V | 1.43 ± 0.24 | S | 128.38 ± 0.32 | 2.5 ± 0.90 | 7000 | (...) | 14-18[29] | UCL (100) | I |
| 141378 | 77464 | (...) | A5IV-V | 2.08 ± 0.30 | S | 53.20 ± 0.17 | 14.69 ± 0.73 | 8361 | 0.82 ± 0.19 | 190-570[35] | Field (100) | N |
| 141518 | (...) | (...) | F3V | 1.43 ± 0.24 | S | 108.19 ± 0.30 | 3.68 ± 0.24 | 6558 | 220.0 ± 16.0 | 14-18[29] | UCL (97) | N |
| 141569 | 77542 | (...) | A2VekB9mB9(_lB) | (...) | A | 111.61 ± 0.37 | 15.15 ± 0.07 | 8446 | 47.0 ± 14.0 | 2-8[11] | Field (100) | I |
| 141943 | (...) | NZ Lupi | G2 | 1.09 ± 0.14 | S | 60.14 ± 0.08 | 2.21 ± 0.05 | 5752 | 1.20 ± 0.43 | 14-18[29] | UCL (53) | I |
| 142446 | 78043 | (...) | F3V | 1.54 ± 0.27 | AB | 135.61 ± 0.37 | 4.02 ± 0.16 | 6529 | 6.10 ± 2.10 | 14-18[29] | UCL (99) | N |
| 145229 | 79165 | (...) | G0 | 1.09 ± 0.14 | S | 33.78 ± 0.02 | 1.02 ± 0.04 | 5935 | 1.20 ± 0.27 | 1300[12] | Field (100) | N |
| 145560 | 79516 | (...) | F5V | 1.49 ± 0.25 | S | 121.23 ± 0.29 | 3.46 ± 0.23 | 6325 | 31.00 ± 8.60 | 14-18[29] | UCL (99) | I, P |
| 146181 | 79742 | (...) | F6V | 1.57 ± 0.27 | S | 127.52 ± 0.26 | 3.00 ± 0.31 | 6407 | 23.00 ± 7.60 | 14-18[29] | UCL (99) | I |
| 146897 | 79977 | (...) | F2/3V | 1.23 ± 0.19 | S | 132.19 ± 0.42 | 3.51 ± 0.16 | 6118 | 76.0 ± 22.0 | 9-13[29] | US (100) | I, P |
| 149914 | 81474 | (...) | B9.5IV | 2.83 ± 0.39 | S | 154.37 ± 0.63 | 41.53 ± 6.10 | 6645 | 10.00 ± 2.20 | 0.5-1[36] | Field (92) | N |
| 153053 | 83187 | (...) | A5IV/V | 1.92 ± 0.30 | S | 52.68 ± 0.13 | 12.54 ± 0.16 | 7872 | 0.85 ± 0.18 | 420-800[37,38] | Field (98) | N |
| 156623 | 84881 | (...) | A0V | 2.33 ± 0.22 | S | 108.33 ± 0.33 | 13.16 ± 0.56 | 8767 | 38.00 ± 8.90 | 9-23[40] | UCL (65) | P |
| 157587 | 85224 | (...) | F5V | 1.26 ± 0.19 | S | 99.87 ± 0.23 | 3.46 ± 0.17 | 6297 | (...) | 165-835[11] | Field (94) | I, P |
| 157728 | 85157 | 73 Her | A7V | 1.77 ± 0.27 | S | 42.72 ± 0.06 | 7.48 ± 0.75 | 7619 | 3.00 ± 0.67 | 200[10] | Field (100) | N |
| 159492 | 86305 | $\pi$ Ara | A5IV/V | 1.88 ± 0.29 | S | 41.01 ± 0.14 | 11.03 ± 0.45 | 7827 | 1.20 ± 0.23 | 50-419[24] | Field (100) | N |
| 160305 | 86598 | (...) | F8/G0V | 1.13 ± 0.15 | S | 65.80 ± 0.10 | 1.78 ± 0.01 | 6016 | 1.40 ± 0.48 | 18-23[1] | $\beta$PMG (96) | I, P |
| 161868 | 87108 | $\gamma$ Oph | A1VnkA0mA0 | 2.36 ± 0.31 | S | 29.75 ± 0.23 | 25.29 ± 4.33 | 8820 | 1.00 ± 0.23 | 435-602[11] | Field (100) | N |
| 164249 | 88399 | (...) | F6V | 1.36 ± 0.22 | A | 49.30 ± 0.06 | 3.20 ± 0.05 | 6340 | 9.40 ± 2.50 | 18-23[1] | $\beta$PMG (100) | N |
| 170773 | 90936 | HR 6948 | F5V | 1.40 ± 0.23 | S | 36.93 ± 0.04 | 3.78 ± 0.22 | 6512 | 5.20 ± 1.10 | 50-200[11] | Field (100) | N |
| 172555 | 92024 | HR 7012 | A7V | 1.87 ± 0.30 | A | 28.79 ± 0.13 | 8.41 ± 0.61 | 7499 | 5.60 ± 1.60 | 18-23[1] | $\beta$PMG (100) | P |
| 174429 | 92680 | PZ Tel | G9IV/K0 | 0.88 ± 0.10 | A | 47.25 ± 0.05 | 0.99 ± 0.02 | 5338 | 0.17 ± 0.01 | 18-23[1] | $\beta$PMG (98) | N |
| 178253 | 94114 | $\alpha$ CrA | A2Va | 2.24 ± 0.32 | S | 36.92 ± 0.49 | 27.22 ± 2.04 | 8706 | 0.21 ± 0.06 | 164-316[24] | Field (100) | N |
| 181296 | 95261 | $\eta$ Tel | A0V | 2.49 ± 0.35 | A | 48.54 ± 0.23 | 21.58 ± 0.04 | 9108 | 1.60 ± 0.28 | 18-23[1] | $\beta$PMG (77) | N |
| 181327 | 95270 | (...) | F6V | 1.32 ± 0.21 | B | 47.78 ± 0.07 | 2.98 ± 0.10 | 6323 | 26.00 ± 7.40 | 18-23[1] | $\beta$PMG (100) | I, P |
| 181869 | 95347 | $\alpha$ Sgr | B8V | 3.01 ± 0.30 | SBC | 55.22 ± 0.90 | 120.1 ± 0.7 | 11721 | 0.01 ± 0.01 | 130-200[7] | ABDMG (81) | N |
| 182681 | 95619 | HR 7380 | B8.5V | 2.57 ± 0.33 | S | 70.69 ± 0.45 | 28.47 ± 0.2 | 9647 | 3.00 ± 0.65 | 18-23[1] | $\beta$PMG (89) | I |
| 183324 | 95793 | c Aql | A0IVp | 2.38 ± 0.33 | S | 60.37 ± 0.16 | 15.21 ± 0.83 | 8561 | 0.20 ± 0.04 | 330[18] | Field (100) | N |
| 188228 | 98495 | $\epsilon$ Pav | A0Va | 2.59 ± 0.34 | S | 31.99 ± 0.20 | 28.25 ± 2.01 | 9740 | 0.05 ± 0.01 | 40-80[8] | ARG (100) | N |
| 191089 | 99273 | (...) | F5V | 1.33 ± 0.21 | SBC | 50.11 ± 0.05 | 2.88 ± 0.14 | 6350 | 16.00 ± 3.40 | 18-23[1] | $\beta$PMG (100) | P |
| 191131 | 99290 | (...) | F0V | 1.50 ± 0.26 | S | 142.25 ± 0.91 | 6.71 ± 0.70 | 6934 | (...) | 1700[12] | Field (100) | N |
| 192263 | 99711 | (...) | K1/2V | 0.82 ± 0.10 | S | 19.63 ± 0.01 | 0.33 ± 0.30 | 5038 | 0.48 ± 0.10 | 2200-11400[41] | Field (100) | N |
| 192425 | 99742 | $\rho$ Aql | A1Va | 2.22 ± 0.33 | S | 47.83 ± 0.31 | 21.39 ± 0.93 | 8718 | 0.34 ± 0.07 | 50-166[24] | Field (98) | N |
| 192758 | (...) | (...) | F0V | 1.57 ± 0.26 | S | 66.50 ± 0.14 | 5.72 ± 0.23 | 6971 | 5.70 ± 1.40 | 40-830[8,42] | ARG (100) | P |
| 197481 | 102409 | AU Mic | M1VeBa1 | 0.66 ± 0.20 | A | 9.71 ± 0.00 | 0.11 ± 0.01 | 3633 | 3.80 ± 0.88 | 18-23[1] | $\beta$PMG (100) | I, P |
| 201219 | 104318 | (...) | G5 | 1.00 ± 0.13 | S | 37.81 ± 0.04 | 0.80 ± 0.01 | 5556 | 1.20 ± 0.27 | 1000[20] | Field (100) | N |
| 202917 | 105388 | (...) | G7V | 0.98 ± 0.13 | S | 46.71 ± 0.03 | 0.71 ± 0.05 | 5537 | 2.79 ± 1.00 | 41-49[7] | THA (100) | P |
| 205674 | 106741 | (...) | F3/5IV | 1.45 ± 0.25 | S | 55.74 ± 0.09 | 3.44 ± 0.04 | 6679 | 3.70 ± 0.79 | 130-200[7] | ABDMG (73) | N |
| 206893 | 107412 | (...) | F5V | 1.36 ± 0.22 | A | 40.77 ± 0.06 | 3.00 ± 0.17 | 6439 | 2.80 ± 0.55 | 66-334[11] | Field (60) | N |
| 216956 | 113368 | Fomalhaut | A4V | (...) | A | 7.70 ± 0.00 | 15.45 ± 1.17 | 8195 | 0.83 ± 0.29 | 400-480[3] | Field (100) | N |
| 218340 | 114236 | (...) | G3V | 1.07 ± 0.14 | S | 56.13 ± 0.06 | 1.14 ± 0.03 | 5834 | 0.68 ± 0.15 | 10-2000[43] | Field (100) | N |

Table E.2: continued.

| HD ID | HIP ID | Alias | SpType | $M_\star$ ($M_\odot$) | Comp.[a] | $d$ (pc) | $L_\star$ ($L_\odot$) | $T_{\rm eff}$ (K) | $L_{\rm IR}/L_\star$ ($10^{-4}$) | Age (Myr) | MG[b] (%) | Detected?[c] |
|---|---|---|---|---|---|---|---|---|---|---|---|---|
| 218396 | 114189 | HR 8799 | F0+VkA5mA5 | 1.61 ± 0.27 | S | 40.88 ± 0.08 | 5.75 ± 0.40 | 7248 | 2.30 ± 0.46 | 38-48 | Field (51) | P |
| 219482 | 114948 | GJ 1282 | F6V | 1.24 ± 0.18 | S | 20.44 ± 0.01 | 2.01 ± 0.13 | 6118 | 0.36 ± 0.07 | 115-385 | Field (100) | N |
| 220825 | 115738 | $\kappa$ Psc | A2VpSrCrSi | 2.40 ± 0.33 | S | 49.22 ± 0.24 | 25.07 ± 2.72 | 9304 | 0.28 ± 0.07 | 130-200[7] | ABDMG (93) | N |
| 221853 | 116431 | (...) | F0 | 1.49 ± 0.25 | S | 65.89 ± 0.12 | 3.93 ± 0.03 | 6842 | 8.60 ± 2.00 | 50-200[11] | Field (99) | N |
| 274255 | 25775 | V⋆ VZ Col | M0V | 0.63 ± 0.08 | AB | 19.16 ± 0.00 | 0.10 ± 0.03 | 4036 | 1.90 ± 820.00 | (...) | Field (100) | N |
| (...) | (...) | TWA 7 | M2Ve | 0.46 ± 0.09 | S | 34.10 ± 0.03 | 0.12 ± 0.01 | 3500 | 21.00 ± 4.50 | 3-5.8[44] | Field (51) | P |
| (...) | (...) | TWA 25 | M0.5 | 0.60 ± 0.08 | S | 53.60 ± 0.07 | 0.27 ± 0.03 | 3803 | (...) | 7-13[7] | TWA (100) | I |
| (...) | (...) | BD-20 951 | K1V(e) | 0.73 ± 0.08 | AB | 62.16 ± 0.08 | 0.48 ± 0.21 | 4550 | 2.15 ± 0.43 | (...) | Field (100) | I, P |

**Notes.** [a] Component with a debris disk in a single or multiple star system which was observed with SPHERE: "S" denotes a single star, "AB" a spectroscopic or eclipsing binary, and "SBC" a SB candidate. [b] MG abbreviations: $\beta$PMG - $\beta$ Pictoris MG, ABDMG = AB Doradus MG, ARG = Argus, CAR - Carina, COL - Columba, ETAC = Eta Chamaeleontis association, LCC - Lower Centaurus Crux, THA - Tucana–Horologium Association, TWA - TW Hydrae Association, UCL - Upper Centaurus Lupus, US - Upper Scorpius, 118TAU - 118 Tauri association. [c] I = disk detected in total intensity, P = disk detected in polarized intensity, N = no detection, m = marginal detection.

**References.** [1] Couture et al. (2023); [2] Lestrade et al. (2012); [3] Mamajek (2012); [4] Marshall et al. (2018); [5] Mamajek & Hillenbrand (2008); [6] Choquet et al. (2016); [7] Bell et al. (2015); [8] Zuckerman (2019); [9] Rhee et al. (2007); [10] Moór et al. (2006); [11] Esposito et al. (2020); [12] Holmberg et al. (2009); [13] Luck (2015); [14] Vican (2012); [15] Metchev & Hillenbrand (2009); [16] Bonavita et al. (2022); [17] Rich et al. (2017); [18] Iliev & Barzova (1995); [19] Zuckerman & Song (2004); [20] Carpenter et al. (2009); [21] Malo et al. (2013); [22] Currie et al. (2017); [23] Moór et al. (2016); [24] Song et al. (2001); [25] Torres et al. (2008); [26] Aguilera-Gómez et al. (2018); [27] Lombart et al. (2020); [28] König (2003); [29] Pecaut et al. (2012); [30] Barrado Y Navascués (2006); [31] Iglesias et al. (2018); [32] Desidera et al. (2021); [33] West et al. (2008); [34] Engler et al. (2020); [35] Lazzoni et al. (2018); [36] Habart et al. (2003); [37] Chen et al. (2006); [38] Rieke et al. (2005); [39] Kennedy et al. (2014); [40] Mellon et al. (2019); [41] Brewer et al. (2016); [42] Chen et al. (2014); [43] Pearce et al. (2022); [44] Herczeg & Hillenbrand (2014); [45] Sissa et al. (2018); [46] Murphy & Lawson (2014); [47] Moór et al. (2011);

Table E.3: Results for one-component MBB model.

| Name | $f_{\rm disk}$ $10^{-4}$ | $R_{\rm MBB}$ (au) | $T_{\rm MBB}$ (K) | $\lambda_0$ ($\mu$m) | $\beta$ | $N_{\rm obs}$ | Name | $f_{\rm disk}$ $10^{-4}$ | $R_{\rm MBB}$ (au) | $T_{\rm MBB}$ (K) | $\lambda_0$ ($\mu$m) | $\beta$ | $N_{\rm obs}$ |
|---|---|---|---|---|---|---|---|---|---|---|---|---|---|
| HIP 11437 | 12.00 ± 2.80 | 7.5 ± 1.5 | 70 | 480 | 0.7 | 6 | HD 104600 | 0.85 ± 0.15 | 27.0 ± 9.7 | 160 | 100 | 0.1 | 3 |
| CPD-72 2713 | 10.00 ± 3.70 | 17.0 ± 6.1 | 45 | 37 | 0.001 | 7 | HD 105850 | 0.30 ± 0.07 | 13.0 ± 5.6 | 170 | 100 | 0.5 | 3 |
| GJ 581 | 0.72 ± 0.14 | 7.4 ± 2.5 | 34 | 69 | 0.6 | 5 | HD 106906 | 14.00 ± 2.90 | 21.0 ± 4.8 | 96 | 69 | 0.7 | 6 |
| GSC 06964-1226 | 1.70 ± 0.40 | 8.1 ± 1.9 | 26 | 170 | 1.0 | 8 | HD 107146 | 11.00 ± 2.90 | 29.0 ± 8.9 | 52 | 350 | 0.7 | 23 |
| HD 105 | 2.80 ± 0.55 | 35.0 ± 9.0 | 50 | 170 | 0.6 | 10 | HD 107301 | 1.20 ± 0.23 | 35.1 ± 16.9 | 117 | 20 | 0.4 | 3 |
| HD 166 | 0.75 ± 0.14 | 6.4 ± 1.5 | 98 | 140 | 1.0 | 9 | HD 107649 | 1.24 ± 0.70 | 8.7 ± 3.4 | 122 | 178 | 1.0 | 5 |
| HD 203 | 1.50 ± 0.29 | 9.0 ± 2.3 | 130 | 42 | 0.2 | 6 | HD 109573 | 27.00 ± 7.00 | 45.0 ± 10.0 | 94 | 190 | 1.2 | 18 |
| HD 377 | 4.08 ± 0.95 | 20.4 ± 4.2 | 64 | 369 | 0.4 | 8 | HD 110058 | 27.00 ± 5.40 | 19.0 ± 4.0 | 110 | 94 | 0.6 | 9 |
| HD 1466 | 0.77 ± 0.15 | 5.2 ± 1.7 | 140 | 170 | 0.4 | 5 | HD 110411 | 0.69 ± 0.14 | 37.0 ± 8.3 | 87 | 160 | 1.5 | 11 |
| HD 3003 | 1.10 ± 0.25 | 10.0 ± 3.3 | 180 | 41 | 1.2 | 7 | HD 111520 | 24.00 ± 5.50 | 19.0 ± 4.2 | 81 | 260 | 0.5 | 7 |
| HD 3670 | 5.30 ± 1.20 | 39.0 ± 8.8 | 56 | 74 | 0.3 | 5 | HD 112810 | 11.00 ± 2.70 | 33.0 ± 6.6 | 65 | 180 | 0.7 | 7 |
| HD 10472 | 3.00 ± 1.10 | 25.0 ± 7.9 | 77 | 150 | 1.2 | 3 | HD 114082 | 38.00 ± 7.80 | 12.0 ± 2.9 | 110 | 180 | 1.0 | 7 |
| HD 10638 | 2.20 ± 0.63 | 35.0 ± 8.6 | 77 | 960 | 1.9 | 3 | HD 115600 | 22.00 ± 4.80 | 16.0 ± 4.1 | 100 | 58 | 0.8 | 7 |
| HD 10700 | 0.24 ± 0.05 | 3.7 ± 0.8 | 120 | 170 | 0.001 | 13 | HD 115617 | 0.27 ± 0.06 | 16.0 ± 4.5 | 67 | 370 | 0.1 | 12 |
| HD 10939 | 0.83 ± 0.23 | 100 ± 23 | 66 | 180 | 1.4 | 6 | HD 117214 | 29.00 ± 6.60 | 15.0 ± 4.2 | 110 | 150 | 1.0 | 5 |
| HD 13246 | 1.60 ± 0.37 | 4.7 ± 1.4 | 150 | 240 | 1.3 | 5 | HD 120326 | 18.00 ± 3.60 | 14.0 ± 3.3 | 110 | 240 | 2.0 | 5 |
| HD 14082 | 2.27 ± 0.49 | 8.9 ± 2.0 | 95 | 306 | 0.2 | 6 | HD 120534 | 3.60 ± 0.80 | 30.0 ± 6.5 | 89 | 91 | 0.7 | 5 |
| HD 15257 | 1.20 ± 0.30 | 68.0 ± 14.0 | 66 | 660 | 1.2 | 5 | HD 121617 | 45.00 ± 10.30 | 25.0 ± 5.6 | 106 | 127 | 0.8 | 7 |
| HD 16743 | 4.20 ± 1.20 | 46.0 ± 10.0 | 62 | 200 | 1.5 | 6 | HD 122652 | 1.14 ± 0.30 | 35.0 ± 12.0 | 54 | 93 | 0.1 | 3 |
| HD 17390 | 2.30 ± 0.60 | 64.0 ± 15.0 | 51 | 440 | 0.8 | 3 | HD 128311 | 0.26 ± 0.06 | 21.0 ± 6.1 | 44 | 64 | 0.2 | 5 |
| HD 17848 | 0.39 ± 0.10 | 97.0 ± 32.9 | 67 | 247 | 0.6 | 7 | HD 129590 | 63.00 ± 18.00 | 17.0 ± 4.5 | 89 | 390 | 0.9 | 5 |

Table E.3: continued.

| Name | $f_{disk}$ $10^{-4}$ | $R_{MBB}$ (au) | $T_{MBB}$ (K) | $\lambda_0$ ($\mu$m) | $\beta$ | $N_{obs}$ |
|---|---|---|---|---|---|---|
| HD 17925 | 0.89 ± 0.19 | 6.2 ± 1.4 | 89 | 28 | 1.4 | 12 |
| HD 20320 | 0.13 ± 0.04 | 79.4 ± 25.8 | 58 | 134 | 1.7 | 7 |
| HD 20794 | 0.02 ± 0.02 | 12.0 ± 6.8 | 74 | 300 | 0.6 | 7 |
| HD 21997 | 6.00 ± 1.30 | 61.0 ± 13.0 | 65 | 190 | 1.2 | 13 |
| HD 22049 | 1.30 ± 0.28 | 6.3 ± 1.3 | 85 | 480 | 0.3 | 23 |
| HD 22179 | 2.99 ± 1.03 | 17.1 ± 4.3 | 70 | 196 | 0.4 | 4 |
| HD 23484 | 0.92 ± 0.19 | 17.0 ± 5.8 | 55 | 400 | 1.1 | 10 |
| HD 24636 | 1.20 ± 0.22 | 10.0 ± 3.0 | 120 | 85 | 1.1 | 5 |
| HD 27290 | 0.23 ± 0.05 | 43.0 ± 11.0 | 68 | 410 | 1.2 | 11 |
| HD 29391 | 0.12 ± 0.06 | 9.0 ± 2.8 | 140 | 100 | 0.5 | 6 |
| HD 30422 | 0.49 ± 0.10 | 38.0 ± 8.5 | 77 | 96 | 1.8 | 6 |
| HD 30447 | 9.60 ± 3.80 | 32.0 ± 8.6 | 68 | 240 | 0.3 | 4 |
| HD 31392 | 1.20 ± 0.24 | 35.0 ± 9.5 | 41 | 68 | 0.6 | 5 |
| HD 35114 | 0.53 ± 0.13 | 6.5 ± 2.8 | 130 | 53 | 0.9 | 3 |
| HD 35650 | 1.00 ± 0.27 | 14.0 ± 7.2 | 44 | 213 | 2.7 | 5 |
| HD 35841 | 17.00 ± 5.90 | 25.0 ± 6.1 | 69 | 220 | 0.8 | 5 |
| HD 36546 | 44.00 ± 9.70 | 14.0 ± 3.2 | 150 | 280 | 0.5 | 5 |
| HD 37484 | 3.30 ± 0.89 | 17.0 ± 4.5 | 92 | 160 | 0.003 | 6 |
| HD 38206 | 1.50 ± 0.37 | 53.0 ± 14.0 | 88 | 280 | 0.8 | 5 |
| HD 38207 | 10.00 ± 3.70 | 36.0 ± 9.4 | 65 | 140 | 1.9 | 4 |
| HD 38397 | 4.80 ± 0.88 | 46.0 ± 13.0 | 44 | 200 | 0.5 | 3 |
| HD 38678 | 1.10 ± 0.30 | 10.0 ± 4.1 | 170 | 54 | 1.3 | 7 |
| HD 38858 | 0.80 ± 0.17 | 19.0 ± 4.1 | 61 | 840 | 0.7 | 7 |
| HD 40540 | 5.30 ± 1.30 | 26.0 ± 6.3 | 86 | 230 | 0.2 | 3 |
| HD 43989 | 0.49 ± 0.46 | 11.0 ± 3.3 | 96 | 200 | 1.7 | 5 |
| HD 48370 | 4.70 ± 0.88 | 44.0 ± 10.0 | 40 | 250 | 0.4 | 8 |
| HD 50571 | 1.10 ± 0.21 | 52.0 ± 15.0 | 53 | 250 | 0.9 | 7 |
| HD 52265 | 0.22 ± 0.06 | 39.3 ± 16.2 | 55 | 58 | 1.6 | 6 |
| HD 53143 | 2.80 ± 0.56 | 11.0 ± 2.6 | 73 | 490 | 0.8 | 6 |
| HD 53842 | 1.26 ± 0.20 | 6.7 ± 3.2 | 140 | 20 | 1.6 | 6 |
| HD 54341 | 2.80 ± 0.65 | 89.0 ± 18.0 | 65 | 270 | 1.4 | 6 |
| HD 60491 | 2.00 ± 0.51 | 7.2 ± 1.6 | 77 | 130 | 1.2 | 3 |
| HD 71155 | 0.30 ± 0.06 | 46.0 ± 10.0 | 100 | 60 | 1.4 | 10 |
| HD 71722 | 1.10 ± 0.23 | 45.0 ± 10.0 | 84 | 120 | 0.9 | 5 |
| HD 73350 | 1.30 ± 0.41 | 21.0 ± 5.6 | 61 | 230 | 1.5 | 5 |
| HD 75416 | 1.10 ± 0.60 | 8.3 ± 4.1 | 320 | 180 | 1.6 | 4 |
| HD 76582 | 2.50 ± 0.55 | 75.0 ± 18.0 | 57 | 210 | 1.0 | 6 |
| HD 80950 | 1.20 ± 0.29 | 13.0 ± 4.6 | 180 | 77 | 0.4 | 5 |
| HD 82943 | 1.10 ± 0.24 | 26.0 ± 5.8 | 61 | 197 | 1.0 | 5 |
| HD 84075 | 2.20 ± 0.64 | 20.0 ± 4.5 | 68 | 610 | 1.8 | 3 |
| HD 90905 | 0.36 ± 0.09 | 17.3 ± 6.5 | 74 | 165 | 1.2 | 6 |
| HD 92945 | 7.90 ± 1.70 | 14.0 ± 3.0 | 59 | 260 | 0.6 | 10 |
| HD 95086 | 14.00 ± 3.10 | 43.0 ± 9.4 | 68 | 370 | 1.0 | 13 |
| HD 98800 | 990 ± 300 | 3.8 ± 1.9 | 140 | 200 | 0.1 | 22 |
| HD 102647 | 0.28 ± 0.05 | 25.0 ± 6.7 | 110 | 23 | 0.6 | 10 |
| HD 133803 | 4.60 ± 0.95 | 5.1 ± 1.9 | 190 | 310 | 0.3 | 5 |
| HD 135379 | 0.53 ± 0.12 | 7.9 ± 2.7 | 200 | 34 | 0.6 | 4 |
| HD 135599 | 1.10 ± 0.32 | 14.0 ± 3.3 | 60 | 170 | 0.6 | 3 |
| HD 136246 | 0.53 ± 0.13 | 60.0 ± 13.0 | 72 | 160 | 0.8 | 5 |
| HD 138965 | 5.24 ± 1.22 | 56.6 ± 11.4 | 72 | 107 | 0.1 | 6 |
| HD 139664 | 1.40 ± 0.32 | 22.0 ± 7.5 | 80 | 170 | 0.8 | 7 |
| HD 140840 | 1.80 ± 0.60 | 40.0 ± 14.0 | 100 | 320 | 1.5 | 3 |
| HD 141378 | 0.82 ± 0.19 | 61.0 ± 13.0 | 71 | 160 | 0.5 | 5 |
| HD 141518 | 220.0 ± 16.0 | 82.0 ± 5.2 | 42 | 5 | 1.0 | 1 |
| HD 141943 | 1.20 ± 0.43 | 15.0 ± 5.7 | 87 | 310 | 0.1 | 3 |
| HD 142446 | 6.10 ± 2.10 | 24.0 ± 6.9 | 80 | 270 | 0.4 | 5 |
| HD 145229 | 1.20 ± 0.27 | 21.0 ± 7.8 | 61 | 70 | 0.3 | 5 |
| HD 145560 | 31.00 ± 8.60 | 21.0 ± 4.9 | 81 | 130 | 0.3 | 5 |
| HD 146181 | 23.00 ± 7.60 | 20.0 ± 5.2 | 81 | 120 | 0.4 | 5 |
| HD 146897 | 76.00 ± 22.00 | 17.0 ± 4.4 | 92 | 320 | 1.1 | 5 |
| HD 149914 | 10.00 ± 2.20 | 42.0 ± 15.0 | 110 | 23 | 1.4 | 4 |
| HD 153053 | 0.85 ± 0.18 | 52.0 ± 11.0 | 72 | 310 | 0.2 | 5 |
| HD 156623 | 38.00 ± 8.90 | 12.0 ± 2.7 | 150 | 140 | 0.7 | 4 |
| HD 157728 | 3.00 ± 0.67 | 11.0 ± 4.8 | 130 | 79 | 0.9 | 3 |
| HD 159492 | 1.20 ± 0.23 | 11.0 ± 4.2 | 150 | 190 | 1.4 | 8 |
| HD 160305 | 1.40 ± 0.48 | 32.0 ± 12.0 | 56 | 180 | 0.6 | 3 |
| HD 161868 | 1.00 ± 0.23 | 64.0 ± 14.0 | 81 | 180 | 1.0 | 9 |
| HD 164249 | 9.40 ± 2.50 | 28.0 ± 6.0 | 70 | 140 | 1.1 | 7 |
| HD 170773 | 5.20 ± 1.10 | 77.0 ± 16.0 | 44 | 140 | 0.9 | 13 |
| HD 172555 | 5.60 ± 1.60 | 5.0 ± 1.6 | 210 | 17 | 0.5 | 8 |
| HD 178253 | 0.21 ± 0.06 | 12.0 ± 4.9 | 190 | 380 | 0.1 | 4 |
| HD 181327 | 26.00 ± 7.40 | 22.0 ± 5.2 | 77 | 300 | 0.6 | 17 |
| HD 182681 | 3.00 ± 0.65 | 56.0 ± 15.0 | 84 | 270 | 1.0 | 7 |
| HD 183324 | 0.15 ± 0.04 | 53.7 ± 24.4 | 74 | 247 | 1.1 | 6 |
| HD 188228 | 0.05 ± 0.01 | 54.0 ± 15.0 | 90 | 190 | 0.5 | 10 |
| HD 191089 | 16.00 ± 3.40 | 15.0 ± 3.1 | 93 | 240 | 0.7 | 12 |
| HD 192263 | 0.48 ± 0.10 | 12.0 ± 4.0 | 59 | 70 | 0.2 | 5 |
| HD 192425 | 0.47 ± 0.10 | 61.6 ± 22.2 | 77 | 181 | 2.6 | 6 |
| HD 192758 | 6.11 ± 1.40 | 46.0 ± 9.3 | 63 | 69 | 0.5 | 5 |
| HD 197481 | 3.80 ± 0.88 | 11.0 ± 2.6 | 48 | 340 | 0.3 | 16 |
| HD 201219 | 1.20 ± 0.27 | 18.0 ± 7.2 | 62 | 90 | 0.9 | 5 |
| HD 202917 | 2.79 ± 1.00 | 9.9 ± 2.4 | 81 | 349 | 1.9 | 7 |
| HD 205674 | 3.70 ± 0.79 | 43.0 ± 9.5 | 58 | 370 | 1.4 | 7 |
| HD 206893 | 2.80 ± 0.55 | 49.0 ± 12.0 | 52 | 150 | 1.1 | 7 |
| HD 216956 | 0.83 ± 0.29 | 60.0 ± 35.0 | 72 | 180 | 0.9 | 17 |
| HD 218340 | 0.68 ± 0.15 | 25.0 ± 8.7 | 57 | 130 | 1.8 | 5 |
| HD 219482 | 0.36 ± 0.07 | 11.0 ± 2.5 | 99 | 78 | 0.4 | 6 |
| HD 220825 | 0.28 ± 0.07 | 11.0 ± 3.6 | 180 | 560 | 1.9 | 5 |
| HD 221853 | 8.60 ± 2.00 | 20.0 ± 4.6 | 87 | 340 | 0.9 | 5 |
| TWA 7 | 20.70 ± 4.51 | 5.3 ± 1.1 | 72 | 540 | 0.4 | 10 |

**Notes.** The columns list target IDs, disk IR excess ($f_{disk}$), BB belt radius ($R_{MBB}$), BB temperature ($T_{MBB}$), characteristic wavelength ($\lambda_0$) and opacity index ($\beta$). The column $N_{obs}$ gives the number of photometric points for wavelengths > 22 $\mu$m which determines whether a fit with more than one component is feasible.

## Appendix F: Companions to the program stars

Table F.1 presents the parameters of confirmed exoplanets with masses below 13 $M_{\rm Jup}$ in the stellar systems of our sample, as of August 1, 2024. The data were retrieved from the NASA Exoplanet archive[17].

To identify companions to the program stars with masses greater than 13 $M_{\rm Jup}$, we followed the methodology outlined in Gratton et al. (2023b,a, 2024, 2025). Our search incorporated both direct observations of visual companions, primarily from Gaia data and HCI, and indirect evidence based on photometry (eclipsing binaries), RV measurements (spectroscopic binaries), and astrometry (mainly from Gaia).

### F.1. Visual binaries

Wide visual companions (separation > 0.7″) can be detected either as separate entries in Gaia DR3 or by HCI. Both are available for all the program stars. Given the age of the stars, Gaia DR3 can reveal companions down to the star-brown dwarfs for separation larger than a few hundred au. HCI is sensitive to massive planetary companions, but only in the separation range from a few tens to about 1000 au.

We considered as physical companions all point sources that have similar parallax and proper motion, and are within 60″ (that is, about 1000-10000 au, depending on distance of the target) of each of the program stars. We also notice that the astrometric solution may be missing for faint sources very close to brighter objects in the Gaia DR3 catalogue. Given the low-density of the background fields, we additionally considered as physical companions the objects listed in Gaia and projected within 2″ of each star, but lacking an astrometric solution.

### F.2. Eclipsing binaries

We searched for eclipsing binaries (EBs) within the TESS dataset but did not identify any among the program stars.

### F.3. Spectroscopic binaries

We inspected the $S_{B^9}$ (Pourbaix et al. (2004), Tokovinin (2018)) and Gaia DR3 binary Gaia Collaboration et al. (2023a) catalogs for spectroscopic and spectrophotometric binaries. In addition, we considered high-precision RV series from the literature and the low-precision one from Gaia DR3. Companions detected using RVs typically have separation less than a few au's.

### F.4. Astrometric binaries

We inspected various catalogs looking for astrometric binaries based on Gaia data: nss_two_body_orbit (Gaia Collaboration et al. 2023a), nss_acceleration_astro (Gaia Collaboration et al. 2023a, Holl et al. 2023).

We considered proper motion acceleration (PMa) from Kervella et al. (2022). The PMa is the difference between the proper motion in Gaia DR3 (baseline of 34 months) and that determined using the position at Hipparcos (1991.25) and Gaia EDR3 (2016.0) epochs. This quantity is available for a vast majority of the program stars. PMa is sensitive to binaries with a projected separation between 1 and 100 au. We consider any value of PMa with a S/N > 3 as an indication of the presence of companions.

We also considered the re-normalised unit weight error (RUWE) as an indication of binarity. This parameter is an indication of the goodness of the 5-parameter solution found by Gaia (Lindegren et al. 2018). Belokurov et al. (2020) showed that when this parameter is > 1.4 the star is likely a binary, at least for stars that are not too bright ($G > 4$) and saturated in the Gaia scans. This method is sensitive to systems that have periods from a few months to a decade (Penoyre et al. 2022). The RUWE parameter is available for the vast majority of the program stars.

### F.5. Parameters for the components

The semimajor axis and the mass for the companions are listed in Table F.2. They are obtained following the methods considered in Gratton et al. (2023b, 2024, 2025), briefly summarised in the following.

For unresolved systems, the sum of the masses is made compatible with the apparent $G$ magnitude of the system, using the mass-luminosity relation for the Gaia $G$ band appropriate for the age of the system. We assumed that the semimajor axis is equal to the projected separation divided by the parallax. On average, this corresponds to the thermal eccentricity distribution considered by Ambartsumian (1937) of $f(e) = 2e$ (see Brandeker et al. 2006). Uncertainties in the masses derived using these recipes are small (well below 10%), while those for the semimajor axes are about 40% (see Figure A.1 in Brandeker et al. 2006).

The indication of binarity for many objects comes from RUWE (> 1.4) or PMa (S/N > 3 objects) or a combination of these techniques. The secondary of these stars is not imaged, and no period or semimajor axis is determined. Since RV variations, RUWE, and PMa have different dependence on the semimajor axis and the mass, we may better constrain the parameters of the companions by combining different methods using all available information rather than considering only a single technique. We did this by means of exploring the semimajor axis – mass ratio plane using a Monte Carlo code. We assumed eccentric orbits, with uniform priors between 0 and 1 in eccentricity (which is in agreement with Hwang et al. 2022 for this range of separations), 0 and 180 degrees in the angle of the ascending node $\Omega$, and 0 and 360 degrees in the periastron angle $\omega$, and left the inclination and phase to assume a random value. In addition, the period was used to fix the solution whenever it was available. Uncertainties are large and only give order-of-magnitude estimates.

## Appendix G: List of symbols

In this section, we provide a list of the symbols employed throughout this work.

| | |
|---|---|
| $A$ | geometrical cross section of particle |
| $A_{\rm disk}$ | vertical disk aspect ratio |
| $a$ | particle radius |
| $a_{\rm blow}$ | blowout grain size |
| $a_{\rm max}$ | maximum grain radius in SD model |
| $a_{\rm min}$ | minimum grain radius in SD model |
| $a_{\rm p}$ | semimajor axis of planet orbit |
| $a_{\rm p\,max}$ | maximum semimajor axis of planet orbit |
| $a_{\rm p\,min}$ | minimum semimajor axis of planet orbit |
| $B_\lambda$ | Planck function |
| $b$ | impact parameter |
| $C^{\rm abs}_\lambda$ | particle spectral absorption cross section |
| $C^{\rm ext}_\lambda$ | particle spectral extinction cross section |
| $C^{\rm sca}_\lambda$ | particle spectral scattering cross section |
| $c$ | speed of light |
| $d$ | distance between the Earth and a star |
| $F_{\star\,\lambda}$ | spectral stellar flux |
| $F_{\rm gr}$ | gravitational force acting on particle |
| $F_{\rm rad}$ | radiation pressure force acting on particle |
| $F^{\rm abs}_\lambda$ | spectral absorbed power |
| $F^{\rm sca}_\lambda$ | spectral scattered power |
| $F_{\rm pol\,\lambda}$ | polarized flux measured from disk image |
| $F_{\rm sca\,\lambda}$ | scattered flux measured from disk image |
| $\langle F_{\rm sca\,\lambda}\rangle$ | scattered disk flux averaged over the full solid angle |

[17] www.exoplanetarchive.ipac.caltech.edu

Table E.4: Results for two-component MBB model.

| Name | Cold component | | | Warm component | | | | | |
|---|---|---|---|---|---|---|---|---|---|
| | $f_{\rm disk}$ $10^{-4}$ | $R_{\rm MBB}$ (au) | $T_{\rm MBB}$ (K) | $f_{\rm disk}$ $10^{-4}$ | $R_{\rm MBB}$ (au) | $T_{\rm MBB}$ (K) | $\lambda_0$ ($\mu$m) | $\beta$ | $N_{\rm obs}$ |
| HD 9672 | 6.8 ± 0.4 | 88 ± 50 | 60 | 1.5 ± 2.0 | 13 ± 19 | 160 | 120 | 0.9 | 16 |
| HD 10647 | 2.7 ± 0.4 | 50 ± 14 | 44 | 0.4 ± 0.2 | 12 ± 5 | 90 | 62 | 0.8 | 15 |
| HD 15115 | 4.8 ± 1.3 | 64 ± 11 | 49 | 0.6 ± 0.2 | 8 ± 5 | 140 | 33 | 0.7 | 11 |
| HD 25457 | 1.0 ± 0.4 | 24 ± 26 | 68 | 1.0 ± 0.5 | 0.9 ± 0.5 | 360 | 103 | 1.3 | 9 |
| HD 31295 | 0.4 ± 0.1 | 140 ± 42 | 47 | 0.3 ± 0.1 | 37 ± 14 | 91 | 35 | 0.5 | 10 |
| HD 32297 | 54 ± 7 | 45 ± 10 | 71 | 12 ± 3 | 10 ± 4 | 150 | 19 | 0.4 | 12 |
| HD 39060 | 30 ± 3 | 82 ± 23 | 53 | 6.6 ± 4.1 | 6 ± 3 | 200 | 100 | 0.8 | 20 |
| HD 61005 | 19 ± 1 | 31 ± 14 | 45 | 6.6 ± 2.7 | 9 ± 3 | 82 | 22 | 0.6 | 13 |
| HD 109085 | 0.2 ± 0.1 | 85 ± 22. | 45 | 2.5 ± 1.7 | 1.2 ± 0.7 | 390 | 20 | 0.1 | 18 |
| HD 131488 | 11 ± 7 | 94 ± 33 | 54 | 21 ± 6 | 5 ± 2 | 240 | 40 | 0.3 | 6 |
| HD 131835 | 28 ± 8 | 46 ± 9 | 72 | 9.1 ± 2.7 | 16 ± 6 | 120 | 410 | 1.4 | 10 |
| HD 141569 | 47 ± 14 | 120 ± 39 | 50 | 53 ± 18 | 7 ± 6 | 210 | 10 | 1.0 | 15 |
| HD 181296 | 1.6 ± 0.3 | 40 ± 9 | 93 | 1.4 ± 1.3 | 7 ± 5 | 230 | 42 | 0.6 | 8 |
| HD 218396 | 2.3 ± 0.5 | 130 ± 33 | 38 | 0.4 ± 0.1 | 8 ± 3 | 160 | 210 | 1.1 | 13 |

**Notes.** The columns list target IDs, disk IR excess ($f_{\rm disk}$), BB belt radius ($R_{\rm MBB}$), BB temperature ($T_{\rm MBB}$), characteristic wavelength ($\lambda_0$) and opacity index ($\beta$). The column $N_{\rm obs}$ gives the number of photometric points for wavelengths > 22 $\mu$m which determines whether a fit with more than one component is feasible.

| | |
|---|---|
| $F_\lambda^{th}$ | spectral flux density of thermal emission |
| $f_{\rm disk}$ | disk fractional luminosity |
| $f_{\rm disk}^{\rm SD}$ | disk fractional luminosity obtained with SD model |
| $f_{{\rm pol}\,\lambda}$ | view factor for polarized flux |
| $f_{{\rm sca}\,\lambda}$ | view factor for scattered flux |
| $G$ | gravitational constant |
| $g$ | HG asymmetry parameter |
| $H$ | disk scale height |
| $H_0$ | disk scale height at reference radius |
| $h$ | disk height coordinate |
| $I_{\star\,\lambda}$ | intensity of the incident radiation |
| $I_\oplus$ | intensity of the solar radiation on Earth |
| $J$ | number of model free parameters |
| $L_\star$ | stellar luminosity |
| $L_{\star\,\lambda}$ | spectral stellar luminosity |
| $L_\odot$ | solar luminosity |
| $L_{\rm IR\,disk}$ | disk IR luminosity |
| $L_{{\rm sca}\,\lambda}$ | spectral disk scattered luminosity |
| $M_\star$ | stellar mass |
| $M_\odot$ | solar mass |
| $M_\oplus$ | Earth mass |
| $M_{\rm B}$ | mass of stellar companion B |
| $M_{\rm CO}$ | carbon monoxide mass |
| $M_{\rm dust}$ | dust mass obtained with MBB model |
| $M_{\rm dust}^{\rm SD}$ | dust mass obtained with SD model |
| $M_{\rm inner}$ | dust mass of inner belt in SD model |
| $M_{\rm outer}$ | dust mass of outer belt in SD model |
| $M_{\rm p}$ | planet mass |
| $M_{\rm total}$ | total dust mass of inner and outer belts in SD model |
| $N_{\rm data}$ | number of data points |
| $N_{\rm obs}$ | number of photometric points for wavelengths > 22 $\mu$m |
| $N_{\rm SED}$ | surface grain number density in the SD model |
| $n(a)$ | differential grain number density |
| $n_{\rm gr}$ | volume grain number density in the disk |
| $P$ | orbital period |
| $p_{\rm max}$ | maximum polarization fraction of scattered light |
| $Q$ | Stokes parameter $Q$ |
| $\langle Q_{\rm pr}\rangle$ | mean radiation pressure coupling coefficient averaged over the stellar flux |
| $Q_r$ | radial Stokes parameter $Q$ |
| $Q_\lambda^{\rm abs}$ | spectral absorption efficiency |
| $Q_\lambda^{\rm ext}$ | spectral extinction efficiency |
| $Q_\lambda^{\rm sca}$ | spectral scattering efficiency |
| $Q_\varphi$ | azimuthal Stokes parameter $Q$ |
| $q$ | SD index |
| $R_\star$ | stellar radius |
| $R_{\rm belt}$ | planetesimal belt radius |
| $R_{\rm belt}^{\rm mes}$ | belt radius measured from the $r^2-$scaled SPHERE images |
| $R_{\rm belt)}^{\rm mod}$ | modeled peak volume density of grains |
| $R_{\rm inner}$ | radius of inner belt in SD model |
| $R_{L_\odot}$ | scaling factor representing the expected radial position of a belt around a star with solar luminosity |
| $R_{\rm MBB}$ | disk radius obtained with MBB model |
| $R_{\rm max(\sigma)}^{\rm mod}$ | radial distance of the modeled peak surface density |
| $R_{\rm outer}$ | radius of outer belt in SD model |
| $R_{\rm p}$ | planet radius |
| $r$ | radial distance from the star |
| $r_0$ | reference radius of planetesimal belt |
| $T_\star$ | stellar temperature |
| $T_{\rm bb}$ | BB temperature of dust grains |
| $T_{\rm dust}^{\rm SD}$ | dust temperature obtained with SD model |
| $T_{\rm grain}$ | grain temperature |
| $T_{\rm eff}$ | effective stellar temperature |
| $T_{\rm eq}$ | equilibrium temperature |
| $T_{\rm MBB}$ | BB temperature of dust grains obtained with MBB model |
| $T_{\rm tr}$ | transit duration |
| $t_{\rm age}$ | stellar age |
| $U$ | Stokes parameter $U$ |
| $U_\varphi$ | azimuthal Stokes parameter $U$ |
| $U_r$ | radial Stokes parameter $U$ |
| $u$ | Heaviside step function |
| $x$ | size parameter |
| $x_{\rm eff}$ | effective size parameter |
| $\alpha$ | index of power-law function for the relation between belt radius and stellar luminosity |
| $\alpha_{\rm in}$ | index of power-law function for the grain number density in the inner region of the disk |
| $\alpha_{\rm mass}$ | index of power-law function for the relation between belt dust mass and stellar mass |
| $\alpha_{\rm out}$ | index of power-law function for the grain number density in the outer region of the disk |

Table F.1: Parameters of confirmed planets in the debris systems of our sample.

| Planet | $R_p$ ($R_{Jup}$) | $M_p$ ($M_{Jup}$) | $T_{eq}$ (K) | $a_p$ (au) | $i$ (deg) | $P$ (days) | Detected by[(a)] |
|---|---|---|---|---|---|---|---|
| HD 10647 b | (...) | $(0.94 \pm 0.18)/\sin i$ | (...) | $2.02 \pm 0.001$ | (...) | $1003 \pm 56$ | RV |
| HD 10700 g | (...) | $(0.006 \pm 0.001)/\sin i$ | (...) | $0.133 \pm 0.002$ | (...) | $20 \pm 1$ | RV |
| HD 10700 h | (...) | $(0.006 \pm 0.002)/\sin i$ | (...) | $0.243 \pm 0.003$ | (...) | $49.4 \pm 0.1$ | RV |
| HD 10700 e | (...) | $(0.012 \pm 0.003)/\sin i$ | (...) | $0.538 \pm 0.060$ | (...) | 163 | RV |
| HD 10700 f | (...) | $(0.012 \pm 0.004)/\sin i$ | (...) | $1.334 \pm 0.044$ | (...) | 636 | RV |
| HD 20794 b | (...) | $(0.009 \pm 0.001)/\sin i$ | 660 | $0.121 \pm 0.002$ | 90 | 18 | RV |
| HD 20794 c | (...) | $(0.008 \pm 0.002)/\sin i$ | 508 | $0.204 \pm 0.003$ | 90 | 40 | RV |
| HD 20794 d | (...) | $(0.015 \pm 0.002)/\sin i$ | 388 | $0.350 \pm 0.006$ | 90 | 90 | RV |
| HD 20794 e | (...) | $(0.015 \pm 0.003)/\sin i$ | (...) | $0.509 \pm 0.006$ | (...) | 147 | RV |
| HD 22049 b | (...) | $0.78 \pm 0.40$ | (...) | $3.5 \pm 0.1$ | 30-89 | 2690 | RV |
| HD 29391 b | (...) | 2 | 700 | $12 \pm 4$ | 43 | 11688 | DI |
| HD 39060 b | $1.5 \pm 0.2$ | $11 \pm 2$ | 1650 | $9.0 \pm 0.4$ | 89 | 8618 | A, DI |
| HD 39060 c | (...) | $10 \pm 1$ | (...) | $2.7 \pm 0.2$ | 89 | 1200 | RV |
| HD 52265 b | (...) | $(1.16 \pm 0.10)/\sin i$ | 405 | $0.50 \pm 0.01$ | (...) | 119 | RV |
| HD 69830 b | (...) | $(0.032 \pm 0.002)/\sin i$ | (...) | $0.079 \pm 0.001$ | (...) | 8.7 | RV |
| HD 69830 c | (...) | $(0.031 \pm 0.003)/\sin i$ | (...) | $0.188 \pm 0.003$ | (...) | 31.6 | RV |
| HD 69830 d | (...) | $(0.044 \pm 0.006)/\sin i$ | (...) | $0.645 \pm 0.010$ | (...) | 200 | RV |
| HD 82943 b | (...) | $1.68 \pm 0.03$ | (...) | $1.183 \pm 0.001$ | 90 | 442 | RV |
| HD 95086 b | (...) | $5.5 \pm 1.5$ | 1000 | $55.7 \pm 2.5$ | (...) | (...) | DI |
| HD 106906 b | (...) | $12.95 \pm 1.85$ | 1820 | 650 | (...) | (...) | DI |
| HD 114082 b | $0.98 \pm 0.03$ | $8 \pm 1$ | (...) | $0.7 \pm 0.4$ | 89.8 | 197 | RV, T |
| HD 115617 b | (...) | $(0.016 \pm 0.002)/\sin i$ | (...) | $0.050 \pm 0.001$ | (...) | 4 | RV |
| HD 115617 c | (...) | $(0.051 \pm 0.004)/\sin i$ | (...) | $0.215 \pm 0.003$ | (...) | 38 | RV |
| HD 115617 d | (...) | $(0.072 \pm 0.004)/\sin i$ | (...) | $0.476 \pm 0.001$ | (...) | 123 | RV |
| HD 128311 b | (...) | $(2.00 \pm 0.16)/\sin i$ | (...) | $1.088 \pm 0.013$ | (...) | 453 | RV |
| HD 128311 c | (...) | $3.8 \pm 0.9$ | (...) | $1.74 \pm 0.01$ | 56 | 922 | RV |
| HD 192263 b | (...) | $(0.658 \pm 0.030)/\sin i$ | 486 | $0.154 \pm 0.002$ | (...) | 24.4 | RV |
| HD 197481 b | $0.36 \pm 0.02$ | $0.053 \pm 0.015$ | $600 \pm 17$ | $0.065 \pm 0.001$ | 89.5 | 8.46 | T, RV, TTV |
| HD 197481 c | $0.25 \pm 0.04$ | $0.045 \pm 0.025$ | $454 \pm 16$ | $0.110 \pm 0.002$ | 89.2 | 18.9 | RV, T |
| HD 197481 d | (...) | $0.003 \pm 0.002$ | (...) | (...) | 89.2 | 12.7 | TTV |
| HD 206893 c | $1.46 \pm 0.18$ | $12.7 \pm 1.2$ | 1182 | $3.53 \pm 0.08$ | 29 | 2090.0 | A, DI, RV |
| HD 218396 b | 0.6 | $7 \pm 4$ | 1200 | $68 \pm 2$ | 26-30 | 166510.0 | A, DI |
| HD 218396 c | 1.0 | $10 \pm 3$ | 1200 | $43 \pm 3$ | 26-30 | 83255.0 | A, DI |
| HD 218396 d | 0.9 | $10 \pm 3$ | 1300 | $27 \pm 3$ | 26-39 | 37000.0 | A, DI |
| HD 218396 e | 0.9 | $10 \pm 4$ | 1300 | $16 \pm 2$ | 26-31 | 18000.0 | A, DI |
| GJ 581 b | (...) | $(0.049 \pm 0.002)/\sin i$ | (...) | $0.041 \pm 0.001$ | (...) | 5.4 | RV |
| GJ 581 c | (...) | $(0.017 \pm 0.002)/\sin i$ | (...) | $0.073 \pm 0.001$ | (...) | 12.9 | RV |
| GJ 581 e | (...) | $(0.005 \pm 0.001)/\sin i$ | (...) | $0.029 \pm 0.001$ | (...) | 3.1 | RV |
| GJ 649 b | (...) | $(0.275 \pm 0.002)/\sin i$ | (...) | $1.133 \pm 0.002$ | (...) | 604.8 | RV |

**Notes.** [(a)] RV = radial velocity; DI = direct imaging; A = astrometry; T = Transit; TTV = transit time variation.

| | |
|---|---|
| $\alpha_R$ | index of power-law function for the relation between belt dust mass and MBB radius |
| $\alpha_t$ | index of power-law function for the evolution of disk fractional IR luminosity |
| $\beta$ | disk flaring index |
| $\beta_{mass}$ | scaling factor of power-law function for the relation between belt dust mass and stellar mass |
| $\beta_{op}$ | dust spectral opacity index |
| $\Delta a$ | width of the cleared zone |
| $\Delta R_{belt}$ | belt width in SD model |
| $\Delta\lambda$ | filter wavelength range |
| $\theta$ | scattering angle |
| $i$ | disk or planet orbit inclination |
| $\Lambda_{pol\,\lambda}$ | $\Lambda$ parameter for polarized flux |
| $\Lambda_{sca\,\lambda}$ | $\Lambda$ parameter for scattered flux |
| $\langle\Lambda_{sca\,\lambda}\rangle$ | $\Lambda$ parameter for disk-averaged scattered flux |
| $\lambda$ | wavelength |
| $\lambda_0$ | SED characteristic wavelength |
| $\lambda_c$ | central wavelength of filter |
| $\varrho$ | grain material density |
| $\Sigma$ | surface density in the SD model |
| $\sigma_\lambda^{ext}$ | spectral total cross section for extinction |
| $\sigma_\lambda^{sca}$ | spectral total cross section for scattering |
| $\nu$ | model degrees of freedom |
| $\Omega$ | solid angle |
| $\omega_\lambda$ | spectral single scattering albedo |

Table F.2: Stellar companions and unconfirmed planets of the program stars

| HD | HIP/Alias | Sep au | $M_B$ $M_\odot$ | Notes |
|---|---|---|---|---|
| (...) | 11437 | 900.98 | 0.552 | (1) |
| (...) | 83043 | 1.11 | 0.003 | (3) |
| 166 | 544 | 45.21 | 0.005 | (2) |
| 203 | 560 | 7.50 | 0.003 | (5) |
| 1466 | 1481 | 4.50 | 0.004 | (6) |
| 3003 | 2578 | 50772.62 | 0.363 | (7) |
| 13246 | 9902 | 2376.46 | 0.834 | (1) |
| 14082 | 10679 | 543.97 | 1.016 | (1) |
| 16743 | 12361 | 12612.24 | 1.150 | (1) |
| 17925 | 13402 | 0.09 | 0.820 | (8,9,10) |
| 20320 | 15197 | 0.18 | 0.450 | (11) |
| 20320 | 15197 | 22.54 | 0.015 | (2) |
| 21997 | 16449 | 12.50 | 0.006 | (2,5) |
| 29391 | 21547 | 2002.52 | 1.100 | (1,12) |
| 30422 | 22192 | 8.24 | 0.019 | (2, 13) |
| 31295 | 22845 | 0.21 | 0.135 | (2) |
| 36546 | 26062 | 5.65 | 0.029 | (2) |
| 37484 | 26453 | 14173.68 | 0.091 | (1) |
| 38678 | 27288 | 0.43 | 0.834 | (2,14) |
| 40540 | 28230 | 3583.28 | 0.310 | (1) |
| 50571 | 32775 | 22.05 | 0.010 | (15) |
| 53143 | 33690 | (...) | (...) | (16,17,18) |
| 53842 | 32435 | 82.74 | 0.386 | (1, 19) |
| 54341 | 34276 | 1.92 | 0.296 | (2) |
| 71155 | 41307 | 1.41 | 0.259 | (2,20,21) |
| 71722 | 41373 | 686.44 | 0.121 | (1) |
| 71722 | 41373 | 18623.50 | 0.795 | (1) |
| 80950 | 45585 | 5.24 | 0.034 | (2) |
| 92945 | 52462 | 7.50 | 0.001 | (2) |
| 98800 | 55505 | 0.86 | 0.290 | (7) |
| 98800 | 55505 | 22.62 | 1.390 | (1,12) |
| 102647 | 57632 | 437.00 | 0.120 | (8) |
| 106906 | 59960 | 0.13 | 1.340 | (22) |
| 107146 | 60074 | 6.06 | 0.009 | (2) |
| 109573 | 61498 | 559.58 | 0.630 | (1) |
| 109573 | 61498 | 12371.94 | 0.480 | (1,23) |
| 111520 | 62657 | 1588.43 | 0.552 | (1) |
| (...) | 63942 | 31.06 | 0.410 | (1,24) |
| 113766 | 63975 | 151.80 | 1.425 | (1) |
| 115600 | 64995 | (...) | (...) | (25) |
| 117214 | 65875 | 1150.40 | 0.556 | (1) |
| 120534 | (...) | 0.03 | 1.327 | (26) |
| 129590 | 72070 | 2.21 | 1.103 | (27) |
| 133803 | 73990 | 17.73 | 0.021 | (28) |
| 133803 | 73990 | 32.00 | 0.022 | (28) |
| 133803 | 73990 | 5205.62 | 0.345 | (1) |
| 135379 | 74824 | 6434.04 | 0.019 | (1) |
| 141011 | 77432 | (...) | (...) | (29) |
| 141378 | 77464 | 4220.89 | 0.098 | (1,30) |
| 141569 | 77542 | 843.77 | 1.290 | (7,31) |
| 161868 | 87108 | 1.02 | 0.136 | (2) |
| 164249 | 88399 | 7.02 | 0.004 | (7) |
| 164249 | 88399 | 320.13 | 0.230 | (1,32) |
| 172555 | 92024 | 2054.09 | 0.800 | (1,7) |
| 174429 | 92680 | 24.50 | 0.046 | (46) |
| 178253 | 94114 | 0.94 | 0.148 | (2) |
| 181296 | 95261 | 197.07 | 0.035 | (33,34) |
| 181296 | 95261 | 20205.78 | 1.228 | (1) |
| 181869 | 95347 | | | (5) |
| 188228 | 98495 | 1.25 | 0.163 | (2) |
| 191089 | 99273 | | | (5) |
| 191131 | 99290 | 3.19 | 0.100 | (2) |
| 206893 | 107412 | 9.60.88 | 0.003 | (36) |
| 216956 | 113368 | 54425.88 | 0.820 | (35) |
| 220825 | 115738 | | | (5) |
| 221853 | 116431 | 23877.86 | 0.031 | (1) |
| (...) | BD-20 951 | 0.07 | 0.802 | (26) |
| (...) | BD-20 951 | 9799.21 | 0.096 | (1) |

**Notes.** (1) Gaia; (2) Astrometric; (3) Pinamonti et al. (2023); (4) Fomalhaut c; (5) Gratton et al. (2024); (6) Mesa et al. (2022); (7) Tokovinin (2018); (8) Rodriguez et al. (2015); (9) Halbwachs et al. (2018); (10) Grandjean et al. (2021); (11) Pourbaix et al. (2004); (12) Secondary, is a binary (Tokovinin 2018); (13) no detection in Lombart et al. (2020); (14) Trifonov et al. (2020); (15) Konopacky et al. (2016); (16) Stark et al. (2023); (17) MacGregor et al. (2022); (18) Constant RV (Trifonov et al. 2020); (19) Bonavita et al. (2022); (20) No detection (Gullikson et al. 2016); (21) X-ray: Schröder & Schmitt (2007); (22) Lagrange et al. (2016); (23) De Rosa et al. (2014); (24) WDS orbit; (25) Gibbs et al. (2019); (26) Gaia SB2; (27) Matthews et al. (2017); Zakhozhay et al. (2022a); Grandjean et al. (2023); overluminous by 0.75 mag. RV amplitude from Gaia assuming the inclination of the disk; (28) Hinkley et al. (2015); (29) Bonnefoy et al. (2021); (30) Waisberg et al. (2023); (31) Secondary is a binary; (32) Tokovinin (2014); (33) Vigan et al. (2021); (34) Rameau et al. (2013); (35) Mamajek et al. (2013); Grandjean et al. (2021); (36) Hinkley et al. (2023); Milli et al. (2017a)